\documentclass[10pt]{article}
\usepackage[lmargin=1in,rmargin=1in,tmargin=1in,bmargin=1in]{geometry}

\usepackage{amsmath,amsthm,amsfonts,amssymb,comment,slashed,mathtools,dsfont,hyperref,xcolor,upgreek}
\usepackage[utf8]{inputenc}
\usepackage[T1]{fontenc}  
\usepackage[english]{babel}

\usepackage{pgfplots}
\pgfplotsset{compat = newest}
\usetikzlibrary{positioning, arrows.meta}
\usepgfplotslibrary{fillbetween}

\usepackage{graphicx}
\usepackage{tikz}
\usetikzlibrary{calc,intersections,patterns,decorations.pathmorphing}

\definecolor{bordeaux}{RGB}{109, 7, 26}
\definecolor{navy}{RGB}{0, 0, 128}

\hypersetup{
  colorlinks = true, %Colours links instead of ugly boxes
  urlcolor   = blue, %Colour for external hyperlinks
  linkcolor  = bordeaux, %Colour of internal links
  citecolor  =  navy %Colour of citations
}

\numberwithin{equation}{section}

\theoremstyle{plain}
\newtheorem{theorem}{Theorem}%[section]
\newtheorem{proposition}{Proposition}[section]
\newtheorem{lemma}[proposition]{Lemma}

\newtheorem*{conjecture*}{Conjecture}
\theoremstyle{definition}
\newtheorem{definition}[proposition]{Definition}
\newtheorem{remark}[proposition]{Remark}

\newcommand{\ve}{\varepsilon}
\newcommand{\ing}{\mathrm{in}}
\newcommand{\out}{\mathrm{out}}
\newcommand{\maxx}{\mathrm{max}}
\newcommand{\les}{\lesssim}
\newcommand{\wtilde}{\widetilde}
\newcommand{\oline}{\overline}
\newcommand{\uline}{\underline}
\newcommand{\Err}{\mathcal{E}rr}
\newcommand{\p}{\partial}
\newcommand{\V}{\Vert}

\newcommand{\R}{\mathcal{R}}
\newcommand{\C}{\mathcal{C}}

\newcommand{\D}{\mathcal{D}}

\title{Nonlinear stability of Schwarzschild black holes for the Einstein--massless Dirac system in spherical symmetry}
\author{Lamberto Tresoldi\thanks{\'Ecole  Polytechnique, Centre de Mathématiques Laurent Schwartz, 91128 Palaiseau Cedex, France.\\\textit{E-mail address}: lamberto.tresoldi@polytechnique.edu}}
\date{\today}

\begin{document}

\maketitle

\begin{abstract}
In this paper, we establish the nonlinear stability of the Schwarzschild exterior for the spherically symmetric Einstein--massless Dirac system. More precisely, we consider sufficiently small perturbations of Schwarzschild characteristic initial data and prove that the corresponding maximal development remains globally controlled in the exterior region \textit{(orbital stability)}, and converges, in a suitable teleological double-null gauge, to a Schwarzschild spacetime with a final mass \textit{(asymptotic stability)}. In particular, the maximal future development contains a complete future null infinity and a regular event horizon, with a nonempty black hole region in its interior.

The main difficulties in the proof arise from the delicate coupling between the Dirac operator and the geometry, compared with the scalar wave counterpart, as well as from the failure of the Dirac stress-energy tensor to satisfy a positive energy condition. To overcome these difficulties, we rely on a bootstrap argument coupling the geometry and the Dirac field. Working in suitable teleological double-null gauges adapted to the asymptotic null boundaries, we establish weighted energy estimates combining integrated local energy decay, red-shift estimates near the event horizon, and an $r^p$-estimates towards null infinity. A core ingredient is the use of suitable reweightings of the spinor components, which allow us to close these estimates despite the nonlinear coupling to the geometry.
\end{abstract}

\tableofcontents

\section{Introduction}
The \textit{Schwarzschild metrics} \cite{schwarzschild1916gravitationsfeld}, which in Boyer--Lindquist coordinates $(t,r,\vartheta,\varphi)$ \cite{boyer1967maximal} take the form
\[
     g_{M}= -\Big(1-\frac{2M}{r}\Big)dt^2+\Big(1-\frac{2M}{r}\Big)^{-1}dr^2+r^2\big(d\vartheta^2+\sin^2\vartheta\,d\varphi^2\big),
\]
constitute the simplest black hole solutions to the \textit{Einstein vacuum equations} \cite{einstein1915feldgleichungen}
\[
    R_{\mu\nu}=0,
\]
where $R_{\mu\nu}$ denotes the Ricci curvature tensor. The parameter $M>0$, representing the black hole mass, labels this one-parameter family of spherically symmetric, static, and asymptotically flat spacetimes, each admitting $T:=\partial_t$ as a stationary Killing vector field. Moreover, Schwarzschild black holes form the non-rotating subfamily of the celebrated two-parameter Kerr family $g_{a,M}$, where the additional parameter $a$ describes the angular momentum of the rotating black holes.

The study of the stability of these solutions is a central problem in general relativity, and substantial developments have been made over the past decades.
In particular, \cite{dafermos2019linear} and \cite{klainerman2020global,dafermos2021non} established linear and nonlinear stability for the exterior of Schwarzschild black holes, employing techniques such as the vector field method, originally introduced by Klainerman \cite{klainerman1985uniform}. More recently, \cite{klainerman2023kerr,klainerman2022construction,klainerman2022effective,giorgi2022wave,shen2023construction} established the nonlinear stability of slowly rotating Kerr black holes, while in \cite{ma2024energy,ma2026energy,szeftel2026kerr,szeftel2026teuk} the nonlinear stability has been extended to the full subextremal range. These results provide important evidence towards the following conjecture.
\begin{conjecture*}[Black Hole Stability Conjecture, \cite{klainerman2022brief}]
"Generic" small perturbations of a Kerr black hole evolve into a spacetime that remains regular outside the event horizon and eventually approaches, globally, a nearby Kerr black hole.
\end{conjecture*}
Noting that the results mentioned above are all established in vacuum, the notion of “generic” in the conjecture refers primarily to the analytic properties of the perturbations. One may, however, interpret the question of stability in a broader sense by also varying the \emph{nature of the perturbation}, namely by considering black holes as solutions of Einstein--matter systems. This leads to the study of the stability of black hole families in the presence of different types of matter fields and, in particular, to the question of whether analogous stability and asymptotic convergence results hold beyond the vacuum setting. It is therefore natural to couple the Einstein equations to, for instance, scalar, Klein--Gordon, Vlasov, Maxwell, or Dirac fields, depending on the physical phenomenon of interest. Among these, the Dirac equation provides the fundamental relativistic model for $1/2$-spin matter and is therefore the natural framework for describing particles such as electrons and neutrinos, complementing the scalar and kinetic models more commonly considered in the literature.

In this work, we shall consider the \textit{Einstein--massless Dirac system}
\begin{align}
    \label{eq:Einstein-eq1}
    R_{\mu\nu}-\frac12 R_g g_{\mu\nu}&=8\pi T_{\mu\nu},\\
    \label{eq:Dirac-eq1}
    \uline{\gamma}^\mu D_\mu \Psi&= 0,  
\end{align}
where $(\mathcal{M},g)$ is a Lorentzian spin manifold, $R_g$ the associated scalar curvature, $\Psi$ is a spinor field, which may locally be represented as a $\mathbb C^4$-valued function, and where the \textit{energy-momentum tensor} is given by
\begin{align*}
    T_{\mu \nu} := \frac{i}{4}\big(\overline{\Psi}\underline{\gamma_\mu} D_\nu \Psi - D_\nu \overline{\Psi} \underline{\gamma_\mu} \Psi\big) + \frac{i}{4}\big(\overline{\Psi}\underline{\gamma_\nu} D_\mu \Psi - D_\mu \overline{\Psi} \underline{\gamma_\nu} \Psi\big).
\end{align*}
Here, $D_\mu$ denotes the \textit{spinorial covariant derivative}, while $\uline{\gamma}^\mu\in M_4(\mathbb C)$ are the \textit{curved Dirac matrices}. Their definition depends on the choice of a \textit{tetrad} and the associated \textit{spin connection}, and is such that the Dirac equation is compatible with changes of coordinates and local Lorentz transformations. We refer to Appendix \ref{app:derivation-Dirac} for a discussion of this construction.

We restrict our attention to the massless case, for which a well-developed decay theory is available on a fixed Schwarzschild background \cite{ma2022sharp} and provides a starting point for our nonlinear analysis. The massive case, on the other hand, entails additional analytical challenges, particularly in controlling the \textit{long-time behaviour} of solutions and the intricate interactions between geometry and matter. We refer to \cite{lefloch2025global}, where these difficulties are addressed in the context of the nonlinear stability of Minkowski spacetime for the Einstein--\textit{massive} Dirac system.

From an analytical perspective, the Dirac field introduces at least two features that distinguish the problem from the corresponding Einstein--scalar field system. First, the Dirac operator is first order, and its square is related to the wave operator through the \textit{Lichnerowicz formula} \cite{schrodinger1932diracsches},
\begin{equation}\label{eq:lichn-identity}
    (\uline{\gamma}^\mu D_\mu)^2 = -D^\mu D_\mu + \frac14 R_g.
\end{equation}
We stress that the second-order operator on the right-hand side may be viewed as a \textit{spinorial wave operator}, since it acts through the spinorial covariant derivatives $D_\mu$. In particular, the passage to a second-order equation introduces additional curvature terms associated with the spin connection.
Second, unlike the energy-momentum tensors associated with other matter models, such as scalar field, Maxwell and Vlasov, the \textit{Dirac energy-momentum tensor does not satisfy the positive energy condition} \cite{dafermos2003time}. This may be related to the fact that the Dirac field is more naturally interpreted in the framework of quantum field theory, after second quantization. In this setting, pointwise energy conditions are replaced by weaker constraints, known as quantum energy inequalities, see \cite{kontou2020energy}. In this context, we also mention \cite{kain2023einstein}, which studies the spherically symmetric Einstein--Dirac system in semiclassical gravity, where the Einstein field equations are sourced by the \textit{expectation value} of the stress-energy tensor.

\subsection{Stability of Schwarzschild for the Einstein--Dirac system in spherical symmetry}
The presence of an external field genuinely complicates the dynamics, as it introduces a backreaction mechanism between matter and geometry. To study this coupling in a tractable setting, we impose spherical symmetry, which reduces the problem to a $(1+1)$-problem. Note that this reduction does not, however, trivialize the dynamics as it does in vacuum, where \textit{Birkhoff's theorem} \cite{jebsen1921allgemeinen} implies that every spherically symmetric vacuum solution is locally Schwarzschild.

The spherical reduction of the Dirac equation requires some care, since a spinor field itself cannot be invariant under rotations in the same sense as a scalar field.
To understand the distinction, recall that on a spherically symmetric background, the wave operator preserves each eigenspace of the orbital angular momentum. On Kerr, by contrast, the equation separates in \textit{spheroidal harmonics} rather than spherical harmonics, leading to coupling between spherical harmonic modes \cite{teukolsky1973perturbations}.
A somewhat reminiscent structure already appears for the Dirac operator in spherical symmetry: the invariant subspaces are labeled by the total angular momentum and mix different orbital modes, so that individual modes do not give rise to a spherically symmetric stress-energy tensor. Thus, to obtain a source compatible with spherical symmetry, one decomposes the angular dependence into
\textit{spinorial spherical harmonics} and restricts to the lowest angular modes.
We refer to Appendix \ref{app:derivation-Dirac} for the explicit reduction used throughout the paper, and to \cite{camporesi1996eigenfunctions,abrikosov2002dirac} for the decomposition of the Dirac operator on spheres. 

\begin{figure}
    \centering
    \begin{tikzpicture}[line cap=rect,inner sep=0pt, x=1cm, y=1cm, scale=1,font=\small]

\coordinate   (U)   at (0,3)      {};
\coordinate  (I)   at (1,2)      {};
\coordinate (II)   at (3,0)      {};
\coordinate  (III)   at (5,2)      {};
\coordinate   (IV)   at (3,4)      {};

\path[name path=parab1, domain=0:3, samples=100, smooth, variable=\x]
    plot ({\x},{1/6*\x*\x - 1/6*\x + 3});

\fill[lightgray] plot[domain=0:3, smooth] (\x,{1/6*\x*\x-1/6*\x+3}) -- (III) -- (II) --(U);

\draw[line width=1pt]  (U) -- (I) 
-- node[pos=0.5, below=2pt, sloped]    {$\underline{\hat{\mathcal{C}}}_{\mathrm{in}}$}
       (II) -- node[pos=0.5, below=2pt, sloped]    {$\hat{\mathcal{C}}_{\mathrm{out}}$} (III); %-- 
            %node[midway, below=2pt, sloped, font=\tiny]    {$u=u_0$}
\draw[line width=1pt,dashed] (III) --
            node[pos=0.5, above=2pt, sloped]    {$\mathcal{I}^+$}
      (IV);
\draw[line width=1pt] (IV)-- node[pos=0.7, below=2pt, sloped]    {$\mathcal{H}^+$}(I);

\node at (1,2.7)          {$\mathcal{BH}$};
\node at ($(IV)+(0,0.25)$)          {$i^+$};
%\node at ($(III)+(0.25,0)$)          {$i^0$};
\node[rotate=60] (D3)     at (2.75,3.4)  {$g\to g_M$};
\node[rotate=60] (D4)     at (2.9,3.1)  {$\psi_\pm\to 0$};   
\filldraw[black] (U) circle (2pt);
\filldraw[fill=white, draw=black] (IV) circle (2pt);
%\filldraw[fill=white, draw=black] (III) circle (2pt);

\end{tikzpicture}
    
    \caption{The Penrose diagram of the maximal development arising from an admissible initial data set as in Theorem \ref{thm:Schwarz-stability1}. For an introduction to Penrose diagrams, see \cite[Appendix C]{dafermos2005proof}.}
    \label{fig:Penrose1-stability}
\end{figure}
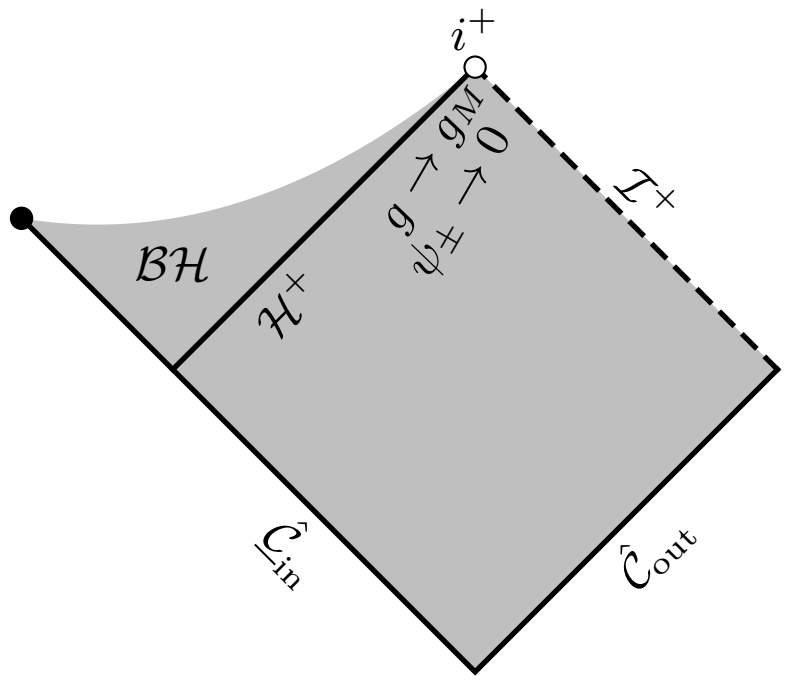

In this setting, the reduced Dirac field is described by two complex-valued functions, denoted by $\psi_+$ and $\psi_-$, giving rise to a \textit{coupled $(1+1)$-dimensional system} consisting of the Dirac equations for the two spinor components and the Einstein equations governing the geometry. See already Section \ref{subsec:Einstein--Dirac system in spherical symmetry} for the precise equations of the model.
In this paper, we establish the nonlinear stability of the Schwarzschild exterior within the spherically symmetric Einstein--massless Dirac system. Let us now state a preliminary version resuming our main result: we refer to Section \ref{sec:stab-schwarzschild-einstein-dirac}, and in particular to Theorem \ref{thm:Schwarz-stability2}, for a detailed statement.
\begin{theorem}[Stability of Schwarzschild for Einstein--massless Dirac in spherical symmetry, first version]\label{thm:Schwarz-stability1} 
Let us consider a characteristic initial data set for the spherically symmetric Einstein--massless Dirac system posed on a bifurcate null hypersurface $\hat{\uline{\C}}_\ing\cup\hat{\C}_\out$, as depicted in Figure \ref{fig:Penrose1-stability}, which lies close to Schwarzschild in an appropriate norm. Then, the initial data evolve into a spacetime with the following properties:
\begin{enumerate}
    \item \emph{Existence of a black hole region}: Future null infinity $\mathcal{I}^+$ is complete and the causal past of future null infinity $J^-(\mathcal{I}^+)$ is bounded by a regular event horizon $\mathcal{H}^+$, which itself bounds a nonempty black hole region $\mathcal{BH}:=\mathcal{M}\setminus J^-(\mathcal{I}^+)$;
    \item \emph{Orbital stability}: The metric remains close to the initial Schwarzschild metric in the domain of outer communication and appropriately defined fluxes and pointwise $C^1$ norms of the Dirac field are bounded in terms of the initial data on $\hat{\uline{\C}}_\ing\cup\hat{\C}_\out$;
    \item \emph{Asymptotic stability}: The metric decays polynomially in $C^1$ (as an appropriate notion of "time" tends to infinity) to a nearby member of the Schwarzschild family, relative to a teleologically defined double-null gauge uniformly in the entire domain of outer communication. The Hawking mass (see \eqref{eq:Hawking-mass}) converges uniformly to the final mass of the black hole and the Dirac field decays polynomially to zero pointwise and in an appropriate energy norm.
\end{enumerate}
\end{theorem} 
We note that the decay rates obtained for the Dirac field are \textit{not expected to be optimal}. Indeed, the argument does not require higher-order regularity estimates for the derivatives of the spinor field, which would allow for a more precise characterization of its decay. Nevertheless, the rates we obtain are consistent with those known for the linear Dirac equation on a Schwarzschild background \cite{ma2022sharp}.

Moreover, the failure of the positive energy condition has a concrete consequence in spherical symmetry: the Hawking mass is not necessarily monotone and may undergo a \textit{small decrease} along the evolution. This contrasts with the standard setting of spherically symmetric Einstein--matter systems (e.g. scalar field, Maxwell, Vlasov), for which this condition holds and monotonicity arguments play a central role in the analysis of black-hole formation and stability, see for instance \cite{dafermos2005spherically,kommemi2013global}. In our model, this failure of monotonicity also means that the apparent horizon is not necessarily contained behind the final event horizon. A related obstruction to standard monotonicity arguments arises in the nonlinear stability analysis of Schwarzschild outside spherical symmetry \cite{dafermos2021non}. Thus, the Einstein--Dirac system provides an interesting model in which the interaction between a black hole and the matter field may lead to a decrease in its mass already in spherical symmetry. This is reminiscent of the semiclassical picture of black hole evaporation through Hawking radiation, see for instance \cite{melnyk2002effet,li2008hawking}.

Recent progress on the formation and stability of extremal black holes \cite{kehle2024extremal,kehle2025gravitational,angelopoulos2026moduli} raises the question of whether the corresponding \textit{gluing constructions} and \textit{threshold results} can be extended to the Einstein--Dirac and Einstein--Dirac--Maxwell systems. Indeed, the techniques developed so far rely in part on energy estimates and monotonicity arguments made possible by the positivity properties of the matter fields. It would therefore be interesting to investigate whether the geometric structure and conservation laws of the Dirac operator can compensate, at least partially, for the lack of positivity and provide alternative mechanisms to establish analogous results.

\subsection{Overview of the proof}
The proof of Theorem \ref{thm:Schwarz-stability1} relies on a bootstrap argument involving both the geometry and the Dirac field. A key ingredient is the construction of teleological gauges together with anchored Schwarzschild backgrounds, which allow us to identify the Schwarzschild black hole towards which the solution converges. In this section, we review the main elements of our strategy.

\subsubsection{Linear Dirac equation on Schwarzschild black holes}
Let us briefly review the approach of \cite{ma2022sharp}, which provides the main source of inspiration for this work.
In that paper, the vector-field method is used to establish a hierarchy of weighted energy estimates yielding boundedness and decay for the Dirac field for any spin-weighted spherical modes on hyperboloidal foliations.
To make the discussion more closely aligned with our framework, we schematically present the corresponding argument for the lowest spherical mode in \textit{double-null Eddington--Finkelstein coordinates} $(u,v)$, adapted to the ingoing and outgoing null directions; see, for instance, \cite{dafermos2019linear} for an introduction to this coordinate system. Our precise conventions are given in Section \ref{sec:preliminaries}.

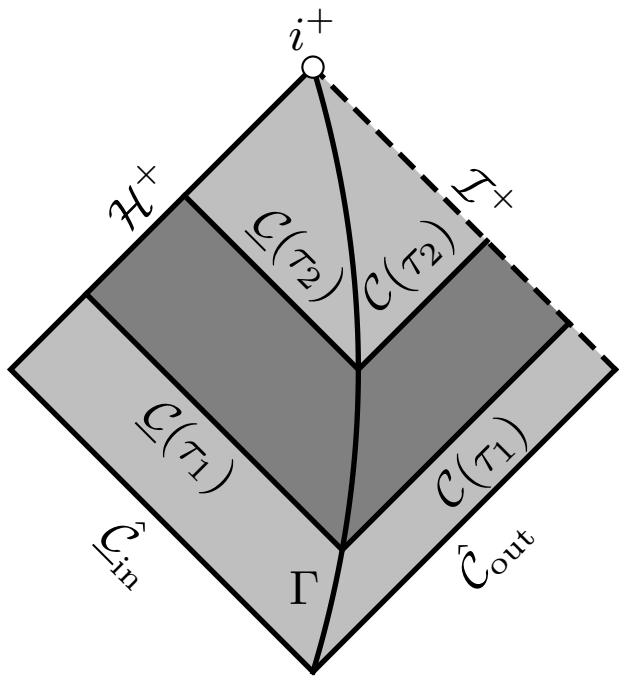
\begin{figure}
\centering
\begin{tikzpicture}[line cap=rect,inner sep=0pt, x=1cm, y=1cm, scale=1,font=\small]

\coordinate  (I)   at (0,2)      {};
\coordinate (II)   at (2,0)      {};
\coordinate  (III)   at (4,2)      {};
\coordinate   (IV)   at (2,4)      {};
\coordinate  (V)   at (0.5,2.5)      {};
\coordinate  (VI)   at (2.19322,0.80678)      {};
\coordinate  (VII)   at (3.69322,2.30678)      {};
\coordinate  (VIII)   at (3.15,2.85)      {};
\coordinate  (IX)   at (2.3,2)      {};
\coordinate  (X)   at (1.15,3.15)      {};

\path[name path=parab1, domain=0:4, samples=100, smooth, variable=\y]
    plot (\y,{-3/40*\y*\y+3/10*\y+2});

\fill[lightgray] (I) -- (II) -- (III) -- (VII) -- (VI) -- (V);
\fill[lightgray] (X) -- (IX) -- (VIII) -- (IV);
\fill[gray] (V) -- (VI) -- (VII) -- (VIII) -- (IX) -- (X);

%\fill[gray] plot[domain=0:1, smooth] (\y,{-3/40*\y*\y+3/10*\y+2}) -- (IV) -- (III) --(I);

\draw[line width=1pt]  (V) 
-- node[pos=0.5, below=2pt, sloped]    {$\underline{\mathcal{C}}(\tau_1)$}
       (VI) -- node[pos=0.5, below=2pt, sloped]    {$\mathcal{C}(\tau_1)$} (VII); %-- 
            %node[midway, below=2pt, sloped, font=\tiny]    {$u=u_0$}
\draw[line width=1pt]  (X) 
-- node[pos=0.5, above=2pt, sloped]    {$\underline{\mathcal{C}}(\tau_2)$}
       (IX) -- node[pos=0.6, above=2pt, sloped]    {$\mathcal{C}(\tau_2)$} (VIII); %-- 
            %node[midway, below=2pt, sloped, font=\tiny]    {$u=u_0$}
            
\draw[line width=1pt,dashed] (III) --
            node[pos=0.5, above=2pt, sloped]    {$\mathcal{I}^+$}
      (IV);
\draw[line width=1pt] (IV) -- node[pos=0.5, above=2pt, sloped]    {$\mathcal{H}^+$} (I) -- node[pos=0.5, below=2pt, sloped]    {$\underline{\hat{\mathcal{C}}}_{\mathrm{in}}$} (II) -- node[pos=0.5, below=2pt, sloped]    {$\hat{\mathcal{C}}_{\mathrm{out}}$} (III);

\draw[line width=1pt,domain=0:4, smooth, variable=\y] plot ({2 + 3/40*\y*(4-\y)}, {\y});

\filldraw[fill=white, draw=black] (IV) circle (2pt);
%\filldraw[fill=white, draw=black] (III) circle (2pt);
\node at ($(IV)+(0,0.25)$)          {$i^+$};
\node at ($(VI)+(-0.25,-0.25)$)          {$\Gamma$};
%\node at ($(III)+(0.25,0)$)          {$i^0$};
\end{tikzpicture}
\caption{The Penrose diagram depicting the foliations $\uline{\mathcal{C}}(\tau)$ and $\mathcal{C}(\tau)$ parametrized by the "time" variable $\tau$, with the shaded region representing the domain of integration.}
\label{fig:Penrose2-foliation}
\end{figure}

Let $\psi_\pm$ be the two Dirac field components obtained after the spherically symmetric reduction. The aim is to establish the following \textit{hierarchy of weighted energy estimates}, consisting of a \textit{flux estimate} along the outgoing and ingoing null hypersurfaces and a corresponding \textit{integrated energy decay estimate}:
\begin{align*}
    \int_{\C(\tau_2)}\big(r^p|\p_v(r\psi_+)|^2 &+ |\psi_-|^2\big)\,dv+ \int_{\uline{\C}(\tau_2)}\Big(\Big|\frac{\p_u\psi_-}{-\p_u r}\Big|^2 + |\psi_+|^2\Big)(-\p_u r)\,du
    \\&+ \int^{\tau_2}_{\tau_1}\int_{\C(\tau)}\big(r^{p-1}|\p_v(r\psi_+)|^2 + |\psi_-|^2\big)\,dv d\tau + \int^{\tau_2}_{\tau_1}\int_{\uline{\C}(\tau)}\Big(\Big|\frac{\p_u\psi_-}{-\p_u r}\Big|^2 + |\psi_+|^2\Big)(-\p_u r)\,du d\tau \\&\les    \int_{\C(\tau_1)}\big(r^p|\p_v(r\psi_+)|^2 + |\psi_-|^2\big)\,dv + \int_{\uline{\C}(\tau_1)}\Big(\Big|\frac{\p_u\psi_-}{-\p_u r}\Big|^2 + |\psi_+|^2\Big)(-\p_u r)\,du + \text{l.o.t.},
\end{align*}
where $\tau$ is a proper time along a timelike curve $\Gamma$ with constant area radius, $\tau_1\le\tau_2$, $p\in[0,3-\delta]$, $\delta>0$. The notation l.o.t. refers to the lower order terms in the hierarchy, while the foliations $C(\tau)$ and $\uline{C}(\tau)$ refer to the \textit{outgoing} and \textit{ingoing null} hypersurfaces, respectively. For a schematic picture summarizing this construction, see Figure \ref{fig:Penrose2-foliation}. 

Since the components $\psi_\pm$ satisfy a \textit{coupled first-order system}, the authors of \cite{ma2022sharp} introduce suitable reweighted variables $\Phi_\pm:=\mathfrak{f}_\pm(r)\psi_\pm$, which recast the equations in the symmetric form
\[
    \p_u\Phi_+ = - \mathfrak{a}(r)\Phi_-,\quad \p_v\Phi_-=\mathfrak{a}(r)\Phi_+,
\]
with $\Phi_-$ now being degenerate at the event horizon.
The derivation of the hierarchy therefore involves two ingredients: the \textit{choice of a vector field}, adapted to the region under consideration, and \textit{suitable reweightings} to deal with the coupling between the two components. Let us briefly illustrate how this interplay enters the argument, focusing in particular on the Morawetz, red-shift and $r^p$ estimates:
\begin{enumerate}
    \item The first step is to establish an \textit{integrated local energy decay estimate} by combining the Killing vector field $\p_t$ together with a Morawetz vector field $f\p_r+q$, where $f(r)$ and $q(r)$ are properly chosen. At this stage, in order to obtain a globally coercive estimate, one exploits \eqref{eq:lichn-identity} expressed in terms of $\Phi_\pm$, which yields a symmetric coupled wave system of the form
    \[
        \p_u\p_v\Phi_\pm +\mathfrak{b}\Phi_\pm+\mathfrak{c}\Phi_\mp= 0.
    \]
    This symmetric structure is essential, as it plays the role of \textit{Chandrasekhar's transformation} \cite{chandrasekhar1975equations}, which reduces the Teukolsky equations for integer spin fields to scalar wave-like equations. Since the resulting estimate degenerates towards the event horizon and has non-optimal $r$-weights towards infinity, one needs the following two steps;
    \item Close to the horizon, a fundamental role is played by the \textit{red-shift estimate}, which captures the well-known \textit{red-shift effect} and provides non-degenerate control of the field towards the event horizon. This mechanism is obtained by considering the vector field $(-\p_u r)^{-1}\p_u$ \cite{dafermos2009red}, which is non-degenerate at the horizon. Indeed, in ingoing Eddington--Finkelstein coordinates $(v,r)$, this vector field is given by $-\p_r$, and is therefore manifestly regular and non-degenerate there. In this region, one first introduces a new reweighting $\widetilde{\Phi}_-$, which removes the degeneracy of $\Phi_-$ at the event horizon and satisfies a decoupled equation with a favourable structure, and then recovers analogous control for the other component;
    \item Towards infinity, the celebrated $r^p$ weighted estimates of Dafermos and Rodnianski \cite{dafermos2010new} are adapted to the Dirac field, obtaining the above hierarchy for $p\in[0,3)$. These estimates are obtained by using the vector field $r^p\p_v$. In the far region, \cite{ma2022sharp} instead start with a suitable reweighting $\wtilde\Phi_+$ satisfying again a decoupled equation, whose zeroth-order coefficient enjoys improved decay in $r$. This allows them to reach the full range of $p$ up to $3$ directly, and to then deal with $\Phi_-$ as a consequence.
\end{enumerate}
Once these estimates are obtained, the pigeonhole principle \cite{dafermos2010new} yields
\[
    \int_{\C(\tau)}\big(r^p|\p_v(r\psi_+)|^2 + |\psi_-|^2\big)\,dv+ \int_{\uline{\C}(\tau)}\Big(\Big|\frac{\p_u\psi_-}{-\p_u r}\Big|^2 + |\psi_+|^2\Big)(-\p_u r)\,du\les\tau^{-3+\delta+p},
\]
where the implicit constant only depends on $\delta$ and the initial data on the initial bifurcate hypersurface. Pointwise decay estimates can then be obtained, while commuting the equations yields corresponding decay estimates for higher-order derivatives.

\subsubsection{The nonlinear case: the bootstrap argument}
The approach described in the previous section, based on identifying and working with appropriate Dirac reweightings, will also be crucial in our nonlinear perturbative setting. 
In this case, however, due to the coupling between the geometry and the Dirac field, we set up a \textit{bootstrap argument}, assuming suitable estimates for both and improving them along the proof, thereby extending these estimates globally.
Our setup is inspired by \cite{dafermos2021non}, and by the works \cite{luk2019strong,angelopoulos2026nonlinear} in the context of Reissner--Nordström black holes in spherical symmetry, in both the subextremal and extremal regimes.

Given a Schwarzschild black hole of mass $M_0>0$, we consider a class of \textit{seed data sets} describing perturbations of the Schwarzschild initial data, prescribed on an initial bifurcate null hypersurface $\uline{\hat\C}_\ing\cup\hat\C_\out$ in the spirit of the characteristic formulation introduced in \cite{klainerman2002evolution}. The bifurcation sphere is located at an area radius $\Lambda:=100M_0$, and the seed data consist of the two Dirac components $\mathring\psi^\out_+$ and $\mathring\psi^\ing_-$, prescribed on $\hat\C_\out$ and $\uline{\hat\C}_\ing$, respectively.
To quantify the distance from the initial data corresponding to the Schwarzschild black hole of mass $M_0$, we consider the \textit{seed data norm}
\begin{equation*}
       \mathfrak D[\mathcal S]\sim
       \sup_{\hat{\uline{\C}}_\ing}\big(|\mathring\psi_-^{\ing}|+|\partial_{\hat u}\mathring\psi_-^{\ing}|\big)\\
       +\sup_{\hat\C_\out} \big(r|\mathring\psi_+^{\out}|+r^2|\partial_{\hat v}(r\mathring\psi_+^{\out})|\big)\le\ve.
\end{equation*}
Since $\ve$ represents the master small parameter measuring the size of the seed data, we shall refer to these data as $\ve$-Schwarzschildean. The coordinates $(\hat u,\hat v)$ are \textit{initial-data-normalized} coordinates, chosen such that $\p_{\hat u}r=-1$ on $\uline{\hat \C}_\ing$ and $\p_{\hat v}r=1$ on $\hat \C_\out$. In particular, $\hat u$ is regular at the event horizon. 

\begin{figure}
\centering
\begin{tikzpicture}[line cap=rect,inner sep=0pt, x=1cm, y=1cm, scale=1,font=\small]

\coordinate  (I)   at (0,2)      {};
\coordinate (II)   at (2,0)      {};
\coordinate  (III)   at (4,2)      {};
\coordinate   (IV)   at (2,4)      {};
\coordinate  (V)   at (0.5,1.5)      {};
\coordinate  (VI)   at (1,2)      {};
\coordinate  (VII)   at (2.19322,3.19322)      {};
\coordinate  (VIII)   at (3.38644,2)      {};
\coordinate  (IX)   at (3.69322,1.69322)      {};
\coordinate  (X)   at (2.19322,0.80678)      {};

\path[name path=parab1, domain=0:4, samples=100, smooth, variable=\y]
    plot (\y,{-3/40*\y*\y+3/10*\y+2});

\fill[lightgray] (I) -- (V) -- (VII) -- (IX) -- (III) -- (IV);
\fill[gray] (V) -- (II) -- (IX) -- (VII);

%\fill[gray] plot[domain=0:1, smooth] (\y,{-3/40*\y*\y+3/10*\y+2}) -- (IV) -- (III) --(I);

\draw[line width=1pt]  (VI) 
-- node[pos=0.5, below=2pt, sloped]    {$\underline{\mathcal{C}}_v$}
       (X) -- node[pos=0.5, below=2pt, sloped]    {$\mathcal{C}_u$} (VIII); %-- 
            %node[midway, below=2pt, sloped, font=\tiny]    {$u=u_0$}
\draw[line width=1pt]  (V) --  (VII) -- (IX); %-- 
            %node[midway, below=2pt, sloped, font=\tiny]    {$u=u_0$}
            
\draw[line width=1pt,dashed] (III) --
            node[pos=0.5, above=2pt, sloped]    {$\mathcal{I}^+$}
      (IV);
\draw[line width=1pt] (IV) -- node[pos=0.6, above=2pt, sloped]    {$\mathcal{H}^+$} (I) -- node[pos=0.5, below=2pt, sloped]    {$\underline{\hat{\mathcal{C}}}_{\mathrm{in}}$} (II) -- node[pos=0.5, below=2pt, sloped]    {$\hat{\mathcal{C}}_{\mathrm{out}}$} (III);

\draw[line width=1pt,domain=0:4, smooth, variable=\y] plot ({2 + 3/40*\y*(4-\y)}, {\y});

\draw[-Stealth, line width=1pt] (1.25,3.5) node[left=2pt] {$\partial_v r = 1-\frac{2m}{r}$} -- (1.875,2.875);
%\draw[-Stealth, line width=1pt] (1.25,3.5) node[left] {$\partial_v r = 1-\frac{2m}{r}$} -- (2.28369,2.46631);
\draw[-Stealth, line width=1pt] (3.43178,2.81822) node[right=2pt] {$\partial_u r = 1-\frac{2m}{r}$} -- (3,2.38644);
\draw[-Stealth, line width=1pt] (2.625,3.625) node[right=2pt] {$\Gamma(\tau_f)$} -- (VII);

\filldraw[fill=white, draw=black] (IV) circle (2pt);
%\filldraw[fill=white, draw=black] (III) circle (2pt);
\node at ($(IV)+(0,0.25)$)          {$i^+$};
\node at ($(X)+(-0.25,-0.25)$)          {$\Gamma$};
%\node at ($(III)+(0.25,0)$)          {$i^0$};
\end{tikzpicture}
\caption{The Penrose diagram showing the bootstrap domain $\D_{\tau_f}:=J^-(\Gamma(\tau_f))$, where $\Gamma:=\{r=\Lambda\}$ is the timelike curve anchoring the bootstrap domain. The double null coordinates $(u,v)$ are teleologically normalized according to the conditions written in the diagram, following the choice introduced in \cite{luk2019strong}.}
\label{fig:Penrose3-gauges}
\end{figure}
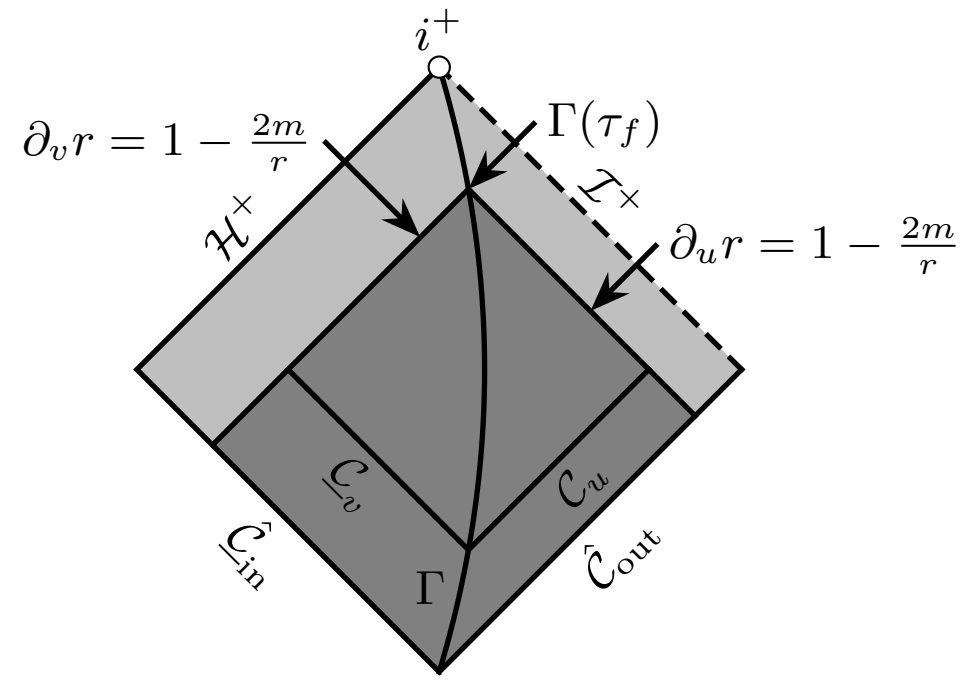

As depicted in Figure \ref{fig:Penrose3-gauges}, we consider the timelike curve $\Gamma:=\{r=\Lambda\}$, parametrized by its proper time $\tau$. For a given $\tau_f$, we define the \textit{bootstrap domain} $\D_{\tau_f}$ as the causal past of the point $\Gamma(\tau_f)$, namely $\D_{\tau_f}:=J^-(\Gamma(\tau_f))$. The bootstrap argument is carried out on $\D_{\tau_f}$, which is progressively extended towards the full future development by letting $\tau_f$ tend to infinity. We also introduce the double-null coordinates $(u,v)$, which are \textit{teleologically normalized} according to the conditions $\p_u r = -(1-\frac{2m}{r})$ on the ingoing future boundary of $\D_{\tau_f}$, where $m$ denotes the Hawking mass \eqref{eq:Hawking-mass}, and $\p_v r = 1-\frac{2m}{r}$ on the outgoing future boundary of $\D_{\tau_f}$. 
Moreover, we extend the proper time $\tau$ on $\Gamma$ to a \textit{time function} on $\D_{\tau_f}$, by setting it to be constant along ingoing cones in the close region and along outgoing cones to the right of $\Gamma$.

To identify the Schwarzschild black hole towards which the solution converges, we introduce a ${\tau_f}$-\textit{anchored Schwarzschild background} $(M_{\tau_f},\bar r_{\tau_f})$, which is updated at each extension of the bootstrap domain. Its reference mass is chosen to coincide with the Hawking mass along $\Gamma$ at the endpoint of the bootstrap domain, namely \[M_{\tau_f}:=m(\Gamma(\tau_f)).\]
We then fix the area radius coordinate $\bar r_{\tau_f}$ of this Schwarzschild solution in double-null Eddington--Finkelstein gauge by setting
\[
\bar r_{\tau_f}(\Gamma(\tau_f))=\Lambda.
\]
Thus, the reference background is updated as the bootstrap domain is extended. The precise construction and its role in the argument are described in Section \ref{sec:anch-solution}.

Motivated by the linear case discussed above, we introduce energy norms such as
\[
    \mathcal{E}_p(\tau):=\int_{\C(\tau)}\big(r^p|\p_v(r\psi_+)|^2  + |\psi_-|^2+...\big)\,dv,\quad \uline{\mathcal{E}}(\tau):= \int_{\uline{C}(\tau)}\Big(\Big|\frac{\p_u\psi_-}{-\p_u \bar r_{\tau_f}}\Big|^2 + |\psi_+|^2 + ...\Big)(-\p_u \bar r_{\tau_f})\,du,
\]
and formulate the corresponding bootstrap assumptions on $\D_{\tau_f}$. For the geometry, we assume, for instance,
\[
    |r-\bar r_{\tau_f}|\les \ve^{3/2}\tau^{-2+\delta},\quad |m-M_{\tau_f}|\les \ve^{3/2}\tau^{-3+\delta},
\]
while for the Dirac field we assume
\[
    \mathcal{E}_p(\tau)\les \ve^2\tau^{-3+\delta+p},\quad \uline{\mathcal{E}}(\tau)\les\ve^2\tau^{-3+\delta},
\]
for $\tau\in[1,\tau_f]$. We refer to Definitions \ref{def-of-energies} and \ref{def:bootstrap-set} for the complete set of assumptions.

\subsubsection{Improving the bootstrap assumptions}
Let us now outline the main elements needed to improve the bootstrap assumptions.

On the geometric side, the double-null gauge together with spherical symmetry reduces the Einstein equations to a \textit{system of transport equations} in terms of geometric quantities such as
\[
    \kappa:=\frac{\p_v r}{1-\frac{2m}{r}},\quad \gamma:=-\frac{\p_u r}{1-\frac{2m}{r}}.
\]
The strategy is then to integrate these equations and compare the dynamical solution with the anchored Schwarzschild background of mass $M_{\tau_f}$, for which 
\[
    \p_v \bar r_{\tau_f}=-\p_u \bar r_{\tau_f}=1-\frac{2M_{\tau_f}}{\bar r _{\tau_f}},\quad \bar \gamma=\bar \kappa=1.
\]
Recalling that the teleological gauges imposed above correspond to $\kappa=1$ on the final outgoing cone and $\gamma=1$  on the final ingoing cone, we therefore note that the perturbations $\tilde\kappa:=\kappa-1$ and $\tilde\gamma:=\gamma-1$ vanish there.
Integrating the corresponding transport equations backwards from these final cones and using the bootstrap assumptions on the Dirac field, yield control over $\tilde\kappa$ and $\tilde\gamma$. Thus, from $\p_v r=\kappa(1-\frac{2m}{r})$ and $\p_u r=-\gamma(1-\frac{2m}{r})$, we recover the corresponding control of the derivatives of the area radius. With this argument, one readily improves the bootstrap assumptions on the geometry.

Let us now emphasize an important point concerning the control of the \textit{null lapse} $\Omega$ associated to the metric \eqref{eq:null-coords}. As already mentioned, once "squared", the Dirac equation contains additional curvature terms that do not appear in the scalar field case and that are expressed in terms of derivatives of $\Omega$. Obtaining pointwise control of $\p_u\log\Omega$ and $\p_v\log\Omega$, however, requires preliminary bounds for the Dirac field. In Lemma \ref{lem:Omega-expansions}, we prove the corresponding estimates and discuss how these quantities are sensitive to the direction of integration relative to the teleological gauge conditions, which are imposed on the final cones.

On the other hand, improving the bootstrap assumptions on the Dirac field consists in adapting the strategy pursued in \cite{ma2022sharp}, discussed above, to the nonlinear perturbative setting. In particular, the reweightings used in the linear setting have to be carefully adapted to the present problem. Indeed, a direct linearization would not in general be sufficient, as the resulting nonlinear error terms may be too large to close the bootstrap argument. We therefore identify suitable Dirac reweightings in each region of interest, allowing us to exploit the same underlying mechanism as in the linear case and derive a hierarchy of the form
\[
    \mathcal{E}_p(\tau_2)+\uline{\mathcal{E}}(\tau_2)\les \mathcal{E}_p(\tau_1)+\uline{\mathcal{E}}(\tau_1) + \Err,
\]
where $\Err$ denotes some decaying nonlinear error, $1\le\tau_1\le\tau_2\le\tau_f$ and $p\in[0,3-\delta]$, and
\[
    \int^{\tau_2}_{\tau_1}\big(\mathcal{E}_{p-1}(\tau)+\uline{\mathcal{E}}(\tau)\big)\,d\tau\les \mathcal{E}_p(\tau_1)+\uline{\mathcal{E}}(\tau_1) + \Err,
\]
for $p\in[1,3-\delta]$. Once this family of estimates is established, the bootstrap assumptions on the Dirac field are improved by a pigeonhole argument, analogously to the linear case.

Finally, once the bootstrap assumptions have been improved and hence established globally in the exterior of the black hole, one last technical step remains: controlling the teleological gauges and proving their convergence to a well-defined \textit{eschatological gauge} as $\tau_f\to+\infty$.
This step requires additional care, since, due to the lack of the positive energy condition, the Hawking mass is not monotone and one cannot a priori guarantee that $1-\frac{2m}{r}>0$ globally. These difficulties are addressed in Section \ref{sec:proof-conclusion}, where we establish the convergence to the asymptotic Schwarzschild black hole with mass $M_\infty$ and conclude the asymptotic stability statement.

\subsection{Related works}
Let us conclude by mentioning some results in the literature that are relevant from the perspective of their setting and techniques. The stability of spherically symmetric black holes as solutions to various Einstein--matter systems has already been studied. We mention, for instance, \cite{luk2019strong,angelopoulos2026nonlinear}, which establish stability results for sub-extremal and extremal Reissner--Nordström black holes, respectively, in the Einstein--Maxwell--neutral scalar field system, as well as \cite{velozo2023linear} for the stability of Schwarzschild in the Einstein--massless Vlasov system.

We next recall some results concerning the Dirac equation on curved spacetimes. The study of Dirac fields on rotating black-hole backgrounds was significantly advanced by the separability results of Teukolsky \cite{teukolsky1973perturbations} for the massless neutrino equation on Kerr and Chandrasekhar \cite{chandrasekhar1976solution} in the massive case. Related results were obtained for Kerr--Newman spacetimes and in alternative coordinate systems  \cite{page1976dirac,roken2017massive}.

From the perspective of dispersive PDEs, the properties of the Dirac operator have been studied on a variety of curved backgrounds. In particular, local smoothing estimates on asymptotically flat and warped product spaces were established in \cite{cacciafesta2019weak}, while Strichartz estimates were obtained on spherically symmetric spaces and asymptotically flat manifolds in \cite{cacciafesta2019strichartz,cacciafesta2022strichartz}. Scattering for both massless and massive Dirac fields on spherically symmetric black holes was studied in \cite{nicolas1995scattering,melnyk2003scattering}, with related results on the exterior of subextremal Kerr \cite{hafner2004scattering} and in the interior of Reissner--Nordström black holes \cite{hafner2021scattering}.

Turning to the coupled Einstein--Dirac system, we mention the works of Finster, Smoller, and Yau, who constructed particle-like solutions for the Einstein--Dirac \cite{finster1999particlelike} and Einstein--Dirac--Maxwell \cite{finster1999particle} systems, which were later studied perturbatively by \cite{nodari2010perturbation1,nodari2010perturbation2}. Finster, Smoller, and Yau \cite{finstera2000non}, as well as Dafermos \cite{dafermos2003time}, also studied the non-existence of time-periodic solutions; we refer to \cite{finster2002non} for a discussion of these results.

More recently, the already mentioned work \cite{ma2022sharp} established Price's laws for the massless Dirac equation on a fixed Schwarzschild background using weighted energy estimates and the vector-field method.
In the nonlinear setting, global stability of Minkowski spacetime for the Einstein--Dirac system has been established in both the massless \cite{chen2022global} and massive \cite{lefloch2025global} cases. We also refer to \cite{dong2021global}, which proves a nonlinear stability result for a coupled system involving a massive Dirac field on Minkowski spacetime.
Local well-posedness for the characteristic initial value problem for the Einstein--Dirac system was studied in \cite{zhao2026local}, and the formation of trapped surfaces has been investigated for the closely related Einstein--Weyl system \cite{zhao2025trapped}.
Finally, \cite{hong2026spinor1,hong2026spinor2} has established global small-data stability for nonlinear Dirac systems near Minkowski spacetime and on fixed Schwarzschild backgrounds.

\subsection{Outline}
The paper is organized as follows. Section \ref{sec:preliminaries} is devoted to presenting all the preliminaries regarding the spherically symmetric Einstein--massless Dirac system, from the explicit formulation of the equations (see also Appendix \ref{app:derivation-Dirac}) to the definition of the characteristic initial value problem and its well-posedness (we refer to Appendix \ref{app:lwp-Einstein-Dirac}). In Section \ref{sec:stab-schwarzschild-einstein-dirac}, we detail the framework of our stability result. In particular, we give a rigorous statement of the main result, describe the proof and the bootstrap argument, and introduce the corresponding teleological gauges and energy norms. Sections \ref{sec:improvement-geometry} to \ref{sec:energy-Dirac} focus on the improvement of the bootstrap assumptions for the geometry and the Dirac field. In Section \ref{sec:proof-conclusion} we finally put all pieces together and conclude the main proof.

\subsection*{Acknowledgments} The author would like to thank his advisors, Anne-Sophie de Suzzoni and Cécile Huneau, for their support and the many helpful discussions. Special thanks also go to the General Relativity Group in Jussieu for the meaningful monthly exchanges and conversations. The author also acknowledges the financial support received through the S. S. Chern Young Faculty Award, funded by AX and awarded to Anne-Sophie de Suzzoni.

\subsection*{Declaration of AI use} The author used ChatGPT (OpenAI) to assist with English language editing, the correction of typos, and the improvement of the clarity and presentation of the manuscript. All AI-assisted suggestions were reviewed and revised by the author, who takes full responsibility for the mathematical content and the final manuscript.
\section{Preliminaries}\label{sec:preliminaries}
In this section we precisely introduce the framework in which we will be working. Section \ref{subsec:Einstein--Dirac system in spherical symmetry} contains the explicit formulation of the spherically symmetric Einstein--massless Dirac system, and we refer to Appendix \ref{app:derivation-Dirac} for a detailed derivation. The main properties of the Schwarzschild solution are recalled in Section \ref{subsec:Schwarzschild-bh}, while in Section \ref{sec:charact-init-value-pb} we set the characteristic initial value problem, whose well-posedness is tackled in Appendix \ref{app:lwp-Einstein-Dirac}.
\subsection{Einstein--Dirac system in spherical symmetry}\label{subsec:Einstein--Dirac system in spherical symmetry}
Let $(\mathcal{M},g)$ be a smooth, connected, time-oriented, four-dimensional Lorentzian manifold, equipped with a metric tensor $g$ of signature $(-,+,+,+)$, endowed with a \textit{spin structure}.
We say that $(\mathcal{M},g)$ is \textit{spherically symmetric} if it is diffeomorphic to $\mathcal{Q}\times\mathbb{S}^2$ and the metric takes the form 
\[
    g=g_\mathcal{Q}+r^2 d\sigma_{\mathbb{S}^2},
\]
where $(\mathcal{Q}, g_\mathcal{Q})$ is a $(1+1)$-dimensional Lorentzian spacetime with boundary (corresponding to the initial data hypersurface), $d\sigma_{\mathbb{S}^2}:=d\vartheta^2+\sin^2\vartheta\,d\varphi^2$ is the round metric on the unit sphere, and $r$ is a nonnegative function on $\mathcal{Q}$, which can be geometrically interpreted as the \textit{area radius} of the orbits of the isometric $SO(3)$ action on $(\mathcal{M},g)$.
We moreover assume that $(\mathcal{Q}, g_\mathcal{Q})$ admits a \textit{global double-null coordinate system} $(u,v)$ such that the metric $g$ takes the form
\begin{equation}\label{eq:null-coords}
    g=-4\Omega^2(u,v)\,dudv + r^2(u,v)\,d\sigma_{\mathbb{S}^2},
\end{equation}
where the \textit{null lapse} $\Omega^2$ is a positive function on $\mathcal{Q}$. We stress that the constant $u$ and $v$ hypersurfaces are null in $(\mathcal{Q},g_\mathcal{Q})$. We will frequently identify the full spacetime with its reduced description $(\mathcal{Q},r,\Omega^2)$.
\begin{remark}\label{rem:null-coords}
The double-null coordinates $(u,v)$ above are not uniquely defined. Indeed, under any strictly increasing smooth reparametrizations $U,V:\mathbb{R}\to\mathbb{R}$, $(\tilde u, \tilde v):=(U(u),V(v))$,  
the metric maintains the same double-null form $g=-4\wtilde\Omega^2\,d\tilde ud\tilde v + r^2\,d\sigma_{\mathbb{S}^2}$, where
\[
    \wtilde\Omega^2(\tilde u,\tilde v)=\frac{1}{U'(U^{-1}(\tilde u))V'(V^{-1}(\tilde v))}\Omega^2(U^{-1}(\tilde u),V^{-1}(\tilde{v})),\quad r(\tilde u,\tilde v)= r(U^{-1}(\tilde u),V^{-1}(\tilde v)).
\]
We refer to such transformations as \emph{double-null gauge transformations}.
\end{remark}

We next introduce the notation for the geometric quantities used throughout the paper. We define the \emph{Hawking mass} $m:\mathcal M  \to\mathbb R$ 
\begin{equation}\label{eq:Hawking-mass}
    m := \frac{r}{2}\Big(1 - g(\nabla r,\nabla r) \Big) = \frac{r}{2}\Big( 1+\frac{\partial_u r\, \partial_vr}{\Omega^2}\Big),
\end{equation}
which is clearly independent of the choice of double-null gauge, and the \emph{mass aspect function}
\begin{equation}
    \mu :=\frac{2m}{r}.
\end{equation}
Moreover, we denote
\begin{equation}\label{eq:geom-quantities}
    \nu:=\partial_ur,\quad\lambda:= \partial_vr, \quad\kappa:=\frac{\Omega^2}{-\nu} = \frac{\lambda}{1-\mu},\quad \gamma:=  \frac{\Omega^2}{\lambda} =\frac{-\nu}{1-\mu}.
\end{equation}

We now turn to the Dirac field. As detailed in Appendix \ref{app:sec:reduction-sph-symm}, we fix the regular \textit{orthonormal frame}
\[
    e_0 := \frac12 \Big( \frac{1}{\kappa}\partial_v + \frac{1}{-\nu}\partial_u\Big), \quad e_1 := \frac{1}{r}\partial_\vartheta, \quad e_2 := \frac{1}{r\sin\vartheta}\partial_\varphi,\quad e_3 := \frac12 \Big( \frac{1}{\kappa}\partial_v - \frac{1}{-\nu}\partial_u\Big),
\]
and restrict the Dirac spinor to the lowest angular mode compatible with spherical symmetry, denoted by $\wtilde{\mathcal{H}}^-_{1/2,1/2}$ (see \eqref{app:eq:lowest-sph-harmonic} and the related discussion). Therefore, the solution $\Psi$ to the Dirac equation \eqref{eq:Dirac-eq1} decomposes into two components $\psi_\pm(u,v)$ satisfying the \textit{reduced Dirac equations}
\begin{align}
    \label{eq:partial_u-psi+}
    \Big(\frac{1}{-\nu}\partial_u +\frac{\partial_u\kappa}{2\Omega^2}\Big)\psi_+  &= -\psi_-,\\
    \label{eq:partial_v-psi-}
    \Big(\frac{1}{\kappa}\partial_v -\frac{\partial_v\nu}{2\Omega^2}+\frac{1-\mu}{r}\Big)\psi_-  &=\frac{1}{r^2} \psi_+.
\end{align}
On the other hand, Einstein equations \eqref{eq:Einstein-eq1} reduce to a hyperbolic-transport system for the metric components $r,\Omega$ given by
\begin{align}
        \label{eq:wave-eq-r}
        \partial_u \partial_v r  &= -\frac{2\Omega^2 m}{r^2} + 4\pi r T_{uv},\\
        \label{eq:wave-eq-Omega}
        \frac{1}{2}\partial_u\partial_v \log \Omega^2
        &=\frac{2\Omega^2 m}{r^3} -  8\pi T_{uv},\\
        \label{eq:Raych-u}
        \partial_u (\Omega^{-2} \partial_u r) &= -\frac{4\pi r}{\Omega^2} T_{uu},\\
        \label{eq:Raych-v}
        \partial_v (\Omega^{-2} \partial_v r) &= -\frac{4\pi r}{\Omega^2} T_{vv},
\end{align}
where the components of the Dirac energy-momentum tensor are given by
\begin{equation}\label{eq:Dirac-energy-momentum-tensor}
    T_{uu}= \frac{-\nu}{2\pi}\Im\big[\oline{\psi_-}\partial_u\psi_-\big],\quad
    T_{vv} = \frac{\kappa}{2\pi r^2} \Im\big[ \oline{\psi_+} \partial_v\psi_+ \big],\quad
    T_{uv} = \frac{\Omega^2}{2\pi r^2}\Im \big[\oline{\psi_-}\psi_+\big].
\end{equation}
We therefore refer to $(r,\Omega,\psi_+,\psi_-)$ as a \textit{spherically symmetric solution} to the Einstein--massless Dirac system if equations \eqref{eq:partial_u-psi+}--\eqref{eq:Raych-v} are satisfied.
\begin{remark}[Lack of monotonicity]
    Equations \eqref{eq:Raych-u}--\eqref{eq:Raych-v} are the so-called \textit{Raychaudhuri equations}, which can be seen as constraint equations for the characteristic initial value problem. In contrast to the scalar field case, the energy-momentum tensor \eqref{eq:Dirac-energy-momentum-tensor} has no definite sign condition, precluding the usual \textit{a priori} monotonicity arguments based on coercivity (see for instance \cite{dafermos2005spherically, kommemi2013global}). The analysis therefore relies on the decay estimates for the Dirac field established throughout the proof.
\end{remark}

\begin{remark}[Gauge freedom]\label{rem:gauge-freedom} A spherically symmetric solution $(r,\Omega,\psi_+,\psi_-)$ to the Einstein--massless Dirac system can be transformed into another solution $(r, \wtilde\Omega,\psi_+,\psi_-)$ of the same system by using a change of variables $(\tilde u,\tilde v):=(U(u),V(v))$, with $U,V$ smooth and strictly increasing, as in Remark \ref{rem:null-coords}.
The new solution under this change of variable is given by
\[
    \wtilde\Omega^2(\tilde u,\tilde v)=\frac{1}{U'(u)V'(v)}\Omega^2(u,v),\quad r(\tilde u,\tilde v)= r(u,v),\quad \psi_\pm(\tilde u,\tilde v)= \psi_\pm(u,v)\footnote{The choice of the orthonormal frame fixed above induces the following \textit{gauge-invariant double-null frame} $\frac{1}{-\nu}\partial_u, \frac{1-\mu}{\lambda}\partial_v,\frac{1}{r}\partial_\vartheta,\frac{1}{r\sin\vartheta}\partial_\varphi$. Consequently, the spinor components are invariant under double-null gauge transformations, as can be verified directly.}.
\]
\end{remark}

From definitions \eqref{eq:Hawking-mass}--\eqref{eq:geom-quantities}, we may rearrange \eqref{eq:wave-eq-r}--\eqref{eq:Raych-v} and derive the equivalent system
\begin{align}
    \label{eq:transport-kappa}
    \partial_u \kappa &= -\frac{4\pi r \kappa^2}{\Omega^2} T_{uu} = -2r\kappa\Im[\overline{\psi_-}\partial_u\psi_-],\\
    \label{eq:transport-gamma}
    \partial_v \gamma &= \frac{4\pi r \gamma^2}{\Omega^2} T_{vv}=\frac{2\gamma}{r(1-\mu)}\Im[\overline{\psi_+}\partial_v\psi_+],\\
    \label{eq:transport-u-varpi}
    \partial_u m &= 2\pi r^2 \Omega^{-2}(T_{uv}\partial_u r- T_{uu}\partial_v r)= \nu \Im[\overline{\psi_-}\psi_+] - r^2(1-\mu) \Im[\overline{\psi_-}\partial_u\psi_-],\\
    \label{eq:transport-v-varpi}
    \partial_v m &= 2\pi r^2 \Omega^{-2}(T_{uv}\partial_v r- T_{vv}\partial_u r)=\lambda \Im[\overline{\psi_-}\psi_+] + \Im[\overline{\psi_+}\partial_v\psi_+].
\end{align}

Throughout the paper, we will employ several reweightings of the spinor components to highlight useful structures and simplify the equations. Of particular importance is the following 
\begin{equation}\label{def:Phi_pm}
    \Phi_+ := \kappa^{1/2} \psi_+ ,\quad \Phi_- := r(-\nu)^{1/2}\psi_-,
\end{equation}
which may be viewed as a dynamical generalization of the transformation introduced in \cite[Definition 3.1]{ma2022sharp} for the Dirac equation on a fixed Schwarzschild background. In terms of these components, \eqref{eq:partial_u-psi+}--\eqref{eq:partial_v-psi-} take the particularly symmetric form
\begin{align}
    \label{eq:Dirac-Phi_+}
    \partial_u \Phi_+ &= -\frac{\Omega}{r}\Phi_-,\\
    \label{eq:Dirac-Phi_-}
    \partial_v \Phi_- &=  \frac{\Omega}{r}\Phi_+.
\end{align}
Applying $\partial_v$ to \eqref{eq:Dirac-Phi_+} and $\partial_u$ to \eqref{eq:Dirac-Phi_-}, one readily finds the following wave-like coupled system
\begin{align}
    \label{eq:wave-Phi_+}
    \partial_u\partial_v\Phi_+  + \frac{\Omega^2}{r^2}\Phi_+ + \frac{\Omega}{r}(\partial_v\log\Omega-\lambda r^{-1})\Phi_-&=0,\\
    \label{eq:wave-Phi_-}
    \partial_u\partial_v\Phi_- + \frac{\Omega^2}{r^2}\Phi_- - \frac{\Omega}{r}(\partial_u\log\Omega-\nu r^{-1})\Phi_+&=0.
\end{align}
\begin{remark} Recalling that the wave equation for a spherically symmetric scalar field $\phi$ in null coordinates writes
\[
    \p_u\p_v\phi+\frac{\lambda}{r}\p_u\phi+\frac{\nu}{r}\p_v\phi=0,
\]
we may view the Einstein--Dirac system, roughly speaking, as the Einstein equations coupled to two nonlinear coupled wave equations. In contrast to the scalar field case, however, the resulting equations contain additional geometric coupling terms, including derivatives of the dynamical null lapse $\Omega$.
\end{remark}

\subsection{The Schwarzschild family of black holes}\label{subsec:Schwarzschild-bh}
Let us now briefly recall the geometric features of the Schwarzschild family that will be relevant for our analysis. As already mentioned, for $M>0$, the Schwarzschild metric is given in Schwarzschild coordinates by
\begin{equation*}
     g_{M}= -X(r)dt^2+X(r)^{-1}dr^2+r^2d\sigma_{\mathbb{S}^2},\quad X(r):=1-\frac{2M}{r}.
\end{equation*}
These coordinates cover only the \textit{domain of outer communication}, corresponding to $(t,r,\vartheta,\varphi) \in \mathbb R \times (2M,+\infty)\times\mathbb{S}^2$. To obtain coordinates regular at the future event horizon $\mathcal H^+:=\{r=2M\}$, we introduce the \textit{tortoise coordinate}
\begin{equation*}
    r_*(r):=\int^r_{3M} \frac{dr'}{X(r')},
\end{equation*}
and the \textit{advanced time coordinate} $v:=\frac12(t+r_*)$. In these new coordinates, called \textit{ingoing Eddington--Finkelstein coordinates}, the metric rewrites
\begin{equation*}
    g_M=-4Xdv^2+4dvdr+r^2 d\sigma_{\mathbb{S}^2},
\end{equation*}
which extends smoothly across $\mathcal H^+$. Introducing in addition the \textit{retarded time coordinate} $u:=\frac12(t-r_*)$, we are able to write the Schwarzschild metric in \textit{double-null Eddington--Finkelstein coordinates}
\begin{equation*}
    g_{M}=-4X(u,v)dudv+r^2(u,v)d\sigma_{\mathbb{S}^2},
\end{equation*} where we stressed that now $X$ and $r$ are implicit functions of $u$ and $v$.
These double-null coordinates cover the domain of outer communication, with the future event horizon $\mathcal{H}^+$ and future null infinity $\mathcal{I}^+$ formally corresponding to $u=+\infty$ and $v=+\infty$, respectively.
The geometric quantities \eqref{eq:Hawking-mass}--\eqref{eq:geom-quantities} in these coordinates are given by
\[
    m=M,\quad\lambda=-\nu=1-\mu=\Omega^2=X,\quad   \kappa=\gamma=1,
\]
while, using $r_*=v-u$, the Killing vector field takes the form $T=\p_t=\p_v+\p_v$.

Although the normalization $r_*(3M)=0$ fixes the tortoise coordinate, and hence the associated double-null coordinates, one may prescribe the value of area radius at any other point. We record this fact in the following lemma, stated for the more general Reissner--Nordström family in \cite[Lemma 2.2]{angelopoulos2026nonlinear}, and apply it here to the Schwarzschild case. The lemma will be useful in Definition \ref{def:anch-solution1} below.
\begin{figure}
\centering
\begin{tikzpicture}[line cap=rect,inner sep=0pt, x=1cm, y=1cm, scale=1,font=\small]

\coordinate  (I)   at (0,2)      {};
\coordinate (II)   at (2,0)      {};
\coordinate  (III)   at (4,2)      {};
\coordinate   (IV)   at (2,4)      {};
\coordinate  (V)   at (1.15,3.15)      {};
\coordinate  (VI)   at (2.3,2)      {};
\coordinate  (VII)   at (3.15,2.85)      {};
\coordinate  (A)   at (0,4)      {};
\coordinate   (U)   at (1,3.3)      {};

\fill[lightgray] (I) -- (II) -- (III) -- (VII) -- (VI) -- (V);
%\fill[lightgray] (I) -- (IV) -- (A);
\fill[gray] (V) -- (VI) -- (VII) -- (IV);
%\path[fill=lightgray, draw=black](I) -- (IV) decorate[decoration={zigzag}]{(IV) -- (A)} -- (I)-- cycle;

\fill[lightgray] (I) -- (IV) decorate[decoration={zigzag}] {(IV) -- (A)} -- (I) -- cycle;
\draw[line width=1pt] (I) -- (IV);
\draw[line width=1pt,decorate,decoration={zigzag}] (IV) -- node[pos=0.5, above=4pt] {$r=0$} (A);

%\fill[gray] plot[domain=0:1, smooth] (\y,{-3/40*\y*\y+3/10*\y+2}) -- (IV) -- (III) --(I);

\draw[line width=1pt]  (U) 
-- node[pos=0.5, below=2pt, sloped]    {$\underline{\hat{\mathcal{C}}}_{\mathrm{in}}$}
       (VI) -- node[pos=0.5, below=2pt, sloped]    {$\hat{\mathcal{C}}_{\mathrm{out}}$} (VII); %-- 
            %node[midway, below=2pt, sloped, font=\tiny]    {$u=u_0$}
%\draw[line width=1pt]  (V) --  (VII) -- (IX); %-- 
            %node[midway, below=2pt, sloped, font=\tiny]    {$u=u_0$}
%\draw[line width=1pt,decorate,decoration=zigzag]  (A) -- node[pos=0.5, above=2pt] {$r=0$} (IV);
    
\draw[line width=1pt,dashed] (II) -- node[pos=0.5, below=2pt, sloped]    {$\mathcal{I}^-$} (III) --
            node[pos=0.5, above=2pt, sloped]    {$\mathcal{I}^+$} (IV);
\draw[line width=1pt] (IV) -- node[pos=0.6, above=2pt, sloped]    {$\mathcal{H}^+$} (I) -- node[pos=0.5, below=2pt, sloped]    {$\mathcal{H}^-$} (II);

%\draw[line width=1pt] (IV) -- node[pos=0.6, above=2pt, sloped]    {$\mathcal{H}^+$} (I) -- node[pos=0.5, below=2pt, sloped]    {$\underline{\hat{\mathcal{C}}}_{\mathrm{in}}$} (II) -- node[pos=0.5, below=2pt, sloped]    {$\hat{\mathcal{C}}_{\mathrm{out}}$} (III);

\draw[line width=1pt,domain=0:4, smooth, variable=\y] plot ({2 + 3/40*\y*(4-\y)}, {\y});
\draw[line width=1pt,domain=1:2, smooth, variable=\x] plot ({\x},{0.2*\x*\x+0.1*\x+3});

\filldraw[fill=white, draw=black] (IV) circle (2pt);
\filldraw[fill=white, draw=black] (III) circle (2pt);
\filldraw[fill=white, draw=black] (II) circle (2pt);
\filldraw[black] (U) circle (2pt);

\node at ($(IV)+(0,0.25)$)          {$i^+$};
\node at ($(II)-(0,0.25)$)          {$i^-$};
\node at ($(II)+(0,0.75)$)      {$\Gamma$};
\node at ($(III)+(0.25,0)$)         {$i^0$};
\node at ($(0.5,3.6)$)         {$\mathcal{BH}$};
\end{tikzpicture}
\caption{The Penrose diagram of a Schwarzschild black hole. Our stability result concerns the exterior region, which is here colored in a darker gray shade.}
\label{fig:Penrose4-schwarzschild}
\end{figure}
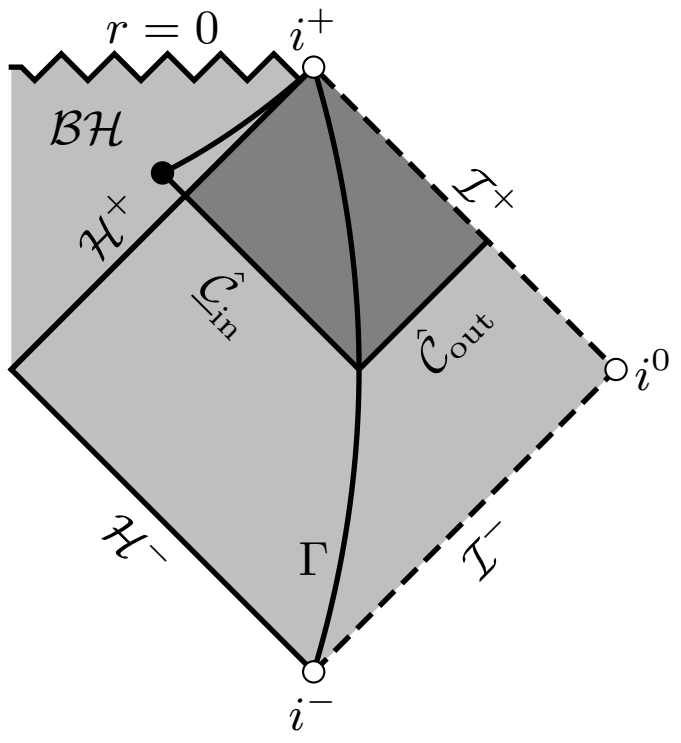

\begin{lemma}\label{lem:existence-anchoring}
    Let $M>0$. Then for any $R_0\in (2M,+\infty)$ and $(u_0,v_0)\in\mathbb R^2$, there exists a unique function $\bar r:\mathbb R^2\to (2M,+\infty)$ such that 
    \begin{equation}\label{eq:Schwarzschild-EF}
        \bar g:= -4X(\bar r)\,dudv+ \bar r^2(u,v)d\sigma_{\mathbb{S}^2}
    \end{equation}
    on $\mathbb{R}^2\times \mathbb{S}^2$ is isometric to the Schwarzschild black hole exterior of mass $M$ and such that 
    \begin{equation*}\oline m=M,\quad
      \partial_v\bar r=-\partial_u\bar r =X(\bar r),\quad\bar r(u_0,v_0)=R_0.
    \end{equation*}
\end{lemma}
Even though, as previously noted, the event horizon $\mathcal{H}^+$ corresponds to the limiting hypersurface $\{u=+\infty\}$ in double-null Eddington--Finkelstein coordinates, one may extend the geometric quantities \eqref{eq:geom-quantities} up to the horizon as follows:
\begin{equation*}
    \bar r(\infty,\cdot)=2M,\quad  \bar \lambda(\infty,\cdot)=0, \quad \bar \mu(\infty,\cdot)=1.
\end{equation*}

\begin{remark}[Dirac on Schwarzschild]\label{rk:Dirac-Phi-Schwarzschild} On a fixed Schwarzschild background, \eqref{eq:wave-Phi_+}--\eqref{eq:wave-Phi_-} therefore reduce to the symmetric hyperbolic system
\[
    \partial_u\partial_v\Phi_\pm  + \frac{X}{\bar r^2}\Phi_\pm + \frac{\sqrt{X}}{\bar r^3}(3M-\bar r)\Phi_\mp=0.
\]
As highlighted in \cite[Remark 4.2]{ma2022sharp}, this feature is crucial in the proof of the Morawetz estimate and it may be regarded as the \textit{analogue of Chandrasekhar's transformation} \cite{chandrasekhar1975equations} for integer spin fields, which recasts the corresponding Teukolsky equations as scalar wave-like equations.
To retain this structure in the dynamical setting, we view \eqref{eq:wave-Phi_+}--\eqref{eq:wave-Phi_-} as a perturbation of the system above, which requires error estimates for the terms involving $\Omega$ and its derivatives, see already Lemma \ref{lem:Mor}.
\end{remark}

\subsection{The characteristic initial value problem}\label{sec:charact-init-value-pb}
We conclude this section by briefly recalling the characteristic initial value problem in general relativity.
Given $U_0<U_1$ and $V_0<V_1$, we define the \textit{bifurcate null hypersurface} and the associated diamond-shaped region by
\begin{align*}
    \C(U_0,U_1,V_0,V_1)  &:= \big(\uline{\C}_{\ing}(U_0,U_1;V_0)\big)\cup \big(\C_{\out}(V_0,V_1;U_0)\big):=\big([U_0,U_1]\times\{V_0\}\big)\cup\big(\{U_0\}\times[V_0,V_1]\big), \\
    \R(U_0,U_1,V_0,V_1) &:= [U_0,U_1]\times [V_0,V_1].
\end{align*}
For simplicity, we will omit the dependence on $(U_0,U_1,V_0,V_1)$ and directly write $\C$ and $\R$ whenever the vertices are clear from the context. We say that a function $f$ is smooth on $\C$ if its restrictions $f_\out:= f|_{\C_{\out}}$ and $f_\ing:= f|_{\C_\ing}$ are smooth on the respective null segments.

\begin{definition}\label{def:init-data-set}
      A smooth \emph{bifurcate characteristic initial data set} for the spherically symmetric Einstein--massless Dirac system consists of a quadruple of smooth functions $(\mathring r,\mathring\Omega,\mathring\psi^\out_+, \mathring\psi^\ing_-)$ with
      \[
        \mathring r:\C\to\mathbb R^+,\quad\mathring\Omega^2:\C\to\mathbb R^+,\quad\mathring\psi^\out_+:\C_\out\to\mathbb C,\quad\mathring\psi^\ing_-:\uline{\C}_\ing\to\mathbb C,
      \]
      such that $\mathring\nu<0$ on $\C$, and \eqref{eq:Raych-u} and \eqref{eq:Raych-v} are satisfied on $\uline{\C}_\ing$ and $\C_\out$ respectively. 
\end{definition}
\begin{remark}
The choice of the free spinor data on each initial hypersurface is dictated by the transport structure of the reduced Dirac equations \eqref{eq:partial_u-psi+}--\eqref{eq:partial_v-psi-}. On the other hand, the assumption $\mathring\nu<0$ on $\C$, which can be interpreted as the \textit{initial absence of anti-trapped spheres}, is needed to define the renormalized components $\Phi_\pm$ in \eqref{def:Phi_pm}, which play a central role in our analysis of the Dirac field.
For small perturbations of Schwarzschild, this condition is in fact automatically satisfied, as shown in Proposition \ref{prop:seed-data-construction} below.
\end{remark}

We next state the local existence result for the characteristic initial value problem for the Einstein--massless Dirac system.

\begin{proposition}\label{prop:slab-existence}
For any $U_0<U_1$, $V_0<V_1$ and $B>0$, there exists a constant $\ve_\mathrm{slab}>0$ with the following property. Let $(\mathring r,\mathring\Omega,\mathring\psi^\out_+, \mathring\psi^\ing_-)$ be a characteristic initial data set for the spherically symmetric Einstein--massless Dirac system on $\mathcal C(U_0,U_1,V_0,V_1)$ satisfying
    \[
        \V\log\mathring r\V_{C^2(\C)} + \V\log\mathring \Omega\V_{C^2(\C)} + \V \mathring\psi^\out_+ \V_{C^0(\C_\out)} + \V \partial_v\mathring\psi^\out_+ \V_{C^0(\C_\out)} + \V \mathring\psi^\ing_- \V_{C^0(\uline{\C}_\ing)} + \V \partial_u\mathring\psi^\ing_- \V_{C^0(\uline{\C}_\ing)}\le B.
    \]
Then there exists a unique smooth solution $(r,\Omega,\psi_+,\psi_-)$ to the spherically symmetric Einstein--Dirac system on the slab
    \begin{equation*}
      \D := \R(U_0,U_0+\min\{\ve_\mathrm{slab},U_1-U_0\},V_0,V_1)\cup\mathcal R(U_0,U_1,V_0,V_0+\min\{\ve_\mathrm{slab},V_1-V_0\})
    \end{equation*}
    satisfying $\nu<0$ on $\mathcal{D}$ and which extends the initial data. Moreover, the norms 
    \[
        \V\log r\V_{C^k(\D)}, \V\log\Omega\V_{C^k(\D)}, \V\psi_+ \V_{C^k(\D)},\V \psi_- \V_{C^k(\D)}
    \] 
    are finite for any $k$ and can be bounded in terms of appropriate higher order initial data norms.
\end{proposition}

\begin{proof} 
We briefly comment on the proof and refer to Appendix \ref{app:lwp-Einstein-Dirac} for the details.
Local existence for the characteristic initial value problem in small rectangles follows from a standard iteration argument, applied to the Einstein--Dirac system \eqref{eq:partial_u-psi+}--\eqref{eq:Raych-v}, after rewriting the Dirac equations in terms of the renormalized variables $\Phi_\pm$. The argument is inspired by \cite{dafermos2005extension}.
The extension to thin slabs is then obtained by working similarly to \cite[Proposition 3.17]{kehle2024extremal}. 
For more general arguments exploiting the underlying null structure, both for the vacuum Einstein equations and the Einstein--Dirac system without spherical symmetry, we refer to \cite{luk2012local,zhao2026local}.
\end{proof}

Thanks to \cite{choquet1969global,sbierski2016existence}, we have a natural notion of maximal globally hyperbolic development. In spherical symmetry, this is a subset of the domain of dependence of $\mathcal C$, viewed as a subset of $(1+1)$-dimensional Minkowski space.

\begin{proposition}
Let $(\mathring r,\mathring\Omega,\mathring\psi^\out_+, \mathring\psi^\ing_-)$ be a characteristic initial data set of the Einstein--massless Dirac system on $\C(U_0,U_1,V_0,V_1)$, where $U_1$ and $V_1$ are allowed to take the value $+\infty$. Then there exists a set $\mathcal Q_\maxx\subset \mathcal R(U_0,U_1,V_0,V_1)$ with the following properties:
\begin{enumerate}
    \item $\mathcal Q_\maxx$ is globally hyperbolic as a subset of $\mathbb{R}^{1+1}_{u,v}$ with Cauchy surface $\mathcal C$.
    \item The solution $(r,\Omega,\psi_+,\psi_-)$ extends uniquely to $\mathcal Q_\maxx$.
    \item $\mathcal Q_\maxx$ is maximal with respect to properties 1.~and 2.
\end{enumerate}
\end{proposition}
\section{Stability of Schwarzschild for the Einstein--Dirac system}\label{sec:stab-schwarzschild-einstein-dirac}
In this section, we present the notation and framework needed to formalize the statement of Theorem \ref{thm:Schwarz-stability1}. In particular, in Section \ref{sec:setup-proof} we set up the main definitions, from the seed data to the energy hierarchies employed in the bootstrap argument. Section \ref{sec:statement-thm} is finally devoted to the detailed statement of our main result, including the corresponding bootstrap assumptions. Our setup is inspired by \cite{angelopoulos2026nonlinear}.

\subsection{Setup for the statement of stability}\label{sec:setup-proof}
Let us start by fixing the mass parameter $M_0>0$, corresponding to the initial mass of the perturbed black hole. We pose our data on the bifurcate null hypersurface
\begin{equation*}
    \hat{\C}:= \uline{\hat{{\C}}}_\ing \cup \hat{\C}_\out = \C(0,U_*,0,+\infty), \quad U_*:=98.5M_0,
\end{equation*}
with null coordinates denoted by $\hat u$ and $\hat v$. 
We normalize the area radius and the Hawking mass at the bifurcation sphere by requiring
\[
    m(0,0)=M_0,\quad r(0,0)=\Lambda:=100M_0.
\]
In particular, the bifurcation sphere lies in the exterior region of the reference Schwarzschild solution, since $1-\frac{2M_0}{\Lambda}>0$.
The constructions and smallness parameters introduced in this section, as well as those appearing throughout the paper, will depend implicitly on the fixed choice of $M_0$.

\subsubsection{Gauge-normalized seed data}
As already remarked, the standard monotonicity arguments that provide \textit{a priori} control of the signs of geometric quantities such as $\lambda$ and $\nu$ are not available here, due to the lack of a definite sign for the Dirac energy-momentum tensor. This issue arises already on the initial bifurcate hypersurface. We therefore restrict to seed data in an almost-Schwarzschildean regime, characterized by smallness with respect to an appropriate norm $\mathfrak{D}$. These quantitative bounds allow us to recover the required sign properties on $\hat{\C}$, which are then recovered in the domain of dependence only \textit{a posteriori} through the bootstrap argument.
\begin{definition}[$\ve$-Schwarzschildean seed data set]
    A \emph{seed data set} on $\hat\C$ is a couple
    \begin{equation*}
        \mathcal S:= (\mathring \psi^\out_+,\mathring \psi^\ing_-),
    \end{equation*}
    where $\mathring\psi^\out_+:\hat{\C}_\out\to\mathbb{C}$ and $\mathring\psi^\ing_-:\hat{\uline{\C}}_\ing\to\mathbb{C}$ are smooth functions.
    Given $\ve>0$, we call \textit{$\ve$-Schwarzschildean} a seed data set satisfying the following condition on the \emph{seed data norm}:
\begin{equation}\label{eq:seed-data-norm}
\begin{aligned}
       \mathfrak D[\mathcal S]:=
       \sup_{\hat{\uline{\C}}_\ing}\Big(|\mathring\psi_-^{\ing}|+|\partial_{\hat u}\mathring\psi_-^{\ing}|\Big)
       +\sup_{\hat\C_\out} \Big((1+\hat v)|\mathring\psi_+^{\out}|+(1+\hat v)^2|\partial_{\hat v}\mathring\psi_+^{\out}|+(1+\hat v)^2|\partial_{\hat v}(\hat v\mathring\psi_+^{\out})|\Big)\le\ve.
\end{aligned}
\end{equation}
\end{definition}

Let us now show that for suitably small $\ve>0$, an $\ve$-Schwarzschildean seed data set properly induces a well-defined characteristic initial data set for the Einstein--massless Dirac system, in the sense of Definition \ref{def:init-data-set}.

\begin{proposition}\label{prop:seed-data-construction}
   There exists an $\ve_\text{loc}>0$ satisfying the following property. Let $\mathcal S=(\mathring \psi^\out_+,\mathring \psi^\ing_-)$ be an $\ve$-Schwarzschildean seed data set on $\hat\C$, with $\ve\le\ve_\text{loc}$. Then, there exists a unique characteristic initial data set $(\mathring r,\mathring\Omega,\mathring\psi_+, \mathring\psi_-)$ on $\hat\C$ such that its maximal development $(\hat{\mathcal Q}_\mathrm{max},r,\Omega,\psi_+,\psi_-)$ has the following properties: 
   \begin{enumerate}
       \item $r(0,0)=\Lambda$,
       \item $m(0,0)=M_0$,
       \item $\nu=-1$ on $\hat{\uline{\C}}_\ing$,
       \item $\lambda= 1$ on $\hat\C_\out$.
   \end{enumerate}
\end{proposition}
We refer to characteristic initial data sets obtained from $\mathcal S$ in this manner as \emph{gauge-normalized}. 

\begin{remark} The seed data norm quantifies the distance from the Schwarzschild black hole with mass $M_0$ and bifurcation sphere area radius $100M_0$, which corresponds to the seed data $\mathcal{S}=(0,0)$. The event horizon for the evolution of $(0,0)$ is located at
\begin{equation}\label{eq:initial-horizon}
    \hat u_{\mathcal H^+,0}:=98M_0
\end{equation}
in the $(\hat u,\hat v)$ gauge determined by Proposition \ref{prop:seed-data-construction}.
\end{remark}

\begin{proof}[Proof of Proposition \ref{prop:seed-data-construction}] For $(\hat u,\hat v)\in\hat\C$, we set
\[
    \mathring r_\ing (\hat u):=\Lambda-\hat u\quad \text{on $\hat{\uline{\C}}_\ing$}, \quad \mathring r_\out(\hat v):=\Lambda+\hat v\quad \text{on $\C_\out$}, \quad \mathring r:=\begin{dcases} \mathring r_\ing &\text{on $\hat{\uline{\C}}_\ing$},\\ \mathring r_\out &\text{on $\C_\out$},
    \end{dcases}
\]
which implies by construction conditions 1., 3., 4. and $\mathring r>0$ as we have
\[
    \min_{\hat \C} \mathring r=\mathring r(U_*,0)=r(0,0)-U_*=\Lambda-U_*=1.5M_0+O(\ve)>0,
\]
for $\ve$ sufficiently small.
Next, on $\hat{\uline{\C}}_\ing$ we set
\[
	\mathring\Omega^2_\ing(\hat u):= 
	\Big(1-\frac{2M_0}{\Lambda}\Big)^{-1}
	\exp\Big(-2\int_0^{\hat u} \big( \mathring r_\ing \Im[\oline{\mathring\psi^\ing_-}\p_{\hat u}\mathring\psi^\ing_-]\big)(\hat u')\,d\hat u'\Big),
 \]
and
\[
    \mathring\psi_+^\ing(\hat u):=\frac{1}{\mathring\Omega_\ing(\hat u)} \Big(\mathring\Omega_\ing(0)\mathring\psi_+^\out(0)-\int^{\hat u}_0 \big(\mathring\Omega_\ing\mathring\psi^\ing_-\big)(\hat u')\,d\hat u'\Big)
\]
which are both well-defined by \eqref{eq:seed-data-norm}.

On the other hand, on $\hat\C_\out$ we note that $\mathring\Omega^2_\out$, $\mathring\nu_\out$, $\mathring m_\out$ and $\mathring\psi^\out_-$ satisfy the coupled system
\begin{equation}\label{eq:syst-seed-data}
\begin{aligned}
     \p_{\hat v} \mathring\Omega^2_\out &=\frac{2\mathring\Omega^2_\out}{\mathring r(1-\mathring\mu_\out)}\Im[\overline{\mathring\psi^\out_+}\p_{\hat v}\mathring\psi^\out_+],& \mathring\Omega^2_\out(0)&=\mathring\Omega^2_\ing(0)=\Big(1-\frac{2M_0}{\Lambda}\Big)^{-1},\\
     \p_{\hat v} \mathring\nu_\out  &= -\frac{2\mathring\Omega^2_\out }{\mathring r^2}\Big( \mathring m_\out-\mathring r\Im \big[\oline{\mathring\psi^\out_-}\mathring\psi^\out_+\big]\Big),& \mathring\nu_\out(0)&=\mathring\nu_\ing(0)=-1,\\
     \p_{\hat v}\mathring m_\out &= \Im[\overline{\mathring\psi^\out_-}\mathring\psi^\out_+] + \Im[\overline{\mathring\psi^\out_+}\p_{\hat v}\mathring\psi^\out_+],& \mathring m_\out(0)&=M_0,\\
     \p_{\hat v} \mathring\psi^\out_-  &=-\Big(\frac{\p_{\hat v}\mathring\nu_\out}{2\mathring \nu_\out}+\frac{1}{\mathring r}\Big)\mathring\psi^\out_- + \frac{1}{\mathring r^2} \mathring\psi^\out_+,& \mathring\psi^\out_-(0)&=\mathring\psi^\ing_-(0),
\end{aligned}
\end{equation}
for which the solution exists locally for $(\Omega^2, \nu, m, \psi_-)$ satisfying $\Omega^2, -\nu,1-\mu >0$.
We now prove that the system admits a global solution on $\hat\C_\out$.
To this end, we assume the following bootstrap assumptions 
\begin{equation}\label{eq:boot-seed-data}
	\frac12 \le 1-\mathring\mu_\out \le 2,\quad (1+\hat v)|\mathring\psi^\out_-|\le 2,
\end{equation}
on $\{0\}\times[0,\hat v_0]$, which are clearly satisfied for $\ve$ and $\hat v_0$ small enough.
Thus, from \eqref{eq:syst-seed-data}, we have
\[
	\mathring\Omega^2_\out(\hat v):=\Big(1-\frac{2M_0}{r_0}\Big)^{-1} \exp\Big(\int^{\hat v}_{0} \frac{2}{\mathring r_\out(1-\mathring\mu_\out)}\Im[\overline{\mathring\psi^\out_+}\p_{\hat v}\mathring\psi^\out_+]\,d\hat v'\Big),
\]
which satisfies $\frac12 \le \mathring\Omega^2_\out(\hat v)\le 2$, by \eqref{eq:seed-data-norm} and \eqref{eq:boot-seed-data}. Furthermore, combining again the $\ve$-Schwarzschildean condition and the bootstrap assumptions, we readily bound for $\hat v\in[0,+\infty)$
\[
    |\mathring m_\out(\hat v)-M_0|\les\int^{\hat v}_{0}\big(\ve (1+\hat v')^{-2}+\ve^2 (1+\hat v')^{-3}\big)\,d\hat v'\les \ve,
\]
which allows to improve and propagate $\frac12 \le 1-\mathring\mu_\out \le 2$ on $\hat\C_\out$. Using  \eqref{eq:syst-seed-data}, we integrate from the bifurcation sphere and bound
\[
    |\mathring\nu_\out(\hat v)+1|\les\int^{\hat v}_{0} \Big|\frac{2\mathring\Omega_\out^2}{\mathring r^2_\out}\Big|(\mathring m_\out+\mathring r_\out|\mathring\psi^\out_-||\mathring\psi^\out_+|)\,d\hat v'\les\frac12,
\]
implying that $-\frac32\le\mathring\nu_\out\le-\frac12$ on $\hat\C_\out$, and hence verifying the condition $\mathring\nu<0$ on $\hat\C$.
Finally, defining \[
\mathring\Phi_-^\out:=\mathring r_\out (-\mathring\nu_\out)^{1/2}\mathring\psi^\out_-\]
and combining \eqref{eq:syst-seed-data} with \eqref{eq:seed-data-norm} and the previous estimates, we readily bound
\[
    (1+\hat v)|\mathring\psi^\out_-(\hat v)|\les |\mathring\Phi_-^\out(\hat v)|\les \Lambda|\mathring\psi_-^\ing(0)|+\int^{\hat v}_0 |\p_{\hat v}\mathring\Phi_-^\out|\,d\hat v'\les \Lambda \ve + \int^{\hat v}_0 (1+\hat v')^{-1}|\mathring\psi_+^\out|\,d\hat v'\le \frac32,
\]
for $\ve$ small enough, improving the remaining bootstrap assumption. 

In this manner, one may assemble the functions built along the proof to a full initial data set, where $\mathring r$ and $\mathring\Omega^2$ are positive (and smooth on the bifurcation sphere), satisfying the constraint equations and $\mathring \nu<0$. Recalling \eqref{eq:Hawking-mass}, one verifies that condition 2. is also attained, concluding the proof.
\end{proof}

\subsubsection{Teleological gauges}\label{subsec:tel-gauges}
\begin{figure}
\centering
\begin{tikzpicture}[line cap=rect,inner sep=0pt, x=1cm, y=1cm, scale=1.5,font=\small]

\coordinate  (I)   at (0,2)      {};
\coordinate (II)   at (2,0)      {};
\coordinate  (III)   at (4,2)      {};
\coordinate   (IV)   at (2,4)      {};
\coordinate  (V)   at (1.15,0.85)      {};
\coordinate  (VI)   at (3.15,1.15)      {};
\coordinate  (VII)   at (2.3,2)      {};
\coordinate  (VIII)   at (3.15,2.85)      {};
\coordinate  (IX)   at (1.15,3.15)      {};

\path[name path=parab1, domain=0:4, samples=100, smooth, variable=\y]
    plot (\y,{-3/40*\y*\y+3/10*\y+2});

\fill[lightgray] (I) -- (II) -- (III) -- (IV);
%\fill[gray] (V) -- (II) -- (IX) -- (VII);
%\fill[gray] plot[domain=0:1, smooth] (\y,{-3/40*\y*\y+3/10*\y+2}) -- (IV) -- (III) --(I);

\draw[line width=1pt]  (IX) 
-- node[pos=0.5, below=2pt, sloped]    {$\underline{\mathcal{C}}^{\tau_f}_v,\underline{\mathcal{E}}^{\tau_f}$}
       (VII) -- node[pos=0.5, above=2pt, sloped]    {$\underline{H}^{\tau_f}_v,\underline{\mathcal{F}}^{\tau_f}$} (VI); %-- 
            %node[midway, below=2pt, sloped, font=\tiny]    {$u=u_0$}
\draw[line width=1pt]  (VIII) 
-- node[pos=0.4, above=2pt, sloped]    {$\mathcal{C}^{\tau_f}_u,\mathcal{E}_p^{\tau_f}$}
       (VII) -- node[pos=0.6, below=2pt, sloped]    {$H^{\tau_f}_u,\mathcal{F}^{\tau_f}$} (V);            %node[midway, below=2pt, sloped, font=\tiny]    {$u=u_0$}

\draw[line width=1pt] (I) -- node[pos=0.5, below=2pt, sloped]  {$v_{\tau_f}=0$} (II) -- node[pos=0.5, below=2pt, sloped]    {$u_{\tau_f}=0$} (III) -- (IV) -- (I);

\draw[line width=1pt,domain=0:4, smooth, variable=\y] plot ({2 + 3/40*\y*(4-\y)}, {\y});

\draw[-Stealth, line width=1pt] (1,3.75) node[left=2pt] {$\kappa_{\tau_f}=1$} -- (1.375,3.375);
%\draw[-Stealth, line width=1pt] (1,3.75) node[left] {$\partial_v r = 1-\frac{2m}{r}$} -- (2.28369,2.46631);
\draw[-Stealth, line width=1pt] (3.68178,3.06822) node[right=2pt] {$\gamma_{\tau_f}=1$} -- (3.30678,2.69322);

\node at ($(IV)+(0.5,0.25)$)          {$M_{\tau_f}=m(\Gamma(\tau_f),\; \bar r_{\tau_f}(\Gamma(\tau_f))=\Lambda$};
\node at ($(II)+(0,0.5)$)      {$\Gamma$};
\filldraw[black] (IV) circle (2pt);

\end{tikzpicture}
\caption{The Penrose diagram of the bootstrap set $\D_{\tau_f}$, illustrating the teleological gauges and the null hypersurfaces on which the relevant energies are defined.}
\label{fig:Penrose5-gauges}
\end{figure}
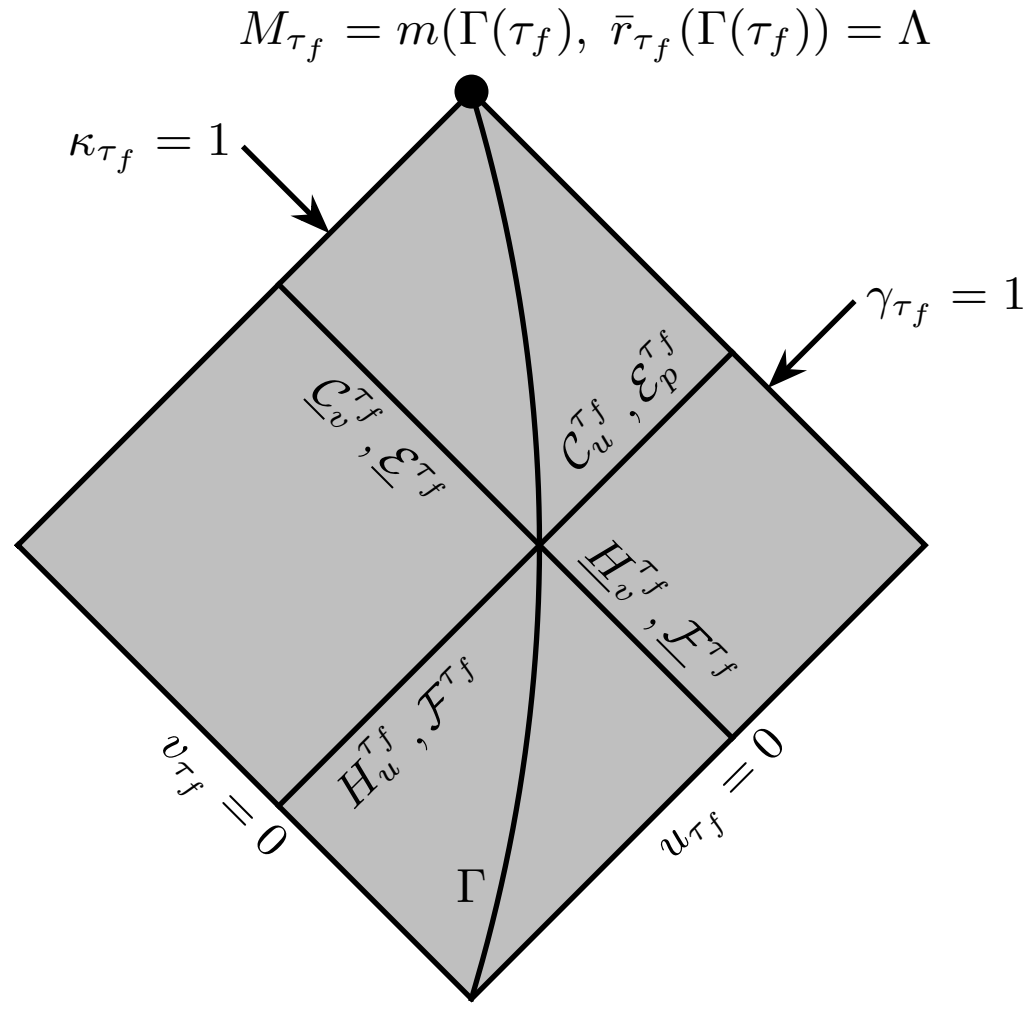

Let us now introduce the first geometric ingredients that formalize how and in which sense the dynamic evolution will converge to a Schwarzschild spacetime, that is the construction of a \textit{teleological gauge} and of a proper \textit{time function}. We follow here the constructions introduced in \cite{angelopoulos2026nonlinear,luk2019strong}.
The assumptions required for these constructions to be well-defined will be later formalized in the definition of the bootstrap set, see already Definition \ref{def:bootstrap-set}. We refer to Figure \ref{fig:Penrose5-gauges} for a schematic overview of the gauge choice and the notation.

Let $\mathcal S$ be an $\ve$-Schwarzschildean seed data set, with $\ve\le\ve_\text{loc}$, and $(\hat{\mathcal Q}_\mathrm{max},r,\Omega,\psi_+,\psi_-)$ its maximal development. We define the set
\begin{equation*}
    \Gamma:=\{r=\Lambda\},
\end{equation*}
which, for $\ve$ sufficiently small, is a timelike curve near $\hat{\mathcal C}$. In the course of the proof of the main theorem, we will verify that $\Gamma$ is an inextendible timelike curve in $\hat{\mathcal Q}_\mathrm{max}$. Assuming this for the moment, we parametrize $\Gamma$ by its proper time $\tau$, normalized so that $\Gamma(1)=(0,0)$. We denote the components of $\Gamma$ by $\Gamma(\tau)=(\Gamma^{\hat u}(\tau),\Gamma^{\hat v}(\tau))$ in the "initial data normalized" $(\hat u,\hat v)$ coordinates on $\hat{\mathcal Q}_\mathrm{max}$.

For $\tau_f\in[1,\infty)$ such that $[1,\tau_f]$ lies in the domain of definition $\Gamma$, we consider the region
\begin{equation*}
    \hat{\mathcal D}_{\tau_f}:=[0,\Gamma^{\hat u}(\tau_f)]\times[0,\Gamma^{\hat v}(\tau_f)].
\end{equation*} 
Assuming that $\hat\gamma>0$ and $\hat\kappa>0$ are both strictly positive on the final ingoing and outgoing cones in $\hat{\mathcal D}_{\tau_f}$ respectively, we may construct the following coordinates.
\begin{definition}[Teleologically normalized coordinates]\label{def:teleolog-coords} Define
$\mathfrak u_{\tau_f}:[0,\Gamma^{\hat u}(\tau_f)]\to\mathbb{R}$ and $\mathfrak v_{\tau_f}:[0,\Gamma^{\hat v}(\tau_f)]\to\mathbb{R}$ by 
\begin{equation}\label{def:gauges}
\begin{aligned}
    \mathfrak u_{\tau_f}(\hat u):= \int_0^{\hat u} \hat\gamma(\hat u',\Gamma^{\hat v}(\tau_f)) \,d\hat u',\quad
    \mathfrak v_{\tau_f}(\hat v):= \int_0^{\hat v} \hat\kappa(\Gamma^{\hat u}(\tau_f),\hat v')\,d\hat v',
\end{aligned}
\end{equation}
which assemble into the map
\begin{equation}\label{def:map-gauges}
\begin{aligned}
   \Upphi_{\tau_f}:\hat{\mathcal D}_{\tau_f}&\to \mathbb{R}^2\\
    (\hat u,\hat v) &\mapsto (\mathfrak u_{\tau_f}(\hat u),\mathfrak v_{\tau_f}(\hat v)),
\end{aligned}
\end{equation}
which is a diffeomorphism onto its image. We denote the image of $\Upphi_{\tau_f}$ by $\mathcal D_{\tau_f}$, which comes equipped with the double-null coordinates $(u_{\tau_f},v_{\tau_f})=\Upphi_{\tau_f}(\hat u,\hat v)$. We refer to these coordinates as \emph{teleologically normalized coordinates}, and we denote by $\hat\Upphi_{\tau_f}$ the inverse of $\Upphi_{\tau_f}$.
\end{definition}
\begin{remark} As illustrated in Figure \ref{fig:Penrose5-gauges}, this choice of coordinates ensures that $\kappa_{\tau_f}=1$ on the final outgoing cone and $\gamma_{\tau_f}=1$ on the final ingoing cone of $\mathcal D_{\tau_f}$.
\end{remark}
From Remark \ref{rem:gauge-freedom}, the solution $(r,\hat\Omega,\psi_+,\psi_-)$ in the coordinate system $(u_{\tau_f},v_{\tau_f})$, is expressed as $(r_{\tau_f},\Omega_{\tau_f}$, $\psi_{+,\tau_f},\psi_{-,\tau_f})$, where
\begin{equation*}
    r_{\tau_f}:= r\circ \hat\Upphi_{\tau_f},\quad \Omega_{\tau_f}:=\frac{1}{\mathfrak u_{\tau_f}'\circ\mathfrak u_{\tau_f}^{-1}}\frac{1}{\mathfrak v'_{\tau_f}\circ\mathfrak v^{-1}_{\tau_f}}\hat\Omega\circ\hat\Upphi_{\tau_f},\quad \psi_{\pm,\tau_f}:= \psi_\pm\circ\hat\Upphi_{\tau_f}.
\end{equation*} 
To simplify the notation, we omit the dependence $\tau_f$ and directly write $(u,v)$ and $(r,\Omega,\psi_{+},\psi_{-})$ when the time $\tau_f$ has been clearly fixed. 

In the $(u_{\tau_f},v_{\tau_f})$ coordinate system, we write the components of $\Gamma$ as $\Gamma(\tau)=(\Gamma^{u_{\tau_f}}(\tau),\Gamma^{v_{\tau_f}}(\tau))$. By a small abuse of notation, we implicitly define a continuous \textit{time function} $\tau$ on $\mathcal{D}_{\tau_f}$ by setting
\begin{equation}\label{def:tau}
    \tau(u_{\tau_f},v_{\tau_f}):=\begin{dcases}
        \tau : \Gamma^{u_{\tau_f}}(\tau)=u_{\tau_f} &\text{if $r(u_{\tau_f},v_{\tau_f})\ge\Lambda$,}\\
        \tau : \Gamma^{v_{\tau_f}}(\tau)=v_{\tau_f} &\text{if $r(u_{\tau_f},v_{\tau_f})<\Lambda$,}
    \end{dcases}
\end{equation}
whose behaviour qualitatively reflects that of Bondi time in the far region and of Eddington--Finkelstein time near the event horizon.
Next, we define the following null segments on $\mathcal{D}_{\tau_f}$:
\begin{align*}
    \uline{C}_v^{\tau_f}&:=\big([0,\Gamma^{u_{\tau_f}}(\tau_f)]\times\{v\}\big)\cap\{r\le\Lambda\},
    &C_u^{\tau_f}:=\big(\{u\}\times[0,\Gamma^{v_{\tau_f}}(\tau_f)]\big)\cap\{r\ge\Lambda\},\\
    \uline{H}_v^{\tau_f}&:=\big([0,\Gamma^{u_{\tau_f}}(\tau_f)]\times\{v\}\big)\cap\{r\ge\Lambda\},
    &H_u^{\tau_f}:=\big(\{u\}\times[0,\Gamma^{v_{\tau_f}}(\tau_f)]\big)\cap\{r\le\Lambda\},
\end{align*}
where again we will omit the dependence on $\tau_f$ when it is clear from the context.
We moreover introduce the following notation for the intersections of $\Gamma=\{r=\Lambda\}$ with the coordinate lines.
For any given $v$, the intersection $(\mathbb{R}\times\{v\})\cap\Gamma$ is either empty or consists of a single point, denoted by $(u^\Lambda(v),v)$. Likewise, for any given $u$, the intersection $(\{u\}\times\mathbb{R})\cap\Gamma$ is either empty or denoted by $(u,v^\Lambda(u))$.
We will also use the following shorthand notation:
\[
    \uline{C}^{\tau_f}(\tau):=\uline{C}_{\Gamma^{v_{\tau_f}}(\tau)}^{\tau_f}, \quad C^{\tau_f}(\tau):=C_{\Gamma^{u_{\tau_f}}(\tau)}^{\tau_f}.
\]

Let us now comment on this construction, giving some intuition for its meaning and emphasizing its teleological aspect in the overall picture.
For the seed data sets under consideration, $\Gamma$ will exist and remain timelike for all $\tau\in[1,+\infty)$. In particular, since $\tau\mapsto\Gamma^{\hat u}(\tau)$ is strictly increasing, the limit
\[
    \hat u_{\mathcal{H}^+}:= \lim_{\tau\to+\infty} \Gamma^{\hat u}(\tau)
\]
is well-defined and satisfies the strict inequality
\[
    \hat u_{\mathcal{H}^+}<U_*,
\]
see already Lemma \ref{lem:thetaU*-bis} below.
We therefore set
\[
\hat{\mathcal{D}}_\infty:=[0,\hat u_{\mathcal{H}^+})\times[0,+\infty),
\]
which represents the final asymptotic black hole exterior. Moreover, there exist surjective and strictly increasing $C^1$ functions $(u_\infty,v_\infty):[0,\hat u_{\mathcal{H}^+})\times[0,+\infty)\to [0,+\infty)\times[0,+\infty)$ such that $\p_{u_\infty} r\to-1$ and $\p_{v_\infty} r\to1$ at null infinity $\mathcal{I}^+$, where $\mathcal{D}_\infty$ denotes $\hat{\mathcal{D}}_\infty$ under this change of coordinates.
These "final teleological" coordinates may therefore be regarded as the \textit{eschatological gauge}. Their construction will be given in Section \ref{sec:control-gauge}.

Furthermore, the geometric estimates established below will imply that, for $R>2M_0+O(\ve)$, $\Gamma_R:=\{r=R\}$ is a timelike curve in $\mathcal{D}_{\tau_f}$ for every $\tau_f\in[1,+\infty]$ (or possibly empty).

\subsubsection{Anchored background solution}\label{sec:anch-solution}
Since we are aiming to converge to a Schwarzschild black hole with a final mass $M_\infty$ only known a posteriori, we define an anchored background solution used to compare and keep track of the evolution of the dynamic spacetime. Let us now detail such a construction.

\begin{definition}\label{def:anch-solution1}
    Assume $\tau_f<\infty$. Given a solution $(r_{\tau_f},\Omega_{\tau_f},\psi_{+,\tau_f},\psi_{-,\tau_f})$ on $\D_{\tau_f}$ with gauge conditions \eqref{def:gauges}, we define the \textit{$\tau_f$-anchored background solution} to be the Schwarzschild metric $(\bar r_{\tau_f},\oline{\Omega}_{\tau_f})$ with mass
\begin{equation}\label{def:anch-solution2}
    M_{\tau_f} :=m_{\tau_f}(\Gamma(\tau_f))
\end{equation} 
in Eddington--Finkelstein double-null form \eqref{eq:Schwarzschild-EF} which is uniquely determined, according to Lemma \ref{lem:existence-anchoring}, by the condition  
\begin{equation*}
        \bar r_{\tau_f}(\Gamma(\tau_f)) = \Lambda.
    \end{equation*} 
\end{definition}

Given the anchored background solution, from now on we denote with barred quantities (e.g. $\oline m_{\tau_f}=M_{\tau_f}$ and $\bar \lambda_{\tau_f}=-\bar\nu_{\tau_f}=\oline{\Omega^2}_{\tau_f}=1-2M_{\tau_f}\bar r^{-1}_{\tau_f}$) the geometric quantities of the anchored background solution and with tilde quantities the differences with the dynamical one, such as 
$\tilde r_{\tau_f}:= r_{\tau_f}-\bar r_{\tau_f}$, $\tilde m_{\tau_f}:= m_{\tau_f}-M_{\tau_f}$, or $\tilde\kappa_{\tau_f} := \kappa_{\tau_f}-1$.

Since in the proof of stability, $\tau_f\to \infty$, we need to extend the above definition to the case when $\tau_f=\infty$. Following the strategy of \cite[Section 3.3]{angelopoulos2026nonlinear}, instead of anchoring at "$\tau_f=\infty$", we anchor the final background solution at $\Gamma\cap\hat{\mathcal C}=\{(0,0)\}$. In particular, after proving in Lemma \ref{lem:final-anch-mass} that the following limit exists
\[
    \lim_{\tau_f\to+\infty}m_{\tau_f}=M_\infty,
\]
we define the \emph{$\infty$-anchored background solution} to be the unique Schwarzschild metric $(\bar r_{\infty},\bar\Omega^2_{\infty})$ with mass $M_\infty$ in Eddington--Finkelstein double-null form \eqref{eq:Schwarzschild-EF} with the following property (see already Proposition \ref{prop:lim-radius-bckground}):
\begin{equation*}
    \bar r_\infty(0,0)=\bar r_\star:=\lim_{\tau_f\to+\infty}\bar r_{\tau_f}(0,0),
\end{equation*} as depicted in Figure \ref{fig:Penrose6-stability}.
Moreover, \eqref{eq:lim-radius-bckground2} will imply that this background solution is indeed anchored "at $\tau_f=\infty$", meaning that
\begin{equation*}
    \lim_{\tau\to+\infty}\bar r_\infty(\Gamma(\tau))=\Lambda.
\end{equation*}

\subsubsection{Energy fluxes for the Dirac field}
We now introduce the final ingredients needed for the precise statement of our main result, namely the quantities used to measure and control the Dirac field along the null hypersurfaces introduced above.
\begin{definition}\label{def-of-energies}
    Let $(r_{\tau_f},\Omega_{\tau_f},\psi_{+,\tau_f},\psi_{-,\tau_f})$ be defined on $\mathcal D_{\tau_f}$ with teleologically normalized coordinates $(u_{\tau_f},v_{\tau_f})$, where $\tau_f\in[1,\infty]$.
    For $\tau,\tau'\in[1,\tau_f]$, $p\in[0,3)$
    and $(u,v)\in\mathcal D_{\tau_f}$, we define:
    \begin{enumerate}
    \item The energy flux to the horizon
    \[
    \uline{\mathcal E}^{\tau_f}(\tau):=\int_{\uline\C^{\tau_f}(\tau)} \Big( -\bar\nu_{\tau_f}|\psi_{+,\tau_f}|^2  
    -\bar\nu_{\tau_f}|\psi_{-,\tau_f}|^2 + \frac{|\partial_{u_{\tau_f}}\psi_{-,\tau_f}|^2}{- \bar\nu_{\tau_f}}\Big)\,du_{\tau_f};
    \] 
    \item The $r^p$-weighted flux to null infinity
    \[
    \mathcal E_p^{\tau_f}(\tau):= \int_{C^{\tau_f}(\tau)}\big( 
    r^p_{\tau_f} |\p_{v_{\tau_f}}(r_{\tau_f}\psi_{+,\tau_f})|^2+c_{p_*}r^{p_*+2}_{\tau_f}|\p_{v_{\tau_f}}\psi_{+,\tau_f}|^2 + c_{p_*}r_{\tau_f}^{p_*}|\psi_{+,\tau_f}|^2 %+c_{p_*}r_{\tau_f}^{p_*}
    +|\psi_{-,\tau_f}|^2\big)\,dv_{\tau_f},
    \]
    where $p_*:=p$ if $p\in[0,1)$ and vanishes otherwise, and $c_{p_*}:=(p_*-1)^2$; 
    \item The flux along ingoing cones in the far region
    \[   
    \underline{\mathcal F}^{\tau_f}(v,\tau'):= \int_{\underline H_v^{\tau_f}\cap\{\tau\ge\tau'\} }
    \big( r^{-2}_{\tau_f}|\psi_{+,\tau_f}|^2 + r^2_{\tau_f}|\psi_{-,\tau_f}|^2 + r^2_{\tau_f}|\p_{u_{\tau_f}}\psi_{-,\tau_f}|^2\big)\,du_{\tau_f};
    \] 
    \item The flux along outgoing cones in the near region
    \[   
    \mathcal F^{\tau_f}(u,\tau'):= \int_{H_u^{\tau_f}\cap\{\tau\ge\tau'\}}\big(|\psi_{+,\tau_f}|^2 + |\psi_{-,\tau_f}|^2 + |\p_{v_{\tau_f}}\psi_{+,\tau_f}|^2\big)\,dv_{\tau_f}.
    \] 
\end{enumerate}
\end{definition}
\begin{remark}
The choice of derivative terms in the above fluxes exploits the transport structure of the reduced Dirac equations \eqref{eq:partial_u-psi+}--\eqref{eq:partial_v-psi-}. In particular, the null derivatives not appearing explicitly in the definitions can be recovered from the Dirac equations in terms of the spinor components and the geometric coefficients. Thus, the fluxes only involve the derivatives needed to control the Dirac field along the corresponding null hypersurfaces.
\end{remark}

\begin{remark} The dependence of $\mathcal{E}^{\tau_f}_p$ on $p_*$ and $c_{p_*}$ is analogous to \cite[Remark 3.6]{angelopoulos2026nonlinear} and we report here the reason for the sake of completeness. For $p\ge1$, the quantity $r^{p+2}_{\tau_f}|\p_{v_{\tau_f}}\psi_{+,\tau_f}|^2 + r_{\tau_f}^{p}|\psi_{+,\tau_f}|^2$ fails to be integrable in $v_{\tau_f}$ uniformly in $\tau_f$, and thus cannot be controlled directly by the energy estimates. The coefficient $c_{p_*}$ is needed because the Hardy inequalities relating this quantity to $r^p_{\tau_f}|\p_{v_{\tau_f}}\wtilde\Phi_{+,\tau_f}|^2$ only hold for $p<1$ and degenerate as $p\nearrow1$ (see already Lemma \ref{lem:Hardy-est1}).
\end{remark}

\subsection{Precise statement of the main result and the bootstrap set}\label{sec:statement-thm}
We are now in position to give the precise formulation of Theorem \ref{thm:Schwarz-stability1} stated in the Introduction.

\begin{theorem}[Stability of Schwarzschild for Einstein--massless Dirac in spherical symmetry]\label{thm:Schwarz-stability2} Let $M_0>0$ and let $0<\delta<0.01$. There exist $\ve_{\mathrm{stab}}(M_0,\delta)>0$ and a constant $C(M_0,\delta)$ (which is implicit in the notation $\les$ below) such that any $\ve$-Schwarzschildean seed data set $\mathcal{S}$, with $0<\ve\le\ve_{\mathrm{stab}}$, satisfies the following properties:
\begin{enumerate}
    \item \emph{Existence of a black hole region:} Denoting by $(\hat{\mathcal{Q}}_{\text{max}},r,\Omega^2,\psi_+,\psi_-)$ the maximal development of $\mathcal{S}$, there exist $\hat u_{\mathcal{H}^+}\in(0,U_*)$ and $M_\infty>0$ such that 
    \[
        [0,\hat u_{\mathcal{H}^+}]\times[0,+\infty)\subset\hat{\mathcal{Q}}_\maxx,\quad \lim_{\hat v \to+\infty}r(\hat u,\hat v)=+\infty\text{ for every $\hat u\in [0,\hat u_{\mathcal{H}^+})$,}\quad \lim_{\hat v \to+\infty}r(\hat u_{\mathcal{H}^+},\hat v)= 2M_\infty.
    \]    
    Therefore, $[0,\hat u_{\mathcal{H}^+})\times\{\hat v=\infty\}$ may be regarded as future null infinity $\mathcal{I}^+$, there exists a non-empty black hole region
    \[
        \mathcal{BH}:=\hat{\mathcal{Q}}_\maxx\setminus J^-(\mathcal{I}^+)
    \]
    and $\mathcal{H}^+:=\p J^-(\mathcal{I}^+)=\{\hat u_{\mathcal{H}^+}\}\times[0,+\infty)$ is the event horizon. Moreover, future null infinity is complete in the sense of Christodoulou \cite{christodoulou1999global}. Moreover, there exist $C^2$ double-null coordinates $(u_\infty,v_\infty)$ on the domain of outer communication $[0,\hat u_{\mathcal{H}^+})\times[0,+\infty)$ such that $u_\infty$ is Bondi normalized, meaning that the event horizon $\mathcal{H}^+$ can be formally regarded as $\{u_\infty=+\infty\}$ and $\p_{u_\infty} r\to-1$ along any outgoing cone in the domain of outer communication.
    \item \emph{Orbital stability:} There exists an $\infty$-anchored Schwarzschild solution $\bar r_\infty$ in the $(u_\infty,v_\infty)$ coordinates with mass
    \begin{equation}\label{eq:orb-stab1}
        |M_\infty-M_0|\les\ve.
    \end{equation}
    Relative to this background solution, the energy of the Dirac field is bounded by its initial value
    \begin{equation}\label{eq:orb-stab2}
        \sup_{\tau\in[1,+\infty)} \big(\uline{\mathcal{E}}^\infty(\tau)+\mathcal{E}_{3-\delta}^\infty(\tau)\big)\les \uline{\mathcal{E}}^\infty(1)+\mathcal{E}_{3-\delta}^\infty(1)\les \ve^2,
    \end{equation}
    and the Dirac field is pointwise bounded in $C^1$ in the domain of outer communication in terms of its initial values
    \begin{equation}\label{eq:orb-stab3}
    \begin{aligned}
        \sup_{J^-(\mathcal{I}^+)}\Big(r|\psi_+|&+r|\psi_-|+r^2|\p_{\hat v}\psi_+|+r^2|\p_{\hat v}(r\psi_+)|+r\Big|\frac{\p_{\hat u}\psi_-}{-\hat \nu}\Big|\Big)
        \\&\les\sup_{J^-(\mathcal{I}^+)\cap \hat\C}\Big(r|\psi_+|+r|\psi_-|+r^2|\p_{\hat v}\psi_+|+r^2|\p_{\hat v}(r\psi_+)|+r\Big|\frac{\p_{\hat u}\psi_-}{-\hat \nu}\Big|\Big)\les\ve.
    \end{aligned}
    \end{equation}
    \item \emph{Asymptotic stability:} The geometry decays towards the $\infty$-anchored Schwarzschild solution in the following sense: 
    \begin{equation}
        |\gamma_\infty-1| \les \ve^2 r^{-2} \tau^{-3+\delta},
    \end{equation}
    in $\mathcal{D}_\infty\cap\{r\ge\Lambda\}$,
    \begin{equation}
        \Big|\frac{\nu_\infty}{\bar\nu_{\infty}}-1\Big|\les\ve^2\tau^{-1+\delta}
    \end{equation}
    in $\mathcal{D}_\infty$, and
    \begin{equation}
        |r -\bar r_\infty|\les\ve^2\tau^{-2+\delta},\quad |\lambda_\infty -\bar\lambda_\infty|\les\ve^2\tau^{-2+\delta},\quad |\kappa_\infty-1|\les\ve^2\tau^{-3+\delta},\quad |m-M_\infty|\les \ve^2\tau^{-3+\delta},
        \end{equation}
        \begin{equation}
        \Big|\p_{v_\infty}\log\Omega_\infty-\frac{M_\infty}{\bar r^{2}_\infty}\Big|\les \ve^2\tau^{-2+\delta},\quad
    \Big|\p_{u_\infty}\log\Omega_\infty+\frac{M_\infty}{\bar r^{2}_\infty}\Big|\les \ve^2 r^{-2}\tau^{-3/4+\delta},
    \end{equation}
    on $\D_\infty$ up to and including the event horizon $\mathcal{H}^+$. The Dirac field decays to zero in energy norm
    \begin{equation}
        \uline{\mathcal{E}}^\infty(\tau)+\mathcal{E}_{p}^\infty(\tau)\les \ve^2\tau^{-3+\delta+p},
    \end{equation}
    for every $\tau\in[1,+\infty)$ and $p\in[0,3-\delta]$, and it decays to zero pointwise
    \begin{equation}
        \bar r^{1/2}_\infty|\psi_+|\les \ve\tau^{-3/2+\delta/2},\quad \bar r_\infty|\psi_-|\les \ve\tau^{-3/2+\delta/2},
    \end{equation}
    on $\D_\infty$ up to and including the event horizon $\mathcal{H}^+$. 
\end{enumerate}
\end{theorem}
\begin{figure}
\centering
\begin{tikzpicture}[line cap=rect,inner sep=0pt, x=1cm, y=1cm, scale=1.25,font=\small]

\coordinate   (U)   at (-1,3)      {};
\coordinate  (I)   at (0,2)      {};
\coordinate (II)   at (2,0)      {};
\coordinate  (III)   at (4,2)      {};
\coordinate   (IV)   at (2,4)      {};

\path[name path=parab1, domain=0:3, samples=100, smooth, variable=\x]
    plot ({\x},{1/6*\x*\x - 1/6*\x + 3});

\fill[lightgray] plot[domain=-1:2, smooth] (\x,{1/6*(\x+1)*(\x+1)-1/6*(\x+1)+3}) -- (III) -- (II) --(U);

\draw[line width=1pt]  (U) -- (I) 
-- node[pos=0.4, below=2pt, sloped]    {$\{\hat v=0\}=\{v_\infty=0\}$}
       (II) -- node[pos=0.55, below=2pt, sloped]    {$\{\hat u=0\}=\{u_\infty=0\}$} (III); %-- 
            %node[midway, below=2pt, sloped, font=\tiny]    {$u=u_0$}
\draw[line width=1pt,dashed] (III) --
            node[pos=0.5, above=2pt, sloped]    {$\mathcal{I}^+=\{\hat v=+\infty\}$} 
            node[pos=0.37, below=2pt, sloped]    {$=\{v_\infty=+\infty\}$}
      (IV);
\draw[line width=1pt] (IV)-- node[pos=0.5, above=2pt, sloped]    {$\mathcal{H}^+=\{\hat u = \hat u_{\mathcal{H}^+}\}$} node[pos=0.37, below=2pt, sloped]    {$=\{u_\infty=+\infty\}$}(I);
\draw[line width=1pt,domain=0:4, smooth, variable=\y] plot ({2 + 3/40*\y*(4-\y)}, {\y});

\node at (-0.35,2.7)          {$\mathcal{BH}$};
\node at ($(IV)+(0,0.25)$)          {$i^+$};
%\node at ($(III)+(0.25,0)$)          {$i^0$}; 
\node at ($(II)+(-0.4,-0.1)$)          {$\bar r= \bar r_\star$};  
\filldraw[black] (U) circle (2pt);
\filldraw[fill=white, draw=black] (IV) circle (2pt);
\filldraw[black] (II) circle (2pt);
\node at ($(II)+(0,0.75)$)      {$\Gamma$};
\end{tikzpicture}
\caption{The Penrose diagram of a Schwarzschild black hole, where we draw the foliations $\uline{\mathcal{C}}(\tau)$ and $\mathcal{C}(\tau)$ parametrized by the "time" variable $\tau$, with the shaded region representing the exterior domain we perturb.}
\label{fig:Penrose6-stability}
\end{figure}
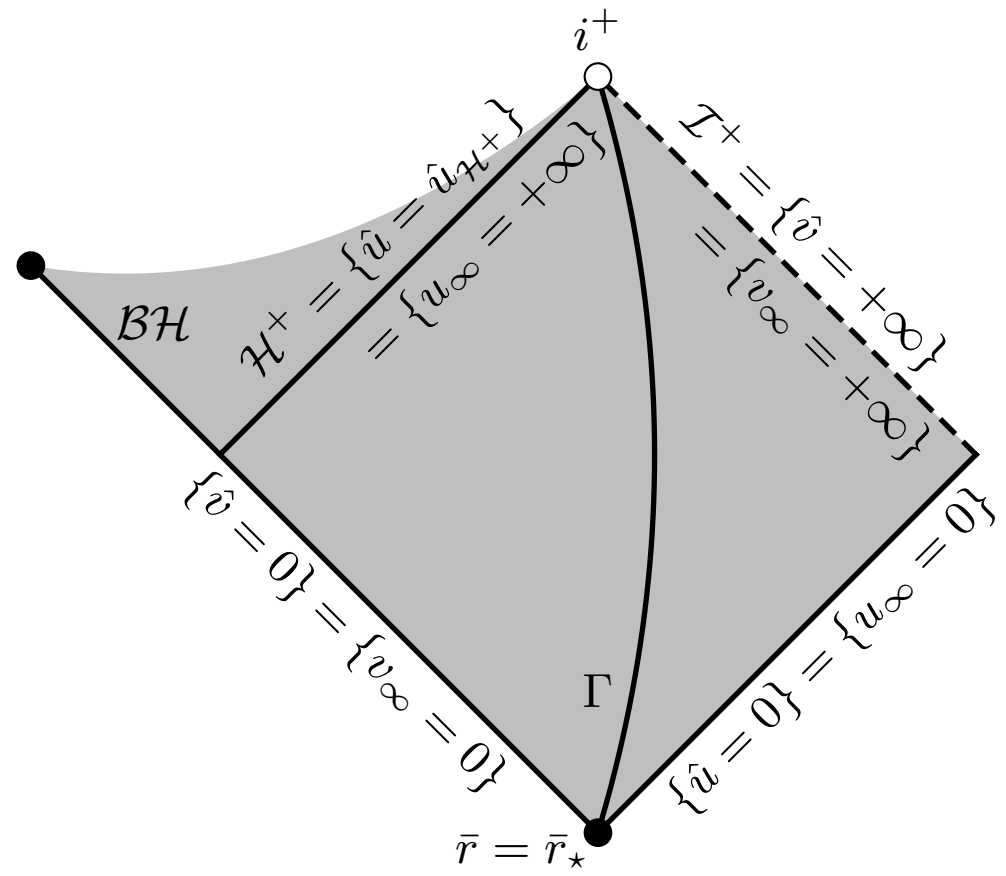
\begin{remark}
The decay rates obtained for the geometric quantities are very similar to those in \cite{angelopoulos2026nonlinear}, which reflects the fact that our bootstrap scheme is adapted from their work. The decay rates obtained for the Dirac field, on the other hand, are not expected to be sharp. As already mentioned, unlike \cite{ma2022sharp} we do not commute the Dirac equation with higher-order differential operators, since this is not needed to close the bootstrap argument, and we therefore do not recover the sharp decay rates established there.
\end{remark}
The proof of Theorem \ref{thm:Schwarz-stability2} is given at the end of Section \ref{sec:proof-conclusion}. Part 1. is shown in Proposition \ref{prop:thm-part1}, while the eschatological gauge $(u_\infty,v_\infty)$ is constructed in Section \ref{sec:control-gauge}. On the other hand, Parts 2.--3. rely on the estimates established in Sections \ref{sec:improvement-geometry} to \ref{sec:energy-Dirac}.

The core of the stability proof is a \textit{continuity argument} implemented through a bootstrap scheme. We therefore begin by introducing the bootstrap set and stating the precise bootstrap assumptions underlying the argument, which are adapted to the Dirac setting from \cite{angelopoulos2026nonlinear}.

\begin{definition}[Bootstrap set]\label{def:bootstrap-set} Let $A\ge1$, $0<\ve\le\ve_{\text{loc}}$. Consider $\mathcal{S}$ an $\ve$-Schwarzschildean seed data set and $(\hat{\mathcal Q}_\mathrm{max},r,\Omega,\psi_+,\psi_-)$ its maximal development in the initial gauge $(\hat u, \hat v)$ determined by Lemma \ref{prop:seed-data-construction}. Then we define the \textit{bootstrap set} $\mathfrak{B}(\mathcal{S},\ve,A)$ as the set of times $\tau_f\in[1,+\infty)$ such that:
\begin{enumerate}
    \item There exists a timelike curve $\Gamma:[1,\tau_f]\to \hat{\mathcal{Q}}_\text{max}$, which is the unique smooth solution to 
    \[
        \frac{d}{d\tau}(\Gamma^{\hat u},\Gamma^{\hat v})= \Big( \frac{\sqrt{1-\hat\mu}}{-2\hat \nu},\frac{\sqrt{1-\hat\mu}}{2\hat\lambda}\Big)\Big|_{\Gamma(\tau)},\quad \Gamma(1)=(0,0).
    \]
    By the choice of the tangent vector to $\Gamma$ and the fact that $r(\Gamma(1))=r(0,0)=\Lambda$, we have $r(\Gamma(\tau))=\Lambda$ for all $\tau\in[1,\tau_f]$.
    Note that, by global hyperbolicity of $\hat{\mathcal{Q}}_{\text{max}}$, it then follows
    \[
        \hat{\mathcal{D}}_{\tau_f}:=[0,\Gamma^{\hat u}(\tau_f)]\times[0,\Gamma^{\hat v}(\tau_f)]\subset\hat{\mathcal{Q}}_{\text{max}};
    \]
    \item The quantities $\hat\gamma$ and $\hat\kappa$ are strictly positive on $[0,\Gamma^{\hat u}(\tau_f)]\times\{\Gamma^{\hat v}(\tau_f)\}$ and $\{\Gamma^{\hat u}(\tau_f)\}\times[0,\Gamma^{\hat v}(\tau_f)]$, respectively. Hence, the teleologically normalized coordinates $(u_{\tau_f},v_{\tau_f})$ in Definition \ref{def:teleolog-coords} are well-defined on $\hat{\mathcal{D}}_{\tau_f}$.
    \item Given the $\tau_f$-anchored Schwarzschild solution $(\bar r_{\tau_f},\oline{\Omega}_{\tau_f})$ on $\mathcal{D}_{\tau_f}$, as in Definition \ref{def:anch-solution1}, the following set of bootstrap assumptions hold: for the geometry 
\begin{align}
    \label{eq:boot-nu}           \Big|\frac{\nu_{\tau_f}}{\bar\nu_{\tau_f}}-1\Big| \le& A^3 \ve^{3/2}\tau^{-1+\delta},\\
    \label{eq:boot-r}
    |\tilde r_{\tau_f}|   \le& A^2 \ve^{3/2} \tau^{-2+\delta},\\
    \label{eq:boot-varpi}|\tilde m_{\tau_f}|  \le& A\ve^{3/2} \tau^{-3+\delta},
\end{align}
on $\D_{\tau_f}$, and for the Dirac field
\begin{align}
    \label{eq:boot-ing-energy}
    \underline{\mathcal E}^{\tau_f}(\tau)&\le A\ve^2\tau^{-3+\delta},\\
    \label{eq:boot-outg-energy}
   \mathcal E_p^{\tau_f}(\tau)&\le A\ve^2\tau^{-3+p+\delta},\\
   \label{eq:boot-ing-flux}
   \underline{\mathcal F}^{\tau_f}(v,\tau)&\le A\ve^2\tau^{-3+\delta},\\
    \label{eq:boot-outg-flux}
    \mathcal F^{\tau_f}(u,\tau)&\le A\ve^2\tau^{-3+\delta},
\end{align} 
for every $\tau\in [1,\tau_f]$, $(u,v)\in\mathcal D_{\tau_f}$, and $p\in [0,3-\delta]$.
\end{enumerate}
\end{definition}

\begin{remark} The bootstrap assumptions on the energy quantities are guided by the non-sharp decay estimates obtained in \cite[Proposition 5.15]{ma2022sharp}. Since these estimates do not capture the optimal decay even on a fixed Schwarzschild background, they are expected to be robust under perturbation, which makes them a natural choice in the nonlinear setting considered here.
\end{remark}

Our aim is therefore to show that $\mathfrak{B}(\mathcal{S},\ve,A)$ is non-empty, open, and closed (see already Proposition \ref{prop:bootstrap-set}), thereby yielding a global solution with the expected decays. The key step is to establish openness by improving the bootstrap assumptions. In particular, the estimates for \eqref{eq:boot-nu}--\eqref{eq:boot-varpi} are obtained in Section \ref{sec:closing-geometry}, while those for \eqref{eq:boot-ing-energy}--\eqref{eq:boot-outg-flux} are proved in Section \ref{sec:improvement-Dirac}.

\section{Estimates for the geometry}\label{sec:improvement-geometry}
The goal of this section is improving the bootstrap assumptions \eqref{eq:boot-nu}--\eqref{eq:boot-varpi}. To this end, in Section \ref{sec:initial-estimates} we establish useful bounds on the main geometric tilde quantities, such as $\tilde\kappa_{\tau_f}, \tilde\gamma_{\tau_f}, \tilde\lambda_{\tau_f}$. In Section \ref{subsec:prop-Gamma-tau}, some important properties on $\Gamma$ and $\tau$ are derived. Finally, in Section \ref{sec:closing-geometry} we prove the following result.
\begin{proposition}[Improving the bootstrap assumptions for the geometry] \label{prop:bootstrap-geometry}
    There exist positive constants $A_0$ and $\ve_0$ depending only on $M_0$ and $\delta$, such that if $A= A_0$, $\ve\le \ve_0$, $\tau_f\in  \mathfrak B(\mathcal S,\ve,A)$, then the estimates 
    \begin{align*}
    \Big|\frac{\nu_{\tau_f}}{\bar\nu_{\tau_f}}-1\Big| &\le \tfrac 12 A^3\ve^{3/2}\tau^{-1+\delta},\\
    |\tilde r_{\tau_f}|   &\le \tfrac 12 A^2 \ve^{3/2} \tau^{-2+\delta}, \\
    |\tilde m_{\tau_f}|  &\le \tfrac 12 A\ve^{3/2} \tau^{-3+\delta}
    \end{align*} hold on $\mathcal D_{\tau_f}$. 
\end{proposition}

From Section \ref{sec:initial-estimates} to Section \ref{sec:rp-est}, we will work on a bootstrap domain $\mathcal{D}_{\tau_f}$ with fixed time $\tau_f$. Thus, to make the exposition simpler, we will omit the subscript $\tau_f$ on $(u_{\tau_f},v_{\tau_f})$, the solution $(r_{\tau_f},\Omega_{\tau_f},\psi_{+,\tau_f},\psi_{-,\tau_f})$, the background Schwarzschild solution $ (\bar r_{\tau_f},\oline{\Omega}_{\tau_f})$, and the energies. In order to keep track of the constant $A$ appearing in the bootstrap assumptions \eqref{eq:boot-nu}--\eqref{eq:boot-outg-flux}, we will use the notations $a\lesssim b$, $a\gtrsim b$, and $a\sim b$, meaning that the bounds do not depend on $u$, $v$, $A$, $\ve$, $\tau$, or $\tau_f$. Finally, we will also often absorb powers of the bootstrap constant $A$ by trading with powers of $\ve$, e.g. $A\ve^{3/2}\les\ve$ for $\ve$ sufficiently small. Note that the constant $A$ will be fixed at the end of Section \ref{sec:improvement-Dirac} depending only on $M_0$ and $\delta$.

\subsection{Initial estimates}\label{sec:initial-estimates}
Since the bootstrap setup is similar to that of \cite{angelopoulos2026nonlinear}, the bounds proved in this section follow a similar strategy. However, the specific features of the Einstein--Dirac system, and in particular the lack of a definite sign for the energy-momentum tensor, require some small modifications to the argument (see for instance Remark \ref{rk:lambda>0}). We therefore include the details of the proofs.

\begin{lemma}\label{lem:basic-consequences} For any $A\ge 1$, $\ve$ sufficiently small, $\tau_f\in\mathfrak B$, it holds that 
    \begin{align}
        \label{eq:rough-bound-varpi} \tfrac 12M  \le m&\le 2M,\\
        \label{eq:nu-ratio} \tfrac 12\le \frac{\nu}{\bar \nu}&\le 2,
    \end{align}
    in $\mathcal D_{\tau_f}$ and 
    \begin{align}
        \label{eq:mu-lower-bound}  
        1-\mu&\ge \tfrac 34,\\
        \label{eq:nu-1}     
        -2\le \nu &\le-\tfrac 12,
    \end{align}
    in $\mathcal D_{\tau_f}\cap\{r\ge \Lambda \}$.
\end{lemma}

\begin{proof}
The bounds \eqref{eq:rough-bound-varpi} and \eqref{eq:nu-ratio} follow directly from \eqref{eq:boot-varpi} and \eqref{eq:boot-nu}, respectively. Regarding \eqref{eq:mu-lower-bound}, we notice that  
\begin{equation*} 
1-\mu \geq 1 - \frac{4M}{\Lambda} \geq \frac 34, 
\end{equation*}
in $\mathcal{D}_{\tau_f}\cap\{r\ge\Lambda\}$. 
Finally, to show \eqref{eq:nu-1}, we use \eqref{eq:boot-nu} together with the fact that $-\bar \nu =  1-2M/\bar r$ and thus $ -1 \leq \bar \nu \leq -\frac 34$ in $\mathcal D_{\tau_f}\cap\{r\ge \Lambda \}$.
\end{proof}

We now use the bootstrap assumptions for the energy decay of the Dirac components to derive bounds for $\tilde\kappa$ and $\tilde\gamma$. 

\begin{lemma}[Estimate for $\tilde\kappa$]\label{lem:kappa-bound}
For any $A\ge 1$, $\ve$ sufficiently small, $\tau_f\in\mathfrak B$, it holds that
    \begin{align}\label{eq:kappa-1}   
        |\tilde\kappa|&\les A\ve^2 \tau^{-3+\delta},\\
        \label{eq:kappa-2}   
       \tfrac 12 &\leq \kappa \leq 2.
    \end{align}
    in $\mathcal D_{\tau_f}$.
\end{lemma}
\begin{proof} 
From \eqref{eq:transport-kappa} and $\kappa = \tilde \kappa +1$, we have 
\begin{equation*}
    \partial_u \tilde \kappa = -2r(\tilde \kappa +1) \Im[\overline{\psi_-} \partial_u\psi_-].
\end{equation*}
Integrating backwards in $u$ from $\{\Gamma^u(\tau_f)\}\times[0,\Gamma^v(\tau_f)]$, where $\tilde\kappa=0$ by our gauge choice, we obtain
\begin{equation*}
    \tilde \kappa(u,v) = -1 + \exp \Big( \int_{\underline C_v \cap \{ u'\ge u\}} 2r \Im[\overline{\psi_-} \partial_u\psi_-]\, du'\Big).
\end{equation*}
Therefore, when $r(u,v)\le\Lambda$, we have
\begin{align*}
   |\tilde \kappa (u,v) | 
   &\lesssim
   \Big| -1 + \exp \Big( \int^{\Gamma^u(\tau_f)}_{u} \big(-\bar\nu|\psi_-|^2 + \frac{|\partial_u \psi_-|^2}{-\bar\nu}\big) \, du'\Big) \Big| 
   \les |-1+\exp(\underline{\mathcal{E}}(\tau))|
   \les A\varepsilon^2\tau^{-3+\delta},
\end{align*}
by the bootstrap assumption \eqref{eq:boot-ing-energy}.
On the other hand, when $r(u,v)>\Lambda$,
\begin{align*}
|\tilde \kappa (u,v)| &\les
   \Big| -1 + 
   \exp \Big( \int_{\uline C_v} 
   \big(-\bar\nu|\psi_-|^2 + \frac{|\partial_u \psi_-|^2 }{-\bar\nu} \big) du'
   +
   \int^{u^\Lambda(v)}_u
   \big(r^2|\psi_-|^2 + r^2|\partial_u \psi_-|^2\big) du'\Big)  \Big| \\&
   \lesssim 
   \big|-1 + \exp\big(\underline{\mathcal{E}}(\tau) + \underline{\mathcal{F}}(v,\tau(u,v))\big)\big|
   \lesssim 
   A\varepsilon^2\tau^{-3+\delta},
\end{align*}
where we used \eqref{eq:boot-ing-energy} and \eqref{eq:boot-ing-flux} to conclude. Finally, \eqref{eq:kappa-2} readily follows from \eqref{eq:kappa-1}.
\end{proof}
\begin{remark}\label{rk:lambda>0}
From \eqref{eq:mu-lower-bound} and \eqref{eq:kappa-2}, and recalling $\lambda = \kappa (1-\mu)$, it follows 
\begin{align}
    \label{eq:lambda-upper-bound} \lambda&\le 2 \quad \text{in $\mathcal D_{\tau_f}$},\\
    \label{eq:lambda-lower-bound}   
    \lambda &\geq \tfrac 38 \quad \text{in $\mathcal D_{\tau_f}\cap\{r\ge \Lambda \}$}.
\end{align}
However, since the Raychaudhuri equation \eqref{eq:Raych-v} has no definite sign, \eqref{eq:lambda-lower-bound} does not by itself imply that $\lambda>0$ throughout $\mathcal{D}_{\tau_f}$, unlike in the scalar field case. 

Similarly, from the identity $\Omega^2= \kappa(-\nu)>0$, we readily deduce
\begin{equation}\label{eq:Omega-bound}
    |\wtilde \Omega^2| = \Big|\bar\nu \Big(\tilde \kappa + \kappa \Big(\frac{\nu}{\bar\nu} - 1\Big)\Big)\Big|\les (-\bar\nu)A^3\ve^{3/2}\tau^{-1+\delta},
\end{equation}
in $\mathcal D_{\tau_f}$.
\end{remark}

\begin{lemma}[Estimate for $\tilde\gamma$]\label{lem:gamma-bound} For any $A\ge 1$, $\ve$ sufficiently small, $\tau_f\in\mathfrak B$, it holds that    
\begin{equation}\label{eq:gamma-1}    
    |\tilde \gamma|\les A\ve^2  r^{-2}\tau^{-3+\delta},
\end{equation}    
    in $\mathcal D_{\tau_f}\cap\{r\ge\Lambda \}$.
\end{lemma}
\begin{proof}
From \eqref{eq:transport-gamma} and $\gamma = \tilde \gamma + 1$, we have  
\begin{equation*}
    \partial_v \tilde \gamma = \frac{2 (\tilde\gamma + 1)}{r\lambda} \Im[\overline{\psi_+}\partial_v \psi_+].
\end{equation*}
Integrating backwards in $v$ from $[0,\Gamma^u(\tau_f)]\times\{\Gamma^v(\tau_f)\}$, where $\tilde\gamma=0$ by our gauge choice, and then multiplying by $r^2$, we obtain for $(u,v) \in \mathcal D_{\tau_f} \cap \{ r \geq \Lambda\}$
\begin{align*}
 |   r^2 \tilde \gamma(u,v)|
 &\lesssim r^2 \int^{\Gamma^v(\tau_f)}_{v}  \frac{  |\tilde \gamma| + 1 }{r} |\psi_+| |\partial_v\psi| \,dv' 
 \lesssim   \int^{\Gamma^v(\tau_f)}_{v} r ( |\tilde \gamma| + 1 )|\psi_+| |\partial_v\psi_+| \, dv' 
 \\
 &\lesssim \int^{\Gamma^v(\tau_f)}_{v} ( |\tilde \gamma| + 1 )(|\psi_+|^2 + r^2  |\partial_v\psi_+|^2) \, dv',
\end{align*}
where we used \eqref{eq:lambda-lower-bound} to move the $r$ weight into the integral. Applying Grönwall's inequality gives
\begin{equation*}
    | r^2 \tilde \gamma(u,v) |\lesssim \mathcal E_0(\tau) \exp \big(\mathcal E_0(\tau)  \big)\les A\ve^2 \tau^{-3+\delta},
\end{equation*}
thanks to the bootstrap assumption \eqref{eq:boot-outg-energy}.
\end{proof}

We now derive Taylor expansions for $\widetilde{1-\mu}$, $\partial_u\tilde r$ and $\partial_v\tilde r$ in terms of tilde quantities. These bounds will be essential to improve the bootstrap assumptions in Section \ref{sec:closing-geometry}.

\begin{lemma}\label{lem:mu}  For any $A\ge 1$, $\ve$ sufficiently small, $\tau_f\in\mathfrak B$, it holds that 
\begin{align}
  \label{eq:mu-1}   
  \left|\widetilde{1-\mu} - \frac{2M\tilde r}{\bar r^2}\right|&\les \frac{|\tilde r|^2}{\bar r^3}+\frac{|\tilde m|}{\bar r},\\
     |\widetilde{1-\mu} |&\les A^2\ve^{3/2}\tau^{-2+\delta},\label{eq:mu-2}
\end{align}
in $\mathcal D_{\tau_f}$.
\end{lemma}
\begin{proof}
From the bootstrap assumption \eqref{eq:boot-r}, we readily obtain the Taylor expansion 
    \begin{equation}
   \label{eq:r-Taylor}     \frac{1}{r^\eta}=\frac{1}{\bar r^\eta}-\frac{\eta\tilde r}{\bar r^{\eta+1}} +O\Big(\frac{|\tilde r|^2}{\bar r^{\eta+2}}\Big)
    \end{equation}
for any $\eta>0$. We therefore compute
\begin{align*}
    \widetilde{1-\mu} &= -\frac{2m}{r}+\frac{2M}{\bar r} = -\frac{2\tilde m}{r}-2M\left(\frac{1}{r}-\frac{1}{\bar r}\right)
    = -\frac{2\tilde m}{r} + \frac{2M\tilde r}{\bar r^2}+O\Big(\frac{|\tilde r|^2}{\bar r ^3}\Big),
\end{align*}
which yields \eqref{eq:mu-1}. Equation \eqref{eq:mu-2} then follows using \eqref{eq:boot-r} and \eqref{eq:boot-varpi}.
\end{proof}

\begin{lemma}\label{lem:nu-lambda-expansion}
   For any $A\ge 1$, $\ve$ sufficiently small, $\tau_f\in\mathfrak B$, we have the expansions
    \begin{align}
    \label{eq:nu-expansion}    
    \partial_u\tilde r &= -\frac{2M\gamma}{\bar r^2}\tilde r+\Err_u \quad\text{in }\mathcal D_{\tau_f}\cap\{r\ge\Lambda  \},\\
    \label{eq:lambda-expansion}    \partial_v\tilde r&= \frac{2M\kappa}{\bar r^2}\tilde r +\Err_v\quad\text{in }\mathcal D_{\tau_f},
    \end{align}
    where the error terms satisfy the bounds
    \begin{align}
      \label{eq:nu-expansion-error}  |\Err_u|&\les A\ve^{3/2}\bar r^{-1}\tau^{-3+\delta},\\
     \label{eq:lambda-expansion-error}   |\Err_v|&\les A\ve^{3/2}\tau^{-3+\delta}.
    \end{align}
\end{lemma}
\begin{proof} Regarding $\partial_u\tilde r$, we compute by \eqref{eq:mu-1},
\begin{equation*}
    \partial_u\tilde r = \tilde\nu 
    = -\gamma(1-\mu)+\bar\gamma(\overline{1-\mu})
    =-\tilde\gamma(\overline{1-\mu}) - \gamma(\widetilde{1-\mu})
    =-\tilde\gamma(\overline{1-\mu})- \gamma\Big[\frac{2M\tilde r}{\bar r^2}+O\Big(\frac{|\tilde r|^2}{\bar r^3}+\frac{|\tilde m|}{\bar r}\Big)\Big],
\end{equation*}
which implies \eqref{eq:nu-expansion}, where the error term satisfies
\begin{equation*}
    |\Err_u|\les \frac{|\tilde r|^2}{\bar r^3}+\frac{|\tilde m|}{\bar r}+|\tilde\gamma|\les A\ve^{3/2}\bar r^{-1}\tau^{-3+\delta},
\end{equation*}
by the bootstrap assumptions and \eqref{eq:gamma-1}. On the other hand, we have
\begin{equation*}
      \partial_v\tilde r=\tilde\lambda = \kappa(1-\mu)-\bar\kappa(\overline{1-\mu})= \tilde\kappa(\overline{1-\mu})+\kappa(\widetilde{1-\mu})
      = \tilde\kappa (\overline{1-\mu}) + \kappa \Big[\frac{2M\tilde r}{\bar r^2}+O\Big(\frac{|\tilde r|^2}{\bar r^3}+\frac{|\tilde m|}{\bar r}\Big)\Big],
  \end{equation*}
  yielding \eqref{eq:lambda-expansion}, where the error term satisfies
  \begin{equation*}
      |\Err_v|\les \frac{|\tilde r|^2}{\bar r^3}+\frac{|\tilde m|}{\bar r}+|\tilde\kappa|\les A\ve^{3/2}\tau^{-3+\delta}
  \end{equation*}
  by the bootstrap assumptions and \eqref{eq:kappa-1}.
\end{proof}

\subsection{Properties of \texorpdfstring{$\Gamma$}{Gamma} and \texorpdfstring{$\tau$}{tau}}\label{subsec:prop-Gamma-tau}

In this section, we derive some basic properties of the timelike curve $\Gamma=\{r=\Lambda \}$ and the associated function $\tau(u,v)$ defined by \eqref{def:tau}, that will be useful throughout the paper. Recall that $\tau$ was also used to denote the proper time parametrization of $\Gamma$, defined via the condition  
\begin{equation}
 \label{def:proper-time} g(\dot\Gamma,\dot\Gamma)=-1,
\end{equation}
where $\dot{}$ denotes $\frac{d}{d\tau}$. Moreover, we recall that $\Gamma^u(\tau)$ and $\Gamma^v(\tau)$ are the components of $\Gamma$ under this parametrization in the teleological coordinates $(u,v)$.

\begin{lemma}\label{lem:gamma-slope} For any $A\ge 1$, $\ve$ sufficiently small, $\tau_f\in\mathfrak B$, it holds that 
\begin{equation}\label{eq:Gamma-slope}
    \dot\Gamma^u(\tau)\sim\dot\Gamma^v(\tau)\sim 1
\end{equation}
for $\tau\in[1,\tau_f]$.
\end{lemma}
\begin{proof} Since $r$ is constant along $\Gamma$, we have $\nu\dot\Gamma^u+\lambda\dot\Gamma^v=0$ on $\Gamma$. Bounds \eqref{eq:mu-lower-bound} and \eqref{eq:kappa-2} yield $\lambda|_{\Gamma}=\kappa(1-\mu)|_{\Gamma}\sim 1$, which implies together with \eqref{eq:nu-1} that $\dot\Gamma^u\sim\dot\Gamma^v$. The proper time condition \eqref{def:proper-time} can be written as 
\begin{equation*}
    \frac{4\nu\lambda}{1-\mu}\dot\Gamma^u\dot\Gamma^v=-1,
\end{equation*}
yielding $\dot\Gamma^u\dot\Gamma^v\sim 1$ by \eqref{eq:mu-lower-bound}. Thus, \eqref{eq:Gamma-slope} readily follows.
\end{proof} 

The following lemma describes the relation between $\tau$ and the coordinates $(u,v)$ in the far and near-horizon regions.
 
\begin{lemma}\label{lem:tau-properties} For any $A\ge 1$, $\ve$ sufficiently small, $\tau_f\in\mathfrak B$, it holds that 
 \begin{equation}\label{eq:tau-difference-u}
        \tau(u_1,v)-\tau(u_2,v)\sim u_1-u_2,
    \end{equation}
    for $(u_1,v),(u_2,v)\in\mathcal D_{\tau_f}\cap\{r\ge\Lambda\}$, and 
\begin{equation}\label{eq:tau-difference-v}
        \tau(u,v_1)-\tau(u,v_2)\sim v_1-v_2,
    \end{equation}
for $(u,v_1),(u,v_2)\in\mathcal D_{\tau_f}\cap\{r\le \Lambda \}$. Moreover, for any $\eta>1$ and $1\le \tau_1\le\tau_2\le \tau_f$, it holds that
\begin{align}
 \label{eq:Hbarv-integral}    \int_{\underline H{}_v\cap\{\tau_1\le\tau\le \tau_2\}}\tau^{-\eta}\,du &\les_\eta \tau_1^{-\eta+1},\\
    \label{eq:Hu-integral}      \int_{H_u\cap\{\tau_1\le\tau\le \tau_2\}}\tau^{-\eta}\,dv &\les_\eta \tau_1^{-\eta+1}.
\end{align}
\end{lemma}
\begin{proof}
    It is a straightforward consequence of \eqref{eq:Gamma-slope}, the fundamental theorem of calculus, and the change of variables formula.
\end{proof}

We conclude this section by showing that the $v$ and $\hat v$ coordinates are comparable.

\begin{lemma}\label{lem:v-est}
    For $\tau_f\in \mathfrak B$, $\ve$ sufficiently small depending on $A$, and $\hat v_1,\hat v_2\in [0,\Gamma^{\hat v}(\tau_f)]$,
    it holds that 
    \begin{equation}
        \mathfrak v(\hat v_2)-\mathfrak v(\hat v_1)\sim \hat v_2-\hat v_1.\label{eq:v-est}
    \end{equation}
    Moreover,  on $\hat{\mathcal D}_{\tau_f}$ it holds that
    \begin{equation}
        \hat\kappa\sim 1.\label{eq:hat-kappa-est}
    \end{equation}
\end{lemma}
\begin{comment}
    you cannot prove an analogue for u coord as the derivative of the inverse would give you a nu on C_in, for which you do not have a statement -c1<nu<-c0<0 (recall: nu here is in double-null)
\end{comment}
\begin{proof} The proof is analogous to \cite[Lemma 5.7]{angelopoulos2026nonlinear}. Recalling the definition of the map $\hat v\mapsto \mathfrak v(\hat v)$ in \eqref{def:gauges}, the derivative of the inverse map $v\mapsto \mathfrak v^{-1}(v)$ is given by $\lambda(0,v)$. Therefore, \eqref{eq:v-est} follows from \eqref{eq:lambda-upper-bound}, \eqref{eq:lambda-lower-bound}, and the fundamental theorem of calculus. 

On the other hand, \eqref{eq:hat-kappa-est} follows from the identity
\begin{equation*}
    \hat\kappa(\hat u,\hat v)=\frac{\kappa(\mathfrak u(\hat u),\mathfrak v(\hat v))}{\lambda(0,\mathfrak v(\hat v))}
\end{equation*}
and \eqref{eq:kappa-2}. 
\end{proof}

\subsection{Improving the bootstrap assumptions for the geometry}\label{sec:closing-geometry}
We are now in position to improve the bootstrap assumptions on the geometry. 
Let us start with $\tilde r$, for which we follow the same strategy as in \cite[Lemma 5.12]{angelopoulos2026nonlinear}, bounding the difference of the area radius function in three steps: along $\Gamma$, and in the regions $\{r\le \Lambda\}$ and $\{r\ge \Lambda\}$.

\begin{proposition}\label{prop:r-estimate} For any $A\ge 1$, $\ve$ sufficiently small, $\tau_f\in\mathfrak B$, it holds that
    \begin{align}
    \label{eq:tilde-r-aux-3}|\tilde r|&\les A\ve^{3/2}\tau^{-2+\delta},\\
    \label{eq:lambda-tilde} |\tilde \lambda |&\lesssim A\varepsilon^{3/2} \tau^{-2+\delta}
    \end{align}
    in $\mathcal D_{\tau_f}$.
\end{proposition}
\begin{proof}
\textsc{Estimate along $\Gamma$}: Using the relation $\nu\dot\Gamma^u+\lambda\dot\Gamma^v=0$ along $\Gamma$ by Definition \ref{def:bootstrap-set}, we obtain 
\begin{equation*}
    \frac{d}{d\tau}\tilde r(\Gamma(\tau))= \big(\nu\dot\Gamma^u+\lambda\dot\Gamma^v\big) -\big(\bar\nu\dot\Gamma^u+\bar\lambda\dot\Gamma^v\big)=-\bar\nu\dot\Gamma^u-\bar\lambda\dot\Gamma^v= \left(\frac{\bar\lambda}{\lambda}\nu-\bar\nu\right)\dot\Gamma^u= -(\overline{1-\mu})\left(\frac{\bar\kappa}{\kappa}-\frac{\bar\gamma}{\gamma}\right)\gamma\dot\Gamma^u.
\end{equation*}
Using \eqref{eq:kappa-1} and \eqref{eq:gamma-1}, we then bound
\begin{equation*}
    \left|\frac{\bar\kappa}{\kappa}-\frac{\bar\gamma}{\gamma}\right|\les A\ve^2\tau^{-3+\delta}
\end{equation*}
along $\Gamma$, and therefore find
    \begin{equation}\label{eq:tilde-r-on-Gamma}
        \big|\tilde r|_\Gamma\big|\les A\ve^2\tau^{-2+\delta},
    \end{equation}
by \eqref{eq:Gamma-slope}, the fundamental theorem of calculus and the anchoring condition $\tilde r(\Gamma(\tau_f))=0$.

\textsc{Estimate for $r\le\Lambda $}: We view \eqref{eq:lambda-expansion} as an ODE for $\tilde r$ in $v$. For $(u,v)\in \mathcal D_{\tau_f}\cap\{r\le\Lambda\}$, we use an integrating factor to write
\begin{align}
   \nonumber  \tilde r(u,v)&=\exp\Big(-\int_{v}^{v^\Lambda(u)}\frac{2M\kappa}{\bar r^2}\,dv'\Big)\Big[\tilde r(u,v^\Lambda(u))-\int_{v}^{v^\Lambda(u)}\exp\Big(\int_{v'}^{v^\Lambda(u)}\frac{2M\kappa}{\bar r^2}\,dv''\Big)\Err_v\,dv'\Big]\\
  \label{eq:tilde-r-near-formula}   &= \exp\Big(-\int_{v}^{v^\Lambda(u)}\frac{2M\kappa}{\bar r^2}\,dv'\Big)\tilde r(u,v^\Lambda(u))- \int_{v}^{v^\Lambda(u)}\exp\Big(-\int_{v}^{v'}\frac{2M\kappa}{\bar r^2}\,dv''\Big)\Err_v\,dv'.
\end{align}
Therefore \eqref{eq:lambda-expansion-error}, \eqref{eq:Hu-integral}, \eqref{eq:tilde-r-on-Gamma} and \eqref{eq:tilde-r-near-formula} yield
\begin{equation}\label{eq:tilde-r-aux-1}
    |\tilde r(u,v)|\les A\ve^2\tau^{-2+\delta}(u,v)+ A\ve^{3/2}\tau^{-2+\delta}(u,v)\les A\ve^{3/2}\tau^{-2+\delta}(u,v),
\end{equation}
after noting that the exponentials are bounded by $1$, as the integrands are non-negative.

\textsc{Estimate for $r\ge \Lambda $}:  We view \eqref{eq:nu-expansion} as an ODE for $\tilde r$ in $u$. For $(u,v)\in \mathcal D_{\tau_f}\cap\{r\geq\Lambda \}$, we use an integrating factor to write
\begin{align}
    \nonumber
    \tilde r(u,v)
    &= \exp\Big(\int_{u}^{u^\Lambda(v)}\frac{2M\gamma}{\bar r^2}\,du'\Big)\Big[\tilde r(u^\Lambda(v),v) -\int_{u}^{u^\Lambda(v)}\exp\Big(-\int_{u'}^{u^\Lambda(v)}\frac{2M\gamma}{\bar r^2} \,du''\Big)\Err_u\,du'\Big]
    \\
    \label{eq:tilde-r-far-formula}
    &= \exp\Big(\int_{u}^{u^\Lambda(v)}\frac{2M\gamma}{\bar r^2}\,du'\Big)\tilde r(u^\Lambda(v),v)- \int_{u}^{u^\Lambda(v)}\exp\Big(\int_{u}^{u'}\frac{2M\gamma}{\bar r^2}\,du''\Big)\Err_u\,du'.
\end{align}
Noting that the integrating factor is bounded by \eqref{eq:nu-1}, as 
\[
    \int_{u}^{u^\Lambda(v)}\frac{2M\gamma}{\bar r^2}\,du'\les \int_{\uline H_v} \bar r^{-2}\,du'\les 1,
\]
we obtain by \eqref{eq:nu-expansion-error}, \eqref{eq:Hbarv-integral}, \eqref{eq:tilde-r-on-Gamma} and \eqref{eq:tilde-r-far-formula} that
\begin{equation}\label{eq:tilde-r-aux-2}
    |\tilde r(u,v)|\les A\ve^2\tau^{-2+\delta}(u,v)+ A\ve^{3/2}\tau^{-2+\delta}(u,v)\les A\ve^{3/2}\tau^{-2+\delta}(u,v).
\end{equation}
Combining \eqref{eq:tilde-r-on-Gamma}, \eqref{eq:tilde-r-aux-1} and \eqref{eq:tilde-r-aux-2} yields \eqref{eq:tilde-r-aux-3} as desired. 
Finally, \eqref{eq:lambda-tilde} readily follows from \eqref{eq:lambda-expansion} and \eqref{eq:lambda-expansion-error}.
\end{proof}

\begin{remark} The favorable sign of the coefficient in \eqref{eq:tilde-r-near-formula} could be exploited by integrating backwards in $v$, and it is a manifestation of the red-shift effect in Schwarzschild. A similar strategy is used by \cite[Lemma 8.19]{luk2019strong} for the subextremal Reissner--Nordström case, and even in \cite[Lemma 5.12]{angelopoulos2026nonlinear} for the extremal case, where the usual red-shift degenerates.
\end{remark}

We now proceed with the estimate for $\tilde\nu$, treating again separately the regions $\{r\le\Lambda\}$ and $\{r\ge\Lambda\}$.
\begin{proposition}\label{prop:nu-estimate}
   For any $A\ge 1$, $\ve$ sufficiently small, $\tau_f\in\mathfrak B$, it holds that
    \begin{equation}
        \label{eq:nu-aux-1} \left|\frac{\nu}{\bar \nu}-1\right|\les A^2\ve^{3/2}\tau^{-1+\delta}
    \end{equation}
    on $\mathcal D_{\tau_f}$.
\end{proposition}
\begin{proof}
    \textsc{Estimate for $r\ge\Lambda $}: From \eqref{eq:nu-expansion} and the bootstrap assumptions, we readily bound
    \begin{equation}\label{eq:nu-aux-2}
        |\tilde\nu|
        \les \bar r^{-2}|\tilde r|+ A\ve^{3/2}\bar r^{-1}\tau^{-3+\delta}
        \les A\ve^{3/2}\bar r^{-1} \tau^{-2+\delta},
    \end{equation}
    which implies \eqref{eq:nu-aux-1} for $r\ge\Lambda$, after dividing by $|\bar\nu|\sim 1$.
    
      \textsc{Estimate for $r\le\Lambda $}: 
      From the wave equation \eqref{eq:wave-eq-r}, we obtain
    \begin{align}
        \nonumber
        \Big|  \partial_v{\log\left(\frac{\nu}{\bar\nu}\right)} \Big|&=\Big|\frac{\p_v\nu}{\nu}-\frac{\p_v\bar\nu}{\bar\nu}\Big| = \Big|\frac{2\kappa m}{r^2}+\frac{4\pi r}{\nu} T_{uv}-\frac{2M}{\bar r^2}\Big|
        \le 
        2 \Big|\frac{\kappa m}{r^2}-\frac{M}{\bar r^2}\Big| + \Big|\frac{4\pi r}{\nu} T_{uv}\Big| \\ \nonumber &\le \frac{2 m}{r^2}|\tilde\kappa|+ \frac{2|\tilde m|}{r^2} + 2M\Big|\frac{1}{r^2}-\frac{1}{\bar r^2}\Big| + \Big|\frac{4\pi r}{\nu} T_{uv}\Big|\les |\tilde\kappa| + |\tilde m|+ |\tilde r| + |\psi_+||\psi_-|
         %\\&\les A^2 \varepsilon^{3/2}\tau^{-2+\delta} + A\ve^2 \tau^{-3+\delta} + |\psi_+|^2+|\psi_-|^2
        \\&
        \label{eq:estimate-partialv-nu}
        \lesssim A^2 \varepsilon^{ 3/2} \tau^{-2+\delta} + |\psi_+|^2+|\psi_-|^2,
      \end{align}
      by \eqref{eq:boot-varpi}, \eqref{eq:kappa-1} and \eqref{eq:tilde-r-aux-3}. Since
      \begin{equation*}
         \left|\left. {\log\left(\frac{\nu}{\bar\nu}\right)}\right|_{\Gamma}\right|\les \big|\tilde\nu|_\Gamma\big|\les A^2\ve^{3/2}\tau^{-2+\delta},
      \end{equation*}
       by \eqref{eq:nu-aux-2}, we conclude by integrating \eqref{eq:estimate-partialv-nu} backwards in $v$ from $\Gamma$, together with \eqref{eq:Hu-integral} and the fact that
       \[
            \int^{v^\Lambda(u)}_v  \big(|\psi_+|^2+|\psi_-|^2\big)\,dv' \le \mathcal{F}(u,\tau)\le A\ve^2\tau^{-3+\delta}.
       \]
\end{proof}

Finally, we bound $\tilde m$, noting that the power of $\ve$ in \eqref{eq:varpi-aux-0} below is strictly better than in the bootstrap assumption \eqref{eq:boot-varpi}.

\begin{proposition}\label{prop:varpi-estimate}
For any $A\ge 1$, $\ve$ sufficiently small, $\tau_f\in\mathfrak B$, it holds that
   \begin{align}
   \label{eq:varpi-aux-0}
       |\tilde m|&\les A\ve^2\tau^{-3+\delta},
   \end{align}
   on $\mathcal D_{\tau_f}$.
\end{proposition}
\begin{proof}
We begin by focusing on the final ingoing cone $[0,\Gamma^u(\tau_f)]\times\{\Gamma^v(\tau_f)\}$. Recalling here \eqref{eq:transport-u-varpi},
\begin{equation*}
    \partial_u\tilde m = \frac{\nu}{\kappa^{1/2}} \Im[\overline{\psi_-}\psi_+] - r^2(1-\mu) \Im[\overline{\psi_-}\partial_u\psi_-],
\end{equation*}
we obtain by integrating backwards in $u$ from $(\Gamma^u(\tau_f),\Gamma^v(\tau_f))$, the anchoring condition \eqref{def:anch-solution2}, \eqref{eq:Hbarv-integral} and the bootstrap assumptions, that
\begin{equation}\label{eq:varpi-aux-3}
    |\tilde m(u,\Gamma^v(\tau_f))|\les \uline{\mathcal{F}}(\Gamma^v(\tau_f),\tau(u,\Gamma^v(\tau_f)))\les A\ve^2 \tau(u,\Gamma^v(\tau_f))^{-3+\delta},
\end{equation}
for every $u\le \Gamma^u(\tau_f)$. Next, for $v<\Gamma^v(\tau_f)$, we consider \eqref{eq:transport-v-varpi}, that we write here
\begin{equation*}
    \partial_v\tilde m = \frac{\lambda}{\kappa^{1/2}} \Im[\overline{\psi_-}\psi_+] + \frac{1}{\kappa} \Im[\overline{\psi_+}\partial_v\psi_+],
\end{equation*}
and we integrate it backwards in $v$. Therefore, in the region $\D_{\tau_f}\cap\{r\ge\Lambda\}$, we obtain
\begin{align}
    \nonumber |\tilde m(u,v)|
    &\les |\tilde m(u,\Gamma^v(\tau_f))| +
    \int^{\Gamma^v(\tau_f)}_v \big(|\psi_-|^2 +|\psi_+|^2+|\partial_v\psi_+|^2\big)\,dv'  
    \\&\les A\ve^2\tau(u,\Gamma^v(\tau_f))^{-3+\delta} +
    \mathcal E_0(\tau(u,v))
    \les A\ve^2\tau(u,v)^{-3+\delta},
    \label{eq:varpi-aux-5}
    \end{align}
where we combined the previous geometric estimates with the bootstrap assumptions. 

Next, when $(u,v)\in \D_{\tau_f}\cap\{r\le\Lambda\}$, we further integrate backwards in $v$ from the point $(u,v^\Lambda(u))$ and use \eqref{eq:varpi-aux-5} to estimate
    \begin{align}
        \nonumber
        |\tilde m(u,v)|&\les A\ve^2\tau(u,v^\Lambda(u))^{-3+\delta} + \int^{v^\Lambda(u)}_v \big(|\psi_+|^2 + |\psi_-|^2 + |\partial_v\psi_+|^2\big)\,dv'\\
        \label{eq:varpi-aux-6}
        &\les
        A\ve^2\tau(u,v^\Lambda(u))^{-3+\delta}
        +
        \mathcal F(u,\tau(u,v))\les A\ve^2\tau(u,v)^{-3+\delta}.
    \end{align}
Combining \eqref{eq:varpi-aux-5} and \eqref{eq:varpi-aux-6}, \eqref{eq:varpi-aux-0} finally follows.
\end{proof}

We now put all of the estimates proved in this section together and improve the bootstrap assumptions for the geometry, \eqref{eq:boot-nu}--\eqref{eq:boot-varpi}.

\begin{proof}[Proof of Proposition \ref{prop:bootstrap-geometry}]
Summarizing Propositions \ref{prop:r-estimate} and \ref{prop:nu-estimate}, we have found on $\mathcal D_{\tau_f}$
\begin{align*}
    |\tilde r|&\les A\ve^{3/2}\tau^{-2+\delta},\\
    \left|\frac{\nu}{\bar\nu}-1\right|&\les A^2\ve^{3/2}\tau^{-1+\delta},
\end{align*}
where the implicit constant does not depend on $A$. Therefore, by choosing $A$ sufficiently large, we obtain
\begin{align*}
    |\tilde r|&\le \tfrac 12 A^2\ve^{3/2}\tau^{-2+\delta},\\
    \left|\frac{\nu}{\bar\nu}-1\right|&\le \tfrac 12 A^3\ve^{3/2}\tau^{-1+\delta}.
\end{align*}
Once $A$ is fixed, by Proposition \ref{prop:varpi-estimate} we obtain
\[
    |\tilde m|\les A\ve^2\tau^{-3+\delta},
\]
which then implies, for $\ve$ sufficiently small,
\[
    |\tilde m|\le \tfrac 12 A\ve^{3/2}\tau^{-3+\delta}
\]
in $\mathcal D_{\tau_f}$, concluding the proof. 
\end{proof}

\section{Preliminary estimates for the coupled system}\label{sec:preliminary-estimates}
The coupling between the Dirac field and the geometry presents additional difficulties compared to the case of the scalar wave equation. 
For instance, as discussed in Remark \ref{rk:Dirac-Phi-Schwarzschild}, exploiting the symmetric structure of the Dirac system on a Schwarzschild background is essential to prove energy estimates. Thus, from \eqref{eq:wave-Phi_+}--\eqref{eq:wave-Phi_-}, we infer that suitable bounds for $\p_u\log\Omega$ and $\p_v\log\Omega$ are crucial in the perturbed setting.

We therefore proceed in four steps. We first establish pointwise estimates for the Dirac field in Lemma \ref{lem:pointwise_estimates_dirac1}, which allow us to control the derivatives of $\log\Omega$ in Lemma \ref{lem:Omega-expansions}. Next, in Lemma \ref{lem:pointwise_estimates_dirac2} we derive pointwise estimates for the derivatives of the Dirac field. Lastly, we employ these bounds to control in Lemma \ref{lem:pointwise-est-Einstein-eq} the derivatives of the geometric quantities (such as $\kappa$, $\gamma$, and $ m$), which arise as error terms in the vector field estimates of Section \ref{sec:energy-Dirac}. We conclude with a final lemma, which follows as a direct consequence of these bounds.

To this end, we adapt the strategy used in \cite[Section 7.2]{angelopoulos2026nonlinear}, where analogous pointwise decay estimates are obtained for the scalar field on a perturbed extremal Reissner–Nordström black hole.

\begin{lemma}\label{lem:pointwise_estimates_dirac1}
For any $A\ge1$, $\ve$ sufficiently small, $\tau_f\in\mathfrak B$, we have the pointwise decay estimates
\begin{align}
\label{est:psi+_bound}
    \bar r^{1/2}|\psi_+|&\les A^{1/2}\ve  \tau^{-3/2+\delta/2},\\
\label{est:psi+_bound-bis}
    \bar r|\psi_+|&\les A^{1/2}\ve  \tau^{-1+\delta/2},\\
\label{est:psi-_bound}
    \bar r|\psi_-|&\les A^{1/2}\ve  \tau^{-3/2+\delta/2},\\
\label{est:T-uv_bound}
    \bar r^{7/2}\frac{|T_{uv}|}{-\nu}&\les A\ve^2  \tau^{-3+\delta},
\end{align}
on $\mathcal D_{\tau_f}$.
\end{lemma}
\begin{proof} 
\textsc{Proof of \eqref{est:psi+_bound}--\eqref{est:psi-_bound} for $r\ge\Lambda$}:  Let $\chi_1=\chi_1(r)$ be a smooth cutoff satisfying $\chi_1(r)=1$ for $r\ge 2\Lambda$ and $\chi_1(r)=0$ for $r\le\Lambda$. By the same argument as in Lemma \ref{lem:gamma-slope}, the set $\{r=2\Lambda\}$ is a timelike curve in $\mathcal D_{\tau_f}$. Moreover, for $(u,v)\in\mathcal D_{\tau_f}\cap\{r\ge2\Lambda\}$, the segments $\{u\}\times[v^\Lambda(u),v]$ and $[u,u^\Lambda(v)]\times\{v\}$ are entirely contained in $\mathcal D_{\tau_f}$. Therefore, for $\beta\in\{1,2\}$ and $(u,v)\in\D_{\tau_f}\cap\{r\ge2\Lambda\}$, we compute
\begin{align*}
    r^{\beta}|\psi_+|^2 &= r^{\beta-2}|r\psi_+|^2 =\Big| \int_{v^\Lambda(u)}^v\big(\chi_1'\lambda r^{\beta}|\psi_+|^2 + \chi_1(\beta-2)\lambda r^{\beta-3}|r\psi_+|^2 + 2\chi_1 r^{\beta-2}\Re[\overline{r\psi_+}\p_v(r\psi_+)]\big)\,dv'\Big|\\
    &\les \int_{v^\Lambda(u)}^v\big(|\beta-2| r^{\beta-1}|\psi_+|^2 + r^{\beta-1}|\psi_+||\partial_v(r\psi_+)|\big)\,dv'
    \les \big(\mathcal E_0(\tau(u,v))\big)^{1/2}\big(\mathcal E_{2\beta-2}(\tau(u,v))\big)^{1/2} \\&\les A\ve^2\tau^{-4+\beta+\delta}(u,v),
   \end{align*}
   \begin{align*}
   r^2|\psi_-|^2 &= \Big| \int^{u^\Lambda(v)}_u\big(\chi_1'\nu r^2|\psi_-|^2 + 2\chi_1 r \nu |\psi_-|^2 + 2\chi_1 r^2\Re[\overline{\psi_-}\partial_u\psi_-]\big)\,du'\Big|\\
   &\les 
   \int^{u^\Lambda(v)}_u\big( r^2|\psi_-|^2 + r^2|\partial_u\psi_-|^2\big)\,du'
   \les 
   \underline{\mathcal{F}}(\tau(u,v))\les A\ve^2\tau^{-3+\delta}(u,v),
\end{align*}
proving \eqref{est:psi+_bound}--\eqref{est:psi-_bound} for $r\geq 2\Lambda$. For $(u,v)\in\{\Lambda \le r\le 2\Lambda\} \cap \mathcal D_{\tau_f}$, define 
\begin{equation*}
    u^*(v):= 
    \begin{dcases}
        u^{2\Lambda}(v) &\text{if $r(0,v)\ge 2\Lambda$,}\\
        0 &\text{otherwise.}
    \end{dcases}
\end{equation*}
Noting that $[u^*(v),u]\times \{v\}\subset \mathcal{D}_{\tau_f}$, $0\le u-u^*(v)\les 1$ and $\tau(u^*(v),v)\sim \tau(u,v)$, we readily obtain
\begin{align*}
    |\psi_-(u,v)| &\leq 
    |\psi_-(u^*(v),v)| + \int_{u^*(v)}^u |\partial_u \psi_-|\,du'\lesssim A^{1/2}\ve \tau(u^*(v),v)^{-3/2+\delta/2} + \big(\underline{\mathcal{F}}(u,\tau(u^*(v),v))\big)^{1/2}\\
    &\les A^{1/2}\ve\tau^{-3/2+\delta/2}(u,v),\\
    |\psi_+(u,v)|&\les |\Phi_+(u,v)|
    \leq 
    |\Phi_+(u^*(v),v)| + \int_{u^*(v)}^u |\partial_u \Phi_+|\,du'\les
    |\psi_+(u^*(v),v)| + \int_{u^*(v)}^u | \psi_-|\,du'\\&\les A^{1/2}\ve \tau(u^*(v),v)^{-3/2+\delta/2} + \big(\underline{\mathcal{F}}(u,\tau(u^*(v),v))\big)^{1/2}\les A^{1/2}\ve\tau^{-3/2+\delta/2}(u,v),
\end{align*}
where we used $\psi_+ = \kappa^{-1/2}\Phi_+\sim \Phi_+$ and $\partial_u \Phi_+=\kappa^{1/2}\nu\psi_-\sim \psi_-$ in $\{\Lambda \le r\le 2\Lambda\}$, by \eqref{eq:nu-1} and \eqref{eq:kappa-2}. This proves \eqref{est:psi+_bound}--\eqref{est:psi-_bound} in the whole region $\{r\ge\Lambda\}$.

\textsc{Proof of \eqref{est:psi+_bound}--\eqref{est:psi-_bound} for $r\le\Lambda$}:  Let $\chi_2=\chi_2(r)$ be a smooth cutoff satisfying $\chi_2(r)=1$ for $r\le \Lambda/2$ and $\chi_2(r)=0$ for $r\ge\Lambda$. By the same argument as in Lemma \ref{lem:gamma-slope}, the set $\{r=\Lambda/2\}$ is a timelike curve in $\mathcal D_{\tau_f}$. Moreover, for $(u,v)\in\mathcal D_{\tau_f}$ with $r(u,v)\le\Lambda/2$, the segment $[u^\Lambda(v),u]\times\{v\}$ is entirely contained in $\mathcal D_{\tau_f}$. Therefore, we have
\begin{align*}
|\psi_+|^2 (u,v)|&\les |\Phi_+|^2 (u,v)
   \les \Big| \int^u_{u^\Lambda(v)}\big(\chi_2'\nu |\Phi_+|^2 + 2\chi_2 \Re[\overline{\Phi_+}\p_u\Phi_+]\big)\,du'\Big|\\
   &\les \int^u_{u^\Lambda(v)} (-\bar\nu)\big( |\psi_+|^2 + |\psi_-|^2\big)\,du'
   \les 
   \underline{\mathcal{E}}(\tau(u,v))\les A\ve^2\tau^{-3+\delta}(u,v),\\
   |\psi_-|^2 (u,v)&=
   \Big| \int^u_{u^\Lambda(v)}\big(\chi_2'\nu |\psi_-|^2 + 2\chi_2 \Re[\overline{\psi_-}\p_u\psi_-]\big)\,du'\Big|\\
   &\les 
   \int^u_{u^\Lambda(v)} \big( -\bar\nu|\psi_-|^2 + \frac{|\partial_u\psi_-|^2}{-\bar\nu}\big)\,du'
   \les 
   \underline{\mathcal{E}}(\tau(u,v))\les A\ve^2\tau^{-3+\delta}(u,v),
\end{align*}
where in the first line we used again $\psi_+\sim \Phi_+$ and $\p_u \Phi_+=\kappa^{1/2}\nu\psi_-\sim \nu\psi_-$. This proves \eqref{est:psi+_bound}--\eqref{est:psi-_bound} for $r\le \Lambda/2$. The region $\Lambda/2\le r\le \Lambda$ can be handled by integrating in $u$, analogously to the case $\Lambda\le r\le 2\Lambda$.

\textsc{Proof of \eqref{est:T-uv_bound}}: This follows immediately by \eqref{eq:Dirac-energy-momentum-tensor} and the previous estimates.
\end{proof}

Let us now proceed with establishing estimates on the derivatives of the null lapse on $\mathcal{D}_{\tau_f}$.
\begin{lemma}\label{lem:Omega-expansions}
   For any $A\ge 1$, $\ve$ sufficiently small, $\tau_f\in\mathfrak B$, we have the expansions
    \begin{align}
    \label{eq:partial_v_Omega-expansion1} 
    |\partial_v \log\Omega - M\bar r^{-2}| &\les A^2\ve^{3/2}\tau^{-2+\delta}\\
    \label{eq:partial_v_Omega-expansion2}
    \bar r|\partial_v \log\Omega - M\bar r^{-2}| &\les A^3\ve^{3/2}\tau^{-1+\delta},\\
    \label{eq:partial_u_Omega-expansion} \bar r^{2}|\partial_u \log\Omega + M\bar r^{-2}| &\les A^3\ve^{3/2} \tau^{-3/4+\delta},
    \end{align}
in $\mathcal D_{\tau_f}.$
\end{lemma}
\begin{proof}
\textsc{Proof of \eqref{eq:partial_v_Omega-expansion1} and \eqref{eq:partial_v_Omega-expansion2}}:
We start by noting that, since $\kappa=1$ on $\{\Gamma^u(\tau_f)\}\times[0,\Gamma^v(\tau_f)]$, we compute
\begin{equation*}
    0=\partial_v\kappa = \partial_u ((-\nu)^{-1}\Omega^2)=(-\nu)^{-1}(\partial_v\Omega^2 +\partial_v\nu)=2\p_v\log\Omega+(-\nu)^{-1}\p_v\nu,
\end{equation*}
yielding
\begin{equation*}
    \partial_v \log\Omega = \frac{ m}{r^2} + 2\pi r\frac{T_{uv}}{\nu}.
\end{equation*}
Therefore, we readily deduce 
\begin{equation}\label{eq:v-Omega-aux1}
    \Big|\p_v \log\Omega- \frac{M}{\bar r^2}\Big|(\Gamma^u(\tau_f),v)\les \frac{|\tilde m|}{r^2} + \frac{|\tilde r|}{r^3} + r\Big|\frac{T_{uv}}{-\nu}\Big|\les A^2\ve^{3/2}\tau^{-2+\delta},
\end{equation}
for $v\in[0,\Gamma^v(\tau_f)]$, where we used \eqref{est:T-uv_bound} and the geometric estimates established in the previous section.
Next, using \eqref{eq:wave-eq-Omega} and \eqref{eq:transport-kappa}, we compute
\begin{align*}
    \partial_u (\partial_v\log\Omega-\kappa M r^{-2}) 
    =2\Omega^2r^{-3}\tilde m - 4\Omega^2 r^{-2}\Im[\oline{\psi_-}\psi_+] + 2Mr^{-1}\Im[\oline{\psi_-}\p_u\psi_-].
\end{align*}
Therefore, integrating backwards in $u$ in $\{r\le\Lambda\}$, we find
\begin{align*}
    |\partial_v\log\Omega -\kappa M r^{-2}|(u,v)
    &\les |\partial_v\log\Omega -\kappa M r^{-2}|(\Gamma^u (\tau_f),v) 
    \\&+ 
    \int^{\Gamma^u (\tau_f)}_u \big(2\Omega^2r^{-3}|\tilde m|+\Omega^2|\psi_-||\psi_+| +|\psi_-||\partial_u\psi_-|\big)\,du'.
\end{align*}
After noting that $\Omega^2=\kappa(-\nu)\sim(-\bar\nu)$ by \eqref{eq:nu-ratio} and \eqref{eq:kappa-2}, we find that
\[
    \int^{\Gamma^u (\tau_f)}_u 2\Omega^2r^{-3}|\tilde m|\,du'\les A\ve^2\tau^{-3+\delta} \int^{\Gamma^u (\tau_f)}_u (-\bar\nu)\,du'\les A\ve^2\tau^{-3+\delta} \int^{\Lambda}_{2M} d\bar r\les A\ve^2\tau^{-3+\delta},
\]
yielding, from \eqref{eq:v-Omega-aux1},
\begin{equation}\label{eq:v-Omega-aux2}
    |\partial_v\log\Omega -\kappa M r^{-2}|(u,v)
    \les A^2\ve^{3/2} \tau^{-2+\delta} +  A\ve^2\tau^{-3+\delta} + \underline{\mathcal{E}}(\tau)\les A^2\ve^{3/2}\tau^{-2+\delta},
\end{equation}
on $\mathcal{D}_{\tau_f}\cap\{r\le\Lambda\}$.
For $(u,v)\in\{r\ge\Lambda\}$, we further integrate backwards in $u$ and readily have 
\begin{align}
    \nonumber |\partial_v\log\Omega -\kappa M r^{-2}|(u,v)
    &\les 
    |\partial_v\log\Omega -\kappa M r^{-2}|(u^\Lambda(v),v) \\ \nonumber&+ \int^{u^\Lambda(v)}_u \big(r^{-3}|\tilde m|+ r^{-2}|\psi_-||\psi_+|+r^{-1}|\psi_-||\partial_u\psi_-| \big)du'
    \\&\label{eq:v-Omega-aux3}
    \les A^2\ve^{3/2} \tau^{-2+\delta} + \underline{\mathcal{F}}(v,\tau)
    \les A^2\ve^{3/2} \tau^{-2+\delta},
\end{align}
where we used \eqref{eq:boot-ing-flux}, \eqref{eq:Hbarv-integral} and \eqref{eq:varpi-aux-0}. By combining \eqref{eq:v-Omega-aux2} and \eqref{eq:v-Omega-aux3} together with the previous geometric estimates on $\kappa$ and $r$, we obtain \eqref{eq:partial_v_Omega-expansion1} as desired,
\begin{equation*}
    |\partial_v \log\Omega - M\bar r^{-2}| \les A^2\ve^{3/2}\tau^{-2+\delta}.
\end{equation*}

To prove \eqref{eq:partial_v_Omega-expansion2}, we proceed as in the previous lemma and consider a smooth cutoff $\chi_1=\chi_1(r)$ satisfying $\chi_1(r)=1$ for $r\ge2\Lambda$ and $\chi_1(r)=0$ for $r\le\Lambda$. Then, for $(u,v)\in\mathcal{D}_{\tau_f}\cap\{r\ge2\Lambda\}$, we bound
\begin{align*}
    \bar r|\partial_v\log\Omega -\kappa M r^{-2}|(u,v)
    &\les \int^{u^\Lambda(v)}_u |\chi'_1\nu\bar r(\partial_v\log\Omega -\kappa M r^{-2})|+ |\chi_1\bar\nu(\partial_v\log\Omega -\kappa M r^{-2})|\\&+\bar r
    \big(r^{-3}|\tilde m|+ r^{-2}|\psi_-||\psi_+|+r^{-1}|\psi_-||\partial_u\psi_-| \big)du'
    \\ &\les \int^{u^\Lambda(v)}_u \big(|\partial_v\log\Omega -\kappa M r^{-2}|+r^{-2}|\tilde m|+ r^{-1}|\psi_-||\psi_+|+|\psi_-||\partial_u\psi_-| \big)du'
    \\&
    \les A^2\ve^{3/2} \tau^{-1+\delta} + \underline{\mathcal{F}}(v,\tau)
    \les A^2\ve^{3/2} \tau^{-1+\delta},
\end{align*}
where in particular we used \eqref{eq:boot-ing-flux}, \eqref{eq:Hbarv-integral} and \eqref{eq:partial_v_Omega-expansion1}. Using the previous geometric estimates on $\kappa$ and $r$, we obtain \eqref{eq:partial_v_Omega-expansion2}.

\begin{remark} Before turning to the estimate for $\p_u\log\Omega$, we briefly explain the strategy underlying the two estimates for $\p_v\log\Omega$.
A direct comparison with $M\bar r^{-2}$ would produce error terms involving $\wtilde\Omega$ and $\tilde r$ whose decay in $\tau$ is too weak to be integrated in $u$ in the far region $r\geq\Lambda$.
Moreover, attempting to obtain \eqref{eq:partial_v_Omega-expansion2} directly with an $\bar r^2$ weight would introduce a term of the form $\int_u^{u^\Lambda(v)} r\big|\p_v\log\Omega-\kappa M r^{-2}\big|,du',$
which cannot be controlled by the estimate \eqref{eq:partial_v_Omega-expansion1}. For this reason, we first establish the unweighted estimate \eqref{eq:partial_v_Omega-expansion1} and subsequently recover the weaker weighted estimate \eqref{eq:partial_v_Omega-expansion2}.
\end{remark}

\textsc{Proof of \eqref{eq:partial_u_Omega-expansion}}:
Proceeding similarly, we start by noting that, since $\gamma=1$ on $[0,\Gamma^u(\tau_f)]\times\{\Gamma^v(\tau_f)\}$, we compute
\begin{equation*}
    0=\partial_u\gamma = \partial_u (\lambda^{-1}\Omega^2)=\lambda^{-1}(\partial_u\Omega^2 - \partial_u\lambda)=2\p_u\log\Omega-\lambda^{-1}\p_u\lambda,
\end{equation*}
yielding
\begin{equation*}
    \partial_u \log\Omega =-\frac{ m}{r^2} + 2\pi r\frac{T_{uv}}{\lambda}= -\frac{m}{r^2} + 2\pi r\frac{T_{uv}}{\Omega^2},
\end{equation*}
where we again used $\gamma=\lambda^{-1}\Omega^2=1$.
Therefore, we readily deduce
\begin{equation*}
    \Big|\p_u \log\Omega + \frac{M}{\bar r^2}\Big|(u,\Gamma^v(\tau_f))\les \frac{|\tilde m|}{r^2} + \frac{|\tilde r|}{r^3} + r\Big|\frac{T_{uv}}{-\nu}\Big|\les A^2\ve^{3/2}\bar r^{-2}\tau^{-2+\delta},
\end{equation*}
for $u\in[0,\Gamma^u(\tau_f)]$, where we used \eqref{est:T-uv_bound} and the geometric estimates established in the previous section.
Next, using \eqref{eq:wave-eq-Omega} and \eqref{eq:transport-gamma}, we compute
\begin{align*}
    \partial_v (\partial_u\log\Omega+\gamma  m r^{-2}) 
    =-\frac{3\kappa^{1/2}(-\nu)}{r^2}\Im[\oline{\psi_-}\psi_+]+\frac{1}{r^2\lambda}\Big(\frac{2\gamma m}{r}-\nu\Big)\Im[\oline{\psi_+}\partial_v\psi_+].
\end{align*}
Therefore, first integrating backwards in $v$ in $\{r\ge\Lambda\}$ and then multiplying by $r^2$, we find
\begin{align}
    \nonumber r^2|\partial_u\log\Omega &+ \gamma m r^{-2}|(u,v) 
    \\\nonumber&\les r^2(u,v)|\partial_u\log\Omega +\gamma m r^{-2}|(u,\Gamma^v (\tau_f)) 
    + 
    r^2(u,v)\int^{\Gamma^v (\tau_f)}_v \big(r^{-2}|\psi_-||\psi_+|+r^{-2}|\psi_+||\partial_v\psi_+|\big)\,dv'
    \\ \nonumber&
    \les A^2\ve^{3/2} \tau^{-2+\delta} + \int^{\Gamma^v (\tau_f)}_v \big(|\psi_-||\psi_+|+|\psi_+||\partial_v\psi_+|\big)\,dv'
    \les A^2\ve^{3/2} \tau^{-2+\delta} + \mathcal{E}_0(\tau)
    \\ \label{eq:u-Omega-aux1}&\les A^2\ve^{3/2} \tau^{-2+\delta},
\end{align}
where we used \eqref{eq:lambda-lower-bound} to bound $r^2(u,v)\le r^2(u,\Gamma^v(\tau_f))$ in the boundary term and to move the $r$ weight into the integral. From the geometric estimates on $\gamma$, $r$ and $ m$, one then obtains \eqref{eq:partial_u_Omega-expansion} in $\D_{\tau_f}\cap\{r\ge\Lambda\}$.

For $\{r\le\Lambda\}$, we define
\[
    \xi:=e^{\bar r} (\partial_{u}\log\Omega+ M\bar r^{-2}),
\]
which satisfies
\[
    \partial_v\xi - \bar\lambda\xi = e^{\bar r}F,\quad F:=\Big(\frac{2\Omega^2 m}{r^3}-8\pi T_{uv} -\frac{2\overline{\Omega^2}M}{\bar r^3}\Big),
\]
indeed
\begin{align*}
    \partial_v\xi &= \partial_v(e^{\bar r} (\partial_{u}\log\Omega+ M\bar r^{-2}))= \bar\lambda e^{\bar r} (\partial_u\log\Omega + M\bar r^{-2})+ e^{\bar r} (\partial_v\partial_u\log\Omega -2M\bar\lambda \bar r^{-3})\\ &=\bar\lambda \xi + e^{\bar r} (2\Omega^2 m r^{-3} -8\pi T_{uv} - 2\oline{\Omega^2}M\bar r^{-3}).
\end{align*}
Multiplying by $\xi^3$ and integrating in $v$ from $\Gamma$, we obtain
\[
    \frac14\xi^4(u,v^\Lambda(u))-\frac14\xi^4(u,v)-\int^{v^\Lambda(u)}_v\bar\lambda\xi^4\,dv' =\int^{v^\Lambda(u)}_v e^{\bar r}F\xi^3\,dv',
\]
hence
\begin{equation*}
    \frac14\xi^4(u,v) + \int^{v^\Lambda(u)}_v \bar\lambda \xi^4 \,dv'\le \frac14\xi^4(u,v^\Lambda) + \int^{v^\Lambda(u)}_v e^{\bar r} |F||\xi|^3\,dv'\le \frac14\xi^4(u,v^\Lambda) + \int^{v^\Lambda(u)}_v \Big(e^{4\bar r}\bar\lambda^{-3}|F|^4 +\frac{\bar \lambda}{2}|\xi|^4\Big)\,dv',
\end{equation*}
yielding
\begin{equation}\label{eq:u-Omega-aux2}
    |\partial_{u}\log\Omega+ M\bar r^{-2}|^4(u,v) \les\frac14\xi^4(u,v) + \int^{v^\Lambda(u)}_v \frac{\bar\lambda}{2} \xi^4\,dv' 
    \les \frac14\xi^4(u,v^\Lambda(u)) + \int^{v^\Lambda(u)}_v \bar\lambda^{-3}|F|^4\,dv',
\end{equation}
where we used the fact that $e^{4\bar r}$ is uniformly bounded in $\{r\le\Lambda\}$. Thus, recalling that $|\wtilde \Omega^2| \les (-\bar\nu)A^3\ve^{3/2}\tau^{-1+\delta}$ (see \eqref{eq:Omega-bound}), 
\[
    |F|\les |\wtilde{\Omega^2}| + \overline{\Omega^2}|\wtilde m| + \overline{\Omega^2} |\wtilde{r}| +
    \Omega^2 |\psi_-||\psi_+|
    \les \overline{\Omega^2}A^3\ve^{3/2}\tau^{-1+\delta}
\]
on $\{r\le\Lambda\}$, we bound the right-hand side in \eqref{eq:u-Omega-aux2} by
\[
    \int^{v^\Lambda(u)}_v \bar\lambda^{-3}|F|^4\,dv'\les (A^3\ve^{3/2})^4\int^{v^\Lambda(u)}_v \oline{\Omega^2}\tau^{-4+4\delta}\,dv'\les (A^3\ve^{3/2})^4\tau^{-3+4\delta}.
\]
Therefore, using \eqref{eq:u-Omega-aux1}, we finally obtain
\[
    |\partial_{u}\log\Omega+ M\bar r^{-2}|(u,v)\les A^3\ve^{3/2}\tau^{-3/4+\delta},
\]
concluding the proof.
\begin{remark}
For $\p_u\log\Omega$, a different obstruction arises in the near region $r\leq\Lambda$. A direct comparison with $\gamma M r^{-2}$ would yield a term $\p_v\gamma\sim(1-\mu)^{-1}$, which degenerates towards the horizon. We therefore introduce the auxiliary quantity $\xi$, whose transport equation contains the favorable linear term $\bar\lambda\xi$. After integrating backwards in $v$, this yields the coercive term in \eqref{eq:u-Omega-aux2}, allowing us to absorb part of the source term by Young's inequality, in the spirit of \cite[Lemma 15.1.1]{dafermos2021non}. This argument allows us to recover a sufficient decay in $\tau$ without explicitly dealing with the degenerating terms arising from $\p_v\gamma$.
\end{remark}
\end{proof}

In order to bound the derivatives of the Dirac field and thus the components $T_{uu}$ and $T_{vv}$, we need an estimate for the initial energy in terms of our seed data norm $\mathfrak D$ in \eqref{eq:seed-data-norm}.
The proof follows \cite[Lemma 7.3]{angelopoulos2026nonlinear}, with the difference that our seed data norm controls only one spinor component on each initial null hypersurface. We exploit the transport structure of the Dirac equations \eqref{eq:Dirac-Phi_+}--\eqref{eq:Dirac-Phi_-} to recover a control of the missing components and thereby bound the energies at the initial time $\tau=1$.

\begin{lemma}\label{lem:initial-energy-est} For any $A\ge 1$, $\ve$ sufficiently small, $\tau_f\in\mathfrak B$, it holds that 
    \begin{equation}\label{eq:data-estimate}
        \mathcal E(1)+\underline{\mathcal E}_p(1)\les \big( \mathfrak D[\mathcal S_0]\big)^2.
    \end{equation}
\end{lemma}
\begin{proof} Let us start with $\uline{\hat{\C}}_\ing$, where $\nu\sim \bar\nu$. From the definition of $\mathfrak D[\mathcal S_0]$ in \eqref{eq:seed-data-norm}, we have on $\uline{\hat{\C}}_\ing$ that
\begin{equation}\label{eq:initial-energy-est1}
    |\psi_-|+\frac{|\partial_u\psi_-|}{-\bar\nu}\les \mathfrak D[\mathcal S_0].
\end{equation}
Thus, it follows
\begin{equation*}
    \int_{\uline{\hat{\C}}_\ing}\Big(-\bar\nu|\psi_-|^2 + \frac{|\partial_u\psi_-|^2}{-\bar\nu}\Big)\, du \le(\mathfrak D[\mathcal S_0])^2\int_{\uline{\hat{\C}}_\ing} (-\bar\nu)\, du\le (\mathfrak D[\mathcal S_0])^2\int_{2M}^\Lambda d\bar r\les (\mathfrak D[\mathcal S_0])^2.
\end{equation*}
Furthermore, recalling that $\kappa\sim1$, we deduce from \eqref{eq:Dirac-Phi_+} that
\[
    |\psi_+|(u,0)\les|\Phi_+|(u,0)\les|\Phi_+|(0,0) +\int_0^u\Big|\frac{\Omega}{r}\Phi_-\Big|\,du'\les |\psi_+|(0,0)+\int_{\uline{\hat{\C}}_\ing}|\psi_-|(-\bar\nu)\,du'\les \mathfrak D[\mathcal{S}_0],
\]
and consequently
\[
    \int_{\uline{\hat{\C}}_\ing}-\bar\nu|\psi_+|^2\, du\les(\mathfrak D[\mathcal S_0])^2\int_{\uline{\hat{\C}}_\ing} (-\bar\nu)\, du\le (\mathfrak D[\mathcal S_0])^2\int_{2M}^\Lambda d\bar r\les (\mathfrak D[\mathcal S_0])^2,
\]
proving $\mathcal{E}(1)\les (\mathfrak D[\mathcal S_0])^2$.
On the other hand, on $\hat\C_\out$, where $\lambda\sim 1$,
we bound
\begin{equation}\label{eq:initial-energy-est2}
    r|\psi_+|+r^2|\partial_v\psi_+|+ r^2|\partial_v(r\psi_+)|\les\mathfrak D[\mathcal S_0]
\end{equation}
on $\hat\C_\out$, which clearly implies 
\[
    \int_{\hat\C_\out}\big(r^p_{\tau_f} |\partial_v(r\psi_+)|^2+r^{p_*+2}_{\tau_f}|\partial_v\psi_+|^2 + r^{p_*}|\psi_+|^2 \big)\,dv\les (\mathfrak D[\mathcal S_0])^2
\]
as $p\in[0,3-\delta]$. Finally, noting that $-\nu\sim1$ on $\hat\C_\out$, we obtain from \eqref{eq:Dirac-Phi_-} that
\[
    r|\psi_-|(0,v)\les |\Phi_-|(0,v)\les |\Phi_-|(0,0)+\int^v_0\Big|\frac{\Omega}{r}\Phi_+\Big|\,dv'\les|\psi_-|(0,0)+\int_{\hat\C_\out}(r|\psi_+|)\frac{1}{r^2}\,dv'\les\mathfrak D[\mathcal S_0],
\]
allowing us to bound the remaining term in $\mathcal{E}_p(1)$.
\end{proof}

We are now ready to establish bounds on the derivatives of the Dirac components.
\begin{lemma}\label{lem:pointwise_estimates_dirac2}
For any $A\ge1$, $\ve$ sufficiently small, $\tau_f\in\mathfrak B$, and $1\le\tau_1\le \tau_2\le\tau_f$, we have the pointwise boundedness estimates
\begin{align}
    \label{est:vpsi+_bound}   \bar r^2|\partial_v\psi_+|&\les A^{1/2}\ve,\\
    \label{est:rvpsi+_bound}   \bar r^2|\partial_v(r\psi_+)|&\les A^{1/2}\ve,\\
    \label{est:upsi-_bound}    \bar  r\frac{|\partial_u\psi_-|}{-\nu}&\les A^{1/2}\ve,\\
    \label{est:T-uu_bound}    \bar r^2\frac{|T_{uu}|}{\nu^2}&\les A\ve^2 \tau^{-3/2+\delta/2},\\
    \label{est:T-vv_bound}    
    \bar r^{7/2}|T_{vv}|&\les A\ve^2 \tau^{-3/2+\delta/2},
\end{align}
on $\mathcal D_{\tau_f}$.
\end{lemma}

\begin{proof}
\textsc{Proof of \eqref{est:vpsi+_bound}}: We consider again the two regions $\{r\le\Lambda\}$ and $\{r\ge\Lambda\}$ separately. Let us begin with the latter and consider $\wtilde\Phi_+:=r(-\nu)^{-1/2}\psi_+$, which satisfies 
\begin{equation*}
    \partial_u \partial_v \widetilde\Phi_+ + \Big(\p_u\log\Omega-\frac{\nu}{r}\Big) \partial_v \widetilde\Phi_+ + \Big(\frac{6\Omega^2 m}{r^3}-12\pi T_{uv}\Big) \widetilde\Phi_+ = 0.
\end{equation*}
Hence, from 
\begin{equation*}
    \partial_u (r^2\partial_v \widetilde\Phi_+) = \Big(\frac{3\nu}{r}-\p_u\log\Omega\Big) (r^2\partial_v \widetilde\Phi_+) -r^2 \Big(\frac{6\Omega^2 m}{r^3}-12\pi T_{uv}\Big) \widetilde\Phi_+,
\end{equation*}
we compute for $(u,v)\in\mathcal{D}_{\tau_f}\cap\{r\ge\Lambda\}$
\begin{equation}\label{eq:bounded-aux-1}
\begin{aligned}
   (r^2\partial_v\wtilde\Phi_+)(u,v) &= \exp\Big(\int_0^{u}\big(\frac{3\nu}{r}-\p_u\log\Omega\big)\,du'\Big)  (r^2\partial_v\wtilde\Phi_+)(0,v) \\&- \int^{u}_0 \exp\Big(\int^{u}_{u'}\big(\frac{3\nu}{r}-\p_u\log\Omega\big)\,du''\Big)r^2 \Big(\frac{6\Omega^2 m}{r^3}-12\pi T_{uv}\Big) \widetilde\Phi_+\,du'.
\end{aligned}
\end{equation}
Using \eqref{eq:nu-1} and \eqref{eq:partial_u_Omega-expansion}, it holds
\begin{equation*}
    \exp\Big(\int^{u_1}_{u_2} \big(\frac{3\nu}{r}-\p_u\log\Omega\big)\,du'\Big)\les \exp\Big(\int^{u_1}_{u_2} \bar r^{-2}(M+A^3\ve^{3/2}\tau^{-3/4+\delta})\,du'\Big)\les1,
\end{equation*}
on $\mathcal{D}_{\tau_f}\cap\{r\ge\Lambda\}$. Next, from \eqref{eq:Hbarv-integral} and \eqref{est:psi+_bound}, we bound
\begin{equation*}
    \int^{u}_0r^2 \Big(\frac{6\Omega^2 m}{r^3}-12\pi T_{uv}\Big) \widetilde\Phi_+\,du'\les  \int^{u}_0 |\psi_+|\,du'\les A^{1/2}\ve \int^{u}_0 \tau^{-3/2+\delta/2}\,du'\les A^{1/2}\ve.
\end{equation*}
Therefore, \eqref{eq:bounded-aux-1}, together with 
\begin{equation*}
    r^2|\partial_v\wtilde\Phi_+|(0,v)\les r^2\big(|\p_v(r\psi_+)|+ r|\psi_+||\p_v\nu|\big)(0,v)\les \big(r^2|\p_v(r\psi_+)|+ r|\psi_+|\big)\les\mathfrak D[\mathcal S_0]\les\ve,
\end{equation*}
which controls the boundary term, yields
\[
    |r^2\partial_v\wtilde\Phi_+|\les A^{1/2}\ve,
\]
and therefore
\[
    r^2|\p_v\psi_+|\les r|\psi_+|+r|\p_v\wtilde\Phi_+|\les A^{1/2}\ve, \quad
    r^2|\p_v(r\psi_+)|\les r|\psi_+|+r^2|\p_v\wtilde\Phi_+|\les A^{1/2}\ve,
\]
on $\mathcal{D}_{\tau_f}\cap\{r\ge\Lambda\}$.
Next, to proceed in the region $\{r\le\Lambda\}$, we instead consider the reweighting $\Phi_+$, which satisfies
\begin{equation*}
    \partial_u\partial_v \Phi_+ = \nu\kappa^{1/2}\Big( \big(\partial_v \log\Omega-\lambda r^{-1}\big)\psi_- + \kappa r^{-2} \psi_+ \Big).
\end{equation*}
Therefore, for $(u,v)\in\mathcal{D}_{\tau_f}\cap \{r\le\Lambda\}$, we readily estimate, by \eqref{est:psi+_bound} and \eqref{est:psi-_bound},
\begin{equation}
\begin{aligned}\label{eq:bounded-aux-2}
    |\p_v\Phi_+|(u,v)&\les|\p_v\Phi_+|(u^\Lambda(v),v)+\int^{u}_{u^\Lambda(v)}\nu\kappa^{1/2}\Big( \big(\partial_v \log\Omega-\lambda r^{-1}\big)\psi_- + \kappa r^{-2} \psi_+ \Big)\,du'\\&\les A^{1/2}\ve + A^{1/2}\ve\int^{u}_{u^\Lambda(v)}(-\nu)\,du'\les A^{1/2}\ve + A^{1/2}\ve\int^\Lambda_{2M}\,d\bar r\les A^{1/2}\ve,
\end{aligned}
\end{equation}
where we controlled the boundary term by noting that
\[
    |\partial_v\Phi_+|(u^\Lambda(v),v)\les \big((\p_v\log\Omega + r^{-1})|\psi_+|+ r^{-1}|\p_v\wtilde\Phi_+|\big)(u^\Lambda(v),v)\les \big(|\psi_+|+|\p_v\wtilde\Phi_+|\big)(u^\Lambda(v),v)\les A^{1/2}\ve,
\]
by \eqref{eq:Omega-bound}, \eqref{eq:lambda-tilde} and \eqref{eq:partial_v_Omega-expansion1}.

Hence, we conclude on $\mathcal{D}_{\tau_f}\cap\{r\le\Lambda\}$ that
\[
    |\partial_v\psi_+|\les |\partial_v\Phi_+| +|\psi_+||\partial_v\kappa|\les A^{1/2}\ve,
\]
where we used \eqref{est:psi+_bound}, \eqref{eq:bounded-aux-2} and the following fact: 
\begin{align*}
    |\partial_v\kappa| &= \Big|2\kappa\Big(\p_v\log\Omega-\kappa  m r^{-2} +2\pi r \frac{T_{uv}}{-\nu}\Big)\Big|\les |\p_v\log\Omega-M\bar r^{-2}|+|\tilde\kappa| \bar r^{-2} + |\tilde m| \bar r^{-2} + |\tilde r|\bar r^{-3} + r\Big|\frac{T_{uv}}{-\nu}\Big|\\&\les A^3\ve^{3/2},%\bar r^{-2}
\end{align*}
again by the geometric estimates, \eqref{est:T-uv_bound} and \eqref{eq:partial_v_Omega-expansion1}.
Moreover, we immediately obtain
\[
    |\p_v(r\psi_+)|\les |\lambda||\psi_+|+ r|\p_v\psi_+|\les |\psi_+|+|\p_v\psi_+|\les A^{1/2}\ve.
\]

\textsc{Proof of \eqref{est:upsi-_bound}}: Let us consider  $\widetilde{\Phi}_-:=\frac{r}{\Omega}\Phi_-=r^2\kappa^{-1/2}\psi_-$, which solves
\begin{equation*}
    \partial_u\partial_v \widetilde \Phi_- = -\Big(\partial_v \log\Omega - \frac{\lambda}{r}\Big)\partial_u \widetilde \Phi_- - \Big(\frac{6\Omega^2 m}{r^3} - 12\pi T_{uv}\Big) \widetilde \Phi_-.
\end{equation*}
Thus, we compute
\begin{equation*}
    \partial_v \Big( \frac{1}{-\nu r} \partial_u \widetilde \Phi_-\Big) 
    =
    \Big( \frac{1}{-\nu r} \partial_u \widetilde \Phi_-\Big) \Big( \frac{\partial_v \nu}{-\nu} - \partial_v \log\Omega \Big) + \frac{1}{\nu r} \Big(\frac{6\Omega^2 m}{r^3} - 12\pi T_{uv}\Big) \widetilde \Phi_-,
\end{equation*}
which can be solved for
\begin{equation}
\begin{aligned}
\label{eq:Ypsi-negative-2}
    \left(\frac{\partial_u \widetilde \Phi_-}{-\nu r} \right)(u,v)
    &=
    \exp\left(\int_0^v \Big(\frac{\partial_v \nu}{-\nu} - \partial_v \log\Omega\Big)dv'\right)
    \left(\frac{\partial_u \widetilde \Phi_-}{-\nu r} \right)(u,0)
    \\&+
    \int_0^v \exp\left(\int_{v'}^{v} \Big(\frac{\partial_v \nu}{-\nu} - \partial_v \log\Omega\Big)\,dv''\right) \frac{1}{\nu r} \Big(\frac{6\Omega^2 m}{r^3} - 12\pi T_{uv}\Big) \widetilde \Phi_-\,dv'.
\end{aligned}
\end{equation}
Let us start by bounding the integrating factor. If $(u,v_1),(u,v_2)\in\D_{\tau_f}\cap\{r\le\Lambda\}$, we have by \eqref{est:T-uv_bound} and \eqref{eq:partial_v_Omega-expansion1},
\begin{align*}
    \exp &\Big(\int_{v_1}^{v_2} \Big(\frac{\partial_v \nu}{-\nu} - \partial_v \log\Omega\Big)dv'\Big)
    = \exp\Big(\int_{v_1}^{v_2} 
    \Big( -\frac{2\kappa  m}{r^2} - \partial_v\log\Omega + 4\pi r \frac{T_{uv}}{-\nu} \Big)dv'\Big)\\
    &\les \exp \Big(\int_{v_1}^{v_2} \Big(-\frac{3M}{\bar r^2} + A^2\ve^{3/2} \tau^{-2+\delta} + A\ve^2 \bar r^{-5/2} \tau^{-3+\delta}\Big)dv'\Big) \les 1 + A^2\ve^{3/2} v_1^{-1+\delta} \les 1;
\end{align*}
on the other hand, when $(u,v_1),(u,v_2)\in\D_{\tau_f}\cap\{r\ge\Lambda\}$, we compute explicitly
\begin{align*}
    \exp &\Big(\int_{v_1}^{v_2} \Big(\frac{\partial_v \nu}{-\nu} - \partial_v \log\Omega\Big)dv'\Big)
    =\exp \left(\int_{v_1}^{v_2} -\p_v\log\big((-\nu)\Omega\big)dv'\right) = \frac{(-\nu)\Omega(u,v_1)}{(-\nu)\Omega(u,v_2)}\les 1,
\end{align*}
by \eqref{eq:nu-1} and \eqref{eq:Omega-bound}. The case when $(u,v_1),(u,v_2)$ lie on different sides of $\Gamma$ may be treated by splitting the integrating factor at $v^\Lambda(u)$ and applying the preceding estimates to each of the two resulting terms.
Next, we control 
\begin{align*}
    \int_0^v \frac{1}{r} \Big(\frac{6\kappa  m}{r^3} + 12\pi \Big|\frac{T_{uv}}{-\nu}\Big|\Big) |\widetilde \Phi_-|\,dv'
    &\les \int_0^v \Big( \frac{ m}{r^3}+ \Big|\frac{T_{uv}}{-\nu}\Big|\Big)|r\psi_-|\,dv'
    \\&\les A^{1/2} \ve\int_0^v \big(r^{-3} \tau^{-3/2+\delta/2} + A\ve^2r^{-7/2} \tau^{-9/2+3/2\delta}
    \big)\,dv'\les A^{1/2}\ve,
\end{align*}
using \eqref{eq:rough-bound-varpi}, \eqref{est:psi-_bound} and \eqref{est:T-uv_bound}. Therefore, from \eqref{eq:Ypsi-negative-2} and the fact that
\[
\Big|\frac{\partial_u \widetilde \Phi_-}{-\nu r} \Big|(u,0)\les \left(|\psi_-|\Big(1+\frac{|\partial_u\kappa|}{-\nu}\Big)+\frac{|\partial_u\psi_-|}{-\nu}\right)(u,0)\les \left(|\psi_-|\Big(1+|\psi_-|\frac{|\partial_u\psi_-|}{-\nu}\Big)+\frac{|\partial_u\psi_-|}{-\nu}\right)(u,0)\les\ve
\]
by \eqref{eq:initial-energy-est1}, we deduce
\begin{equation}\label{est:upsi-_bound-aux}
    \Big| \frac{\partial_u \widetilde \Phi_-}{-\nu r}\Big| \les A^{1/2}\ve.
\end{equation}
Then, \eqref{est:upsi-_bound} follows by $\psi_-= \kappa^{1/2} r^{-2}\widetilde \Phi_-$, \eqref{est:psi-_bound} and \eqref{est:upsi-_bound-aux}, indeed
\begin{equation*}
    r\frac{|\partial_u\psi_-|}{-\nu}
    \les 
    \Big(1+r\frac{|\partial_u \kappa|}{-\nu}\Big) |\psi_-| + \Big|\frac{\partial_u \widetilde \Phi_-}{-\nu r}\Big|
    \les \Big( 1+\Big|\frac{\partial_u \widetilde \Phi_-}{-\nu r}\Big||r\psi_-|\Big)|\psi_-|+ \Big|\frac{\partial_u \widetilde \Phi_-}{-\nu r}\Big|\les A^{1/2}\ve,
\end{equation*}
where we rewrote \eqref{eq:transport-kappa} as $\partial_u\kappa=-2r\kappa\Im[\oline{\psi_-}\partial_u\psi_-]=-2r^{-1}\kappa^{3/2}\Im[\oline{\psi_-}\partial_u\wtilde\Phi_-]$.

\textsc{Proof of \eqref{est:T-uu_bound} and \eqref{est:T-vv_bound}}: These follow immediately by \eqref{eq:Dirac-energy-momentum-tensor} and the previous estimates.
\end{proof}
\begin{remark}
The meaning and role of the reweightings $\wtilde\Phi_\pm$ introduced above, as well as the advantages of the equations they satisfy, will become clearer in the next section. In particular, $\wtilde\Phi_-$ will be particularly suited to the near-horizon region, where it will be used to establish the red-shift estimates in Proposition \ref{prop:red-shift}, while $\wtilde\Phi_+$ will be more relevant in the far region, where it will be essential in the $r^p$-estimates in Proposition \ref{prop:rp-est}.
\end{remark}
As a consequence, we finally establish pointwise boundedness estimates for the Einstein equations.
\begin{lemma}\label{lem:pointwise-est-Einstein-eq}
For any $A\ge1$, $\ve$ sufficiently small, $\tau_f\in\mathfrak B$, and $1\le\tau_1\le \tau_2\le\tau_f$, we have the pointwise boundedness estimates
\begin{align}
    \label{est:partial_uv-r}
    \Big|\frac{\partial_u\partial_v r}{-\nu}
    +\frac{2M}{\bar r^2}\Big| &\les A^2\ve^{3/2}\bar r^{-2}\tau^{-2+\delta},\\
    \label{est:partial_uv-Omega}
    \Big|\frac{\partial_u\partial_v\log \Omega}{-\nu}
    -\frac{2M}{\bar r^3}\Big| &\les A^2\ve^{3/2}\bar r^{-2}\tau^{-2+\delta},\\
    \label{est:partial_u-varpi}
    \Big| \frac{\partial_u m}{-\nu}\Big|&\les A\ve^2\tau^{-3/2+\delta/2},\\
    \label{est:partial_v-varpi}
    |\partial_v  m| &\les A\ve^2\bar r^{-3/2}\tau^{-3/2+\delta/2},\\
    \label{est:partial_u-kappa}
    \Big| \frac{\partial_u \kappa}{-\nu}\Big| &\les  A\ve^2 \bar r^{-1}\tau^{-3/2+\delta/2}
\end{align}
in $\mathcal D_{\tau_f}$, and 
\begin{align}
    \label{est:partial_v-gamma}
    | \partial_v \gamma| &\les A\ve^2 \bar r^{-7/2}\tau^{-3/2+\delta/2}
\end{align}
in $\mathcal D_{\tau_f}\cap\{r\ge\Lambda\}$.
\end{lemma}
\begin{proof} The bounds follow readily by \eqref{eq:wave-eq-r}, \eqref{eq:wave-eq-Omega} and \eqref{eq:transport-kappa}--\eqref{eq:transport-v-varpi}, together with Lemmas \ref{lem:pointwise_estimates_dirac1} and \ref{lem:pointwise_estimates_dirac2}.
\end{proof}

We conclude this section with an argument that has been adapted from \cite[Proposition 5.12]{luk2019strong}, which relies on the previous lemma. This result will be crucial in controlling the gauge map $\Upphi_{\tau_f}$ in Lemma \ref{lem:inv-est}.
\begin{lemma}\label{lem:u-v-gamma} For any $A\ge 1$, $\ve$ sufficiently small, $\tau_f\in\mathfrak B$, it holds that 
\begin{equation}\label{eq:nu-exp}
    -\nu \sim e^{-(u-v)}
\end{equation}
on $\D_{\tau_f}\cap\{r\le\Lambda\}$.
\end{lemma}
\begin{proof} \textsc{Step 1.} We start by proving
\begin{equation}\label{eq:u-v-gamma}
    \sup_{\Gamma\cap\mathcal{D}_{\tau_f}}|u-v|\les A\ve^2.
\end{equation}
    Recalling Definition \ref{def:bootstrap-set} and denoting by $T:=\frac{1}{\kappa}\p_v+\frac{1}{\gamma}\p_u$ the Kodama vector field \cite{kodama1980conserved}, we note that
    \[(u-v)(\Gamma(1))=(u-v)(0,0)=0,\quad \dot\Gamma(\tau)=\frac{1}{2\sqrt{1-\mu}}T|_{\Gamma(\tau)}.\] Thus, on $\Gamma(\tau)$ with $\tau\in[1,\tau_f]$, we compute
    \[
    \Big|\frac{d}{d\tau}(u-v)\Big|= \frac{1}{2\sqrt{1-\mu}}|T(u-v)|=\frac{1}{2\sqrt{1-\mu}}\Big|\frac{1}{\gamma}-\frac{1}{\kappa}\Big|\les \frac{1}{2\sqrt{1-\mu}}\big(|\tilde\gamma|+|\tilde\kappa|\big),
    \]
    where we used that $\kappa$ and $\gamma$ are uniformly bounded on $\Gamma$ by \eqref{eq:kappa-1} and \eqref{eq:gamma-1}. Hence,
    \begin{align*}
        (u-v)(\Gamma(\tau))&\les \int^\tau_1\frac{1}{2\sqrt{1-\mu}}\big(|\tilde\gamma|+|\tilde\kappa|\big)\,d\tau'\\&\les \big(\sup_{\Gamma}|\gamma|\big)\int^{\Gamma^u(\tau)}_0|\tilde\gamma|(u',v^\Lambda(u'))\,du'+\big(\sup_{\Gamma}|\kappa|\big)\int^{\Gamma^v(\tau)}_0|\tilde\kappa|(u^\Lambda(v'),v')\,dv'.
    \end{align*}
    Using again the uniform bounds for $\kappa$ and $\gamma$ along $\Gamma$, it remains to estimate the two integrals. Regarding the first term, we use \eqref{eq:transport-gamma} and \eqref{est:partial_v-gamma} to write
    \begin{align*}
        \int^{\Gamma^u(\tau_f)}_0 |\tilde\gamma|(u',v^\Lambda(u'))\,du'&\les\int^{\Gamma^u(\tau_f)}_0 \int^{\Gamma^v(\tau_f)}_{v^\Lambda(u')} A\ve^2 r^{-7/2}\tau^{-3/2+\delta/2}\,dv'du'
        \\&\les A\ve^2\int^{\Gamma^u(\tau_f)}_0 (u')^{-3/2+\delta/2}\int^{\Gamma^v(\tau_f)}_{v^\Lambda(u')}  r^{-7/2}\,dv'du'\les A\ve^2,
    \end{align*}
    where in the last line we used the definition of $\tau$ in \eqref{def:tau}, together with the integrable decay in both $\tau$ and $r$, to obtain the bound \(A\varepsilon^2\). On the other hand, we compute similarly, thanks to \eqref{eq:transport-kappa} and \eqref{est:partial_u-kappa},
    \begin{align*}
        \int^{\Gamma^v(\tau_f)}_0 |\tilde\kappa|(u^\Lambda(v'),v')\,dv'&\les\int^{\Gamma^v(\tau_f)}_0 \int^{\Gamma^u(\tau_f)}_{u^\Lambda(v')} A\ve^2 (-\nu)\bar r^{-1}\tau^{-3/2+\delta/2}\,du'dv'
        \\&\les A\ve^2\int^{\Gamma^v(\tau_f)}_0 (v')^{-3/2+\delta/2}\int^{\Gamma^u(\tau_f)}_{u^\Lambda(v')}  (-\nu)\,dv'du'\les A\ve^2,
    \end{align*}
    where in the last line we used again \eqref{def:tau}, together with the integrable decay in $\tau$ and the integrability of the \(-\nu\) factor along the $u$-direction, to obtain the bound $A\varepsilon^2$.

\textsc{Step 2.} Let us now prove \eqref{eq:nu-exp} on $\D_{\tau_f}\cap\{r\le\Lambda\}$.
From \eqref{eq:wave-eq-r}, we readily deduce 
    \[
        \p_v\log(-\nu)=\frac{2\kappa m}{r^2}+\frac{4\pi r}{\nu}T_{uv}.
    \]
    Noting by \eqref{est:T-uv_bound} and the previous geometric estimates established in Section \ref{sec:improvement-geometry} that 
    \[
        0<C\le \Big|\frac{2\kappa m}{r^2}+\frac{4\pi r}{\nu}T_{uv}\Big|\le C',
    \]
    for some strictly positive constants $C,C'$, we may integrate in $v$ and estimate, for $(u,v)\in\D_{\tau_f}\cap\{r\le\Lambda\}$,
    \[
        (-\nu)(u,v)=(-\nu)(u,v^\Lambda(u))\exp\Big(-\int^{v^\Lambda(u)}_v \big(\frac{2\kappa m}{r^2}+\frac{4\pi r}{\nu}T_{uv}\big)dv'\Big)\sim \exp\big(-(v^\Lambda(u)-u)\big)\sim \exp(-(u-v))
    \]
    where we used \eqref{eq:nu-1} to control the boundary term and \eqref{eq:u-v-gamma} to compare $v^\Lambda(u)-u$ with $u-v$.
\end{proof}
\section{Energy estimates for the Dirac field}\label{sec:energy-Dirac}
The aim of this section is to improve the bootstrap assumptions on the Dirac field, namely \eqref{eq:boot-ing-energy}--\eqref{eq:boot-outg-flux}, by employing the vector-field method. In particular, in Section \ref{subsec:iled} we establish an integrated local energy decay estimate by combining a suitable Morawetz multiplier and the current conservation for the Dirac field.
Since this estimate degenerates towards the horizon and does not have optimal decay in the far region, we establish red-shift estimates in Section \ref{sec:red-shift} and $r^p$-weighted energy estimates in Section \ref{sec:rp-est}.
Finally, in Section \ref{sec:improvement-Dirac} we prove the following result via a pigeonhole argument.

\begin{proposition}[Improving the bootstrap assumptions for the Dirac field] \label{prop:bootstrap-Dirac}
There exist positive constants $A_0$ and $\ve_0$ depending only on $M_0$ and $\delta$, such that if $A= A_0$, $\ve\le \ve_0$, $\tau_f\in  \mathfrak B(\mathcal S,\ve,A)$, then the estimates
\begin{align}
    \label{eq:boot-improved1}
    \underline{\mathcal{E}}^{\tau_f} (\tau) &\le \tfrac 12A\ve^2 \tau^{-3+\delta}, \\
    \mathcal E_{p}^{\tau_f}(\tau) &\le \tfrac 12A\ve^2 \tau^{-3+p+\delta}, \\
    \underline{\mathcal F}^{\tau_f}(v,\tau)&\le \tfrac 12A\ve^2\tau^{-3+\delta},\\
    \label{eq:boot-improved4}
    \mathcal F^{\tau_f}(u,\tau)&\le \tfrac 12A\ve^2 \tau^{-3+\delta},
    \end{align}
hold for every $\tau\in [1,\tau_f]$, $(u,v)\in\mathcal D_{\tau_f}$, and $p\in [\delta,3-\delta]$.
\end{proposition}

\begin{figure}
\centering
\begin{tikzpicture}[line cap=rect,inner sep=0pt, x=1cm, y=1cm, scale=1,font=\small]

\coordinate  (I)   at (0,2)      {};
\coordinate (II)   at (2,0)      {};
\coordinate  (III)   at (4,2)      {};
\coordinate   (IV)   at (2,4)      {};
\coordinate  (V)   at (1.15,0.85)      {};
\coordinate  (VI)   at (3.15,1.15)      {};
\coordinate  (VII)   at (2.3,2)      {};
\coordinate  (VIII)   at (3.15,2.85)      {};
\coordinate  (IX)   at (1.15,3.15)      {};

\path[name path=parab1, domain=0:4, samples=100, smooth, variable=\y]
    plot (\y,{-3/40*\y*\y+3/10*\y+2});

\fill[lightgray] (I) -- (II) -- (III) -- (VIII) -- (VII) -- (IX);
%\fill[gray] (V) -- (II) -- (IX) -- (VII);
%\fill[gray] plot[domain=0:1, smooth] (\y,{-3/40*\y*\y+3/10*\y+2}) -- (IV) -- (III) --(I);

\draw[line width=1pt]  (IX) 
-- node[pos=0.5, above=2pt, sloped]    {$\mathrm{V}$}
       (VII) -- node[pos=0.5, above=2pt, sloped]    {$\mathrm{IV}$} (VIII) -- node[pos=0.5, above=2pt, sloped]    {$\mathrm{III}$}
       (III) -- node[pos=0.5, below=2pt, sloped]    {$\mathrm{II}$} (II) -- node[pos=0.5, below=2pt, sloped]    {$\mathrm{I}$} (I) -- node[pos=0.5, above=2pt, sloped]    {$\mathrm{VI}$} (IX);

\draw[line width=1pt,domain=0:2, smooth, variable=\y] plot ({2 + 3/40*\y*(4-\y)}, {\y});

\node at ($(II)+(0.4,0.75)$)      {$\Gamma$};
\node at ($(1.5,1.4)$)      {$\mathcal{R}_{\le\Lambda}$};
\node at ($(2.8,1.4)$)      {$\mathcal{R}_{\ge\Lambda}$};
\node at ($(II)+(-0.7,0)$)      {$(u_1,v_1)$};
\node at ($(VII)+(-0.7,0)$)      {$(u_2,v_2)$};
\filldraw[draw=black] (II) circle (2pt);
\filldraw[draw=black] (VII) circle (2pt);

\end{tikzpicture}
\caption{A Penrose diagram of the region $\mathcal{R}$, together with the labelled hypersurfaces $\mathrm{I}$--$\mathrm{VI}$ on which the energy estimates will be established.}
\label{fig:Penrose7-energydomain}
\end{figure}
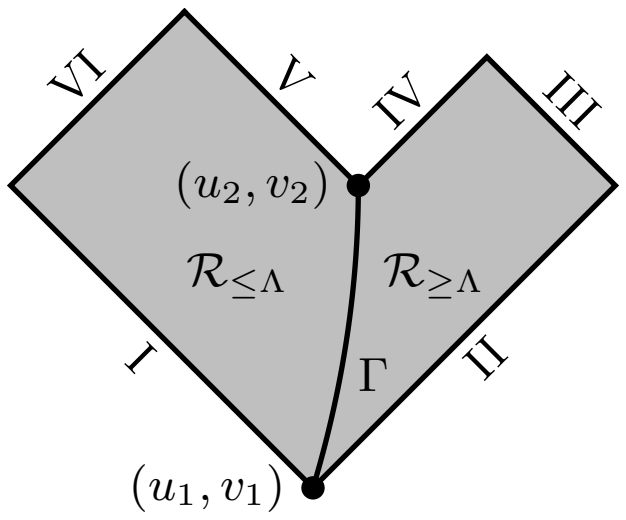

The energy estimates in this section (except in Lemma \ref{lem:T-cons-law-bis}) will be carried out in a region $\R\subset\mathcal{D}_{\tau_f}$ defined as follows. Let $\tau_f\in\mathfrak{B}$, $(u_1,v_1),(u_2,v_2)\in\Gamma\cap\mathcal{D}_{\tau_f}$, where $(u_2,v_2)$ is to the future of $(u_1,v_1)$. Consider $u'_2>u_2$ and $v'_2>v_2$ and define
\[
    \R:=\big([u_1,u_2]\times[v_1,v'_2]\big)\cup\big([u_1,u'_2]\times[v_1,v_2]\big),\quad \R_{\le\Lambda}:=\R\cap\{r\le\Lambda\},\quad \R_{\ge\Lambda}:=\R\cap\{r\ge\Lambda\}.
\]
The null segments constituting $\p\R$ are labelled $\mathrm{I}$--$\mathrm{VI}$, as depicted in Figure \ref{fig:Penrose7-energydomain}. We define $\tau_1$ to be the value of $\tau$ on $\mathrm{I}\cup\mathrm{II}$ and $\tau_2$ to be the value of $\tau$ on $\mathrm{IV}\cup\mathrm{V}$.

\subsection{Integrated local energy decay (ILED) estimate}\label{subsec:iled}
We begin with the first estimate for the Dirac field. The argument combines the Dirac energy current, which provides control of the boundary terms, with a Morawetz multiplier yielding the required coercivity in the bulk.
\begin{proposition}\label{prop:T+Mor} 
For any $A\ge 1$, $\ve$ sufficiently small, $\tau_f\in\mathfrak B$, $(u_1,v_1),(u_2,v_2)\in\Gamma\cap\mathcal D_{\tau_f}$,  $u_2'>u_2$, and $v_2'>v_2$,  where $(u_2,v_2)$ is to the future of $(u_1,v_1)$, it holds that
    \begin{equation}\label{eq:T+Mor}
    \begin{aligned}
        &\int_\mathrm{III} \big(r^2|\psi_-|^2 + r^{-2}|\psi_+|^2 + r^2|\partial_u\psi_-|^2\big)\, du 
        + \int_\mathrm{IV} \big(|\psi_+|^2+|\psi_-|^2+|\partial_v\psi_+|^2\big)\,dv 
        \\ &+ \int_\mathrm{V} -\bar\nu\big(|\psi_-|^2+|\psi_+|^2+|\partial_u\psi_-|^2\big)\, du + \int_\mathrm{VI} \big(|\psi_+|^2 +\bar\nu^2|\psi_-|^2+|\partial_v\psi_+|^2\big)\,dv 
        \\& +\iint_\mathcal{R} -\bar\nu \,\bar r^{-4} (|\psi_+|^2 + \bar r^2 |\partial_v\psi_+|^2) + \bar\nu^2\bar r^{-2}(|\psi_-|^2 + \bar r^2|\partial_u\psi_-|^2 )\,dudv
        \\&\les \int_\mathrm{I} -\bar\nu\big(|\psi_-|^2+|\psi_+|^2+|\partial_u\psi_-|^2\big)\, du
        + \int_\mathrm{II} \big(|\psi_+|^2+|\psi_-|^2+|\partial_v\psi_+|^2\big)\,dv
        +\ve^3\tau_1^{-7/2+3\delta}.
\end{aligned}
\end{equation}
\end{proposition}
At this stage, before turning to the proof, we highlight the degeneracy of certain coefficients near the horizon, such as $\p_u\psi_-$ in $\mathrm{V}$ and $\psi_-$ in $\mathrm{VI}$, together with the absence of $r$ weights on the hypersurface $\mathrm{IV}$.

\subsubsection{Proof of the ILED estimate}
Let us begin with the following lemma, which yields a control on the boundary terms.

\begin{lemma}[Degenerate energy estimate -- I]\label{lem:T-cons-law}
With hypotheses as in Proposition \ref{prop:T+Mor}, it holds that
\begin{equation}\label{eq:T-cons}
    \begin{aligned}
        &\int_\mathrm{III} \big( r^{-2}|\psi_+|^2 + r^2|\psi_-|^2 + r^2|\partial_u\psi_-|^2\big)\, du 
        + \int_\mathrm{IV} \big(|\psi_+|^2+|\psi_-|^2+|\partial_v\psi_+|^2\big)\,dv 
        \\ &+ \int_\mathrm{V} -\bar\nu\big(|\psi_+|^2+|\psi_-|^2+|\partial_u\psi_-|^2\big)\, du + \int_\mathrm{VI} \big(|\psi_+|^2 +\bar\nu^2|\psi_-|^2+|\partial_v\psi_+|^2\big)\,dv 
        \\&\les \int_\mathrm{I} -\bar\nu\Big(|\psi_+|^2+|\psi_-|^2+\frac{|\partial_u\psi_-|^2}{\bar\nu^2}\Big)\, du
        + \int_\mathrm{II} \big(|\psi_+|^2+|\psi_-|^2+|\partial_v\psi_+|^2\big)\,dv
        + \Err^{T} +\ve^3\tau_1^{-7/2+3\delta} ,
    \end{aligned}
\end{equation}
where
\begin{equation}\label{eq:T-cons-err}
    \Err^{T} :=A^3\ve^{3/2} \iint_\mathcal{R} \oline{\Omega^2}\bar r^{-4}(|\Phi_+|^2 +|\Phi_-|^2 + \bar r^2|\partial_v\Phi_+|^2 + \bar r^2|\partial_u\Phi_-|^2 )\,dudv.
\end{equation}
\end{lemma}

In the proof, we follow the strategy in \cite[Proposition 3.9]{ma2022sharp}, which crucially relies on the structure of the Dirac equations in the form \eqref{eq:Dirac-Phi_+}--\eqref{eq:Dirac-Phi_-}. To control the first-order derivatives, we commute the equations with the vector field $\oline{T}:=\partial_v+\partial_u$, which corresponds to the background time vector field. While Killing on Schwarzschild, in the perturbed setting $\oline{T}$ gives rise to a bulk error term, which is later absorbed by Lemma \ref{lem:Mor}. Alternatively, one could use the \textit{Kodama vector field} \cite{kodama1980conserved}
\[
K:=\Big(\frac{1}{\kappa}\p_v+\frac{1}{\gamma}\p_u\Big),
\]
and treat the resulting error terms in an analogous manner. Indeed, error terms such as $\p_u\kappa$ and $\p_v\gamma$ can be controlled by Lemma \ref{lem:pointwise-est-Einstein-eq}, while $\p_u\gamma$ and $\p_v\kappa$ can be expressed in terms of derivatives of the null lapse and thus bounded by Lemma \ref{lem:Omega-expansions}.
We do not pursue this approach here in order to avoid lengthy computations. 

Finally, to obtain full control of the null hypersurface terms, we follow \cite[Proposition 6.1]{angelopoulos2026nonlinear} and introduce an additional "null Lagrangian" current \cite{christodoulou2000action}. This allows us to recover the missing control of $\Phi_-$ on $u$-constant hypersurfaces and of $\Phi_+$ on $v$-constant hypersurfaces.

\begin{proof}[Proof of Lemma \ref{lem:T-cons-law}]
    We begin by multiplying \eqref{eq:Dirac-Phi_+} by $\overline{\Phi_+}$ and \eqref{eq:Dirac-Phi_-} by $\overline{\Phi_-}$. By taking the real part and adding the two identities together, one obtains
    \begin{equation}\label{eq:T-cons-aux1}
        \partial_u(|\Phi_+|^2) + \partial_v(|\Phi_-|^2) = 0.
    \end{equation}
    
Let us now commute the Dirac equations \eqref{eq:Dirac-Phi_+}-\eqref{eq:Dirac-Phi_-} with the vector field $\oline{T}=\partial_v+\partial_u$, yielding
    \begin{align*}
        \partial_u(\oline{T}\Phi_+) &=-\frac{\Omega}{r}\big(\partial_v\log\Omega+\partial_u\log\Omega - (\lambda+\nu)r^{-1}\big)\Phi_-
        - \frac{\Omega^2}{r^2}\Phi_+ - \frac{\Omega}{r}\partial_u\Phi_- ,\\
        \partial_v(\oline{T}\Phi_-) &= \frac{\Omega}{r}\big(\partial_v\log\Omega+\partial_u\log\Omega - (\lambda+\nu)r^{-1}\big)\Phi_+
        - \frac{\Omega^2}{r^2}\Phi_- + \frac{\Omega}{r}\partial_v\Phi_+.
    \end{align*}
Similarly as before, we multiply by $\oline{\oline{T}\Phi_\pm}$, take the real parts and sum the two resulting equations. By using that $\oline{T}\Phi_+=\p_v\Phi_+-(\Omega/r)\Phi_-$ and $\oline{T}\Phi_-=\p_u\Phi_-+(\Omega/r)\Phi_+$, we find
\begin{equation}\label{eq:T-cons-aux2}
    \frac12\big(\partial_u(|\oline{T}\Phi_+|^2)+\partial_v(|\oline{T}\Phi_-|^2)\big)= \mathcal{A}\Big(\Re[\oline{\Phi_+}\partial_u\Phi_-]-\Re[\oline{\Phi_-}\partial_v\Phi_+] +\frac{\Omega}{r}(|\Phi_-|^2 +|\Phi_+|^2)\Big),
\end{equation}
where
\[
    \mathcal{A}:=\frac{\Omega}{r}\Big(\partial_v\log\Omega+\partial_u\log\Omega - (\lambda+\nu)r^{-1}\Big).
\]
Recalling \eqref{eq:Omega-bound}, \eqref{eq:partial_v_Omega-expansion2}, \eqref{eq:partial_u_Omega-expansion}, and the fact that
\[
|\lambda+\nu| = |(1-\bar\mu)(\kappa-\nu/\bar\nu)+(\wtilde{1-\mu})\kappa|\les A^3\ve^{3/2}\tau^{-1+\delta},
\]
we have
\[
    |\mathcal{A}|\les A^3\ve^{3/2}(-\bar\nu)^{1/2}\bar r^{-2} \tau^{-3/4+\delta}.
\]
Thus, we integrate \eqref{eq:T-cons-aux2} over $\mathcal{R}$, and treat the RHS as an error term. For instance, we bound the first term in the right-hand side as follows
\begin{align*}
    \iint_\R |\mathcal{A}||\Phi_+||\p_u\Phi_-|\,dudv&\les A^3\ve^{3/2}\Big(\iint_{\R_{\le\Lambda}} \big(-\bar\nu|\Phi_+|^2+\tau^{-3/2+2\delta}|\p_u\Phi_-|^2\big)\,dudv \\&+ \iint_{\R_{\ge\Lambda}}\big(\bar r^{-4}|\p_u\Phi_-|^2 + \tau^{-3/2+2\delta}|\Phi_+|^2\big)\,dudv\Big)
    \\&\les A^3\ve^{3/2}\Big(\iint_{\R_{\le\Lambda}} -\bar\nu\big(|\Phi_+|^2+\tau^{-3/2+2\delta}(|\psi_-|^2+|\p_u\psi_-|^2)\big)\,dudv \\&+ \iint_{\R_{\ge\Lambda}}\big(\bar r^{-4}|\p_u\Phi_-|^2 + \tau^{-3/2+2\delta}|\psi_+|^2\big)\,dudv\Big)
    \\&\les \Err^T + A^3\ve^{3/2} \Big( \int^{\tau_2}_{\tau_1} \tau^{-3/2+2\delta}\underline{\mathcal{E}}(\tau)\,d\tau + \int^{\tau_2}_{\tau_1}\tau^{-3/2+2\delta} \mathcal{E}_0(\tau)\,d\tau\Big) \\& \les \Err^T + \ve^3\tau_1^{-7/2+3\delta},
\end{align*}
by the bootstrap assumptions \eqref{eq:boot-ing-energy}, \eqref{eq:boot-outg-energy} and the definition of $\Err^T$ in \eqref{eq:T-cons-err}.
In this manner, we obtain
\begin{align*}
    \Big|\iint_\mathcal{R} &\text{(RHS of \eqref{eq:T-cons-aux2})}\,dudv\Big|\les A^3\ve^{3/2}\Big(\iint_\mathcal{R} -\bar\nu\bar r^{-4}(|\Phi_+|^2 +|\Phi_-|^2 + r^2|\partial_v\Phi_+|^2 + r^2|\partial_u\Phi_-|^2 )\,dudv \\&+ \iint_{\R_{\le\Lambda}}\tau^{-3/2+2\delta}(-\bar\nu)(|\psi_-|^2+|\partial_u\psi_-|^2+|\psi_+|^2)\,dudv
    +\iint_{\R_{\ge\Lambda}} \tau^{-3/2+2\delta}(|\psi_+|^2 + |\psi_-|^2 )\,dudv\Big)\\
    &\les \Err^T + \ve^3\tau_1^{-7/2+3\delta}.
\end{align*}

Next, we consider the trivial identity $\partial_u\partial_v(\oline\sigma |\Phi_\pm|^2) - \partial_v\partial_u(\oline\sigma |\Phi_\pm|^2)=0$ with $\oline\sigma:=\oline{\Omega^2}=1-2M/\bar r$, which writes  
\begin{equation}\label{eq:T-cons-aux3}
\begin{aligned}
    &\partial_u \Big( 2 M\bar\lambda \bar r^{-2} \big(|\Phi_+|^2+|\Phi_-|^2\big) + 2\oline{\Omega^2}\big(\Re[\overline{\Phi_+}\partial_v\Phi_+] + \Omega r^{-1}\Re[\overline{\Phi_-}\Phi_+]\big)\Big)
    \\
    &- \partial_v \Big( 2 M\bar\nu \bar r^{-2} \big(|\Phi_+|^2+|\Phi_-|^2\big) + 2\oline{\Omega^2}\big(\Re[\overline{\Phi_-}\partial_u\Phi_-] -\Omega r^{-1}\Re[\overline{\Phi_-}\Phi_+]\big)\Big)= 0.
\end{aligned}
\end{equation}
Finally, adding \eqref{eq:T-cons-aux1}--\eqref{eq:T-cons-aux2}--\eqref{eq:T-cons-aux3} yields
\begin{align*}
    \iint_\mathcal{R}\partial_u \Big( |\Phi_+|^2 + \oline{\Omega^2}\bar r^{-2} |\Phi_-|^2 + |\partial_v\Phi_+|^2\Big) +\partial_v \Big( \oline{\Omega^2}\bar r^{-2} |\Phi_+|^2 + |\Phi_-|^2 +|\partial_u\Phi_-|^2\Big)\les \Err^T + \ve^3\tau_1^{-7/2+3\delta}.
\end{align*}
Recalling the definition \eqref{def:Phi_pm} relating $\Phi_\pm$ to $\psi_\pm$, the statement follows.
\end{proof}

\begin{figure}
\centering
\begin{tikzpicture}[line cap=rect,inner sep=0pt, x=1cm, y=1cm, scale=0.75,font=\small]

\coordinate  (I)   at (0,2)      {};
\coordinate (II)   at (2,0)      {};
\coordinate  (III)   at (4,2)      {};
\coordinate   (IV)   at (2,4)      {};

\path[name path=parab1, domain=0:4, samples=100, smooth, variable=\y]
    plot (\y,{-3/40*\y*\y+3/10*\y+2});

\fill[lightgray] (I) -- (II) -- (III) -- (IV);
%\fill[gray] (V) -- (II) -- (IX) -- (VII);
%\fill[gray] plot[domain=0:1, smooth] (\y,{-3/40*\y*\y+3/10*\y+2}) -- (IV) -- (III) --(I);

\draw[line width=1pt]  (I) 
-- node[pos=0.5, below=2pt, sloped]    {$\mathrm{i}$}
       (II) -- node[pos=0.5, below=2pt, sloped]    {$\mathrm{ii}$} (III) -- node[pos=0.5, above=2pt, sloped]    {$\mathrm{iii}$}
       (IV) -- node[pos=0.5, above=2pt, sloped]    {$\mathrm{iv}$} (I);

\draw[line width=1pt,domain=0:4, smooth, variable=\y] plot ({2 + 3/40*\y*(4-\y)}, {\y});

\node at ($(II)+(0.1,2)$)      {$\Gamma$};
\node at ($(II)+(-1,0)$)      {$(u_1,v_1)$};
\node at ($(IV)+(-1,0)$)      {$(u_2,v_2)$};
\filldraw[draw=black] (II) circle (2pt);
\filldraw[draw=black] (IV) circle (2pt);

\end{tikzpicture}

\caption{A Penrose diagram of the region considered in Lemma \ref{lem:T-cons-law-bis}.}
\label{fig:Penrose8-energydomain}
\end{figure}
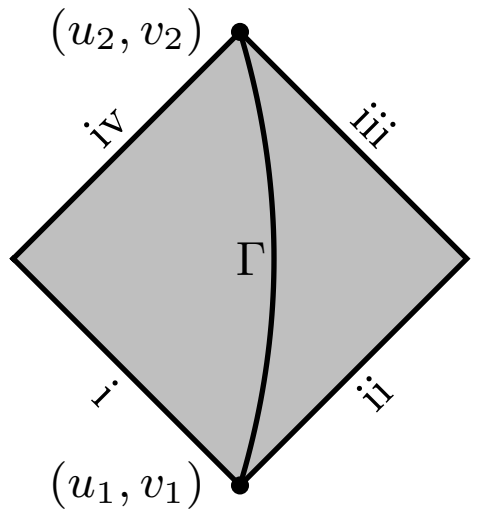

With the same argument, one also proves the following.
\begin{lemma}[Degenerate energy estimate -- II]\label{lem:T-cons-law-bis}
With hypotheses as in Proposition \ref{prop:T+Mor}, it holds that
\begin{equation}\label{eq:T-cons-bis}
    \begin{aligned}
        &\int_\mathrm{iii} \big( r^{-2}|\psi_+|^2 + r^2|\psi_-|^2 + r^2|\partial_u\psi_-|^2\big)\, du 
        + \int_\mathrm{iv} \big(|\psi_+|^2 +\bar\nu^2|\psi_-|^2+|\partial_v\psi_+|^2\big)\,dv 
        \\&\les \int_\mathrm{i} -\bar\nu\Big(|\psi_+|^2+|\psi_-|^2+\frac{|\partial_u\psi_-|^2}{\bar\nu^2}\Big)\, du
        + \int_\mathrm{ii} \big(|\psi_+|^2+|\psi_-|^2+|\partial_v\psi_+|^2\big)\,dv
        + \Err^{T} +\ve^3\tau_1^{-7/2+3\delta} ,
    \end{aligned}
\end{equation}
where $\Err^{T}$ is as in Lemma \ref{lem:T-cons-law} and the hypersurfaces $\mathrm{i}$--$\mathrm{iv}$ are as depicted in Figure \ref{fig:Penrose8-energydomain}.
\end{lemma}

As a second step, we now prove the estimate that allows us to recover a globally coercive bulk term. 
\begin{lemma}[Morawetz estimate]\label{lem:Mor} With hypotheses as in Proposition \ref{prop:T+Mor}, it holds that
    \begin{equation}\label{eq:Mor}
        \iint_\mathcal{R} -\bar\nu \,\bar r^{-4} (|\psi_+|^2 + \bar r^2 |\partial_v\psi_+|^2) + \bar\nu^2\bar r^{-2}(|\psi_-|^2 + \bar r^2|\partial_u\psi_-|^2 )\,dudv \les (\text{LHS of \eqref{eq:T-cons})} + \ve^3\tau_1^{-7/2+3\delta}.
    \end{equation}
\end{lemma}

As noted in Remark \ref{rk:Dirac-Phi-Schwarzschild}, the symmetry between the equations for the two Dirac components is crucial for establishing the Morawetz estimate in the linear case. Our strategy is therefore to regard the equations for $\Phi_\pm$ as perturbations of their Schwarzschild counterparts, combining the pointwise estimates of the previous section and the bootstrap assumptions to control the resulting error terms. This allows us to adapt the multiplier used in \cite[Theorem 4.1]{ma2022sharp} to our double-null gauge and carry out an analogous analysis of the bulk terms. 

\begin{proof}[Proof of Lemma \ref{lem:Mor}] Thanks to \eqref{eq:Omega-bound}, \eqref{eq:partial_v_Omega-expansion2}, \eqref{eq:partial_u_Omega-expansion} and the geometric estimates established in Section \ref{sec:improvement-geometry}, we can linearize equations \eqref{eq:wave-Phi_+}--\eqref{eq:wave-Phi_-} as follows
\begin{equation}\label{eq:dirac-system}
\begin{aligned}
    \partial_u\partial_v\Phi_\pm + \Big(\frac{\overline{\Omega^2}}{\bar r^2} + \Err^{\pm\pm}\Big)\Phi_\pm + \Big(\frac{\overline{\Omega}(3M-\bar r)}{\bar r^3}+\Err^{\pm\mp}\Big)\Phi_\mp &=0,
\end{aligned}
\end{equation}
where
\begin{align}
    |\Err^{\pm\pm}| \les A^3\ve^{3/2}\oline{\Omega^2} \bar r^{-2} \tau^{-1+\delta},\quad
    |\Err^{\pm\mp}|\les A^3\ve^{3/2}\oline{\Omega} \bar r^{-2} \tau^{-3/4+\delta}.
\end{align}
Let us apply to \eqref{eq:dirac-system} the following Morawetz multiplier
\[
    \mathfrak{X}\Phi_\pm := f (\partial_v-\partial_u)\Phi_\pm + q\Phi_\pm,
\]
where $f=f(\bar r)$, $q=q(\bar r)$ to be chosen, and sum the two resulting equations. We proceed by computing explicitly all the terms:
\begin{equation}\label{eq:Morawetz1}
\begin{aligned}
    \Re[ \overline{\mathfrak{X}\Phi_+} \partial_u\partial_v\Phi_+]&= \partial_u\Big(\frac{f}{2}|\partial_v\Phi_+|^2 + q\Re[\oline{\Phi_+}\partial_v\Phi_+] \Big) - \partial_v\Big(\frac{f\Omega^2}{2r^2}|\Phi_-|^2 - \frac{\oline{\Omega^2}q'}{2}|\Phi_+|^2\Big)\\
    &+ \frac{\oline{\Omega^2}f'}{2}\Big(|\partial_v\Phi_+|^2+\frac{\Omega^2}{r^2}|\Phi_-|^2\Big)-\frac{\oline{\Omega^2}(\oline{\Omega^2}q')'}{2}|\Phi_+|^2 + \frac{q\Omega}{r}\Re[\oline{\Phi_-}\partial_v\Phi_+],\\
%\end{align*}
%\begin{align*}
    \Re[ \overline{\mathfrak{X}\Phi_-} \partial_u\partial_v\Phi_-]&= \partial_u\Big(\frac{f\Omega^2}{2r^2}|\Phi_+|^2 -\frac{\oline{\Omega^2}q'}{2}|\Phi_-|^2 \Big) - \partial_v\Big(\frac{f}{2}|\partial_u\Phi_-|^2 - q\Re[\oline{\Phi_-}\partial_u\Phi_-] \Big)\\
    &+ \frac{\oline{\Omega^2}f'}{2}\Big(|\partial_u\Phi_-|^2+\frac{\Omega^2}{r^2}|\Phi_+|^2\Big)-\frac{\oline{\Omega^2}(\oline{\Omega^2}q')'}{2}|\Phi_-|^2 - \frac{q\Omega}{r}\Re[\oline{\Phi_+}\partial_u\Phi_-].
\end{aligned}
\end{equation}
Next,
\begin{equation}\label{eq:Morawetz2}
\begin{aligned}          
    \sum_\pm\Big(\frac{\overline{\Omega^2}}{\bar r^2} + \Err^{\pm\pm}\Big) \Re[ \overline{\mathfrak{X}\Phi_\pm} \Phi_\pm]&=\sum_\pm \partial_v\Big( \frac{\oline{\Omega^2}f}{2\bar r^2}|\Phi_\pm|^2 \Big)-\partial_u\Big( \frac{\oline{\Omega^2}f}{2 \bar r^2}|\Phi_\pm|^2 \Big) + \Big( -\oline{\Omega^2} \Big(\frac{\oline{\Omega^2}f}{\bar r^2}\Big)'+\frac{\oline{\Omega^2}q}{\bar r^2}\Big)|\Phi_\pm|^2 \\&+ \Err^{\pm\pm}\Re[ \overline{\mathfrak{X}\Phi_\pm} \Phi_\pm].
\end{aligned}
\end{equation}
Regarding the last term, we compute
\begin{align}
    \nonumber
    \sum_\pm\Big(\frac{\overline{\Omega}(3M-\bar r)}{\bar r^3}&+ \Err^{\pm\mp}\Big) \Re[ \overline{\mathfrak{X}\Phi_\pm} \Phi_\mp] 
    \\\nonumber
    &= \frac{\overline{\Omega}(3M-\bar r)}{\bar r^3}\Big[f\Big(\Re[\oline{\partial_v\Phi_+} \Phi_--\oline{\partial_u\Phi_-}\Phi_+]+\frac{\Omega}{r}(|\Phi_-|^2+|\Phi_+|^2)\Big) + 2q\Re[\oline{\Phi_-}\Phi_+]\Big] \\\nonumber
    &+ \Err^{+-}\Re[\oline{\mathfrak{X}\Phi_+}\Phi_-]+ \Err^{-+}\Re[\oline{\mathfrak{X}\Phi_-}\Phi_+]
%\end{align*}
%\begin{equation*}
    %\sum_\pm\Big(\frac{\overline{\Omega}(3M-\bar r)}{\bar r^3}+ \Err^{\pm\mp}\Big) \Re[ \overline{\mathfrak{X}\Phi_\pm} \Phi_\mp] 
    \\\nonumber
    &= \partial_v\Big(\frac{f\overline{\Omega}(3M-\bar r)}{\bar r^3}\Re[\oline{\Phi_+}\Phi_-]\Big)-\partial_u\Big(\frac{f\overline{\Omega}(3M-\bar r)}{\bar r^3}\Re[\oline{\Phi_+}\Phi_-]\Big) 
    \\\label{eq:Morawetz3}
    &+ 2\Big(-\oline{\Omega^2} \Big(\frac{f\oline{\Omega}(3M-\bar r)}{\bar r^3}\Big)'+ \frac{q\overline{\Omega}(3M-\bar r)}{\bar r^3}\Big)\Re[\oline{\Phi_+}\Phi_-] 
    \\\nonumber
    &+ \Err^{+-}\Re[\oline{\mathfrak{X}\Phi_+}\Phi_-]+ \Err^{-+}\Re[\oline{\mathfrak{X}\Phi_-}\Phi_+],
\end{align}
where we used the Dirac equations \eqref{eq:Dirac-Phi_+}--\eqref{eq:Dirac-Phi_-} to reduce the bulk to total derivatives terms plus one cross contribution $\Re[\oline{\Phi_+}\Phi_-]$.
Denoting
\[
    \mathcal{F}:=2\Big(-\oline{\Omega^2} \Big(\frac{f\oline{\Omega}(3M-\bar r)}{\bar r^3}\Big)'+ \frac{q\overline{\Omega}(3M-\bar r)}{\bar r^3}\Big),
\]
and writing
\[
    1= \frac{\bar r}{\oline{\Omega}}\frac{\Omega}{r}  + \Err,\quad |\Err|\les A^3\ve^{3/2}\tau^{-1+\delta},
\]
we further rewrite the term $\Re[\overline{\Phi_+}\Phi_-]$ in \eqref{eq:Morawetz3} as
\begin{align*}
    \mathcal{F}\Re[\overline{\Phi_+}\Phi_-] &= \mathcal{F}\Big(\frac{\bar r}{\oline{\Omega}}\frac{\Omega}{r} + \Err\Big)\Re[\overline{\Phi_+}\Phi_-] = \frac{\mathcal{F}\bar r}{2\oline{\Omega}}\Big(\Re[\overline{\partial_v\Phi_-}\Phi_-] - \Re[\overline{\Phi_+}\partial_u \Phi_+]\Big) + \mathcal{F}\Err\Re[\overline{\Phi_+}\Phi_-]\\&= \partial_v\Big( \frac{\mathcal{F} \bar r}{4\oline{\Omega}}|\Phi_-|^2 \Big)-\partial_u\Big( \frac{\mathcal{F}\bar r }{4\oline{\Omega}}|\Phi_+|^2 \Big) - \partial_v\Big(\frac{\mathcal{F}\bar r }{4\oline{\Omega}}\Big)|\Phi_-|^2 + \partial_u\Big(\frac{\mathcal{F}\bar r }{4\oline{\Omega}}\Big)|\Phi_+|^2+ \mathcal{F}\Err\Re[\overline{\Phi_+}\Phi_-]\\&=
    \partial_v\Big( \frac{\mathcal{F} \bar r}{4\oline{\Omega}}|\Phi_-|^2 \Big)-\partial_u\Big( \frac{\mathcal{F}\bar r }{4\oline{\Omega}}|\Phi_+|^2 \Big) - \oline{\Omega^2}\Big(\frac{\mathcal{F}\bar r }{4\oline{\Omega}}\Big)'(|\Phi_-|^2 +|\Phi_+|^2) + \mathcal{F}\Err\Re[\overline{\Phi_+}\Phi_-].
\end{align*}
In this manner, we are able to exploit, also in the perturbed setting, Ma and Zhang's trick to transform the mixed term $\Re[\overline{\mathfrak{X}\Phi_\pm} \Phi_\mp]$ into a sum of boundary terms and bulk terms involving $|\Phi_\pm|^2$.

Therefore, combining \eqref{eq:Morawetz1}, \eqref{eq:Morawetz2} and \eqref{eq:Morawetz3} yields
\begin{equation}\label{eq:Morawetz4}
\begin{aligned}
    \frac{\oline{\Omega^2}f'}{2}&\Big[\Big(|\partial_v\Phi_+|^2 + \frac{\Omega^2}{r^2}|\Phi_-|^2 + \frac{2\Omega}{r}\frac{q}{\oline{\Omega^2}f'}\Re[\oline{\Phi_-}\partial_v\Phi_+]\Big)
    + \Big(|\partial_u\Phi_-|^2 + \frac{\Omega^2}{r^2}|\Phi_+|^2 - \frac{2\Omega}{r}\frac{q}{\oline{\Omega^2}f'}\Re[\oline{\Phi_+}\partial_u\Phi_-]\Big) \Big]\\
    &+\oline{\Omega^2}\Big(-\frac{(\oline{\Omega^2}q')'}{2}- \Big(\frac{\oline{\Omega^2}f}{\bar r^2}\Big)'+\frac{q}{\bar r^2} - \Big(\frac{\mathcal{F}\bar r }{4\oline{\Omega}}\Big)'\Big)(|\Phi_+|^2+|\Phi_-|^2) + \mathcal{E}rrors = -\partial_u F_u - \partial_v F_v,
\end{aligned}
\end{equation}
where 
\begin{align*}
    F_u:=&\frac{f}{2}|\partial_v\Phi_+|^2 + \Big[ \frac{f}{2}\Big(\frac{\Omega^2}{r^2}- \frac{\oline{\Omega^2}}{ \bar r^2}\Big) + \frac{\bar r\oline{\Omega}}{2} \Big(\frac{f\oline{\Omega}(3M-\bar r)}{\bar r^3}\Big)'- \frac{q(3M-\bar r)}{\bar 2r^2}\Big]|\Phi_+|^2\\
    -&\frac{\oline{\Omega^2}}{2}(q' + f\bar r^{-2})|\Phi_-|^2 + q\Re[\oline{\Phi_+}\partial_v\Phi_+]-\frac{f\overline{\Omega}(3M-\bar r)}{\bar r^3}\Re[\oline{\Phi_+}\Phi_-],\\
%\end{align*}
%\begin{align*}
    F_v:=&- \frac{f}{2}|\partial_u\Phi_-|^2 + \Big[ \frac{f}{2}\Big(\frac{\oline{\Omega^2}}{\bar r^2}-\frac{\Omega^2}{r^2}\Big)  - \frac{\bar r\oline{\Omega}}{2} \Big(\frac{f\oline{\Omega}(3M-\bar r)}{\bar r^3}\Big)'+ \frac{q(3M-\bar r)}{2\bar r^2} \Big]|\Phi_-|^2 
    \\&+\frac{\oline{\Omega^2}}{2}( q'+ f\bar r^{-2})|\Phi_+|^2+ q\Re[\oline{\Phi_-}\partial_u\Phi_-] + \frac{f\overline{\Omega}(3M-\bar r)}{\bar r^3}\Re[\oline{\Phi_+}\Phi_-],\\
    \mathcal{E}rrors:=&\sum_{\pm,\pm'}\Err^{\pm\pm'}\Re[\oline{\mathfrak{X\Phi_\pm}}\Phi_{\pm'}]+\mathcal{F}\Err\Re[\oline{\Phi_+}\Phi_-].
\end{align*}
By integrating over $\mathcal{R}$ and choosing 
\[
    f:=1-\frac{3M}{\bar r},\quad q:=\oline{\Omega^2}f'=\frac{3M\oline{\Omega^2}}{\bar r^2},
\]
the LHS of \eqref{eq:Morawetz4} becomes
\begin{equation}\label{eq:Morawetz_bulk1}
    \iint_\mathcal{R}\frac{3M\oline{\Omega^2}}{2\bar r^2}\Big[|\partial_v\Phi_+ - \partial_u\Phi_+|^2 + |\partial_v\Phi_- - \partial_u\Phi_-|^2 \Big] + \frac{M\oline{\Omega^2}}{\bar r^6}(9\bar r^2 - 48M\bar r + \frac{135}{2}M^2)(|\Phi_+|^2+|\Phi_-|^2)\,dudv,
\end{equation}
which is strictly positive on $\mathcal{D}_{\tau_f}$.
On the other hand, the flux terms satisfy
\begin{align*}
    |F_u| & \les r^{-2}|\Phi_+|^2 - \nu r^{-2}|\Phi_-|^2 + |\partial_v\Phi_+|^2\les r^{-2}|\psi_+|^2 + \nu^2|\psi_-|^2 + |\partial_v\psi_+|^2 ,\\
    |F_v| &\les \Omega^2 r^{-2}|\Phi_+|^2 + r^{-2}|\Phi_-|^2 + |\partial_u\Phi_-|^2 \les -\bar\nu( r^{-2}|\psi_+|^2 + |\psi_-|^2 + r^2|\partial_u\psi_-|^2 ),
\end{align*}
and thus are bounded by the LHS of \eqref{eq:T-cons}.

On the other hand, by choosing 
\[
    f:=1-\frac{3M}{\bar r},\quad q:=-\oline{\Omega^2}f'=-\frac{3M\oline{\Omega^2}}{\bar r^2},
\]
the LHS of \eqref{eq:Morawetz4} becomes
\begin{equation}\label{eq:Morawetz_bulk2}
    \iint_\mathcal{R} \frac{3M\oline{\Omega^2}}{2\bar r^2}\Big[|\partial_v\Phi_+ + \partial_u\Phi_+|^2 + |\partial_v\Phi_- + \partial_u\Phi_-|^2 \Big] + C(\bar r)(|\Phi_+|^2+|\Phi_-|^2)\,dudv,
\end{equation}
where 
\[
    |C(\bar r)|=\Big|\frac{3M\oline{\Omega^2}}{2\bar r^6}(20\bar r^2-144M\bar r+225M^2)\Big|\les \oline{\Omega^2}\bar r^{-4},
\]
with boundary terms analogous to those above. Thus, combining \eqref{eq:Morawetz_bulk1} and \eqref{eq:Morawetz_bulk2}, we obtain % just take a small multiple of the second one to absorb the eventual indefinite coefficient C(bar r)
\[
    \iint_\mathcal{R} \oline{\Omega^2}\bar r^{-4} (|\Phi_+|^2 +|\Phi_-|^2+\bar r^2 |\partial_v\Phi_+ |^2 + \bar r^2 |\partial_u\Phi_- |^2)\,dudv + \mathcal{E}rrors \les (\text{LHS of \eqref{eq:T-cons})}.
\]

We conclude the proof by briefly discussing the error terms $\mathcal{E}rrors$ in the same fashion as in the proof of Lemma \ref{lem:T-cons-law}. Therefore, we limit ourselves to treat explicitly only $\Err^{-+}$, since the remaining error terms are analogous or simpler.  Thus, one obtains
\begin{align*}
    |\Err^{-+}\Re[ \overline{\mathfrak{X}\Phi_-} \Phi_+]|&\les |\Err^{-+}|(\Omega r^{-1}|\Phi_+|^2 + |\partial_u\Phi_-||\Phi_+| +\Omega^2 r^{-2}|\Phi_-||\Phi_+|)\\
    &\les A^3\ve^{3/2}\Big(\mathds{1}_{r\le\Lambda}\big(\oline{\Omega^2}(|\Phi_+|^2 + |\Phi_-|^2) + \tau^{-3/2+2\delta}|\partial_u\Phi_-|^2\big)
    \\&+\mathds{1}_{r\ge\Lambda}\big( r^{-4}( |\Phi_+|^2 + |\Phi_-|^2+r^{2}|\partial_u\Phi_-|^2 )+\tau^{-2+2\delta}|\Phi_+|^2\big) \Big).
\end{align*}
Noting that, with the choices $f\sim 1$ and $q\sim \oline{\Omega^2}\bar r^{-2}$, the coefficient $\mathcal{F}$ appearing in $\Err$ satisfies $\mathcal{F}\sim \oline{\Omega}\bar r^{-3}$, one finds
\begin{align*}
    \iint_\mathcal{R}\mathcal{E}rrors\,dudv&\les A^3\ve^{3/2}\iint_\mathcal{R} \Big(\oline{\Omega^2}\bar r^{-4}(|\Phi_+|^2 +|\Phi_-|^2 + r^2|\partial_v\Phi_+|^2 + r^2|\partial_u\Phi_-|^2 ) \\&+ \mathds{1}_{r\le\Lambda}\tau^{-3/2+2\delta}(|\partial_u\Phi_-|^2+|\Phi_-|^2)
    +\mathds{1}_{r\ge\Lambda}  \tau^{-2+2\delta}(|\Phi_+|^2 + |\partial_v\Phi_+|^2 +r^{-2}|\Phi_-|^2 )\Big)dudv,
\end{align*}
where the first term can be absorbed into the main coercive bulk term for $\ve$ small enough, while the remaining contributions are controlled using the bootstrap assumptions \eqref{eq:boot-ing-energy}--\eqref{eq:boot-outg-energy}:
\begin{align*}
    A^3\ve^{3/2} \iint_{\mathcal{R}}\Big(\mathds{1}_{r\le\Lambda}\tau^{-3/2+2\delta}(|\partial_u\Phi_-|^2&+|\Phi_-|^2)
    +\mathds{1}_{r\ge\Lambda}  \tau^{-2+2\delta}(|\Phi_+|^2 + |\partial_v\Phi_+|^2 +r^{-2}|\Phi_-|^2 )\Big)dudv\\
    &\les A^3\ve^{3/2} \Big( \int^{\tau_2}_{\tau_1} \tau^{-3/2+2\delta}\underline{\mathcal{E}}(\tau)\,d\tau + \int^{\tau_2}_{\tau_1}\tau^{-2+2\delta} \mathcal{E}_0(\tau)\,d\tau\Big) \les \ve^3\tau_1^{-7/2+3\delta}.
\end{align*}
\end{proof}

\begin{proof}[Proof of Proposition \ref{prop:T+Mor}]
    The result follows by adding \eqref{eq:T-cons} and \eqref{eq:Mor} together. Indeed, for $\ve$ suitably small, one can absorb the bulk error $\Err^T$ appearing in the RHS of \eqref{eq:T-cons} into the strictly coercive LHS of \eqref{eq:Mor}.
\end{proof}

\subsection{Red-shift estimate near the horizon}\label{sec:red-shift}
As illustrated in the previous proposition, the ILED estimate \eqref{eq:T+Mor} degenerates towards the event horizon. The following red-shift estimate allows us to remove this degeneracy.
\begin{proposition}\label{prop:red-shift} For any $A\ge 1$, $\ve$ sufficiently small, $\tau_f\in\mathfrak B$, $(u_1,v_1),(u_2,v_2)\in\Gamma\cap\mathcal D_{\tau_f}$,  $u_2'>u_2$, and $v_2'>v_2$,  where $(u_2,v_2)$ is to the future of $(u_1,v_1)$, let $r_1:=2.5M$ and $r_2:=2.7M$. Then, it holds
\begin{equation}\label{est:red-shift-est}
\begin{aligned}
    \int_{\mathrm{VI}_{\le r_1}} &\left( |\psi_-|^2 + |\psi_+|^2 +|\partial_v \psi_+|^2\right) dv 
    + 
    \int_{\mathrm{V}_{\le r_1}} -\bar\nu\left(|\psi_-|^2  +|\psi_+|^2 + \frac{|\partial_u\psi_-|^2}{\bar\nu^2}\right)du 
    \\
    &+ \iint_{\mathcal{R}_{\le r_1}}-\bar\nu\left( |\psi_-|^2 + |\psi_+|^2 + |\partial_v\psi_+|^2 + \frac{|\partial_u\psi_-|^2}{\bar\nu^2} \right)dudv
    \\
    &\les \int_\mathrm{I} -\bar\nu\Big(|\psi_-|^2+|\psi_+|^2+\frac{|\partial_u\psi_-|^2}{\bar\nu^2}\Big)\, du
    + \int_\mathrm{II} \big(|\psi_+|^2+|\psi_-|^2+|\partial_v\psi_+|^2\big)\,dv
    +\ve^3\tau_1^{-7/2+3\delta}.
\end{aligned}
\end{equation}
\end{proposition}

The proof of this estimate will be given in the next section after a series of lemmas. 

\subsubsection{Proof of the red-shift estimate}\label{subsubsec:proof-red-shift} 

To improve the ILED estimate, we exploit once again the coupled system satisfied by the spinorial components $\Phi_\pm$, \eqref{eq:wave-Phi_+}--\eqref{eq:wave-Phi_-}. In contrast to Lemma \ref{lem:Mor}, where the coupling between the two components plays a crucial role, the red-shift mechanism can be initiated from an equation for a single component. To this end, we introduce a suitable reweighting $\wtilde{\Phi}_-$ of $\Phi_-$, which satisfies a decoupled equation up to nonlinear error terms. This allows us first to establish a red-shift estimate for $\widetilde{\Phi}_-$ and then to obtain a control for $\Phi_+$ in a second step. Since $\wtilde{\Phi}_-$ and $\Phi_+$ are both solutions to an equation with structure as in \eqref{eq:rs-prototype} below, we start by proving the following red-shift identities.
 
\begin{lemma}[Near-horizon identities]\label{lem:red-shift-ids} Let $\mathcal{O}$ be an open set such that $\overline{\mathcal{O}}\subseteq\mathcal{D}_{\tau_f}$ and $\varphi\in C^2(\overline{\mathcal{O}})$ be a solution to
\begin{equation}\label{eq:rs-prototype}
    \partial_u \partial_v \varphi + b_u \partial_u \varphi + b_0 \varphi = 0,
\end{equation}
where the coefficients $b_u\in C^0(\mathcal{O})$, $b_0\in C^1(\mathcal{O})$. Let $F$ and $\chi$ be $C^1$ functions on $(0,\infty)$. Denoting $\chi=\chi(r)$ and $\chi'= \frac{d}{dr}\chi(r)$, the followings hold:
\begin{equation}\label{eq:red-shift-id1}
\begin{aligned}
    \partial_u \Big(\frac{\chi F b_0}{2(-\nu)}|\varphi|^2\Big)
    &+ \partial_v \Big( \frac{\chi F}{2(-\nu)}|\partial_u \varphi|^2\Big)  
    + \frac{\chi |\varphi|^2}{2\nu}  \Big[ b_0  \partial_u F+ F\Big(\partial_u b_0 - \frac{b_0\partial_u\nu}{\nu}\Big)\Big]
    \\&+  \Big[ F\Big( b_u + \frac{\partial_v \nu}{2\nu}\Big) - \frac{\partial_v F}{2}
    \Big]\frac{\chi|\partial_u \varphi|^2}{-\nu}
    = - \frac{\chi' F}{2} \Big( b_0 |\varphi|^2 
    + \frac{ \lambda}{\nu} |\partial_u \varphi|^2\Big),
\end{aligned}
\end{equation}
and
\begin{equation}\label{eq:red-shift-id2}
	\begin{aligned}
		&\partial_u \Big( \frac{\chi r^2}{2}|\partial_v \varphi|^2\Big)+\partial_v \Big(\frac{\chi r^2 b_0}{2}|\varphi|^2\Big) + \chi(-\nu)r|\partial_v\varphi|^2 +\chi r^2 b_u \Re[\oline{\partial_v\varphi}\partial_u\varphi]
		\\
		&= -  \partial_u \Big(\frac{\chi r^2 b_0}{2}|\varphi|^2\Big) - \partial_v \Big( \frac{\chi r^2}{2}|\partial_u \varphi|^2\Big) + \Big(\partial_v\Big( \frac{\chi r^2}{2}\Big) -\chi r^2 b_u\Big)|\partial_u\varphi|^2 \\&+ \chi' \frac{\nu r^2}{2}|\partial_v\varphi|^2 + (\partial_u+\partial_v)\Big( \frac{\chi r^2 b_0}{2}\Big)|\varphi|^2.
	\end{aligned}
\end{equation}
\end{lemma}
\begin{proof} We start by multiplying \eqref{eq:rs-prototype} by $\chi F(-\nu)^{-1}\overline{\partial_u\varphi}$ and take the real part. Computing separately, we obtain
\begin{align*}
    \frac{\chi F}{-\nu}\Re[\overline{\partial_u\varphi}\partial_v \partial_u \varphi] &= \partial_v \Big( \frac{\chi F}{2(-\nu)}|\partial_u \varphi|^2\Big) - \partial_v\Big( \frac{\chi F}{2(-\nu)}\Big)|\partial_u\varphi|^2,\\
    \frac{\chi F}{-\nu}\Re[\overline{\partial_u\varphi}b_u \partial_u \varphi] &=  \frac{\chi F b_u}{(-\nu)}|\partial_u \varphi|^2,\\
    \frac{\chi F}{-\nu}\Re[\overline{\partial_u\varphi}b_0\varphi] &= \partial_u \Big(\frac{\chi F b_0}{2(-\nu)}|\varphi|^2\Big) - \partial_u\Big( \frac{\chi F b_0}{2(-\nu)}\Big)|\varphi|^2.
\end{align*}
Expanding the last term of the first and third lines by the product rule and rearranging terms, we arrive at \eqref{eq:red-shift-id1}. 

The proof of \eqref{eq:red-shift-id2} follows similarly by considering $\chi r^2 \oline{\oline{T}\varphi}=\chi r^2(\oline{\partial_v\varphi}+\oline{\partial_u\varphi})$. With analogous computations, one obtains
\begin{align*}
	\chi r^2\Re[\oline{T\varphi}\partial_v \partial_u \varphi] &= \partial_v \Big( \frac{\chi r^2}{2}|\partial_u \varphi|^2\Big)+\partial_u \Big( \frac{\chi r^2}{2}|\partial_v \varphi|^2\Big) - \partial_v\Big( \frac{\chi r^2}{2}\Big)|\partial_u\varphi|^2- \partial_u\Big( \frac{\chi r^2}{2}\Big)|\partial_v\varphi|^2,\\
	\chi r^2\Re[\oline{T\varphi}b_u \partial_u \varphi] &=  \chi r^2 b_u (|\partial_u \varphi|^2 + \Re[\oline{\partial_v\varphi}\partial_u\varphi]),\\
	\chi r^2\Re[\overline{\partial_u\varphi}b_0\varphi] &= \partial_u \Big(\frac{\chi r^2 b_0}{2}|\varphi|^2\Big)+\partial_v \Big(\frac{\chi r^2 b_0}{2}|\varphi|^2\Big) - (\partial_u+\partial_v)\Big( \frac{\chi r^2 b_0}{2}\Big)|\varphi|^2.
\end{align*}
Rearranging the terms yields \eqref{eq:red-shift-id2}. 
\end{proof}

We are now in position to state the main ingredient for the proof of Proposition \ref{prop:red-shift}.
\begin{lemma}\label{lem:red-shift} With hypotheses as in Proposition \ref{prop:red-shift}, it holds that
\begin{equation}\label{est:red-shift1}
    \begin{aligned}
    &\int_{\mathrm{VI}_{\le r_1}} (|\psi_-|^2+ |\psi_+|^2)\,dv
    + 
    \int_{\mathrm{V}_{\le r_1}} \Big(-\bar\nu|\psi_-|^2 + \frac{|\partial_u\psi_-|^2}{-\bar\nu}\Big)du 
    \\&+ 
    \iint_{\mathcal{R}_{\le r_1}}
    \Big( -\bar\nu|\psi_-|^2 - \bar\nu|\psi_+|^2 +\frac{|\partial_u\psi_-|^2}{-\bar \nu}\Big)dudv
    \\
    &\les 
    \int_{\mathrm{I}} \Big(-\bar\nu|\psi_-|^2 + \frac{|\partial_u\psi_-|^2}{-\bar\nu}\Big)du
    +
    \iint_{\mathcal{R}_{r_1,r_2}} \Big( -\bar\nu|\psi_-|^2 -\bar\nu|\psi_+|^2 + \frac{|\partial_u\psi_-|^2}{-\bar\nu}\Big)dudv 
    + \ve^3 \tau^{-7/2+3\delta}_1.
    \end{aligned}
\end{equation}    
\end{lemma}

\begin{proof}\textsc{Step 1.} Let $\chi_{\mathcal{H}^+}$ be a smooth non-negative function, with $0\le\chi\le 1$, which equals $1$ on $\{r\le r_1\}$ and $0$ on $\{r\ge r_2\}$. We note that the sets $\{r=r_1\}$ and $\{r=r_2\}$ are both timelike curves in $\D_{\tau_f}$ by the same argument as in Lemma \ref{lem:gamma-slope}, and noting that $\lambda|_{\{r=r_i\}}\sim\bar\lambda_{\{r=r_i\}}\sim1$ by \eqref{eq:lambda-tilde}. Let us consider \[\wtilde\Phi_-:=\frac{r}{\Omega}\Phi_-,\] 
which solves 
\begin{equation*}
    \partial_u \partial_v \widetilde\Phi_- + b^-_u \partial_u \widetilde\Phi_- + b^-_0 \widetilde\Phi_- = 0,
\end{equation*}
where 
\begin{equation*}
    b^-_u := \partial_v\log\Omega-\frac{\lambda}{r},\quad b^-_0 := \frac{6\Omega^2 m}{r^3} - 12\pi T_{uv}.
\end{equation*}
Choosing  $\chi=\chi_{\mathcal{H}^+}$ and \[F:=1+C_0 (1-\mu),\] as in \cite[Proposition 3.15]{ma2022sharp}, with $C_0$ to be fixed, we use \eqref{eq:red-shift-id1} and obtain
\begin{align*}
    \partial_u \Big(\frac{\chi_{\mathcal{H}^+}(1+C_0(1-\mu))b^-_0}{2(-\nu)}  |\widetilde\Phi_-|^2\Big) 
    &+
    \partial_v \Big( \frac{\chi_{\mathcal{H}^+} (1+C_0(1-\mu))}{2(-\nu)} |\partial_u\widetilde\Phi_-|^2\Big) 
    -
    \chi_{\mathcal{H}^+}\nu\Big(\mathcal{B}^-_0  |\widetilde\Phi_-|^2 + \mathcal{B}^-_u\frac{|\partial_u\widetilde\Phi_-|^2}{\nu^2}\Big)
    \\&= - \frac{\chi'_{\mathcal{H}^+}(1+C_0(1-\mu))}{2}  \Big( b^-_0 |\widetilde\Phi_-|^2 
    + \frac{\lambda}{\nu}  |\partial_u \widetilde\Phi_-|^2\Big),
\end{align*}
where 
\begin{align*}
    \mathcal{B}^-_0 &:= \mathcal{B}^-_{0,\mathrm{I}} - \mathcal{B}^-_{0,\mathrm{II}}:= \frac{C_0b^-_0}{r(-\nu)}\Big(\frac{ m}{r}+\frac{\partial_u m}{-\nu}\Big) - \frac{(1+C_0(1-\mu))}{2\nu^2}\Big(\partial_u b^-_0 - \frac{b^-_0\partial_u \nu}{\nu}\Big),
    \\
    \mathcal{B}^-_u &:=(1+C_0(1-\mu))\Big( b^-_u + \frac{\partial_v \nu}{2\nu}\Big) - \frac{C_0}{r}\Big(\frac{ m\lambda}{r}-\partial_v m\Big).
\end{align*}

Let us start with the analysis of the boundary terms. We readily compute by \eqref{eq:kappa-1}, \eqref{eq:mu-2}, \eqref{est:T-uv_bound} and the other geometric estimates,
\begin{equation}\label{eq:redshift-aux1}
\begin{aligned}
	\frac{\chi_{\mathcal{H}^+}(1+C_0(1-\mu))b^-_0}{2(-\nu)} &= \chi_{\mathcal{H}^+}\Big( \frac{3M}{\bar r^3}(1+C_0(1-\bar\mu))+C_0A^2\ve^{3/2}\tau^{-2+\delta}\Big),\\
	\frac{\chi_{\mathcal{H}^+} (1+C_0(1-\mu))}{2(-\nu)} &= \frac{\chi_{\mathcal{H}^+}}{-\nu}\Big(\frac12(1+C_0(1-\bar\mu))+C_0A^2\ve^{3/2}\tau^{-2+\delta}\Big),
\end{aligned}
\end{equation}
which are both strictly coercive for $\ve$ sufficiently small.

We now focus on the bulk coefficients. Regarding $\mathcal{B}^-_u$, by \eqref{eq:partial_v_Omega-expansion1}, \eqref{est:partial_uv-r}, \eqref{est:partial_v-varpi}, we have 
\begin{align}
    \nonumber \mathcal{B}^-_u &\sim \Big(1+C_0(1-\bar\mu)\Big)\Big( \frac{4M-\bar r}{\bar r^2}\Big) + C_0A^2\ve^{3/2}\bar \tau^{-2+\delta} - \frac{C_0 M\bar\lambda}{\bar r^2} + C_0A\ve^{3/2}\bar r^{-2}\tau^{-3/2+\delta/2}
    \\\label{eq:redshift-aux2}
    &=\frac{4M-\bar r}{\bar r^2} + C_0\Big(1-\frac{2M}{\bar r}\Big)\frac{3M-\bar r}{\bar r^2} + C_0 A^2\ve^{3/2}\tau^{-3/2+\delta/2},
\end{align}
which is strictly positive for $r\le r_1$ and $\ve$ sufficiently small.
Next, we expand the two summands of $\mathcal{B}^-_0$ separately:
\begin{equation*}
    \mathcal{B}^-_{0,\mathrm{I}} \sim \frac{6C_0M^2}{\bar r^5} + C_0A^2\ve^{3/2}\tau^{-3/2+\delta/2},
\end{equation*}
by \eqref{est:partial_u-varpi} and the estimates established in the previous section;
on the other hand,
\begin{align*}
    \mathcal{B}^-_{0,\mathrm{II}} &= (1+C_0(1-\mu))\Big(\frac{9\kappa m}{r^4} - \frac{6\pi}{\nu^2}(\partial_uT_{uv}-\partial_u\log\Omega^2 T_{uv}) + \frac{3\kappa}{r^3}\frac{\partial_u m}{-\nu} -\frac{12\pi\kappa m}{r^2}\frac{T_{uu}}{\nu^2} + 24\pi^2r\frac{T_{uv}}{-\nu}\frac{T_{uu}}{\nu^2}\Big)\\
    &\sim (1+C_0(1-\bar\mu))\frac{9M}{\bar r^4} + C_0A^2\ve^{3/2}\tau^{-3/2+\delta/2} - 6\pi(1+C_0(1-\bar\mu))\Big[\frac{\partial_uT_{uv}-\partial_u\log\Omega^2 T_{uv}}{\nu^2}\Big],
\end{align*}
by \eqref{est:T-uv_bound}, \eqref{est:T-uu_bound},  \eqref{est:partial_u-varpi}.
Computing explicitly the last term in square brackets, one finds that it is equal to
\begin{equation*}
    \frac{\kappa}{-\nu} \partial_u\Big( \frac{1}{2\pi r^2\kappa^{1/2}}\Im[\overline{\psi_-}\psi_+]\Big)\sim \Big(-1+\frac{\partial_u\kappa}{-\nu}\Big)\Im[\overline{\psi_-}\psi_+] + \Im\Big[\frac{\overline{\partial_u\psi_-}}{-\nu}\psi_+\Big],
\end{equation*}
which can be estimated thanks to Lemmas \ref{lem:pointwise_estimates_dirac1}, \ref{lem:pointwise_estimates_dirac2} and \ref{lem:pointwise-est-Einstein-eq}, yielding
\begin{equation*}
    \mathcal{B}^-_{0,\mathrm{II}}\sim(1+C_0(1-\bar\mu))\frac{9M}{\bar r^4}+ C_0A^2\ve^{3/2}\tau^{-3/2+\delta/2}.
\end{equation*}
Therefore, we find
\begin{equation}\label{eq:redshift-aux3}
    \mathcal{B}^-_0 = \mathcal{B}^-_{0,\mathrm{I}} - \mathcal{B}^-_{0,\mathrm{II}} \sim \frac{3M}{\bar r^5}(C_0(8M-3\bar r) - 3\bar r) + C_0 A^2\ve^{3/2}\tau^{-3/2+\delta/2},
\end{equation}
which is strictly positive in $\{r\le r_1\}$ for $C_0$ large enough (e.g. $C_0=20$) and $\ve$ sufficiently small. 
Therefore, we integrate over $\mathcal{R}$ and find
\begin{equation}\label{est:red-shift-step1}
\begin{aligned}
	\int_{\mathrm{VI}_{\le r_1}} &|\wtilde\Phi_-|^2\,dv
	+ 
	\int_{\mathrm{V}_{\le r_1}} \frac{|\partial_u\wtilde\Phi_-|^2}{-\nu}\,du 
	+ 
	\iint_{\mathcal{R}_{\le r_1}}
	\Big( -\nu|\wtilde\Phi_-|^2 + \frac{|\partial_u\wtilde\Phi_-|^2}{-\nu}\Big)\,dudv
	\\
	&\les 
	\int_{\mathrm{I}} \frac{|\partial_u\wtilde\Phi_-|^2}{-\nu}\,du
	+
	\iint_{\mathcal{R}_{r_1,r_2}}
	\Big( -\nu|\wtilde\Phi_-|^2 + \frac{|\partial_u\wtilde\Phi_-|^2}{-\nu}\Big)\,dudv
	+C_0A^2\ve^{3/2}\Err^{\text{red}},
\end{aligned}
\end{equation}
where the error appearing in \eqref{eq:redshift-aux1}--\eqref{eq:redshift-aux2}--\eqref{eq:redshift-aux3} is explicitly given by
\begin{align*}
    \Err^{\text{red}}:=&\iint_\mathcal{R} \chi_{\mathcal{H}^+} \tau^{-3/2+\delta/2} \Big( -\nu |\widetilde\Phi_-|^2 + \frac{|\partial_u\widetilde\Phi_-|^2}{-\nu}\Big)\,dudv
    +
    \int_{\mathrm{VI}} \chi_{\mathcal{H}^+} \tau^{-2+\delta} |\wtilde\Phi_-|^2\,dv 
     \\+& \int_{\mathrm{V}} \chi_{\mathcal{H}^+}\tau^{-2+\delta}|\partial_u\wtilde\Phi_-|^2\,du 
   +\int_{\mathrm{I}} \chi_{\mathcal{H}^+} \tau^{-2+\delta}|\partial_u\wtilde\Phi_-|^2\,du,
\end{align*}
while the cutoff terms in $\chi'$ have been absorbed into the spacetime bulk in the RHS, being localized in $\{r_1\le r\le r_2\}$.
Thanks to the bootstrap assumption \eqref{eq:boot-ing-energy} and \eqref{eq:boot-outg-flux}, we bound
\begin{align*}
    |\Err^{\text{red}}|&\les
    \iint_{\mathcal{R}_{r\le r_2}} \tau^{-3/2+\delta/2} \Big(-\nu|\psi_-|^2 + \frac{|\partial_u\psi_-|^2}{-\nu}\Big)\,dudv
    + \int_{\mathrm{VI}_{\le r_2}} \tau^{-2+\delta}|\psi_-|^2dv 
    \\&+ 
    \int_{\mathrm{V}_{\le r_2}} \tau^{-2+\delta}\Big(-\nu|\psi_-|^2 + \frac{|\partial_u\psi_-|^2}{-\nu}\Big)du
    +
    \int_{\mathrm{I}_{\le r_2}} \tau^{-2+\delta}\Big(-\nu|\psi_-|^2 + \frac{|\partial_u\psi_-|^2}{-\nu}\Big)du
    \\& \les \int^{\tau_2}_{\tau_1} \tau^{-3/2+\delta/2} \underline{\mathcal{E}}(\tau)\,d\tau + \tau^{-2+\delta}_2 \mathcal{F}(\tau_2)  +  \tau^{-2+\delta}_2\underline{\mathcal{E}}(\tau_2) +\tau^{-2+\delta}_1\underline{\mathcal{E}}(\tau_1) \les A\ve^2\tau^{-7/2+3/2\delta}_1.
\end{align*}

\textsc{Step 2.} Let us now look at the component $\Phi_+$, which is a solution to
\begin{equation}\label{eq:Phi_+-aux}
    \partial_u \partial_v \Phi_+ + b^+_u \partial_u \Phi_+ + b^+_0 \Phi_+ = 0,
\end{equation}
where 
\begin{equation*}
    b^+_u := -\partial_v\log\Omega+\frac{\lambda}{r},\quad b^+_0 := \frac{\Omega^2}{r^2}.
\end{equation*}
Relying again on \eqref{eq:red-shift-id1} with $\chi=\chi_{\mathcal{H}^+}$ and $F:=1+C_0 (1-\mu)$, we obtain
\begin{align*}
	\partial_u \Big(\frac{\chi_{\mathcal{H}^+}(1+C_0(1-\mu))b^+_0}{2(-\nu)}  |\Phi_+|^2\Big) 
	&+
	\partial_v \Big( \frac{\chi_{\mathcal{H}^+} (1+C_0(1-\mu))}{2(-\nu)} |\partial_u\Phi_+|^2\Big) 
	+
	\chi_{\mathcal{H}^+}(-\nu)\Big(\mathcal{B}^+_0  |\Phi_+|^2 + \mathcal{B}^+_u\frac{|\partial_u\Phi_+|^2}{\nu^2}\Big)
	\\&= - \frac{\chi'_{\mathcal{H}^+}(1+C_0(1-\mu))}{2}  \Big( b^+_0 |\Phi_+|^2 
	+ \frac{\lambda}{\nu}  |\partial_u \Phi_+|^2\Big),
\end{align*}
where 
\begin{align*}
    \mathcal{B}^+_0 &:= \mathcal{B}^+_{0,\mathrm{I}} - \mathcal{B}^+_{0,\mathrm{II}}:= \frac{C_0b^+_0}{r(-\nu)}\Big(\frac{ m}{r}+\frac{\partial_u m}{-\nu}\Big) - \frac{(1+C_0(1-\mu))}{2\nu^2}\Big(\partial_u b^+_0 + \frac{b^+_0\partial_u \nu}{-\nu}\Big),
    \\
    \mathcal{B}^+_u &:=(1+C_0(1-\mu))\Big( b^+_u + \frac{\partial_v \nu}{2\nu}\Big) - \frac{C_0}{r}\Big(\frac{ m\lambda}{r}-\partial_v m\Big).
\end{align*}

As before, the boundary terms rewrite
\begin{align*}
	\frac{\chi_{\mathcal{H}^+}(1+C_0(1-\mu))b^+_0}{2(-\nu)} &=\chi_{\mathcal{H}^+}\Big( \frac{1+C_0(1-\bar\mu)}{2\bar r^2}+C_0A^2\ve^{3/2}\tau^{-2+\delta}\Big),\\
	\frac{\chi_{\mathcal{H}^+} (1+C_0(1-\mu))}{2(-\nu)} &= \frac{\chi_{\mathcal{H}^+}}{-\nu}\Big(\frac12(1+C_0(1-\bar\mu))+C_0A^2\ve^{3/2}\tau^{-2+\delta}\Big).
\end{align*}
Proceeding with the bulk terms, we have that
\begin{align*}
    \mathcal{B}^+_u &\sim \Big(1+C_0(1-\bar\mu)\Big)\Big( \frac{\bar r-2M}{\bar r^2}\Big) + C_0A^2\ve^{3/2}\bar \tau^{-2+\delta} - \frac{C_0 M\bar\lambda}{\bar r^2} + C_0A\ve^{3/2}\bar r^{-2}\tau^{-3/2+\delta/2}
    \\
    &=\frac{\bar r-2M}{\bar r^3}(\bar r+ C_0(\bar r-3M)) + C_0A^2\ve^{3/2}\tau^{-3/2+\delta/2},
\end{align*}
which is degenerate towards the horizon.
Therefore, we use the Dirac equation to rewrite $(-\nu)^{-1}|\partial_u\Phi_+|^2 \sim (-\nu)|\psi_-|^2$ and bound this bulk term by the RHS of \eqref{est:red-shift-step1}.
Regarding $\mathcal{B}^+_0$, we estimate as before the two summands separately:
\begin{align*}
    \mathcal{B}^+_{0,\mathrm{I}} &\sim \frac{C_0M}{\bar r^4} + C_0A^2\ve^{3/2}\tau^{-3/2+\delta/2};\\
    \mathcal{B}^+_{0,\mathrm{II}} &= (1+C_0(1-\mu))\Big(\frac{\kappa}{r^3} - \frac{2\pi\kappa}{r}\frac{T_{uu}}{\nu^2}\Big)
    \sim \frac{1+C_0(1-\bar\mu)}{\bar r^3} + C_0A^2\ve^{3/2}\tau^{-3/2+\delta/2}.
\end{align*}
Thus, we find
\begin{equation*}
    \mathcal{B}^+_0 = \mathcal{B}^+_{0,\mathrm{I}} - \mathcal{B}^+_{0,\mathrm{II}} \sim \frac{1}{\bar r^4}(C_0(3M-\bar r) - \bar r) + C_0A^2\ve^{3/2}\tau^{-3/2+\delta/2}.
\end{equation*}
which is strictly positive in $\{r\le r_1\}$ for $C_0$ large enough (e.g. $C_0=20$ as before) and $\ve$ small enough. 

Therefore, we integrate over $\mathcal{R}$ and find
\begin{equation}\label{est:red-shift-step2}
	\begin{aligned}
		\int_{\mathrm{VI}_{\le r_1}} &|\Phi_+|^2\,dv
		+ 
		\int_{\mathrm{V}_{\le r_1}} -\nu |\wtilde\Phi_-|^2\,du 
		+ 
		\iint_{\mathcal{R}_{\le r_1}}
		-\nu|\Phi_+|^2 \,dudv
		\\
		&\les 
		\iint_{\mathcal{R}_{r_1,r_2}}
		-\nu|\Phi_+|^2 \,dudv
		+ \text{(RHS of \eqref{est:red-shift-step1})} + \ve^3\tau^{-7/2+3/2\delta}_1,
	\end{aligned}
\end{equation}
where the error terms are treated as in the previous step. 

Finally, adding \eqref{est:red-shift-step1} and \eqref{est:red-shift-step2}, the proof is completed. Indeed, recalling that $\wtilde\Phi_-=r^2\kappa^{-1/2}\psi_-$ and $\Phi_+=\kappa^{1/2}\psi_+$, the following equivalences hold on $\R_{\le \Lambda}$
\[
|\wtilde\Phi_-|\sim|\psi_-|, \quad |\p_u\wtilde\Phi_-|\sim |\psi_-|+|\p_u\psi_-|,\quad |\Phi_+|\sim|\psi_+|,
\]
thanks to the geometric estimates.
\end{proof}
The next lemma provides control of the remaining terms involving $\psi_+$ by making use of \eqref{eq:red-shift-id2}, thereby completing the proof of Proposition \ref{prop:red-shift}.

\begin{lemma} With hypotheses as in Proposition \ref{prop:red-shift}, it holds that
    \begin{equation}\label{est:red-shift2}
    \begin{aligned}
    &\int_{\mathrm{VI}_{\le r_1}} |\partial_v\psi_+|^2dv
    + 
    \int_{\mathrm{V}_{\le r_1}} -\bar\nu|\psi_+|^2 du 
    + 
    \iint_{\mathcal{R}_{\le r_1}}
    -\bar\nu|\partial_v\psi_+|^2 dudv
    \\
    &\les 
    \int_{\mathrm{I}} -\bar\nu|\psi_+|^2 du
    +
    \iint_{\mathcal{R}_{r_1,r_2}} -\bar\nu|\partial_v\psi_+|^2 dudv
    + \text{(RHS of \eqref{est:red-shift1})}.
    \end{aligned}
\end{equation}  
\end{lemma}
\begin{proof} Since $\Phi_+$ is a solution to \eqref{eq:Phi_+-aux}, we apply identity \eqref{eq:red-shift-id2} with $\chi_{\mathcal{H}^+}$ as in the previous lemma. Thus, we find
\begin{equation*}
	\begin{aligned}
		&\partial_u \Big( \frac{\chi_{\mathcal{H}^+} r^2}{2}|\partial_v \Phi_+|^2\Big)+\partial_v \Big(\frac{\chi_{\mathcal{H}^+} r^2 b^+_0}{2}|\Phi_+|^2\Big) + \chi_{\mathcal{H}^+}(-\nu)r|\partial_v\Phi_+|^2 +\chi_{\mathcal{H}^+} r^2 b^+_u \Re[\oline{\partial_v\Phi_+}\partial_u\Phi_+]
		\\
		&= -  \partial_u \Big(\frac{\chi_{\mathcal{H}^+} r^2 b^+_0}{2}|\Phi_+|^2\Big) - \partial_v \Big( \frac{\chi_{\mathcal{H}^+} r^2}{2}|\partial_u \Phi_+|^2\Big) + \Big(\partial_v\Big( \frac{\chi_{\mathcal{H}^+} r^2}{2}\Big) -\chi_{\mathcal{H}^+} r^2 b^+_u\Big)|\partial_u\Phi_+|^2 \\&+ \chi_{\mathcal{H}^+}' \frac{\nu r^2}{2}|\partial_v\Phi_+|^2 + (\partial_u+\partial_v)\Big( \frac{\chi_{\mathcal{H}^+} r^2 b^+_0}{2}\Big)|\Phi_+|^2.
	\end{aligned}
\end{equation*}

Let us start with the analysis of the LHS. Combining the flux terms with \eqref{est:red-shift1}, we readily obtain 
\begin{align*}
	\chi_{\mathcal{H}^+}\big(\frac{r^2}{2}|\partial_v\Phi_+|^ 2 +|\psi_+|^2\big) \gtrsim \chi_{\mathcal{H}^+} |\partial_v\psi_+|^2,\quad
	\frac{\chi_{\mathcal{H}^+} r^2 b^+_0}{2}|\Phi_+|^2 \sim \chi_{\mathcal{H}^+} (-\bar\nu)|\psi_+|^2.
\end{align*}
Proceeding analogously with the bulk term $|\partial_v\Phi_+|^2$, we use again the previous bound on $|\psi_+|$ obtained in \eqref{est:red-shift1}, to estimate
\begin{equation*}
    \chi_{\mathcal{H}^+}(-\nu)\big( |\partial_v\Phi_+|^2 + |\psi_+|^2\big)\gtrsim \chi_{\mathcal{H}^+}(-\bar\nu)|\partial_v\psi_+|^2.
\end{equation*}
On the other hand, we split the cross term $\Re[\oline{\partial_v\Phi_+}\partial_u\Phi_+]=-\Omega r^{-1}\Re[\oline{\partial_v\Phi_+}\Phi_-]$ as
\begin{align*}
	\left|\chi_{\mathcal{H}^+} r \Omega \Big(\partial_v\log\Omega-\frac{\lambda}{r}\Big)\Re[\oline{\partial_v\Phi_+}\Phi_-]\right| 
&\les \chi_{\mathcal{H}^+}(-\bar\nu) \Big( \frac{\epsilon}{2}|\partial_v\psi_+|^2 + \frac{r^4}{2\epsilon}\Big|\partial_v\log\Omega-\frac{\lambda}{r}\Big|^2 |\psi_-|^2\Big)
\\&
\les \chi_{\mathcal{H}^+}(-\bar\nu)\Big(\frac{\epsilon}{2}|\partial_v\psi_+|^2 + \frac{1}{2\epsilon} |\psi_-|^2\Big),
\end{align*}
so that the first summand can be absorbed on the LHS, while the second can be controlled by the RHS of \eqref{est:red-shift1}.
Regarding the right-hand side, the bulk $|\partial_u\Phi_+| \sim (-\nu)|\psi_-|$ and the remaining terms may be bounded again by Lemma \ref{lem:red-shift}, concluding the proof.
\end{proof}

\begin{proof}[Proof of Proposition \ref{prop:red-shift}]
    Combining \eqref{est:red-shift1} and \eqref{est:red-shift2}, one obtains 
    \begin{equation*}
    	\begin{aligned}
    	\int_{\mathrm{VI}_{\le r_1}} (|\psi_-|^2+ |\psi_+|^2 &+ |\partial_v\psi_+|^2)\,dv + 
    	\int_{\mathrm{V}_{\le r_1}} \Big(-\bar\nu|\psi_-|^2  -\bar\nu|\psi_+|^2 + \frac{|\partial_u\psi_-|^2}{-\bar\nu}\Big)du 
    		\\
    		&+ \iint_{\mathcal{R}_{\le r_1}}
    				\Big( -\bar\nu|\psi_-|^2 - \bar\nu|\psi_+|^2 -\bar\nu|\partial_v\psi_+|^2 +\frac{|\partial_u\psi_-|^2}{-\bar \nu}\Big)dudv
    				\\
    				&\les 
    				\int_{\mathrm{I}} \Big(-\bar\nu|\psi_-|^2 -\bar\nu|\psi_+|^2 + \frac{|\partial_u\psi_-|^2}{-\bar\nu}\Big)du
    				\\&+
    				\iint_{\mathcal{R}_{r_1,r_2}} \Big( -\bar\nu|\psi_-|^2 -\bar\nu|\psi_+|^2 + \frac{|\partial_u\psi_-|^2}{-\bar\nu} -\bar\nu|\partial_v\psi_+|^2\Big)dudv 
    				+ \ve^3 \tau^{-7/2+3\delta}_1.
    	\end{aligned}
    \end{equation*}  
The estimate \eqref{est:red-shift-est} then follows by bounding the spacetime term on the right-hand side by the LHS of \eqref{eq:T+Mor}.
\end{proof}

\subsection{\texorpdfstring{$r^p$}{rp}-estimates}\label{sec:rp-est}
This section is devoted to upgrading \eqref{eq:T+Mor} in the far region by obtaining additional decay estimates for the energy. In particular, we establish the celebrated Dafermos--Rodnianski hierarchy of $r^p$-weighted energy estimates for the Dirac field.
\begin{proposition}\label{prop:rp-est}
		For any $p\in[0,3-\delta]$, $A\ge 1$, $\ve$ sufficiently small, $\tau_f\in\mathfrak B$, $(u_1,v_1),(u_2,v_2)\in\Gamma\cap\mathcal D_{\tau_f}$,  $u_2'>u_2$, and $v_2'>v_2$,  where $(u_2,v_2)$ is to the future of $(u_1,v_1)$, it holds that 
		\begin{equation}\label{est:rp-est}
			\begin{aligned}
			&\int_\mathrm{IV}\big(r^p |\partial_v(r\psi_+)|^2+ c_{{p}_*}r^{p_*+2}|\partial_v\psi_+|^2+c_{{p}_*}r^{p_*}|\psi_+|^2 \big)\,dv
			\\&+\iint_{\mathcal R_{\ge\Lambda}}\big(r^{p-1} |\partial_v\wtilde\Phi_+|^2+ c_{{(p-1)}_*}r^{(p-1)_*+2}|\partial_v\psi_+|^2+c_{{(p-1)}_*}r^{(p-1)_*}|\psi_+|^2 + |\psi_-|^2\big)\,dudv \\
			&\les \int_\mathrm{I} -\bar\nu\Big(|\psi_-|^2+|\psi_+|^2+\frac{|\partial_u\psi_-|^2}{\bar\nu^2}\Big)\, du
			+ \int_\mathrm{II} \big(r^p|\partial_v\wtilde{\Phi}_+|^2+r^2|\partial_v\psi_+|^2+|\psi_+|^2+|\psi_-|^2\big)\,dv
			+\ve^3\tau_1^{-7/2+3\delta},
			\end{aligned}
		\end{equation}
        where we recall from Definition \ref{def-of-energies} that $p_*:=p$ if $p\in[0,1)$ and vanishes otherwise, and $c_{p_*}:=(p_*-1)^2$.
\end{proposition}
The proof of this estimate will be given in the next section after a series of lemmas. 

\subsubsection{Proof of the \texorpdfstring{$r^p$}{rp}-estimates}
To prove the following estimates, we proceed in a manner analogous to the red-shift argument, with the roles of the two spinorial components interchanged. Here we start with the $\Phi_+$ component and introduce the reweighting
\[
\widetilde\Phi_+:=\frac{r}{\Omega}\Phi_+=\frac{r}{(-\nu)^{1/2}}\psi_+,
\]
which satisfies a decoupled equation up to nonlinear error terms. As in Section \ref{sec:red-shift}, we first establish a preliminary lemma for the $r^p$-identities associated with the prototype equation \eqref{eq:rp-prototype}, which is satisfied both by $\widetilde\Phi_+$ and by $\Phi_-$.

\begin{lemma}[$r^p$-identities]\label{lem:r^p identity} Let $\mathcal{O}$ be an open set such that $\overline{\mathcal{O}}\subseteq\mathcal{D}_{\tau_f}$ and $\varphi\in C^2(\overline{\mathcal{O}})$ be a solution to
\begin{equation}\label{eq:rp-prototype}
    \partial_u \partial_v \varphi + b_v \partial_v \varphi + b_0 \varphi = 0,
\end{equation}
where the coefficients $b_v\in C^0(\mathcal{O})$, $b_0\in C^1(\mathcal{O})$. Let $\chi$ be a $C^1$ function on $(0,\infty)$. Denoting $\chi=\chi(r), \chi'= \frac{d}{dr}\chi(r)$, we have
\begin{equation}\label{eq:r^p-id}
\begin{aligned}
    \partial_u \Big(\frac12 \chi r^p|\partial_v\varphi|^2\Big)
    &+
    \partial_v \Big( \frac12 \chi r^p b_0 |\varphi|^2\Big) 
    + \frac12\chi r^{p-1}\Big( (2r b_v -p\nu)|\partial_v\varphi|^2  
    - (p \lambda b_0 + r \partial_v b_0)|\varphi|^2\Big)
    \\
    &= \frac12 \chi' r^p ( \nu |\partial_v\varphi|^2 
    + \lambda b_0 |\varphi|^2).
\end{aligned}
\end{equation}
\end{lemma}
\begin{proof} We multiply \eqref{eq:rp-prototype} by $\chi r^p \overline{\partial_v\varphi}$ and take the real part. Computing separately, we obtain
\begin{align*}
    \chi r^p\Re[\overline{\partial_v\varphi}\partial_v \partial_u \varphi] &= \partial_u \Big( \frac12 \chi r^p |\partial_v\varphi|^2\Big) - \partial_u\Big(\frac12\chi r^p\Big)|\partial_v\varphi|^2,\\
    \chi r^p \Re[\overline{\partial_v\varphi}b_v \partial_v \varphi] &= \chi r^p b_v |\partial_v \varphi|^2,\\
    \chi r^p \Re[\overline{\partial_v\varphi}b_0\varphi] &= \partial_v \Big( \frac12 \chi r^p b_0 |\varphi|^2\Big) - \partial_v\Big(\frac12\chi r^p b_0\Big)|\varphi|^2.
\end{align*}
Expanding the last term of the first and third lines by the product rule and rearranging terms, we arrive at \eqref{eq:r^p-id}.
\end{proof}
We are now ready to state the main ingredient for the proof of Proposition \ref{prop:rp-est}.
\begin{lemma}\label{lem:rp-1}
With hypotheses as in Proposition \ref{prop:rp-est}, it holds that
    \begin{equation}\label{eq:rp-aux1}
    \begin{aligned}
    \int_\mathrm{IV_{\ge4\Lambda}}r^p|\partial_v\widetilde\Phi_+|^2\,dv&+\iint_{\mathcal R_{\ge4\Lambda}} \big(r^{p-1}|\partial_v\widetilde\Phi_+|^2 +|\psi_-|^2\big)\,dudv  
    \\&\les \iint_{\mathcal R_{\Lambda,4\Lambda}} \big(
    |\partial_v\widetilde\Phi_+|^2 + 
    |\psi_+|^2+|\psi_-|^2\big)\,dudv +\int_\mathrm{II}\big(r^p|\partial_v\widetilde\Phi_+|^2 +|\psi_+|^2\big)\,dv.
    \end{aligned}
    \end{equation}
\end{lemma}

The reweighted field $\wtilde\Phi_+$ introduced above can be viewed as a \textit{reweighted radiation field} for $\psi_+$, since the additional factor is precisely what allows the two Dirac components to decouple and the $r^p$-hierarchy to be established. This construction is part of a broader decoupling scheme for non-zero spin fields, introduced in \cite{andersson2025stability,ma2022almost} and adapted to the Dirac field on Schwarzschild in \cite[Definitions 5.5 and 5.14]{ma2022sharp}. In the latter setting, they introduced reweighted fields $\wtilde\Phi^{(i)}_{\pm1/2}$, indexed by the spherical mode $i$, each satisfying an equation with a favorable structure to extend the $r^p$-hierarchy.
The function $\wtilde\Phi_+$ above corresponds to the first member $\wtilde\Phi^{(1)}_{+1/2}$ of that family, here adapted to our dynamical and perturbed framework.

\begin{proof}[Proof of Lemma \ref{lem:rp-1}] \textsc{Step 1.} Let $\chi$ be a smooth non-negative function, with $0\le\chi\le 1$, which equals $1$ on $\{r\ge 4\Lambda\}$ and $0$ on $\{r\le \Lambda\}$. Again, one can prove that $\{r=4\Lambda\}$ is a timelike curve by following the argument in Lemma \ref{lem:gamma-slope}.
Let us consider $\widetilde\Phi_+=\frac{r}{\Omega}\Phi_+$, 
which solves 
\begin{equation*}
    \partial_u \partial_v \widetilde\Phi_+ + b^+_v \partial_v \widetilde\Phi_+ + b^+_0 \widetilde\Phi_+ = 0,
\end{equation*}
where 
\begin{equation*}
    b^+_v := \partial_u\log\Omega-\frac{\nu}{r},\quad b^+_0 := \frac{6\Omega^2 m}{r^3} - 12\pi T_{uv}.
\end{equation*}
Therefore, from Lemma \ref{lem:r^p identity}, we obtain
\begin{equation}\label{eq:rp-aux1-step1-1}
	\begin{aligned}
		\partial_u \Big(\frac12 \chi r^p|\partial_v\wtilde{\Phi}_+|^2\Big)
    &+
    \partial_v \Big( \frac12 \chi r^p b^+_0 |\wtilde{\Phi}_+|^2\Big) 
    + \frac12\chi r^{p-1}\Big( (2r b^+_v -p\nu)|\partial_v\wtilde{\Phi}_+|^2  
    - (p \lambda b^+_0 + r \partial_v b^+_0)|\wtilde{\Phi}_+|^2\Big)
    \\
    &= \frac12 \chi' r^p ( \nu |\partial_v\wtilde{\Phi}_+|^2 
    + \lambda b^+_0 |\wtilde{\Phi}_+|^2).
    	\end{aligned}
\end{equation}
As the analysis of the boundary terms is straightforward, we directly focus on the bulk terms. Looking at $|\partial_v\wtilde{\Phi}_+|^2$, we compute
\begin{align*}
    2r b^+_v -p\nu = -\nu(2+p) + 2r\partial_u\log\Omega  \sim p+2>0,
\end{align*}
for $r\ge 2\Lambda$ and $\ve$ sufficiently small.
Regarding $|\wtilde{\Phi}_+|^2$, we have by \eqref{eq:gamma-1}, \eqref{est:T-uv_bound}, \eqref{eq:partial_v_Omega-expansion2}, \eqref{est:partial_v-varpi},
\begin{align*}
    - p \lambda b^+_0 - r \partial_v b^+_0&= 6r^{-3}\Big(\lambda^2\gamma m(3-p)+r\Omega^2(\partial_v m-2 m\partial_v\log\Omega)+2\pi r^3(p\lambda T_{uv} - r\partial_v T_{uv})\Big)\\
    &\sim Mr^{-3} (3-p - M\bar r^{-1}  + A^2\ve^{3/2}\tau^{-3/2+\delta/2}),
\end{align*}
which is strictly positive for $p\in[0,3-\delta]$, $r\ge 4\Lambda$ and $\ve$ sufficiently small. In the previous estimate, we used the explicit bound
\begin{align*}
    |\partial_vT_{uv}| &= \Big|\Big(\partial_v\log\Omega-3\frac{\lambda}{r}\Big)T_{uv} + \frac{\kappa^{1/2}(-\nu)}{2\pi r^2}\Im[\overline{\psi_-}\partial_v\psi_+]\Big|\les A\ve^2 r^{-9/2}\tau^{-3+\delta} + A\ve^2 r^{-5}\tau^{-3/2+\delta/2} \\&\les A\ve^2 r^{-9/2}\tau^{-3/2+\delta/2}.
\end{align*}
Noticing that the RHS of \eqref{eq:rp-aux1-step1-1} is localized in $\mathcal{R}_{\Lambda,4\Lambda}$, we find
    \begin{equation}\label{eq:rp-aux1-step1-2}
	\begin{aligned}
		\int_\mathrm{IV_{\ge4\Lambda}}r^p|\partial_v\widetilde\Phi_+|^2\,dv&+\iint_{\mathcal R_{\ge4\Lambda}} r^{p-1}|\partial_v\widetilde\Phi_+|^2 \,dudv  
		\les \iint_{\mathcal R_{\Lambda,4\Lambda}} \big(
		|\partial_v\widetilde\Phi_+|^2 + 
		|\psi_+|^2\big)\,dudv +\int_\mathrm{II}r^p|\partial_v\widetilde\Phi_+|^2 \,dv,
	\end{aligned}
\end{equation}
for $p\in[0,3-\delta]$.

\textsc{Step 2.} Let us now recall that $\Phi_-$ is a solution to
\begin{equation*}
    \partial_u \partial_v \Phi_- + b^-_v \partial_v \Phi_- + b^-_0 \Phi_- = 0,
\end{equation*}
where 
\begin{equation*}
    b^-_v := -b^+_v= -\partial_u\log\Omega+\frac{\nu}{r},\quad b^-_0 := \frac{\Omega^2}{r^2}.
\end{equation*}
Therefore, with $\chi$ as in \textsc{Step 1}, we obtain
\begin{align*}
    \partial_u \Big(\frac12 \chi r^p|\partial_v\Phi_-|^2\Big)
    &+
    \partial_v \Big( \frac12 \chi r^p b^-_0 |\Phi_-|^2\Big) 
    + \frac12\chi r^{p-1}\Big( (2r b^-_v -p\nu)|\partial_v\Phi_-|^2  
    - (p \lambda b^-_0 + r \partial_v b^-_0)|\Phi_-|^2\Big)
    \\
    &= \frac12 \chi' r^p ( \nu |\partial_v\Phi_-|^2 
    + \lambda b^-_0 |\Phi_-|^2).
\end{align*}
Proceeding as before, one finds that the bulk $|\partial_v\Phi_-|^2$ is negative: indeed,
\[
    2r b^-_v -p\nu= (2-p)\nu-2r\partial_u\log\Omega\sim -(2-p)+M\bar r^{-1}<0,
\]
for $r\ge4\Lambda$.
Thus, we use the Dirac equation, to rewrite it as  $r^{p-1}|\partial_v\Phi_-|^2\sim r^{p-3}|\Phi_+|^2$, which is controlled by the previous step. On the other hand, regarding $|\Phi_-|^2$ we have by arguing as before
\begin{align*}
    - p \lambda b^-_0 - r \partial_v b^-_0&= r^{-2}\Big(\lambda^2\gamma (2-p) -2r\Omega^2\partial_v\log\Omega\Big)\sim r^{-2}(2-p - M\bar r^{-1} + A^2\ve^{3/2}\bar r^{-1}\tau^{-2+\delta}),
\end{align*}
which is strictly positive for $p\in[0,2-\delta]$, $r\ge4\Lambda$ and $\ve$ sufficiently small. Therefore, by choosing $p=1$ and recalling $|\Phi_-|^2\sim r^2|\psi_-|^2$, we obtain 
    \begin{equation}\label{eq:rp-aux1-step1-3}
	\begin{aligned}
	   \iint_{\mathcal R_{\ge4\Lambda}} |\psi_-|^2 \,dudv  
		\les \iint_{\mathcal R_{\Lambda,4\Lambda}} \big(
		|\psi_+|^2 +
		|\psi_-|^2\big)\,dudv +\int_\mathrm{II}r^{-1}|\psi_+|^2 \,dv.
	\end{aligned}
\end{equation}
Putting together \eqref{eq:rp-aux1-step1-2} and \eqref{eq:rp-aux1-step1-3}, we conclude the proof.
\end{proof}

To obtain control over $\psi_+$, we use the following Hardy-type inequalities \cite[Lemma 6.21]{angelopoulos2026nonlinear}, which are a consequence of \cite[Lemma 8.30]{luk2019strong}. These estimates are also the source of the degenerating exponents and constants $p_*$ and $c_{p_*}$ appearing in Proposition \ref{prop:rp-est}.

\begin{lemma}[$r^p$-weighted Hardy inequalities]\label{lem:Hardy-est1} Let $\alpha<-1$.  With hypotheses as in Proposition \ref{prop:rp-est}, it holds that
\begin{align*}
    (\alpha+1)^2  \iint_{\mathcal R_{\ge 4\Lambda}} r^{\alpha+2}|\psi_+|^2\,dudv &\les\iint_{\mathcal R_{\ge 4\Lambda}} r^{\alpha+2}|\partial_v\widetilde{\Phi}_+|^2\,dudv+\iint_{\mathcal R_{\Lambda,4\Lambda}} \big(|\partial_v\widetilde{\Phi}_+|^2+|\psi_+|^2\big)\,dudv,
    \\ 
    (\alpha+1)^2\int^{v'_2}_{v^\Lambda} r^{\alpha+2}|\psi_+|^2 \,dv &\les \int^{v'_2}_{v^\Lambda} \big(r^{\alpha+2}|\partial_v\wtilde\Phi_+|^2 + |\psi_+|^2\big)\,dv.
   \end{align*}
\end{lemma}
\begin{proof}
It suffices to redo the proof of \cite[Lemma 6.21]{angelopoulos2026nonlinear} with $\wtilde\Phi_+$, which satisfies $\wtilde\Phi_+\sim r\psi_+$ in $\{r\ge\Lambda\}$. In particular, for the first inequality, an additional cutoff allows us to replace the boundary integral along $\Gamma$ appearing there by the localized spacetime integral on the right-hand side.
\begin{comment}
    \p_v(\wtilde\Phi_+ \chi) gives \wtilde\Phi_+ \chi' that is localized and \chi\p_v\wtilde\Phi_+ which is controlled by r^p
\end{comment}
\begin{comment}
From \eqref{eq:hardy_v}, we derive
    \begin{equation}\label{eq:hardy_v1}
        (\alpha+1)^2\int^v_{v^\Lambda} r^\alpha f^2\,dv'\les \int^v_{v^\Lambda} r^{\alpha+2}(\partial_v f)^2\,dv' + (r^{\alpha+1}f^2)(u,v^\Lambda(u)),
    \end{equation}
    where we have used that the boundary term at $v$ has a favorable sign. Applying this inequality to $f=\chi_1|\widetilde{\Phi}_+|$, where $\chi_1=\chi_1(r)$ be a cutoff function such that $\chi(r)=0$ for $r\le \Lambda$ and $\chi(r)=1$ for $r\ge 2\Lambda$ and integrating in $u$, yields \eqref{eq:Hardy-I-1}.
    To prove \eqref{eq:Hardy-I-1}, let $\chi_1=\chi_1(r)$ be a cutoff function such that $\chi(r)=0$ for $r\le \Lambda$ and $\chi(r)=1$ for $r\ge 2\Lambda$ and apply \eqref{eq:hardy_v1} to $f=\chi_1|\widetilde{\Phi}_+|$. The cutoff kills the boundary term along $\Gamma$, yielding
    \begin{equation*}
        (\alpha+1)^2  \int_{S^{2\Lambda}_v} r^{\alpha+2}|\psi_+|^2\,dv' \les \int_{S^{2\Lambda}_v} r^{\alpha+2}|\partial_v\widetilde{\Phi}_+|^2\,dv' +  \int^{v^{2\Lambda}}_{v^\Lambda} \big(|\partial_v\widetilde{\Phi}_+|^2 + |\psi_+|^2\big)\,dv'.
    \end{equation*}
    Integrating this inequality in $u$ leads to \eqref{eq:Hardy-I-1}.
\end{comment}
\end{proof}
\begin{proof}[Proof of Proposition \ref{prop:rp-est}]
	The proof follows from Lemmas \ref{lem:rp-1} and \ref{lem:Hardy-est1}, Proposition \ref{prop:T+Mor}, and the relation $\psi_+=r^{-1}(-\nu)^{1/2}\wtilde{\Phi}_+$, together with the straightforward bounds
\[
    |\partial_v\psi_+|^2\les r^{-2}(|\partial_v\wtilde\Phi_+|^2+|\psi_+|^2),\quad |\partial_v(r\psi_+)|^2\les |\partial_v\wtilde\Phi_+|^2+r^{-2}|\psi_+|^2,
\]
    in $\mathcal R_{\ge\Lambda}$.
\begin{comment}
    1. ILED controls the |\psi_+|^2 in the RHS of the second equation in Lemma 6.14 + it absorbs all the bulk terms
    2. r^p|\partial_v(r\psi_+)|^2\les r^p|\partial_v\wtilde\Phi_+|^2+r^{p-2}|\psi_+|^2 is fine as p<3 and therefore p-2<1
\end{comment}
\end{proof}

\subsection{Improving the bootstrap assumptions for the Dirac field}\label{sec:improvement-Dirac}
We are now in position to prove Proposition \ref{prop:bootstrap-Dirac} using a pigeonhole principle argument. 
Recalling the definition of the energies $\uline{\mathcal E}$, $\mathcal E_p$, $\underline{\mathcal F}$ and $\mathcal F$ (see Definition \ref{def-of-energies}), we define  
\begin{equation*}
    \mathcal{G}_p(\tau):= \underline{\mathcal E}(\tau) + \mathcal E_p(\tau),
\end{equation*}
and summarize the content of Propositions \ref{prop:T+Mor}, \ref{prop:red-shift} and \ref{prop:rp-est} as follows.
\begin{proposition}
     For any $A\ge 1$, $\ve$ sufficiently small, $\tau_f\in\mathfrak B$, and $1\le\tau_1\le \tau_2\le\tau_f$, it holds that 
     \begin{equation}\label{eq:final-bdd}
        \sup_{\tau\in[\tau_1,\tau_2]} \mathcal{G}_p(\tau) \les \mathcal{G}_p(\tau_1)+\ve^3\tau_1^{-7/2+3\delta+p},
     \end{equation}
     for every $p\in [0,3-\delta]$ and
     \begin{equation}\label{eq:final-integral-bdd}
          \int_{\tau_1}^{\tau_2}\mathcal{G}_{p-1}(\tau)\,d\tau \les \mathcal{G}_{p}(\tau_1)+\ve^3\tau_1^{-7/2+3\delta+p},
     \end{equation}
     for every $p\in[1,3-\delta]$.
\end{proposition}

Before proceeding with the proof, we need the following interpolation property, which is an immediate consequence of Hölder's inequality.
\begin{lemma}\label{lem:interpolation}
    Let $0\le p_1<p<p_2\le 3-\delta$ and $\tau\in[1,\tau_f]$. Then, it holds that \begin{align*}
       \mathcal G_p(\tau)&\les \big(\mathcal G_{p_1}(\tau))^\frac{p_2-p}{p_2-p_1}\big(\mathcal G_{p_2}(\tau))^\frac{p-p_1}{p_2-p_1}.
    \end{align*}
\end{lemma}

\begin{proof}[Proof of Proposition \ref{prop:bootstrap-Dirac}] We start by noting that for any $p\in[0,3-\delta]$, it holds
\begin{equation}\label{eq:decay-aux-1}
    \sup_{\tau\in[1,\tau_f]}\mathcal{G}_p(\tau) \les \mathcal{G}_p(1)+\ve^3\les\big(\mathfrak D[\mathcal S_0]\big)^2+\ve^3\les \ve^2,
\end{equation} 
where we combined \eqref{eq:final-bdd} applied on $[1,\tau_f]$, and \eqref{eq:data-estimate}.
Let us now split $[1,\tau_f]$ in dyadic intervals $[2^i,2^{i+1}]$ where $i\in\{0,...,I_{\tau_f}-1\}$, with $I_{\tau_f}$ being the largest integer satisfying $2^{I_{\tau_f}}\le\tau_f$. We assume $I_{\tau_f}\ge 1$, else \eqref{eq:decay-aux-1} already suffices. 
By the pigeonhole principle applied on each sub-interval, there exists $\tau'_i\in[2^i,2^{i+1}]$, for $i\in \{0,\dotsc,I_{\tau_f}-1\}$, such that
\begin{equation}\label{eq:decay-aux-2}
    \mathcal{G}_{2-\delta}(\tau_i')\les \frac{1}{2^{i}}\int^{2^{i+1}}_{2^i}\mathcal{G}_{2-\delta}(\tau)\,d\tau\les 2^{-i}\big(\mathcal{G}_{3-\delta}(2^i)+\ve^3 2^{i(-1/2+2\delta)}\big)\les \ve^2 2^{-i},
\end{equation}
where we combined \eqref{eq:final-integral-bdd} and \eqref{eq:decay-aux-1} with $p=3-\delta$. Therefore, thanks to dyadicity and \eqref{eq:final-bdd} with $p=2-\delta$, we are able to upgrade \eqref{eq:decay-aux-2} to 
\begin{equation}\label{eq:decay-aux-3}
    \mathcal{G}_{2-\delta}(\tau)\les \ve^2 \tau^{-1},
\end{equation}
for every $\tau\in[1,\tau_f]$. Using Lemma \ref{lem:interpolation}, we interpolate between \eqref{eq:decay-aux-1} for $p=3-\delta$ and \eqref{eq:decay-aux-3}, yielding
\begin{equation}\label{eq:decay-aux-4}
     \mathcal G_{2}(\tau) \les \ve^2\tau^{-1+\delta},
\end{equation}
for every $\tau\in[1,\tau_f]$. Applying again the pigeonhole principle, there exists $\tau_i''\in[2^i,2^{i+1}]$
for $i\in \{0,\dotsc,$ $I_{\tau_f}-1\}$, such that
\begin{equation*}
    \mathcal{G}_{1}(\tau_i'')\les \frac{1}{2^{i}}\int^{2^{i+1}}_{2^i}\mathcal{G}_{1}(\tau)\,d\tau\les 2^{-i}\big(\mathcal{G}_{2}(2^i)+\ve^3 2^{i(-3/2+3\delta)}\big)\les \ve^2 (2^{i})^{-2+\delta},
\end{equation*}
where we combined \eqref{eq:decay-aux-4} and \eqref{eq:final-integral-bdd} with $p=2$, and which can be again upgraded to
\begin{equation*}
    \mathcal G_{1}(\tau)\les \ve^2\tau^{-2+\delta},
\end{equation*}
for every $\tau\in[1,\tau_f]$. Repeating this argument once more, one finds 
\begin{equation*}
    \mathcal G_{0}(\tau)\les \ve^2\tau^{-3+\delta},
\end{equation*}
for every $\tau\in[1,\tau_f]$. Thus, interpolating between this estimate and \eqref{eq:decay-aux-1} for $p=3-\delta$, we obtain
\begin{equation}
    \mathcal G_{p}(\tau)\les \ve^2\tau^{-3+\delta+p}\label{eq:decay-aux-5},
\end{equation}
for every $p\in[0,3-\delta]$ and $\tau\in[1,\tau_f]$.
Next, recalling Lemma \ref{lem:T-cons-law-bis}, we deduce
\begin{equation}\label{eq:decay-aux-7}
    \mathcal F(u,\tau)\les \ve^2\tau^{-3+\delta},\quad \underline{\mathcal F}(v,\tau)\les \ve^2\tau^{-3+\delta},
\end{equation}
for any $(u,v)\in\mathcal D_{\tau_f}$ and $\tau\in[1,\tau_f]$. Therefore, for $A=A_0$ large enough, \eqref{eq:decay-aux-5} and \eqref{eq:decay-aux-7} imply \eqref{eq:boot-improved1}--\eqref{eq:boot-improved4}.
\end{proof}
\section{Proof of Theorem \ref{thm:Schwarz-stability2}}\label{sec:proof-conclusion}
This section is devoted to the proof of Theorem \ref{thm:Schwarz-stability2}, concerning the stability of Schwarzschild for the spherically symmetric Einstein--Dirac system.

In particular, in Section \ref{subsec:continuity-argument} we prove by a continuity argument that the bootstrap set satisfies $\mathfrak{B}=[1,+\infty)$. In Section \ref{sec:control-gauge}, we control the teleological gauges introduced in Section \ref{subsec:tel-gauges}, showing that the sequence of gauges obtained through successive updates converges to an eschatological gauge. Moreover, we show that the corresponding anchored backgrounds, introduced in Section \ref{sec:anch-solution}, converge to a Schwarzschild black hole with mass $M_\infty$ and area radius function $\bar r_\infty$. Finally, in Section \ref{sec:completing-proof}, we complete the proof of Theorem \ref{thm:Schwarz-stability2}.

Since in this part of the paper we will work with gauges and background solutions corresponding to different bootstrap times, we will no longer suppress the $\tau_f$ subscripts in the bootstrap quantities. Moreover, from now on, we fix $A=A_0$, so that Propositions \ref{prop:bootstrap-geometry} and \ref{prop:bootstrap-Dirac} apply.

\subsection{The continuity argument}\label{subsec:continuity-argument}
We start by showing that the bootstrap set $\mathfrak{B}(\mathcal{S},\ve,A)$ in Definition \ref{def:bootstrap-set} coincides with the full interval $[1,+\infty)$.

\begin{proposition}\label{prop:bootstrap-set} Let $M_0$ and $\delta_0$ be as in Theorem \ref{thm:Schwarz-stability2}. Then, there exists an $\ve_{\mathrm{stab}}(M_0,\delta)>0$ such that, for any $\ve$-Schwarzschildean seed data set $\mathcal{S}$, with $0<\ve\le\ve_{\mathrm{stab}}$, it holds $\mathfrak{B}(\mathcal{S},\ve,A_0)=[1,+\infty)$. 
\end{proposition}
Before proving the result, we first show that the bootstrap region remains strictly away from the future boundary of the initial data hypersurface $\hat{\C}$.

\begin{lemma}\label{lem:thetaU*} There exists a constant $\theta\in(0,1)$ such that for $\ve$ sufficiently small and $\tau_f\in\mathfrak{B}(\mathcal{S},\ve,A_0)$, it holds that $\Gamma^{\hat u}(\tau_f)\le\theta U_*$.
\end{lemma}
\begin{proof} We begin by noting that 
\[
    |M_{\tau_f}-M_0| = |M_{\tau_f}-m_{\tau_f}(0,0)|=|\tilde m_{\tau_f}(0,0)|\les \ve^{3/2}\les\ve,
\]
by Proposition \ref{prop:seed-data-construction} and \eqref{eq:boot-varpi}. Moreover, since $\p_{\hat u} r=-1$ on $\hat{\C}_{\ing}$, we have that $r(\Gamma^{\hat u}(\tau_f),0)=\Lambda - \Gamma^{\hat u}(\tau_f)$. Therefore, using \eqref{eq:boot-r}, we bound
\begin{align*}
    \Gamma^{\hat u}(\tau_f)&= \Lambda-r(\Gamma^{\hat u}(\tau_f),0)=\Lambda - \bar r_{\tau_f}\circ\Upphi_{\tau_f}(\Gamma^{\hat u}(\tau_f),0)-\tilde r\circ\Upphi_{\tau_f}(\Gamma^{\hat u}(\tau_f),0) \le 100M_0- 2M_{\tau_f} + O(\ve)\\&= 98M_0+O(\ve),
\end{align*}
which is strictly smaller than $U_*=98.5M_0$ for $\ve$ sufficiently small.
\end{proof}

\begin{proof}[Proof of Proposition \ref{prop:bootstrap-set}]
    Let us prove that $\mathfrak{B}(\mathcal{S},\ve,A_0)$ is non-empty, open and closed for $\ve$ sufficiently small. 
    
    \textsc{Non-emptiness:} It is a straightforward consequence of local well-posedness and continuous dependence on the initial data. Indeed, a solution arising from a seed data set $\mathcal{S}$ extends to a neighborhood of the initial data hypersurface $\hat{\mathcal C}$ of uniform size near $(0,0)$. Therefore, as the solution stays close to Schwarzschild in this small region by Cauchy stability, the bootstrap assumptions 1.--3. are automatically satisfied for $\ve$ sufficiently small and $\tau_f$ sufficiently close to $1$. Note that, at this stage, the choice $A=A_0$ plays no role.
    
    \textsc{Openness:}
    Let us now prove it is open, that is, proving that if $\tau_f\in\mathfrak{B}(\mathcal{S},\ve,A_0)$, then  $\tau_f+\eta\in \mathfrak{B}(\mathcal{S},\ve,A_0)$ for $\eta>0$ small enough. As $\tau_f\in\mathfrak{B}$, the solution $(r,\hat\Omega,\psi_+,\psi_-)$ exists in the region \[\hat{\mathcal{D}}_{\tau_f}:=[0,\Gamma^{\hat u}(\tau_f)]\times[0,\Gamma^{\hat v}(\tau_f)]\subset\hat{\mathcal{Q}}_{\text{max}}.\] By Proposition \ref{prop:slab-existence} and Lemma \ref{lem:thetaU*}, there exists $\sigma>0$ such that
    \[
    \hat{\mathcal{D}}_{\tau_f}^\sigma:=[0,\Gamma^{\hat u}(\tau_f)+\sigma]\times[0,\Gamma^{\hat v}(\tau_f)+\sigma]\subset\hat{\mathcal{Q}}_{\text{max}}.
    \] 
    
    Our aim is to show that there exists $\eta>0$ sufficiently small such that the region
    \[
        \hat{\mathcal{D}}_{\tau_f+\eta}:=[0,\Gamma^{\hat u}(\tau_f+\eta)]\times[0,\Gamma^{\hat v}(\tau_f+\eta)]
    \]
    satisfies
    \[
    \hat{\mathcal{D}}_{\tau_f}\subset\hat{\mathcal{D}}_{\tau_f+\eta} \subseteq\hat{\mathcal{D}}_{\tau_f}^\sigma \subset\hat{\mathcal{Q}}_{\text{max}},
    \]
    and, in addition, properties 1.--3. of Definition \ref{def:bootstrap-set} hold on $\hat{\mathcal{D}}_{\tau_f+\eta}$.  
    
    Since $\Gamma$ being a timelike curve is an open condition, $\Gamma|_{[0,\tau_f]}$ extends to an inextendible timelike curve in $\hat{\mathcal{D}}_{\tau_f}^\sigma$ for $\sigma$ sufficiently small (see also Lemma \ref{lem:gamma-slope}). Moreover, \eqref{eq:nu-1}, \eqref{eq:mu-lower-bound} and \eqref{eq:kappa-2} imply, by continuity, that $\hat\kappa>0$ in $\hat{\mathcal{D}}_{\tau_f}^\sigma$ and $\hat\gamma>0$ in $\hat{\mathcal{D}}_{\tau_f}^\sigma\cap\{r\ge\Lambda\}$, for $\sigma$ small enough. Hence, it follows that $\hat{\mathcal{D}}_{\tau_f+\eta} \subset\hat{\mathcal{Q}}_{\text{max}}$ for $\eta$ sufficiently small. We may therefore define the map $\Upphi_{\tau_f+\eta}$ and the associated teleological coordinates $(\mathfrak{u}_{\tau_f+\eta},\mathfrak{v}_{\tau_f+\eta})$, thereby establishing points 1.--2.
    
    It remains to verify point 3. By Definition \ref{def:teleolog-coords}, we note that $\mathfrak{u}_{\tau_f+\eta}\to\mathfrak{u}_{\tau_f}$ smoothly on $[0,\Gamma^{\hat u}(\tau_f)]$ and $\mathfrak{v}_{\tau_f+\eta}\to\mathfrak{v}_{\tau_f}$ smoothly on $[0,\Gamma^{\hat v}(\tau_f)]$, as $\eta\to 0$. Furthermore, recalling \eqref{def:anch-solution2} and denoting by $\bar r_{\tau_f+\eta}$ and $\bar r_{\tau_f}$ the corresponding anchored Schwarzschild backgrounds, we have $\bar r_{\tau_f+\eta}\circ\Upphi_{\tau_f+\eta}\to \bar r_{\tau_f}\circ\Upphi_{\tau_f}$ smoothly on $\hat{\mathcal{D}}_{\tau_f}$ as $\eta\to0$.
    Now, thanks to Propositions \ref{prop:bootstrap-geometry} and \ref{prop:bootstrap-Dirac}, the constants of the bootstrap assumptions \eqref{eq:boot-nu}--\eqref{eq:boot-outg-flux} have been strictly improved on $\mathcal{D}_{\tau_f}$. Combining these improved estimates with the smooth convergence of the coordinate charts and of the anchored backgrounds above, we conclude that, for $\eta>0$ sufficiently small, the bootstrap assumptions \eqref{eq:boot-nu}--\eqref{eq:boot-outg-flux} continue to hold on $\D_{\tau_f+\eta}$. Thus, $\tau_f+\eta\in \mathfrak{B}(\mathcal{S},\ve,A_0)$.

    \textsc{Closedness:} We now show that the bootstrap set is closed, by proving that if $\tau_f^n\to\tau^\infty_f$ as $n\to+\infty$ with $\tau_f^n$ strictly increasing sequence in $\mathfrak{B}(\mathcal{S},\ve,A_0)$ and $\tau^\infty_f$ finite, then $\tau^\infty_f\in\mathfrak{B}(\mathcal{S},\ve,A_0)$.

    We start by noting that, as $\Gamma$ is a timelike curve, $\Gamma^{\hat u}(\tau^n_f)$ and $\Gamma^{\hat v}(\tau^n_f)$ are strictly monotone increasing.
    By Lemma \ref{lem:thetaU*}, $\Gamma^{\hat u}(\tau^n_f)$ converges to a number $\hat u_\star<U_*$. On the other hand, by Lemma \ref{lem:v-est} and \eqref{eq:tau-difference-v}, $\Gamma^{\hat v}(\tau^n_f)$ converges to a finite number $\hat v_\star$. Hence, the set $\hat{\mathcal{D}}_\star:=[0,\hat u_\star)\times[0,\hat v_\star)$ is contained in $\hat{\mathcal{Q}}_\maxx$. Let us now prove that 
    \[
    \hat{\D}_{\tau^\infty_f}=\oline{\hat{\D}_\star}\subset\hat{\mathcal{Q}}_{\text{max}},
    \]
    that is showing that the solution $(r,\hat{\Omega},\psi_+,\psi_-)$ extends smoothly to $\oline{\hat{\D}_\star}$. By the bootstrap assumptions, Lemma \ref{lem:v-est}, \eqref{eq:lambda-upper-bound}, \eqref{eq:lambda-lower-bound}, \eqref{est:psi+_bound}, \eqref{est:psi-_bound}, \eqref{est:vpsi+_bound}, we find
    $1\les r\les 1+\tau^{\infty}_f$, $|\hat\lambda|\les1$, $\hat\kappa\sim 1$, $|\psi_\pm|\les1$ and $|\p_{\hat v}\psi_+|\les1$ on $\hat{\D}_\star$. Combining these estimates with \eqref{eq:wave-eq-r} yields $|\log(-\hat\nu)|\les 1+\tau^\infty_f$ on $\hat{\D}_\star$, and hence $|\p_{\hat u}\psi_-|\les 1+\tau^\infty_f$ by \eqref{est:upsi-_bound}. Recalling $\hat\Omega^2=-\hat\kappa\hat\nu$, it follows $1\les\hat\Omega^2\les1+\tau^\infty_f$. Moreover, by Lemma \ref{lem:Omega-expansions}, $|\p_{\hat u}\log\Omega|+|\p_{\hat v}\log\Omega|\les 1+(\tau^\infty_f)^2$ on $\hat{\D}_\star$. These bounds imply that the solution extends to $\oline{\hat{\D}_\star}$ in $C^1$, and by standard propagation of regularity results one verifies that the extension is actually $C^\infty$. Conditions 1.--3. can then be proved by similar continuity arguments as in the previous step, proving that $\tau^\infty_f\in\mathfrak{B}(\mathcal{S},\ve,A_0)$.
\end{proof}
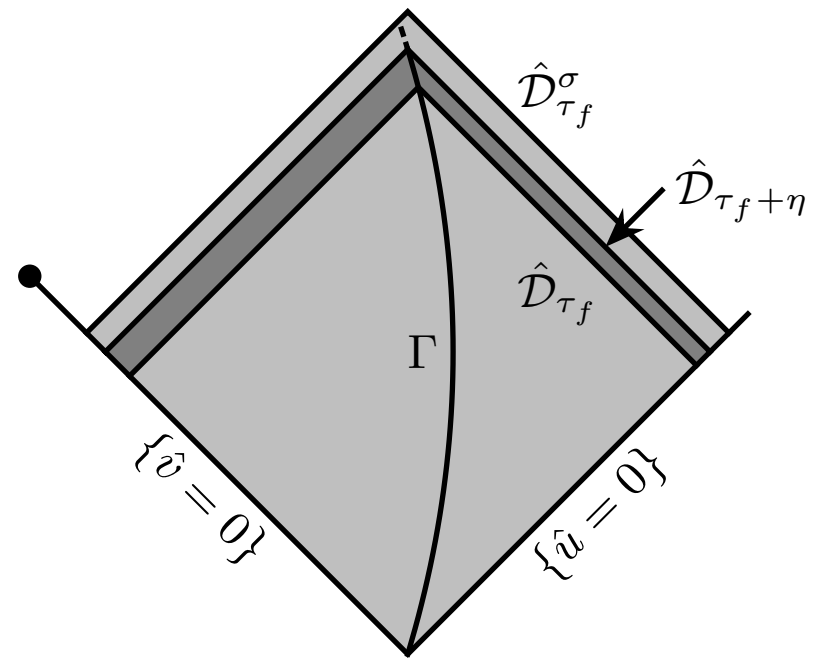
\begin{figure}
\centering
\begin{tikzpicture}[line cap=rect,inner sep=0pt, x=1cm, y=1cm, scale=1,font=\small]

\coordinate   (U)   at (-0.5,2.5)      {};
\coordinate   (W)   at (4.25,2.25)      {};
\coordinate  (I)   at (0,2)      {};
\coordinate (II)   at (2,0)      {};
\coordinate  (III)   at (4,2)      {};
\coordinate   (IV)   at (2,4)      {};
\coordinate   (V)   at (0.16016,1.83984)      {};
\coordinate   (VI)   at (2.07031,3.75)      {};
\coordinate   (VII)   at (3.91016,1.91016)      {};
\coordinate   (IV-bis)   at (2,4.25)      {};
\coordinate   (I-bis)   at (-0.125,2.125)      {};
\coordinate   (III-bis)   at (4.125,2.125)      {};

\path[name path=parab1, domain=0:3, samples=100, smooth, variable=\x]
    plot ({\x},{1/6*\x*\x - 1/6*\x + 3});

\fill[lightgray] (V) -- (II) -- (VII) -- (VI) -- (V);
\fill[lightgray] (I-bis) -- (IV-bis) -- (III-bis) -- (III) -- (IV) -- (I);
\fill[gray] (I) -- (V) -- (VI) -- (VII) -- (III) -- (IV) -- (I);

\draw[line width=1pt]  (V) -- (VI) -- (VII); 
\draw[line width=1pt]  (I-bis) -- (IV-bis) -- (III-bis); 
\draw[line width=1pt]  (III) -- (W); 
\draw[line width=1pt]  (U) -- (I) 
-- node[pos=0.4, below=2pt, sloped]    {$\{\hat v=0\}$}
       (II) -- node[pos=0.55, below=2pt, sloped]    {$\{\hat u=0\}$} (III) -- (IV) -- (I); %-- 
            %node[midway, below=2pt, sloped, font=\tiny]    {$u=u_0$}

\draw[line width=1pt,domain=0:4, smooth, variable=\y] plot ({2 + 3/40*\y*(4-\y)}, {\y});
\draw[line width=1pt,dotted,domain=4:4.19022, smooth, variable=\y] plot ({2 + 3/40*\y*(4-\y)}, {\y});

\draw[-Stealth, line width=1pt] (3.68178,3.06822) node[right=2pt] {$\hat{\mathcal{D}}_{\tau_f+\eta}$} -- (3.30678,2.69322);

\filldraw[black] (U) circle (2pt);
\node at ($(II)+(0.1,2)$)      {$\Gamma$};
\node at ($(3,3.7)$) {$\hat{\mathcal{D}}_{\tau_f}^\sigma$};
\node at ($(3,2.38644)$) {$\hat{\mathcal{D}}_{\tau_f}$};
\end{tikzpicture}

\caption{We depict here the Penrose diagram of the region considered in the openness argument of Proposition \ref{prop:bootstrap-set}. Within the extended region $\hat{\D}^\sigma_{\tau_f}$, we can find $\eta>0$ sufficiently small such that $\hat{\mathcal{D}}_{\tau_f}\subset\hat{\mathcal{D}}_{\tau_f+\eta} \subseteq\hat{\mathcal{D}}_{\tau_f}^\sigma \subset\hat{\mathcal{Q}}_{\text{max}}$.
Note that, by Lemma \ref{lem:thetaU*}, the region $\hat{\mathcal{D}}^\sigma{\tau_f}$ must terminate before reaching the solid black point.}
\label{fig:Penrose9-bootstrap}
\end{figure}
\begin{remark} Let us note that $\ve_{\mathrm{stab}}$ will be eventually reduced one more time in Proposition \ref{prop:thm-part3} below.
\end{remark}
\subsection{Control of the gauge and anchored backgrounds}\label{sec:control-gauge}
Since we proved $\mathfrak{B}(\mathcal{S},\ve,A_0)=[1,+\infty)$ in the previous section, we may define
\[
    \hat{\mathcal{D}}_\infty := \bigcup_{\tau_f\ge1} \hat{\mathcal{D}}_{\tau_f}\subset \hat{\mathcal{Q}}_{\maxx}.
\]
The aim of this section is to properly define the eschatological gauge $\Upphi_\infty$ on $\hat{\mathcal{D}}_\infty$ as the limit of the teleological gauges $\Upphi_{\tau_f}=(u_{\tau_f},v_{\tau_f})$.

The arguments and techniques are analogous to those employed in \cite[Section 8.2]{angelopoulos2026nonlinear}. In our setting, however, there are two additional points that need to be addressed. First, due to our different choice of teleological gauge, we are required to control the behaviour of the coordinate component $\mathfrak{v}_{\tau_{f}}$; on the other hand, we need to prove the convergence of the mass $M_{\tau_f}$ of the anchored background, which is updated at each bootstrap time, rather than being fixed and adjusted by modulation as in \cite{angelopoulos2026nonlinear}. For these reasons, we provide the details of the argument below.

\begin{proposition}\label{prop:gauge-diffeo} For $(\hat u,\hat v)\in\hat{\mathcal{D}}_{\infty}$, the limit
    \[
        \Upphi_\infty(\hat u,\hat v):=\lim_{\tau_f\to+\infty} \Upphi_{\tau_f}(\hat u,\hat v)
    \]
exists and defines a $C^2$ diffeomorphism $\Upphi_\infty:\hat{\mathcal{D}}_{\infty}\to[0,+\infty)\times[0,+\infty)$.
\end{proposition}
The proof is given after a series of preliminary lemmas.
Let us begin with a straightforward consequence of Lemma \ref{lem:thetaU*}.
\begin{lemma}\label{lem:thetaU*-bis} The limit 
\begin{equation}\label{eq:thetaU*-bis}  
    \hat u_{\mathcal{H}^+}:=\lim_{\tau\to+\infty}\Gamma^{\hat u}(\tau)
\end{equation}

exists and satisfies $\hat u_{\mathcal{H}^+}\le\theta U_*$, where $\theta\in(0,1)$ is the constant from Lemma \ref{lem:thetaU*}. Furthermore, \[
\hat{\mathcal{D}}_\infty=[0,\hat u_{\mathcal{H}^+})\times[0,+\infty).\]
\end{lemma}
We now prove that the family $\{\Upphi_{\tau_f}\}_{\tau_f}$ is Cauchy in $C^2_\mathrm{loc}$. %on any region bounded away from $\hat u_{\mathcal{H}^+}$.
\begin{lemma}\label{lem:cauchy-prop} For any $\eta>0$ and $(\hat u_0,\hat v_0)\in [0,\hat u_{\mathcal{H}^+})\times[0,+\infty)$, there exists $\tau_{f,1}$ sufficiently large such that for all $\tau_{f,2}>\tau_{f,1}$, it holds
    \begin{equation}\label{eq:cauchy-prop}
        \Vert \Upphi_{\tau_{f,2}} -\Upphi_{\tau_{f,1}}\Vert_{C^2([0,\hat u_0]\times[0,\hat v_0])} \les_{\hat u_0,\hat v_0
        }\ve^2\eta .
    \end{equation}
\end{lemma}
\begin{remark}
In the proof of the lemma, we will need an additional construction that stays away from the event horizon. This is necessary because, in the absence of the positive energy condition, we cannot establish an a priori sign for $\lambda$, and hence cannot guarantee that $\lambda>0$ globally. Indeed, since $\kappa\sim 1$ and $1-\mu=\lambda/\kappa$, a global positivity estimate for $\lambda$ would imply $1-\mu>0$.
\end{remark}
\begin{remark} Note that the Cauchy estimate \eqref{eq:cauchy-prop} is weaker than \cite[Lemma 8.10]{angelopoulos2026nonlinear}, where a stronger explicit decay in $\tau_{f,1}$ is achieved. This is due to the need of controlling $\mathfrak{v}_{\tau_{f}}$ and the fact that, by Lemma \ref{lem:thetaU*-bis}, we do not have an explicit rate of convergence for the sequence $\{\Gamma^{\hat u}(\tau_f)\}_{\tau_f}$, which therefore prevents us from obtaining a quantitative bound on the late-tail integrals over $[\Gamma^{\hat u}(\tau_{f,1}),\Gamma^{\hat u}(\tau_{f,2})]$.
\end{remark}

\begin{proof}[Proof of Lemma \ref{lem:cauchy-prop}] 
Let us recall \eqref{eq:transport-gamma}, which, due to gauge invariance, retains the same form
\[ 
\partial_{\hat v} \hat\gamma =\frac{2\hat \gamma}{r(1-\mu)}\Im[\overline{\psi_+}\partial_{\hat v}\psi_+].
\]
Therefore, from \eqref{def:gauges} we compute
\begin{align*}
        \mathfrak{u}'_{\tau_{f,2}}(\hat u) - \mathfrak{u}'_{\tau_{f,1}}(\hat u) &= \hat \gamma(\hat u, \Gamma^{\hat v}(\tau_{f,2}))-\hat \gamma(\hat u,\Gamma^{\hat v}(\tau_{f,1}))\\
        &= \hat \gamma(\hat u,\Gamma^{\hat v}(\tau_{f,1}))\Big( \exp\Big( \int^{\Gamma^{\hat v}(\tau_{f,2})}_{\Gamma^{\hat v}(\tau_{f,1})} \frac{2}{r(1-\mu)}\Im[\overline{\psi_+}\partial_{\hat v}\psi_+]\,d\hat v\Big)-1\Big),
\end{align*}
for $\hat u\in[0,\hat u_0]$. Hence, as depicted in Figure \ref{fig:Penrose10-gauge-proof}, by defining
\[
\hat v^*(\hat u):=\begin{dcases}
    \hat v^{3M}(\hat u) &\text{if $r(\hat u,0)\le3M_{\tau_f}$}\\
    0 &\text{otherwise}
\end{dcases}
\]
we may express $\hat \gamma(\hat u,\Gamma^{\hat v}(\tau_{f,1}))$ by integrating the transport equation for $\hat\gamma$ in $\hat v$ from $\hat v^*(\hat u)$, and write
\begin{equation}\label{eq:gauge-est-aux1}
\begin{aligned}
        &\mathfrak{u}'_{\tau_{f,2}}(\hat u) - \mathfrak{u}'_{\tau_{f,1}}(\hat u) \\
        &= \frac{1}{1-\mu(\hat u,\hat v^*(\hat u))}\exp\Big( \int^{\Gamma^{\hat v}(\tau_{f,1})}_{\hat v^*(\hat u)} \frac{2}{r(1-\mu)}\Im[\overline{\psi_+}\partial_{\hat v}\psi_+]\,d\hat v\Big)\Big[ \exp\Big( \int^{\Gamma^{\hat v}(\tau_{f,2})}_{\Gamma^{\hat v}(\tau_{f,1})} \frac{2}{r(1-\mu)}\Im[\overline{\psi_+}\partial_{\hat v}\psi_+]\,d\hat v\Big)-1\Big],
\end{aligned}
\end{equation}
As the solution exists and is smooth on $\hat\D_{\tau_{f,1}}$, it holds
\[
    \Big|\frac{1}{1-\mu(\hat u,\hat v^*(\hat u))}\exp\Big( \int^{\Gamma^{\hat v}(\tau_{f,1})}_{\hat v^*(\hat u)} \frac{2}{r(1-\mu)}\Im[\overline{\psi_+}\partial_{\hat v}\psi_+]\,d\hat v\Big)\Big|\le C(\hat u_0).
\]
Indeed, the choice of $\hat v^*(\hat u)$ ensures that the reference point $(\hat u,\hat v^*(\hat u))$ lies in the region $\{r\geq 3M_{\tau_f}\}$, where $1-\mu$ remains strictly positive for $\hat u\in[0,\hat u_0]$ and $\ve$ small enough, thanks to \eqref{eq:mu-2}. The integral can then be subdivided into $\{r\leq\Lambda\}$ and $\{r\geq\Lambda\}$: the former is readily bounded by the compactness of the region $\{0\le\hat u\le \hat u_0\}\cap\{r\le\Lambda\}$, while the latter is controlled by \eqref{eq:boot-outg-energy}. Regarding the second factor in \eqref{eq:gauge-est-aux1}, we obtain by \eqref{est:psi+_bound-bis} and \eqref{est:vpsi+_bound} that
\begin{align}
    \Big| \exp\Big( \int^{\Gamma^{\hat v}(\tau_{f,2})}_{\Gamma^{\hat v}(\tau_{f,1})} \frac{2}{r(1-\mu)}\Im[\overline{\psi_+}\partial_{\hat v}\psi_+]\,d\hat v\Big)-1\Big|\les \int^{\Gamma^{\hat v}(\tau_{f,2})}_{\Gamma^{\hat v}(\tau_{f,1})} \ve^2 r(\hat u,\hat v)^{-4}\,d\hat v
    \les_{\hat u_0} \ve^2 \tau_{f,1}^{-3},
\end{align}
\begin{comment}
    analogous to AKU
\end{comment}
for $\tau_{f,1}$ large enough. Hence, these bounds yield $|\mathfrak{u}'_{\tau_{f,2}}(\hat u) - \mathfrak{u}'_{\tau_{f,1}}(\hat u)|\les_{\hat u_0} \ve^2 \tau_{f,1}^{-3}$, which therefore gives, after integrating in $\hat u$, 
\[
|\mathfrak{u}_{\tau_{f,2}}(\hat u) - \mathfrak{u}_{\tau_{f,1}}(\hat u)|\les_{\hat u_0} \ve^2 \tau_{f,1}^{-3}\les\ve^2\eta,
\]
for $\tau_{f,1}$ sufficiently large and $\hat u\in[0, \hat u_0]$.
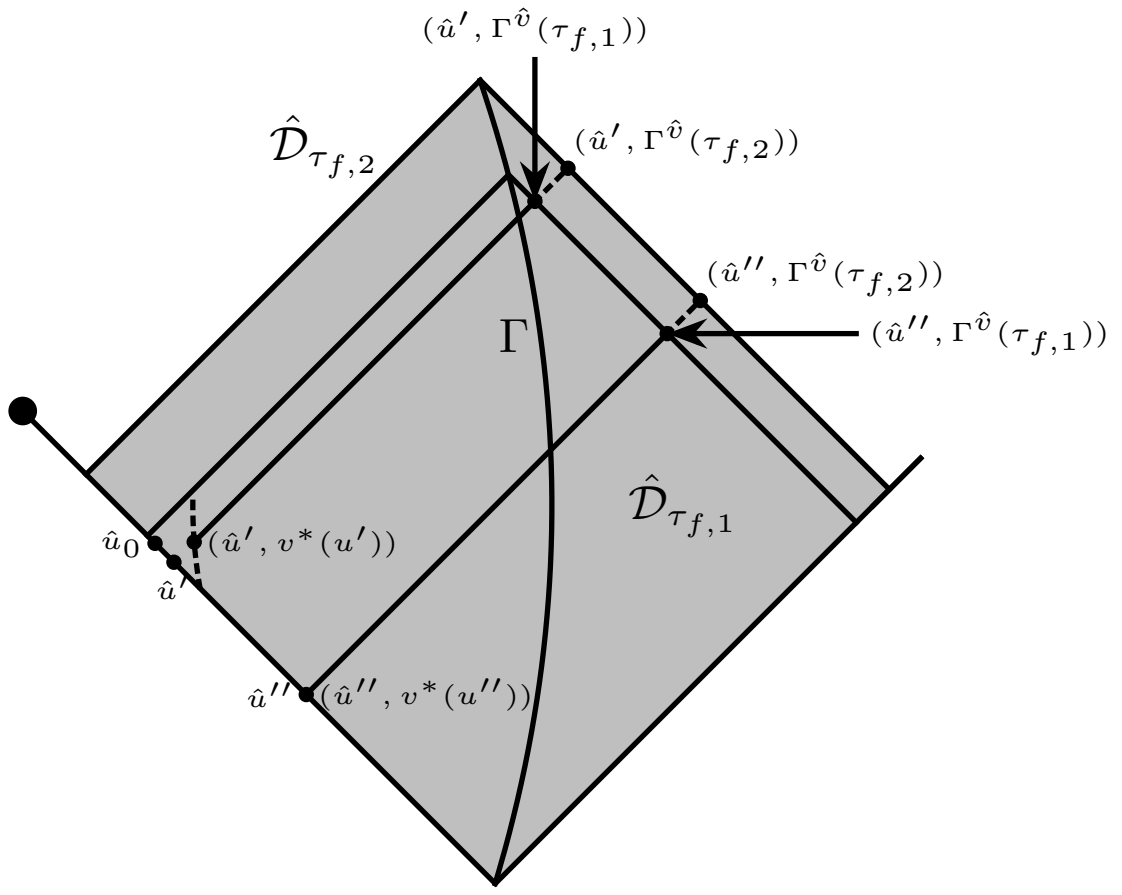
\begin{figure}
\centering
\begin{tikzpicture}[line cap=rect,inner sep=0pt, x=1cm, y=1cm, scale=1.25,font=\small]

\coordinate   (U)   at (-0.5,2.5)      {};
\coordinate   (W)   at (4.25,2.25)      {};
\coordinate   (A)   at (0.2,1.8)      {};
\coordinate   (B)   at (0.3,1.7)      {};
\coordinate   (C)   at (1,1)      {};
\coordinate  (I)   at (0,2)      {};
\coordinate (II)   at (2,0)      {};
\coordinate  (III)   at (4,2)      {};
\coordinate   (IV)   at (2,4)      {};
\coordinate   (V)   at (0.16016,1.83984)      {};
\coordinate   (VI)   at (2.07031,3.75)      {};
\coordinate   (VII)   at (3.91016,1.91016)      {};
\coordinate   (IV-bis)   at (1.92031,4.25)      {};
\coordinate   (I-bis)   at (-0.16484,2.16484)      {};
\coordinate   (III-bis)   at (4.08515,2.08516)      {};

\path[name path=parab1, domain=0:3, samples=100, smooth, variable=\x]
    plot ({\x},{1/6*\x*\x - 1/6*\x + 3});

\fill[lightgray] (I-bis) -- (IV-bis) -- (III-bis) -- (II) --(I-bis)  ;
%\fill[gray] (I) -- (V) -- (VI) -- (VII) -- (III) -- (IV) -- (I);
%\fill[lightgray] (V) -- (II) -- (VII) -- (VI) -- (V);

\draw[line width=1pt]  (U) -- (II) -- (W); 
\draw[line width=1pt]  (V) -- (VI) -- (VII); 
\draw[line width=1pt]  (I-bis) -- (IV-bis) -- (III-bis); 
\draw[line width=1pt]  (III) -- (W); 
%\draw[line width=1pt]  (U) -- (I) 
%-- node[pos=0.4, below=2pt, sloped]    {$\{\hat v=0\}$}
%       (II) -- node[pos=0.55, below=2pt, sloped]    {$\{\hat u=0\}$} (III) -- (IV) -- (I); %-- 
            %node[midway, below=2pt, sloped, font=\tiny]    {$u=u_0$}

\draw[line width=1pt,domain=0:4.25, smooth, variable=\y] plot ({2 + 3/40*\y*(4-\y)}, {\y});
%\draw[line width=1pt,dotted,domain=4:4.25, smooth, variable=\y] plot ({2 + 3/40*\y*(4-\y)}, {\y});
\draw[line width=1pt,dotted,domain=1.57:2.05, smooth, variable=\y] plot ({1/5*(\y-2)^2 + 2/5}, {\y});

\draw[line width=1pt]  (0.407418,1.807418) -- (2.210155,3.61016);

\draw[line width=1pt]  (C) -- (2.91016,2.91016);

\draw[line width=1pt,dotted] (2.210155, 3.61016)--(2.385155,3.78516);

\draw[line width=1pt,dotted] (2.910155,2.910155)--(3.085155,3.085155);
%\draw[-Stealth, line width=1pt] (3.68178,3.06822) node[right=2pt] {$\hat{\mathcal{D}}_{\tau_f+\eta}$} -- (3.30678,2.69322);

\filldraw[black] (U) circle (2pt);
\filldraw[black] (A) circle (1pt);
\filldraw[black] (B) circle (1pt);
\filldraw[black] (C) circle (1pt);
\filldraw[black] (0.407418,1.807418) circle (1pt);

\node[left=2pt] at (A)     {\tiny $\hat u_0$};
\node[below=2pt] at (B)     {\tiny $\hat u'$};
\node[right=2pt] at (0.407418,1.807418)     {\tiny $(\hat u',v^*(u'))$};
\node[right=2pt] at (C)     {\tiny $(\hat u'',v^*(u''))$};
\node[left=2pt] at (C)     {\tiny $\hat u''$};

\node at (3,2) {$\hat{\mathcal{D}}_{\tau_{f,1}}$};
\node at (2.1,2.9) {$\Gamma$};
\node at (1.1,3.9) {$\hat{\mathcal{D}}_{\tau_{f,2}}$};

\filldraw[black] (2.385155,3.78516) circle (1pt);
\node[above right=1pt] at (2.385155,3.78516)     {\tiny $(\hat u',\Gamma^{\hat v}(\tau_{f,2}))$};
\filldraw[black] (2.210155, 3.61016) circle (1pt);
\draw[-Stealth, line width=1pt] (2.210155, 4.36016) node[above=1pt] {\tiny $(\hat u',\Gamma^{\hat v}(\tau_{f,1}))$} -- (2.210155, 3.61016);

\filldraw[black] (3.085155,3.085155) circle (1pt);
\node[above right=1pt] at (3.085155,3.085155)     {\tiny $(\hat u'',\Gamma^{\hat v}(\tau_{f,2}))$};

\filldraw[black] (2.910155,2.910155) circle (1pt);
\draw[-Stealth, line width=1pt] (3.910155,2.910155) node[right=2pt] {\tiny $(\hat u'',\Gamma^{\hat v}(\tau_{f,1}))$} -- (2.910155,2.910155);
\end{tikzpicture}

\caption{We depict the Penrose diagram of the spacetime region and the null hypersurfaces considered in the argument controlling $\mathfrak{u}'_{\tau_{f,2}}(\hat u) - \mathfrak{u}'_{\tau_{f,1}}(\hat u)$. For $\tau_{f,1}$ sufficiently large, it holds $\Gamma^{\hat u}(\tau_{f,2})>\Gamma^{\hat u}(\tau_{f,1})>\hat u_0$. We illustrate the two possible configurations of $\hat v^*(\hat u)$, depending on the position of $\hat u$ relative to the curve $\{r=3M_{\tau_{f}}\}$. This distinction ensures that the resulting hypersurface remains in the region $\{r\ge3M_{\tau_{f}}\}$, allowing us to invoke \eqref{eq:mu-2} to deduce the positivity of $1-\mu$ for sufficiently small $\ve$. The late-time tail region from which the error estimate arises is also indicated by dotted lines.}
\label{fig:Penrose10-gauge-proof}
\end{figure}
Proceeding analogously with $\mathfrak{v}$, we use the gauge-invariant equation \eqref{eq:transport-kappa}, which reads
\[
    \partial_{\hat u} \hat \kappa  = -2r\hat\kappa\Im[\overline{\psi_-}\partial_{\hat u}\psi_-].
\]
Thus, from \eqref{def:gauges} we obtain
\begin{align}
        \nonumber
        \mathfrak{v}'_{\tau_{f,2}}(\hat v) &- \mathfrak{v}'_{\tau_{f,1}}(\hat v) = \hat \kappa(\Gamma^{\hat u}(\tau_{f,2}),\hat v)-\hat \kappa(\Gamma^{\hat u}(\tau_{f,1}),\hat v)\\
        \nonumber
        &= \hat \kappa(\Gamma^{\hat u}(\tau_{f,1}),\hat v)\Big( \exp\Big( \int^{\Gamma^{\hat u}(\tau_{f,2})}_{\Gamma^{\hat u}(\tau_{f,1})} -2r \Im[\oline{\psi_-}\partial_{\hat u}\psi_-]\,d\hat u\Big)-1\Big)
        \\\label{eq:gauge-est-aux2}&= \frac{1}{1-\mu(0,\hat v)}\exp\Big( \int^{\Gamma^{\hat u}(\tau_{f,1})}_0 -2r \Im[\oline{\psi_-}\partial_{\hat u}\psi_-]\,d\hat u\Big)\Big( \exp\Big( \int^{\Gamma^{\hat u}(\tau_{f,2})}_{\Gamma^{\hat u}(\tau_{f,1})} -2r \Im[\oline{\psi_-}\partial_{\hat u}\psi_-]\,d\hat u\Big)-1\Big),
\end{align}
for $\hat v\in [0,\hat v_0]$. Similarly as before, we infer 
\[
    \Big|\frac{1}{1-\mu(0,\hat v)}\exp\Big( \int^{\Gamma^{\hat u}(\tau_{f,1})}_0 -2r \Im[\oline{\psi_-}\partial_{\hat u}\psi_-]\,d\hat u\Big)\Big|\les C(\hat v_0),
\]
by treating separately the regions $\{r\le\Lambda\}$ and $\{r\ge\Lambda\}$, and using \eqref{eq:mu-lower-bound}, \eqref{eq:boot-ing-energy} and \eqref{eq:boot-ing-flux}. Regarding the last factor, we use \eqref{est:psi-_bound}, \eqref{est:upsi-_bound} and the fact that $\{\Gamma^{\hat u}(\tau_f)\}_{\tau_f}$ is a Cauchy sequence by \eqref{eq:thetaU*-bis}, to bound
\[
\Big| \exp\Big( \int^{\Gamma^{\hat u}(\tau_{f,2})}_{\Gamma^{\hat u}(\tau_{f,1})} -2r \Im[\oline{\psi_-}\partial_{\hat u}\psi_-]\,d\hat u\Big)-1\Big|\les
\ve^2 |\Gamma^{\hat u}(\tau_{f,2})-\Gamma^{\hat u}(\tau_{f,1})|\les
\ve^2\eta,
\]
for $\tau_{f,1}$ sufficiently large. Thus, these bounds yield $|\mathfrak{v}'_{\tau_{f,2}}(\hat v) - \mathfrak{v}'_{\tau_{f,1}}(\hat v)|\les_{\hat v_0} \ve^2 \eta$, which therefore gives, after integrating in $\hat v$, 
\[
|\mathfrak{v}_{\tau_{f,2}}(\hat v) - \mathfrak{v}_{\tau_{f,1}}(\hat v)|\les_{\hat v_0}\ve^2\eta,
\]
for $\tau_{f,1}$ sufficiently large and $\hat v\in[0, \hat v_0]$.

Let us now focus on the second derivatives. Differentiating \eqref{eq:gauge-est-aux1} yields
\begin{align*}
    \mathfrak{u}''_{\tau_{f,2}}(\hat u) &- \mathfrak{u}''_{\tau_{f,1}}(\hat u) = \Big[\frac{\partial_{\hat u}\mu(\hat u,\hat v^*(\hat u))}{1-\mu(\hat u,\hat v^*(\hat u))} + \int^{\Gamma^{\hat v}(\tau_{f,1})}_{\hat v^*(\hat u)} \partial_{\hat u}\Big( \frac{2}{r(1-\mu)}\Im[\overline{\psi_+}\partial_{\hat v}\psi_+]\Big)\,d\hat v\Big] (\mathfrak{u}'_{\tau_{f,2}}(\hat u) - \mathfrak{u}'_{\tau_{f,1}}(\hat u))\\ & + \frac{1}{1-\mu(\hat u,\hat v^*(\hat u))} \exp\Big( \int^{\Gamma^{\hat v}(\tau_{f,2})}_{0} \frac{2}{r(1-\mu)}\Im[\overline{\psi_+}\partial_{\hat v}\psi_+]\,d\hat v\Big)\int^{\Gamma^{\hat v}(\tau_{f,2})}_{\Gamma^{\hat v}(\tau_{f,1})}\partial_{\hat u}\Big( \frac{2}{r(1-\mu)}\Im[\overline{\psi_+}\partial_{\hat v}\psi_+]\Big)\,d\hat v.  
\end{align*}
Using \eqref{eq:partial_u-psi+}--\eqref{eq:partial_v-psi-}, we explicitly bound
\begin{align*}
    \Big|\partial_{\hat u}\Big( \frac{2}{r(1-\mu)}\Im[\overline{\psi_+}\partial_{\hat v}\psi_+]\Big)\Big|&= \Big|\frac{2}{r(1-\mu)}\Big[\Big(\frac{\hat\gamma}{r} +\frac{2\partial_{\hat u} m}{r(1-\mu)} -\frac{\partial_{\hat u}\hat\kappa}{\hat\kappa}\Big)\Im[\oline{\psi_+}\partial_{\hat v}\psi_+]+\hat \nu\Im[\oline{\psi_-}\partial_{\hat v}\psi_+] \\&+ \Big(\frac12\partial_{\hat v}\hat \nu + \frac{\hat \Omega^2(1-\mu)}{r}\Big)\Im[\oline{\psi_+}\psi_-]\Big]\Big|\\&\les_{\hat u_0} \ve^2r^{-4}\tau^{-3/2+\delta/2}(1+ \tau^{-3/2+\delta/2}) \les_{\hat u_0}\ve^2r^{-4}\tau^{-3/2+\delta/2}.
\end{align*}
Therefore, with analogous arguments as before, we obtain $|\mathfrak{u}''_{\tau_{f,2}} - \mathfrak{u}''_{\tau_{f,1}}|\les_{\hat u_0} \ve^2 \tau^{-3}_{f,1}\les \ve^2 \eta$ for $\tau_{f,1}$ large enough.

Similarly, we differentiate \eqref{eq:gauge-est-aux2} and find
\begin{align*}
    \mathfrak{v}''_{\tau_{f,2}}(\hat v) &- \mathfrak{v}''_{\tau_{f,1}}(\hat v) = \Big[\frac{\partial_{\hat v}\mu(0,\hat v)}{1-\mu(0,\hat v)} + \int^{\Gamma^{\hat u}(\tau_{f,1})}_0 \partial_{\hat v}\big(-2r \Im[\overline{\psi_-}\partial_{\hat u}\psi_-]\big)\,d\hat u\Big] (\mathfrak{v}'_{\tau_{f,2}}(\hat v) - \mathfrak{v}'_{\tau_{f,1}}(\hat v))\\ & + \frac{1}{1-\mu(0,\hat v)} \exp\Big( \int^{\Gamma^{\hat u}(\tau_{f,2})}_{0} \big(-2r \Im[\overline{\psi_-}\partial_{\hat u}\psi_-]\big)\,d\hat u\Big)\int^{\Gamma^{\hat u}(\tau_{f,2})}_{\Gamma^{\hat u}(\tau_{f,1})}\partial_{\hat v} \big(-2r \Im[\overline{\psi_-}\partial_{\hat u}\psi_-]\big)\,d\hat u.  
\end{align*}
Again, we substitute \eqref{eq:partial_u-psi+}--\eqref{eq:partial_v-psi-} to compute
\begin{align*}
    \partial_{\hat v}\big(-2r \Im[\oline{\psi_-}\partial_{\hat u}\psi_-]\big) = 2r\Big(\frac{\partial_{\hat v}\hat\nu}{-\hat\nu} -\frac{\hat \lambda}{r}\Big)\Im[\oline{\psi_-}\partial_{\hat u}\psi_-] + \frac{\hat \kappa}{r^2}\Im[\oline{\psi_+}\partial_{\hat u}\psi_-] + \Big(\frac{\partial_{\hat u}\hat \kappa}{2r^2}-\frac{2\hat \kappa \hat \nu}{r^3}\Big)\Im[\oline{\psi_-}\psi_+],
\end{align*}
which can be uniformly bounded as before by Lemmas \ref{lem:pointwise_estimates_dirac1}, \ref{lem:pointwise_estimates_dirac2} and \ref{lem:pointwise-est-Einstein-eq}: 
\[
\big|\partial_{\hat v}\big(-2r \Im[\oline{\psi_-}\partial_{\hat u}\psi_-]\big)\big|\les \ve^2.
\]
Therefore, using again the fact that $\{\Gamma^{\hat u}(\tau_f)\}_{\tau_f}$ is Cauchy, we deduce
\begin{align*}
    \Big|\int^{\Gamma^{\hat u}(\tau_{f,2})}_{\Gamma^{\hat u}(\tau_{f,1})}\partial_{\hat v} \big(-2r \Im[\overline{\psi_-}\partial_{\hat u}\psi_-]\big)\,d\hat u\Big| \les_{\hat v_0}
    \ve^2 |\Gamma^{\hat u}(\tau_{f,2})- \Gamma^{\hat u}(\tau_{f,1})|\les_{\hat v_0} \ve^2\eta,
\end{align*}
for $\tau_{f,1}$ large enough. Thus, by the same arguments as before, we obtain $|\mathfrak{v}''_{\tau_{f,2}} - \mathfrak{v}''_{\tau_{f,1}}|\les_{\hat v_0} \ve^2 \eta$ for $\tau_{f,1}$ large enough.
\end{proof}

Let us now prove Proposition \ref{prop:gauge-diffeo}.
\begin{proof}[Proof of Proposition \ref{prop:gauge-diffeo}] Existence and regularity of $\Upphi_\infty=(\mathfrak{u}_\infty,\mathfrak{v}_\infty)$ are now consequences of the previous lemma. It remains to prove that it is a diffeomorphism. First, $\mathfrak{u}_\infty$ and $\mathfrak{v}_\infty$ are strictly increasing as
    \begin{align*}
        \frac{d\mathfrak{u}_\infty}{d\hat u} &= \lim_{\tau_f\to+\infty} \hat \gamma(\hat u,\Gamma^{\hat v}(\tau_f))= \frac{1}{1-\mu(\hat u,\hat v^*(\hat u))}\exp\Big( \int^{+\infty}_{\hat v^*(\hat u)} \frac{2}{r(1-\mu)}\Im[\oline{\psi_+}\partial_{\hat v}\psi_+]\,d\hat v\Big)>0
    \end{align*}
by \eqref{eq:mu-2}, and 
    \begin{align*}
            \frac{d\mathfrak{v}_\infty}{d\hat v} &= \lim_{\tau_f\to+\infty} \hat \kappa(\Gamma^{\hat u}(\tau_f),\hat v)>0,
    \end{align*}
    by Lemma \ref{lem:v-est}. Moreover, $\mathfrak{u}_\infty$ is surjective onto $[0,+\infty)$, since $\mathfrak{u}_\infty(0)=0$ and $\mathfrak{u}_{\tau_f}(\Gamma^{\hat u}(\tau_f))\sim \tau_f$ as $\tau_f\to+\infty$ by Lemma \ref{lem:tau-properties}, while  $\mathfrak{v}_\infty$ being surjective is again a consequence of Lemma \ref{lem:v-est}.
\end{proof}

From now on, we denote the eschatological gauge by $(u_\infty,v_\infty):=\Upphi_\infty(\hat u,\hat v)$ and by $\mathcal{D}_\infty$ its domain. We stress that, since $\Upphi_\infty$ is only $C^2$, the solution $(r,\hat\Omega,\psi_+,\psi_-)$ expressed in these coordinates is, accordingly, only $C^2$. 

Since the previous estimate is not uniform up to the horizon, and therefore problematic to uniformly control the anchored backgrounds, we prove the following proposition, which bounds the mapping $\Upphi_{\tau_{f,1}}\mapsto\Upphi_{\tau_{f,2}}$ in the entire domain of outer communication as $\tau_{f,1}\to+\infty$.

\begin{lemma}\label{lem:inv-est} For $1\le\tau_{f,1} <\tau_{f,2}\le+\infty$ and $\tau_{f,1}$ sufficiently large, the following bounds hold
\begin{align}
    \label{eq:inv1-est}
    \sup_{J^-(\Gamma(\tau_{f,1}))\cap\mathcal{D}_{\tau_{f,2}}}|\Upphi_{\tau_{f,1}}\circ\hat\Upphi_{\tau_{f,2}}-\mathrm{Id}|&\les \ve^2\tau_{f,1}^{-1},\\
    \label{eq:inv2-est}
    \sup_{J^-(\Gamma(\tau_{f,1}))\cap\mathcal{D}_{\tau_{f,2}}}|d(\Upphi_{\tau_{f,1}}\circ\hat\Upphi_{\tau_{f,2}}-\mathrm{Id})|&\les \ve^2\tau_{f,1}^{-2}.
\end{align}
\end{lemma}
\begin{proof} From Definition \ref{def:map-gauges}, we write $\Upphi_{\tau_{f,1}}\circ\hat\Upphi_{\tau_{f,2}}=\mathfrak{g}_{\tau_{f,1};\tau_{f,2}}\times\mathfrak{h}_{\tau_{f,1};\tau_{f,2}}$, where
    \begin{equation*}
        \mathfrak{g}_{\tau_{f,1};\tau_{f,2}}(u_{\tau_{f,2}}):=\int^{u_{\tau_{f,2}}}_0 \gamma_{\tau_{f,2}}(u'_{\tau_f,2},\Gamma^{v_{\tau_{f,2}}}(\tau_{f_1}))\,du'_{\tau_{f,2}} = u_{\tau_{f,2}}+\int^{u_{\tau_{f,2}}}_0 \big(\gamma_{\tau_{f,2}}(u'_{\tau_f,2},\Gamma^{v_{\tau_{f,2}}}(\tau_{f_1}))-1\big)\,du'_{\tau_{f,2}},
    \end{equation*}
    \begin{equation*}
        \mathfrak{h}_{\tau_{f,1};\tau_{f,2}}(v_{\tau_{f,2}}):=\int^{v_{\tau_{f,2}}}_0 \kappa_{\tau_{f,2}}(\Gamma^{u_{\tau_{f,2}}}(\tau_{f_1}),v'_{\tau_f,2})\,dv'_{\tau_{f,2}} = v_{\tau_{f,2}}+\int^{v_{\tau_{f,2}}}_0 \big(\kappa_{\tau_{f,2}}(\Gamma^{u_{\tau_{f,2}}}(\tau_{f_1}),v'_{\tau_f,2})-1\big)\,dv'_{\tau_{f,2}}.
    \end{equation*}
Hence, by \eqref{eq:gamma-1}, we bound
\begin{align*}
    \sup_{u_{\tau_{f,2}}\in[0,\Gamma^{u_{\tau_{f,2}}}(\tau_{f,1})]}|\mathfrak{g}_{\tau_{f,1};\tau_{f,2}}'(u_{\tau_{f,2}})-1|&\les \sup_{u_{\tau_{f,2}}\in[0,\Gamma^{u_{\tau_{f,2}}}(\tau_{f,1})]}|\tilde\gamma_{\tau_{f,2}}(u_{\tau_{f,2}},\Gamma^{v_{\tau_{f,2}}}(\tau_{f,1}))|\\&\les \sup_{u_{\tau_{f,2}}\in[0,\Gamma^{u_{\tau_{f,2}}}(\tau_{f,1})]}\ve^2 r^{-2} \tau^{-3+\delta}(u_{\tau_{f,2}},\Gamma^{v_{\tau_{f,2}}}(\tau_{f,1}))\les \ve^2\tau_{f,1}^{-2},
\end{align*}
where we have split $[0,\Gamma^{u_{\tau_{f,2}}}(\tau_{f,1})]\times\{\Gamma^{v_{\tau_{f,2}}}(\tau_{f,1})\}$ in the cases $\{\tau\le \frac12\tau_{f,1}\}$, where $r\gtrsim\tau_{f,1}$, and $\{\tau\ge \frac12\tau_{f,1}\}$, where $r\les\tau_{f,1}$. 

On the other hand, we analogously need to estimate
\begin{align*}
    \sup_{v_{\tau_{f,2}}\in[0,\Gamma^{v_{\tau_{f,2}}}(\tau_{f,1})]}|\mathfrak{h}_{\tau_{f,1};\tau_{f,2}}'(v_{\tau_{f,2}})-1|&\les \sup_{v_{\tau_{f,2}}\in[0,\Gamma^{v_{\tau_{f,2}}}(\tau_{f,1})]}|\tilde\kappa_{\tau_{f,2}}(\Gamma^{u_{\tau_{f,2}}}(\tau_{f,1}),v_{\tau_{f,2}})|.
\end{align*}
We proceed by subdividing again $\{\Gamma^{u_{\tau_{f,2}}}(\tau_{f,1})\}\times[0,\Gamma^{v_{\tau_{f,2}}}(\tau_{f,1})]$ in the cases $\{\tau\le\frac12 \tau_{f,1}\}$ and $\{\tau\ge\frac12\tau_{f,1}\}$.
Let us begin with the latter: since $u\les \tau_{f,1}$ there, \eqref{eq:kappa-1} readily yields
\[
    |\mathfrak{h}_{\tau_{f,1};\tau_{f,2}}'(v_{\tau_{f,2}})-1|\les \ve^2 \tau^{-3+\delta}(\Gamma^{u_{\tau_{f,2}}}(\tau_{f,1}),v_{\tau_{f,2}})\les\ve^2 \tau_{f,1}^{-3+\delta},
\]
for $v_{\tau_{f,2}}\in[0,\Gamma^{v_{\tau_{f,2}}}(\tau_{f,1})]\cap\{\tau\ge\frac12\tau_{f,1}\}$.
To treat the case $\{\tau\le\frac12 \tau_{f,1}\}$, we need to improve the current bound on $\tilde\kappa$ by exploiting that now $u\gtrsim \tau_{f,1}$. To this end, we combine \eqref{est:partial_u-kappa} with the exponential decay estimate \eqref{eq:nu-exp} to obtain
\begin{align*}
    |\tilde\kappa(\Gamma^{u_{\tau_{f,2}}}(\tau_{f,1}),v_{\tau_{f,2}})|&\les \int^{\Gamma^{u_{\tau_{f,2}}}(\tau_{f,2})}_{\Gamma^{u_{\tau_{f,2}}}(\tau_{f,1})}|\p_u\tilde\kappa|\,du'\les\ve^2\int^{\Gamma^{u_{\tau_{f,2}}}(\tau_{f,2})}_{\Gamma^{u_{\tau_{f,2}}}(\tau_{f,1})}(-\nu) \bar r^{-1}\tau^{-3/2+\delta/2}\,du'\\&\les \ve^2 \int^{\Gamma^{u_{\tau_{f,2}}}(\tau_{f,2})}_{\Gamma^{u_{\tau_{f,2}}}(\tau_{f,1})}(-\nu)\,du'\les \ve^2 e^{-\Gamma^{u_{\tau_{f,2}}}(\tau_{f,1})},
\end{align*}
where we used that $\tilde\kappa(\Gamma^{u_{\tau_{f,2}}}(\tau_{f,2}),v_{\tau_{f,2}})=0$, and that $\bar r^{-1}\tau^{-3/2+\delta/2}$ is bounded on the region of integration, since $v_{\tau_{f,2}}\in[0,\Gamma^{v_{\tau_{f,2}}}(\tau_{f,1})]\cap\{\tau\le\frac12\tau_{f,1}\}$. As the resulting bound decays faster than $\ve^2\tau_{f,1}^{-3+\delta}$ for $\tau_{f,1}$ sufficiently large,
combining it with the previous case we conclude
\[
\sup_{u_{\tau_{f,2}}\in[0,\Gamma^{u_{\tau_{f,2}}}(\tau_{f,1})]}|\mathfrak{g}_{\tau_{f,1};\tau_{f,2}}'(u_{\tau_{f,2}})-1|\les \ve^2\tau_{f,1}^{-2}, \quad \sup_{v_{\tau_{f,2}}\in[0,\Gamma^{v_{\tau_{f,2}}}(\tau_{f,1})]}|\mathfrak{h}_{\tau_{f,1};\tau_{f,2}}'(v_{\tau_{f,2}})-1|\les\ve^2 \tau_{f,1}^{-3+\delta},
\]
which readily implies \eqref{eq:inv1-est} and \eqref{eq:inv2-est}.
\end{proof}

We next turn to determining the final mass of the anchored Schwarzschild background.
\begin{lemma}\label{lem:final-anch-mass} The following limit exists 
\begin{equation}\label{eq:anch-mass-bound}
    M_\infty:=\lim_{\tau_f\to+\infty}M_{\tau_f}.
\end{equation}
\end{lemma}
\begin{proof} It is a straightforward consequence of \eqref{def:anch-solution2} and \eqref{eq:boot-varpi}. Indeed, one can note that the sequence $\{M_{\tau_{f}}\}_{\tau_f}$ is Cauchy
\begin{equation}\label{eq:anch-mass-bound2}
    |M_{\tau_{f,2}}- M_{\tau_{f,1}}|=|M_{\tau_{f,2}}- m_{\tau_{f,2}}(\Gamma(\tau_{f,1}))|\les|\tilde m_{\tau_{f,2}}(\Gamma(\tau_{f,1}))|\les \ve^{3/2}\tau^{-3+\delta}_{f,1},
\end{equation}
and therefore denote by $M_\infty$ its limit.
\end{proof}

Thanks to Lemmas \ref{lem:inv-est} and \ref{lem:final-anch-mass}, we are now able to compare $\bar r_{\tau_{f,1}}$ and $\bar r_{\tau_{f,2}}$. Indeed, as they are associated to two different bootstrap times, and therefore defined in different coordinate systems, one needs to pull them back to the same domain. Rather than pulling both functions back to $\hat\D_{\infty}$, which would require us to rely on the weak Cauchy estimate \eqref{eq:cauchy-prop}, we instead pull $\bar r_{\tau_{f,1}}$ back to the later $(u_{\tau_{f,2}},v_{\tau_{f,2}})$ coordinates, allowing us to exploit the stronger estimate \eqref{eq:inv1-est}.
\begin{lemma} For $1\le\tau_{f,1} <\tau_{f,2}\le+\infty$ and $\tau_{f,1}$ sufficiently large, it holds that
    \begin{equation}\label{eq:pull-back-r}
        \sup_{J^-(\Gamma(\tau_{f,1}))\cap\mathcal{D}_{\tau_{f,2}}}|\bar r_{\tau_{f,2}} - \bar r_{\tau_{f,1}}\circ (\Upphi_{\tau_{f,1}}\circ\hat\Upphi_{\tau_{f,2}})|\les \ve^{3/2}\tau^{-1/2}_{f,1}.
    \end{equation}
\end{lemma}
\begin{proof} 
Let us define $\varrho:=\bar r_{\tau_{f,2}}-\bar r_{\tau_{f,1}}\circ(\Upphi_{\tau_{f,1}}\circ\hat\Upphi_{\tau_{f,2}})$. We proceed by differentiating in $u_{\tau_{f,2}}$ and obtain
\begin{align*}
    \partial_{u_{\tau_{f,2}}}\varrho &= \bar \nu_{\tau_{f,2}} - (\bar\nu_{\tau_{f,1}}\circ(\Upphi_{\tau_{f,1}}\circ\hat\Upphi_{\tau_{f,2}}))\mathfrak{g}'_{\tau_{f,1};\tau_{f,2}} = \Big(1-\frac{2M_{\tau_{f,1}}}{\bar r_{\tau_{f,1}}\circ(\Upphi_{\tau_{f,1}}\circ\hat\Upphi_{\tau_{f,2}})}\Big)\mathfrak{g}'_{\tau_{f,1};\tau_{f,2}}-\Big(1-\frac{2M_{\tau_{f,2}}}{\bar r_{\tau_{f,2}}}\Big) \\&= \Big(1-\frac{2M_{\tau_{f,1}}}{\bar r_{\tau_{f,1}}\circ(\Upphi_{\tau_{f,1}}\circ\hat\Upphi_{\tau_{f,2}})}\Big)-\Big(1-\frac{2M_{\tau_{f,2}}}{\bar r_{\tau_{f,2}}}\Big)+ O(\ve^2 \tau^{-2}_{f,1}),
\end{align*}
where we used \eqref{eq:inv2-est} in the last equality.
We now sum and subtract the term $2M_{\tau_{f,1}}/\bar r_{\tau_{f,2}}$ and use \eqref{eq:anch-mass-bound2}, and deduce
\begin{align*}
    \partial_{u_{\tau_{f,2}}}\varrho &=\frac{2(M_{\tau_{f,2}}-M_{\tau_{f,1}})}{\bar r_{\tau_{f,2}}}+ 2M_{\tau_{f,1}}\Big(\frac{1}{\bar r_{\tau_{f,2}}}-\frac{1}{\bar r_{\tau_{f,1}}\circ(\Upphi_{\tau_{f,1}}\circ\hat\Upphi_{\tau_{f,2}})}\Big)+ O(\ve^2 \tau^{-2}_{f,1})
    \\&= 2M_{\tau_{f,1}}\Big(\frac{1}{\bar r_{\tau_{f,2}}}-\frac{1}{\bar r_{\tau_{f,1}}\circ(\Upphi_{\tau_{f,1}}\circ\hat\Upphi_{\tau_{f,2}})}\Big)+ O(\ve^{3/2} \tau^{-2}_{f,1}),
\end{align*}
for $\tau_{f,1}$ sufficiently large. 
Under the bootstrap assumption $|\varrho|\le \ve \tau_{f,1}^{-1/3}$ with $\tau_{f,1}$ large, we may Taylor expand and write
\begin{equation}\label{eq:varrho-aux-u}
    \partial_{u_{\tau_{f,2}}}\varrho = - \frac{2M_{\tau_{f,1}}}{\bar r^2_{\tau_{f,2}}}\varrho+ O(\bar r^{-3}_{\tau_{f,2}}\varrho^2)+O(\ve^{3/2} \tau^{-2}_{f,1}).
\end{equation}
From \eqref{eq:tilde-r-on-Gamma}, we note that $|\varrho(\Gamma(\tau_{f,1}))|\les \ve^2\tau^{-2+\delta}_{f,1}$, which shows that the bootstrap assumption is satisfied near $\Gamma(\tau_{f,1})$ for $\tau_{f,1}$ large enough. Integrating \eqref{eq:varrho-aux-u} backwards in $u$ from $\Gamma(\tau_{f,1})$ (in the same fashion as in Proposition \ref{prop:r-estimate}), we bound
\begin{align}\label{eq:varrho-aux-u1}
    |\varrho(u_{\tau_{f,2}},\Gamma^{v_{\tau_{f,2}}}(\tau_{f,1}))|\les \ve^2\tau_{f,1}^{-2+\delta} + \int^{\Gamma^{u_{\tau_{f,2}}}(\tau_{f,1})}_{u_{\tau_{f,2}}}(\bar r^{-3}_{\tau_{f,1}}\varrho^2 +\ve^{3/2} \tau^{-2}_{f,1})\,du'_{\tau_{f,2}}\les \ve^2 \tau_{f,1}^{-2/3}+ \ve^{3/2}\tau_{f,1}^{-1} \les \ve^2 \tau_{f,1}^{-2/3},
\end{align}
which improves the bootstrap assumption on $\varrho$ for $\tau_{f,1}$ large enough. Hence, \eqref{eq:varrho-aux-u1} holds for $u_{\tau_{f,2}}\in[0,\Gamma^{u_{\tau_{f,2}}}(\tau_{f,1})]$.
Similarly to \eqref{eq:varrho-aux-u}, we may derive in $v_{\tau_{f,2}}$ and obtain
\[
    \partial_{v_{\tau_{f,2}}}\varrho = \frac{2M_{\tau_{f,2}}}{\bar r^2_{\tau_{f,2}}}\varrho+ O(\bar r^{-3}_{\tau_{f,2}}\varrho^2)+O(\ve^{3/2} \tau_{f,1}^{-2}).
\]
This can be integrated backwards from $[0,\Gamma^{u_{\tau_{f,2}}}(\tau_{f,1})]\times\{\Gamma^v(\tau_{f,1})\}$, and by using \eqref{eq:varrho-aux-u1}, together with the good sign of the coefficient of the linear term on the right-hand side, one finally obtains \eqref{eq:pull-back-r}.
\end{proof}

\subsection{Completing the proof}\label{sec:completing-proof}
Thanks to the bounds on the gauges and anchored backgrounds established in the previous sections, we are now in a position to rigorously pass to the limit $\tau_f\to+\infty$. This allows us to derive the final decay estimates on $\D_\infty$ and complete the proof of Theorem \ref{thm:Schwarz-stability2}.

We first extract the limiting anchored Schwarzschild background, with area radius function $\bar r_\infty$ and mass $M_\infty$, as described in \eqref{eq:anch-mass-bound}.
\begin{proposition}\label{prop:lim-radius-bckground} The limit 
\begin{equation}\label{eq:lim-radius-bckground1}
    \bar r_\star := \lim_{\tau_f\to+\infty} \bar r_{\tau_f}(0,0)
\end{equation}
exists and satisfies 
\begin{equation}\label{eq:lim-radius-bckground2}
    |\bar r_\star-\Lambda|\les \ve^2.
\end{equation}
Let us denote by $\bar r_\infty$ the $\infty$-anchored background solution on $\mathcal{D}_\infty$ with mass $M_{\infty}$ and bifurcation sphere area radius $\bar r_\infty(0,0)=\bar r_\star$, as defined in Section \ref{sec:anch-solution}. Then, it holds that
\begin{equation}\label{eq:lim-radius-bckground3}
    \sup_{J^{-}(\Gamma({\tau_f}))\cap \mathcal{D}_\infty}|\bar r_{\infty}-\bar r_{\tau_f}\circ (\Upphi_{\tau_f}\circ \hat\Upphi_\infty)|\les \ve^2\tau_f^{-1/2}.
\end{equation}
\end{proposition}
\begin{proof} The argument is analogous to \cite[Proposition 8.13]{angelopoulos2026nonlinear}. The existence of the limit \eqref{eq:lim-radius-bckground1} follows from \eqref{eq:pull-back-r} evaluated at $(0,0)$, whereas \eqref{eq:lim-radius-bckground2} is a consequence of \eqref{eq:tilde-r-on-Gamma}.
To prove \eqref{eq:lim-radius-bckground3}, we first note that, by \eqref{eq:pull-back-r}, the limit 
\[
    \mathfrak{r}(\hat u,\hat v):=\lim_{\tau_f\to+\infty} \bar r_{\tau_f}\circ \Upphi_{\tau_f} 
\]
exists on $\hat{\mathcal{D}}_\infty$ and satisfies
\begin{equation}\label{eq:lim-radius-bckground4}
    \sup_{J^-(\Gamma(\tau_f))\cap\hat{\mathcal{D}}_\infty}|\mathfrak{r}-\bar r_{\tau_f}\circ \Upphi_{\tau_f}|\les \ve^{3/2}\tau^{-1/2}_f.
\end{equation}
Claiming that $\bar r_\infty = \mathfrak{r}\circ\hat\Upphi_\infty$, we then infer \eqref{eq:lim-radius-bckground3} from \eqref{eq:lim-radius-bckground4}. As $\bar r_\infty(0,0)=\bar r_\star=\mathfrak{r}\circ\hat\Upphi_\infty(0,0)$, we are left to prove that $\mathfrak{r}\circ\hat\Upphi_\infty$ satisfies the Eddington--Finkelstein gauge conditions in the coordinates $(u_\infty,v_\infty)$. Let us compute explicitly
\begin{align*}
    \partial_{u_\infty}(\bar r_{\tau_f}\circ (\Upphi_{\tau_f}\circ \hat\Upphi_{\infty}))&= -\Big(1-\frac{2M_{\tau_f}}{\bar r_{\tau_f}\circ (\Upphi_{\tau_f}\circ \hat\Upphi_{\infty})}\Big)\mathfrak{g}'_{\tau_f,\infty},\\
    \partial_{v_\infty}(\bar r_{\tau_f}\circ (\Upphi_{\tau_f}\circ \hat\Upphi_{\infty}))&= \Big(1-\frac{2M_{\tau_f}}{\bar r_{\tau_f}\circ (\Upphi_{\tau_f}\circ \hat\Upphi_{\infty})}\Big)\mathfrak{h}'_{\tau_f,\infty}.
\end{align*}
Thanks to Lemmas \ref{lem:inv-est} and \ref{lem:final-anch-mass}, we infer that the two quantities above converge uniformly as $\tau_f\to+\infty$ to $-(1-2M_\infty/\mathfrak{r}\circ\hat\Upphi_\infty)$ and $(1-2M_\infty/\mathfrak{r}\circ\hat\Upphi_\infty)$, respectively. Thus, since this verifies the Eddington--Finkelstein gauge conditions for $\partial_{u_\infty}(\mathfrak{r}\circ\hat\Upphi_\infty)$ and $\partial_{v_\infty}(\mathfrak{r}\circ\hat\Upphi_\infty)$, we conclude.
\end{proof}
As noted before, we stress that $\bar r_\infty$ is smooth on $\D_\infty$ with the coordinates $(u_\infty,v_\infty)$, whereas $\bar r_\infty\circ\Upphi_\infty$ is only $C^2$ on $\hat\D_\infty$ with $(\hat u,\hat v)$.

\begin{proposition}\label{prop:energy-convergence} For any $\tau\in[1,+\infty)$ and $p\in[0,3-\delta]$, the following limits hold
    \begin{align*}
        \lim_{\tau_f\to+\infty} \uline{\mathcal{E}}^{\tau_f}(\tau)= \uline{\mathcal{E}}^{\infty}(\tau),\\
        \lim_{\tau_f\to+\infty} \mathcal{E}^{\tau_f}_p(\tau)= \mathcal{E}^{\infty}_p(\tau).
    \end{align*}
\end{proposition}
\begin{proof} The argument relies on the dominated convergence theorem. We refer to \cite[Proposition 8.15]{angelopoulos2026nonlinear} for more details on the proof, as the energies defined in this work are very similar to those considered there.
\end{proof}

Let us now prove the first part of Theorem \ref{thm:Schwarz-stability2}, that is showing the existence of a black hole region and of a regular event horizon in the maximal development of the seed data.
\begin{proposition}\label{prop:thm-part1} Given an $\ve$-Schwarzschildean seed data set with $\ve\le\ve_{\mathrm{stab}}$ and $\hat{\mathcal{Q}}_{\mathrm{max}}$ its maximal development, there exists a $\hat u_{\mathcal{H}^+}\in(0,U_*)$ satisfying
\begin{equation}\label{eq:u-H+}
    |\hat u_{\mathcal{H}^+}-\hat u_{\mathcal{H}^+,0}|\les \ve,
\end{equation}
where $\hat u_{\mathcal{H}^+,0}$ is defined in \eqref{eq:initial-horizon}, such that $[0,\hat u_{\mathcal{H}^+}]\times[0,+\infty)\subset\hat{\mathcal{Q}}_{\mathrm{max}}$, 
\begin{equation}
    \lim_{\hat v\to+\infty}r(\hat u,\hat v)=+\infty
\end{equation}
for every $\hat u\in[0,\hat u_{\mathcal{H}^+})$, and
\begin{equation}\label{eq:final-H+}
    \lim_{\hat v\to+\infty}r(\hat u_{\mathcal{H}^+},\hat v)=\lim_{\hat v\to+\infty} 2m(\hat u_{\mathcal{H}^+},\hat v)=2M_{\infty}.
\end{equation}
Therefore, $[0,\hat u_{\mathcal{H}^+})\times\{\hat v=+\infty\}$ may be regarded as future null infinity $\mathcal{I}^+$ and the future boundary of $J^-(\mathcal{I}^+)$ is the event horizon $\mathcal{H}^+:=\{\hat u_{\mathcal{H}^+}\}\times [0,+\infty)$. The black hole region is 
    \[
        \mathcal{BH}:=\hat{\mathcal{Q}}_{\mathrm{max}}\setminus J^-(\mathcal{I}^+)=[\hat u_{\mathcal{H}^+},U_*]\times[0,+\infty)\neq\emptyset,
    \]
    and future null infinity is complete in the sense of Christodoulou \cite{christodoulou1999global}.
\end{proposition}
\begin{proof} Thanks to the pointwise estimates for the Dirac field in Section \ref{sec:preliminary-estimates} and the extension principle proved in Proposition \ref{app:prop:slab-existence}, we have $\oline{\hat{\mathcal{D}}_\infty}\subset\hat{\mathcal{Q}}_\maxx$, where the closure is taken in the $(\hat u,\hat v)$-plane. By Lemma \ref{lem:thetaU*-bis} $\oline{\hat{\mathcal{D}}_\infty}\setminus\hat{\mathcal{D}}_\infty=\{\hat u_{\mathcal{H}^+}\}\times[0,+\infty)$. By \eqref{eq:boot-r} and \eqref{eq:lim-radius-bckground2}, it holds that 
\begin{equation}\label{eq:lim-radius}
    |r-\bar r_\infty\circ\Upphi_{\infty}|\les \ve^{3/2}\tau^{-2+\delta}
\end{equation}
on $\hat{\mathcal{D}}_\infty$, whereas, by the geometry of Schwarzschild, we have
\begin{align*}
    \lim_{\hat u\to \hat u_{\mathcal{H}^+}}\bar r_\infty\circ\Upphi_\infty(\hat u,\hat v) &= \lim_{u_\infty\to +\infty}\bar r_{\infty}(u_\infty,v_\infty(\hat v))=2M_\infty,
\end{align*}
for $\hat v\in[0,+\infty)$. Therefore, we deduce
\begin{align}
    \label{eq:final-H+-bis}
    |r(\hat u_{\mathcal{H}^+},\hat v)-2M_\infty|&\les \ve^{3/2}\tau^{-2+\delta}(\hat u_{\mathcal{H}^+},\hat v),
\end{align}
for $\hat v\in[0,+\infty)$.
This last fact and 
\begin{equation}\label{eq:final-mass}
    | m-M_\infty|\les \ve^{3/2}\tau^{-3+\delta}
\end{equation}
on $\oline{\hat{\mathcal{D}}}_\infty$, which follows by passing to the limit in \eqref{eq:boot-varpi}, together imply \eqref{eq:final-H+}.
Next, \eqref{eq:u-H+} follows by \eqref{eq:final-H+-bis} evaluated at $\hat v=0$, by \eqref{eq:final-mass} evaluated at $\tau=1$, and by the gauge condition $\hat\nu(\hat u,0)=-1$. Indeed,
\[
    \hat u_{\mathcal{H}^+} - \hat u_{\mathcal{H}^+,0} = (\Lambda-r(\hat u_{\mathcal{H}^+},0)) - \hat u_{\mathcal{H}^+,0}=(\Lambda-2M_\infty+O(\ve))- \hat u_{\mathcal{H}^+,0} =2(M_0-M_\infty)+O(\ve)=O(\ve).
\]
Moreover, noting that for $\hat u\in[0,\hat u_{\mathcal{H}^+})$,
\begin{align*}
    \lim_{\hat v\to +\infty}\bar r_\infty\circ\Upphi_\infty(\hat u,\hat v) &= \lim_{v_\infty\to +\infty}\bar r_{\infty}(u_\infty(\hat u),v_\infty)=+\infty,
\end{align*}
by the geometry of Schwarzschild, we deduce again by \eqref{eq:lim-radius} that
\begin{align*}
    \lim_{\hat v\to +\infty}r(\hat u,\hat v)&=+\infty ,
\end{align*}
for $\hat u\in[0,\hat u_{\mathcal{H}^+})$.
Finally, by the fact that $r$ is uniformly bounded on $\mathcal{H}^+$, the completeness of $\mathcal{I}^+$ follows by \cite[Section 5.4]{dafermos2005spherically}.
\end{proof}

Before proving Theorem \ref{thm:Schwarz-stability2}, we collect the final geometric estimates in the eschatological gauge $(u_\infty,v_\infty)$. Note that only certain estimates below follow directly from the bounds established in Sections \ref{sec:improvement-geometry} and \ref{sec:preliminary-estimates}, whereas others carry improved exponents in $\ve$.

Although the event horizon $\mathcal{H}^+$ is not covered by the coordinates $(u_\infty,v_\infty)$, we may formally attach it as the limiting hypersurface $\{u_\infty=+\infty\}$. With this interpretation, we shall consider the extended region $\D_\infty\cup\mathcal{H}^+$ and define certain geometric quantities continuously up to the horizon.

\begin{proposition}\label{prop:thm-part3} For $\ve_{\mathrm{stab}}$ sufficiently small, the following estimates, expressed in the eschatological gauge $(u_\infty,v_\infty)$, hold:
\begin{equation}\label{eq:gamma-2}
    |\gamma_\infty-1|\les\ve^2r^{-2}\tau^{-3+\delta}
\end{equation}
on $\mathcal{D}_\infty\cap \{r\ge\Lambda\}$, 
\begin{equation}\label{eq:nu-final}
    \Big|\frac{\nu_\infty}{\bar\nu_\infty}-1\Big|\les \ve^2\tau^{-1+\delta}
\end{equation}
on $\mathcal{D}_\infty$, and
\begin{align}
    \label{eq:r-final}
    |r-\bar r_\infty| &\les \ve^2\tau^{-2+\delta},\\
    |\lambda_\infty-\bar\lambda_\infty|&\les \ve^2\tau^{-2+\delta},\\
    \label{eq:kappa-3}
    |\kappa_\infty-1|&\les \ve^2\tau^{-2+\delta},\\
    \label{eq:varpi-final}
    | m-M_\infty|&\les \ve^2\tau^{-3+\delta},\\
    |(1-\mu)-(1-\bar\mu_\infty)|&\les \ve^2\tau^{-2+\delta},\\
    \Big|\p_{v_\infty}\log\Omega_\infty-\frac{M_\infty}{\bar r^{2}_\infty}\Big|&\les \ve^2\tau^{-2+\delta},\\
    \Big|\p_{u_\infty}\log\Omega_\infty+\frac{M_\infty}{\bar r^{2}_\infty}\Big|&\les \ve^2 r^{-2}\tau^{-3/4+\delta},
\end{align}
on $\mathcal{D}_\infty\cup\mathcal{H}^+$.
\end{proposition}
\begin{proof} Estimates \eqref{eq:gamma-2} and \eqref{eq:kappa-3} are simply obtained by passing to the limit in \eqref{eq:gamma-1} and \eqref{eq:kappa-1}. On the other hand, \eqref{eq:boot-varpi} is upgraded to \eqref{eq:varpi-final} by passing to the limit in \eqref{eq:varpi-aux-0}.

Let us now prove the remaining estimates by exploiting the refined estimate \eqref{eq:varpi-final} via a bootstrap argument in the eschatological gauge on $\D_\infty$. We denote by $\mathfrak{B}'$ the set of the times $\tau_f\in[1,+\infty)$ such that 
\[
    |r-\bar r_\infty|\les A'\ve^2\tau^{-2+\delta},\quad \Big|\frac{\nu_\infty}{\bar\nu_\infty}-1\Big|\les A'^2\ve^2\tau^{-1+\delta},
\]
where $A'\ge1$ is a constant to be determined. 

For $\tau_f\in\mathfrak{B}'$, we may use the bootstrap assumptions and \eqref{eq:varpi-final} to improve the Taylor expansions in Lemma \ref{lem:nu-lambda-expansion} to
    \begin{align*}
    \partial_{u_\infty}(r-\bar r_\infty) &= -\frac{2M_\infty\gamma_\infty}{\bar r^2_\infty}(r-\bar r_\infty)+O(\ve^2\bar r^{-1}_\infty \tau^{-3+\delta}) &\text{in }&\mathcal D_{\infty}\cap\{\tau\le\tau_f\}\cap\{r\ge\Lambda  \},\\
    \partial_v(r-\bar r_\infty)&= \frac{2M\kappa_\infty}{\bar r^2_\infty}(r-\bar r_\infty) +O(\ve^2\tau^{-3+\delta})&\text{in }&\mathcal D_{\infty}\cap\{\tau\le\tau_f\}.
    \end{align*}
Therefore, combining these estimates with $|(r-\bar r_\infty)|_\Gamma|\les \ve^2\tau^{-2+\delta}$, which follows by passing to the limit in \eqref{eq:tilde-r-on-Gamma}, one may repeat the arguments in Propositions \ref{prop:r-estimate} and \ref{prop:nu-estimate} and improve the above bootstrap assumptions for $A'$ chosen sufficiently large. In this manner, $\mathfrak{B}'=[1,+\infty)$, and \eqref{eq:nu-final} and \eqref{eq:r-final} hold. Finally, thanks to these upgraded bounds, one may improve the remaining ones by repeating the arguments in Lemmas \ref{lem:mu} and  \ref{lem:Omega-expansions}, and Proposition \ref{prop:r-estimate}.
\end{proof}

Let us now finally put everything together and conclude with the proof of Theorem \ref{thm:Schwarz-stability2} about the nonlinear stability of Schwarzschild as a solution to the Einstein--Dirac system. 

\begin{proof}[Proof of Theorem \ref{thm:Schwarz-stability2}] 
\textsc{Existence of a black hole region:} It is a consequence of Propositions \ref{prop:gauge-diffeo} and \ref{prop:thm-part1}. In addition, $\p_u r_\infty\to-1$ follows by \eqref{eq:gamma-2}.

\textsc{Orbital stability:} \eqref{eq:orb-stab1} is a consequence of the definition of $\mathfrak{D}$ and \eqref{eq:final-mass} evaluated at $\tau=1$. On the other hand, \eqref{eq:orb-stab2} does not immediately follow by Proposition \ref{prop:energy-convergence} with $p=3-\delta$. Indeed, recalling Lemma \ref{lem:initial-energy-est}, we see that the smallness parameter $\ve\ge\mathfrak{D}[\mathcal{S}]$ only bounds the energy $\uline{\mathcal{E}^\infty}(1)+\mathcal{E}^\infty_{3-\delta}(1)$ from above. To solve this issue, we proceed as in the previous proposition via an additional bootstrap argument on $\D_\infty$. 

We introduce $\mathfrak{B}'$ the set of times $\tau_f\in[1,+\infty)$ such that the bootstrap assumptions \eqref{eq:boot-nu}--\eqref{eq:boot-outg-flux} hold, where $A$ is replaced by a new constant $A'$ and $\ve$ by $\ve':=\big(\uline{\mathcal{E}}^\infty(1)+\mathcal{E}^\infty_{3-\delta}(1)\big)^{1/2}$. Therefore, by repeating all the arguments from Section \ref{sec:improvement-geometry} with $\tau_f$ replaced by $\infty$ and $\ve$ by $\ve'$, one shows that $\mathfrak{B}'=[1,+\infty)$. In this manner, one obtains for every $\tau\in[1,+\infty)$ and $p\in[0,3-\delta]$ that
\[
    \uline{\mathcal{E}}^\infty(\tau)+\mathcal{E}^\infty_{p}(\tau)\les \ve'^2\tau^{-3+\delta+p} = \big(\uline{\mathcal{E}}^\infty(1)+\mathcal{E}^\infty_{3-\delta}(1)\big)\tau^{-3+\delta+p},
\]
from which \eqref{eq:orb-stab2} follows. Regarding \eqref{eq:orb-stab3}, one may revisit the proofs of Lemmas \ref{lem:pointwise_estimates_dirac1} and \ref{lem:pointwise_estimates_dirac2} and observe that the estimates only depend on the energies and pointwise norms coming from initial data of the solution restricted to $J^-(\mathcal{I}^+)\cap\hat{\C}$. Hence, the right-hand side of the pointwise estimates can be replaced by the right-hand side of \eqref{eq:orb-stab3}.

\textsc{Asymptotic stability:} It follows by Propositions \ref{prop:energy-convergence}, \ref{prop:thm-part3} and Lemma \ref{lem:pointwise_estimates_dirac1}.
\end{proof}
\appendix
\section{The Dirac equation in spherical symmetry}\label{app:derivation-Dirac}
The construction of the Dirac operator on a Lorentzian manifold is delicate, but fairly classical, and relies on the introduction of a spin structure together with an orthonormal frame. Section \ref{subsec:Dirac equation in curved spacetime} is devoted to recalling the main ingredients needed for the definition of the Dirac operator, while in Section \ref{app:sec:reduction-sph-symm} the reduction to spherical symmetry is shown. 

\subsection{The Dirac equation in curved spacetime}\label{subsec:Dirac equation in curved spacetime}

We briefly recall the definition of the Dirac operator on a
Lorentzian manifold. We shall refer to
\cite{woit2017quantum} for an
introduction to spinorial formalism, and to \cite[Section 5.6]{parker2009quantum} and \cite{cacciafesta2019weak} for more details on the geometric construction of the Dirac operator. We stress that in what follows we will commit an abuse of notation by identifying spinor fields with their $\mathbb C^4$-valued
representatives.

Let $(\mathcal M,g)$ be a smooth, connected, time-oriented
four-dimensional Lorentzian manifold with signature $(-,+,+,+)$, and
suppose that $(\mathcal M,g)$ is endowed with a \textit{spin structure}. 
We consider a \textit{vierbein}, that is a time-oriented local orthonormal frame $\{e_a\}_{a=0,...,3}$ satisfying
\[
    e_a\,^\mu g_{\mu\nu}e_b\,^\nu = \eta_{ab},
\]
where $\eta:=\text{diag}(-1,1,1,1)$ is the Minkowski metric. Greek indices refer to spacetime coordinates and are raised and lowered
using $g$, while Latin indices refer to the orthonormal frame and are
raised and lowered using $\eta$. %Intuitively, the vierbein permits to relate the tangent space of the manifold to the Minkowski tangent space.

The Dirac equation provides the relativistic description of spin-$1/2$ matter and takes the form
\begin{equation}\label{eq:Dirac-curved}
    i\uline{\gamma}^\mu D_\mu\Psi=\mathfrak{m}\Psi,
\end{equation}
where $\mathfrak{m}\geq0$ is the matter mass, $\Psi$ is a Dirac spinor field,
$\uline{\gamma}^\mu$ are the \textit{curved Dirac matrices} and $D_\mu$ denotes
the \textit{spinorial covariant derivative}.
The matrices $\uline{\gamma}^\mu\in M_4(\mathbb{C})$ satisfy the
Clifford anticommutation relations
\[
    \{\uline{\gamma}^\mu,\uline{\gamma}^\nu\}
    =-2g^{\mu\nu}\text{Id}_4,
\]
and are obtained from the flat Dirac matrices $\gamma^a$ through the
vierbein according to
\[
    \uline{\gamma}^\mu=e_a\,^\mu\gamma^a,
\]
where the flat Dirac matrices satisfy
$\{\gamma^a,\gamma^b\}=-2\eta^{ab}\text{Id}_4$.
Throughout this work, we use the
\textit{Weyl representation} \cite{collas2019dirac}
\[
    \gamma^0 := \begin{pmatrix}
        0_2 & \text{Id}_2\\
        \text{Id}_2 & 0_2
    \end{pmatrix},
    \quad
    \gamma^j := \begin{pmatrix}
        0_2 & \sigma_j\\
        -\sigma_j & 0_2
    \end{pmatrix},
\]
where the Pauli matrices $\sigma_j$ write
\[
    \sigma_1 =\begin{pmatrix}
        0 & 1\\
        1 & 0
    \end{pmatrix}
    ,\quad
    \sigma_2 =\begin{pmatrix}
        0 & -i\\
        i & 0
    \end{pmatrix}
    ,\quad 
    \sigma_3 = \begin{pmatrix}
        1 & 0\\
        0 & -1
    \end{pmatrix}.
\]
The \textit{spinorial covariant derivative} $D_\mu$ is given by
\[
    D_\mu\Psi
    :=
    \partial_\mu\Psi
    -\frac18\omega_\mu\,^{ab}
    [\gamma_a,\gamma_b]\Psi,
\]
where $\omega_\mu{}^{ab}=-\omega_\mu{}^{ba}$ are the coefficients of the
\textit{spin connection}. In terms of the vierbein and the Christoffel symbols,
the spin connection is given by
\[
    \omega_\mu\,^{ab}:= e^a\,_\nu\big(\Gamma^\nu_{\mu\sigma} e^{\sigma b}+\partial_\mu e^{\nu b}\big).
\]
Equivalently, its coefficients are characterized by \textit{Cartan's
structure equation}
\begin{equation}\label{eq:spin-connection}
    de^a + \omega^a\,_b \wedge e^b = 0,
\end{equation}
where $e^a:=e_\mu\,^a\,dx^\mu$ and $\omega^a\,_b:=\omega_\mu\,^a\,_b\,dx^\mu$. Note that the spinorial covariant derivative is covariant under both coordinate transformation and local Lorentz transformations (see for instance \cite{yepez2011einstein}).
Strictly speaking, $\Psi$ is a section of the spinor bundle associated
to the spin structure of the manifold. As already said, by slight abuse of notation, we identify such
sections with $\mathbb{C}^4$-valued functions.

For later use, we define the \textit{adjoint spinor} $\oline\Psi:=\Psi^*\gamma^0$, which is a solution to the \textit{adjoint Dirac equation}
\[
    iD_\mu\oline\Psi\uline{\gamma}^\mu =-\mathfrak{m}\oline\Psi,
\]
where the covariant derivative now acts as
\[
    D_\mu \oline\Psi = \partial_\mu \oline\Psi + \frac{1}{8}\oline\Psi\omega_\mu\,^{ab}[\gamma_a,\gamma_b].
\]
Finally, we conclude this section by giving the formula of the \textit{Dirac energy-momentum tensor}
\begin{align*}
    T_{\mu \nu} := \frac{i}{4}\big(\overline{\Psi}\uline{\gamma_\mu} D_\nu \Psi - D_\nu \overline{\Psi} \uline{\gamma_\mu} \Psi\big) + \frac{i}{4}\big(\overline{\Psi}\uline{\gamma_\nu} D_\mu \Psi - D_\mu \overline{\Psi} \uline{\gamma_\nu} \Psi\big).
\end{align*}
\begin{remark}\label{app:rk:trace-T} If $\Psi$ solves the Dirac equation, and consequently $\oline\Psi$ satisfies the adjoint Dirac equation, the trace of the energy-momentum tensor is given by
\begin{equation}\label{eq:trace-energy-momentum-tensor}
    \text{tr}_g T= g^{\mu\nu}T_{\mu\nu}= \frac{i}{2}\Big(\oline\Psi \uline{\gamma}^\mu D_\mu\Psi - D_\mu\oline\Psi\gamma^\mu\Psi\Big)=\mathfrak{m}\oline\Psi \Psi.
\end{equation}
Therefore, we note that when $\mathfrak{m}=0$, the energy-momentum tensor is traceless, implying that solutions to the Einstein–massless Dirac system have vanishing scalar curvature.
\end{remark}

\subsection{Reduction to spherical symmetry}\label{app:sec:reduction-sph-symm}
We proceed here with the derivation of the Dirac equation in spherical
symmetry. We follow the approach of
\cite{camporesi1996eigenfunctions,cacciafesta2019weak},
which is closer to the perspective of dispersive PDEs. For a derivation
more closely related to the Newman--Penrose formalism, we refer to
\cite[Appendix A]{ma2022sharp}.

Analogously to the Einstein equations, for which a coordinate gauge is fixed, the derivation of the Dirac equation requires the choice of a frame gauge, that is, the choice of a tetrad.
Since we consider black hole spacetimes which are perturbations of Schwarzschild, we can make such a choice globally in the exterior region.
From Section \ref{subsec:Einstein--Dirac system in spherical symmetry}, let $(\mathcal{M},g)$ be a spherically symmetric Lorentzian manifold,
written in double-null coordinates as
\[
    g=-4\Omega^2(u,v)\,dudv + r^2(u,v)\,d\sigma_{\mathbb{S}^2}.
\]
Recalling the geometric quantities defined in \eqref{eq:Hawking-mass}--\eqref{eq:geom-quantities}, we consider the following \textit{gauge invariant double-null tetrad}
\[
    Y := \frac{1}{-\nu}\partial_u,\quad V:=\frac{1-\mu}{\lambda}\partial_v,\quad e_1 := \frac{1}{r}\partial_\vartheta,\quad e_2 := \frac{1}{r\sin\vartheta}\partial_\varphi,
\]
which is regular away from $\{r = 0\}$. We therefore associate the following orthonormal frame
$e_a=e_a\,^\mu\,\partial_\mu$ 
\begin{align*}  
    e_0 = \frac12 (V+Y)= \frac12 \Big( \frac{1-\mu}{\lambda}\partial_v + \frac{1}{-\nu}\partial_u\Big), \quad e_1, \quad e_2,\quad e_3 = \frac12 (V-Y)= \frac12 \Big( \frac{1-\mu}{\lambda}\partial_v - \frac{1}{-\nu}\partial_u\Big),
\end{align*}
with associated dual frame given by $e^a = \eta^{ab} g_{\mu\nu} e_b\,^\nu\, dx^\mu$ 
\begin{align*}
    e^0 =  -\nu\, du + \kappa \,dv,\quad 
    e^1 = r\,d\vartheta,\quad 
    e^2 = r\sin\vartheta\,d\varphi,\quad 
    e^3 &= \nu\, du + \kappa\, dv.
\end{align*}
Computing preliminarily
\begin{align*}
    de^0 &= (\partial_u\kappa + \partial_v\nu)\,du\wedge dv = \frac{\partial_u\kappa + \partial_v\nu}{2\Omega^2}\,e^0\wedge e^3,\\
    de^1 &= -\frac{1}{r}\left(\frac{ m}{r}\,e^0\wedge e^1 + \Big(1-\frac{ m}{r}\Big)\,e^1\wedge e^3\right),\\
    de^2 &= -\frac{1}{r}\left( \frac{ m}{r} \,e^0\wedge e^2 + \Big(1-\frac{ m}{r}\Big)\, e^2\wedge e^3 \right) + \frac{\cot\vartheta}{r}\,e^1\wedge e^2,\\
    de^3 &= (\partial_u\kappa - \partial_v\nu)\,du\wedge dv = \frac{\partial_u\kappa - \partial_v\nu}{2\Omega^2}e^0\wedge e^3,
\end{align*}
we obtain by Cartan's formula \eqref{eq:spin-connection} the spin connection coefficients:
\begin{align*}
    \omega_v\,^{03}&= \frac{\partial_v\nu}{\nu},\\
    \omega_u\,^{03} &= -\frac{\partial_u \kappa}{\kappa},\\
    \omega_\vartheta\,^{01} &= -\frac{ m}{r}, \quad &\omega_\vartheta\,^{13} &= 1-\frac{ m}{r},\\
    \omega_\varphi\,^{02} &= -\frac{ m}{r}\sin\vartheta, \quad &\omega_\varphi\,^{23} &= (1-\frac{ m}{r})\sin\vartheta, \quad &\omega_\varphi\,^{12} = -\cos\vartheta.
\end{align*}
Thus, recalling $D_\mu = \partial_\mu - \frac{1}{8}\omega_\mu\,^{ab}[\gamma_a,\gamma_b]$, we find
\begin{align*}
    D_0 &= D_v = \partial_v -\frac{\partial_v\nu}{2\nu}\gamma_0\gamma_3,\\
    D_1 &= D_\vartheta = \partial_\vartheta +\frac12\left(\frac{ m}{r}\gamma_0\gamma_1-\Big( 1-\frac{ m}{r}\Big)\gamma_1\gamma_3\right),\\
    D_2 &= D_\varphi = \partial_\varphi +\frac{\sin\vartheta}{2}\left(\frac{ m}{r}\gamma_0\gamma_2-\Big( 1-\frac{ m}{r}\Big)\gamma_2\gamma_3\right) + \frac{\cos\vartheta}{2}\gamma_1\gamma_2,\\
    D_3 &= D_u= \partial_u + \frac{\partial_u \kappa}{2\kappa}\gamma_0\gamma_3,
\end{align*}
while the curved Gamma matrices  $\uline{\gamma}^\mu = e_a\,^\mu \gamma^a$ are given by
\begin{align*}
    \uline{\gamma}^0 = \uline{\gamma}^v = \frac{1}{2\kappa}(\gamma^0 + \gamma^3),\quad
    \uline{\gamma}^1 = \uline{\gamma}^\vartheta = \frac{1}{r}\gamma^1,\quad
    \uline{\gamma}^2 = \uline{\gamma}^\varphi = \frac{1}{r\sin\vartheta}\gamma^2,\quad
    \uline{\gamma}^3 = \uline{\gamma}^u = \frac{1}{-2\nu}(\gamma^0 - \gamma^3).
\end{align*}

Therefore, we are ready to explicitly write the Dirac equation \eqref{eq:Dirac-curved}:
\begin{equation}\label{eq:Dirac-explicit1}
\begin{aligned}
    i\left[ (\gamma^0 +\gamma^3)\Big(\frac{1}{2\kappa}\partial_v - \frac{\partial_v\nu}{4\Omega^2}\Big) + (\gamma^0-\gamma^3)\Big(\frac{1}{-2\nu}\partial_u+\frac{\partial_u \kappa}{4\Omega^2}\Big) + \frac{1}{r}\Big(1-\frac{ m}{r}\Big)\gamma^3 -\frac{ m}{r^2}\gamma^0   
    \right. \\ \left.
    +\frac{1}{r}\Big(\gamma^1\Big( \partial_\vartheta + \frac{\cot\vartheta}{2} \Big) + \frac{\gamma^2}{\sin\vartheta}\partial_\varphi\Big)    \right]\Psi = \mathfrak{m}\Psi.
\end{aligned}
\end{equation}
Multiplying \eqref{eq:Dirac-explicit1} by $\gamma^0$ yields
\begin{align*}
    \Big[(\text{Id}_4 - \gamma^0\gamma^3)\Big(\frac{1}{-2\nu}\partial_u + \frac{\partial_u \kappa}{4\Omega^2}\Big) &+ (\text{Id}_4+ \gamma^0\gamma^3)\Big(\frac{1}{2\kappa}\partial_v - \frac{\partial_v\nu }{4\Omega^2}\Big) \\&+ \frac{\gamma^0}{r}\mathcal{D}_{\mathbb{S}^2} -\frac{ m}{r^2}\text{Id}_4 + \frac1r\Big(1-\frac{ m}{r}\Big)\gamma^0\gamma^3\Big]\Psi = -i\mathfrak{m}\gamma^0\Psi,
\end{align*}
where we denoted
\[
    \mathcal{D}_{\mathbb{S}^2}:= \gamma^1\Big( \partial_\vartheta + \frac{\cot\vartheta}{2} \Big) + \frac{\gamma^2}{\sin\vartheta}\partial_\varphi.
\]
To better highlight the block structure of the matrices, we further
transform the equation by applying the orthogonal matrix $P = \begin{psmallmatrix}
    1 & & &  \\ & & & 1 \\ & & 1 & \\ & -1 
\end{psmallmatrix}$, which yields
\begin{equation}\label{eq:Dirac-explicit2}
\begin{aligned}
\Big[\Big(\frac{1}{-\nu}\partial_u +\frac{\partial_u \kappa}{2\Omega^2}\Big)\begin{pmatrix}
    \text{Id}_2 & 0_2 \\ 0_2 & 0_2 
\end{pmatrix} &+
\Big(\frac{1}{\kappa}\partial_v - \frac{\partial_v\nu }{2\Omega^2}\Big)\begin{pmatrix}
    0_2 & 0_2 \\ 0_2 & \text{Id}_2
\end{pmatrix} \\&+ \frac1r \begin{pmatrix}
    0_2 & \hat\nabla\\ \hat\nabla & 0_2
\end{pmatrix} + \frac1r\begin{pmatrix}
    - \text{Id}_2 & 0_2 \\ 0_2 & (1-\mu)\text{Id}_2
\end{pmatrix}\Big] P\Psi =
    -i\mathfrak{m}\begin{pmatrix}
    0_2 & \sigma_3 \\ \sigma_3 & 0_2
\end{pmatrix}P\Psi,
\end{aligned}
\end{equation}
where
\[
    \widehat{\nabla}_{\mathbb{S}^2}:=\sigma_1\Big( \partial_\vartheta + \frac{\cot\vartheta}{2} \Big) + \frac{\sigma_2}{\sin\vartheta}\partial_\varphi
\]
is the Dirac operator on the sphere.

The operator $\widehat{\nabla}_{\mathbb S^2}$ admits a \textit{partial wave
decomposition} (see \cite{abrikosov2002dirac,cacciafesta2019strichartz,thaller2013dirac}), allowing us to simplify the angular dependence in terms of \textit{spinorial spherical harmonics} $\mathcal E^\pm_{j,m_j}$, satisfying
\[
    \widehat{\nabla}_{\mathbb{S}^2} \mathcal{E}^\pm_{j,m_j}=\mp\Big(\frac12+j\Big)\mathcal{E}^\mp_{j,m_j} \quad \text{ and } \quad \sigma^3 \mathcal{E}^\pm_{j,m_j} = \mathcal{E}^\mp_{j,m_j},
\]
where $j\in\frac12+\mathbb{N}$, $m_j\in\frac12+\mathbb{Z}$ with $-j\le m_j\le j$. Each $\mathcal E^\pm_{j,m_j}$ can be expressed, up to a local rotation, as a combination of \textit{standard spherical harmonics} $Y_{j^\pm,m^\pm_j}$, where $j^\pm:=j\pm\frac12$ and $m^\pm_j:=m_j\pm\frac12$.

Moreover, these spinorial spherical harmonics can then be combined into $\mathbb C^4$-valued spinors, which generate the partial wave subspaces $\widetilde{\mathcal H}^\pm_{j,m_j}$ for the full Dirac operator.
However, there is a crucial subtlety in the reduction to spherical symmetry. In general, the Dirac flow does not preserve arbitrary angular modes. As shown in \cite[Section 5]{cacciafesta2011global},the partial wave subspaces are preserved by the nonlinear dynamics provided that the initial datum has an angular part belonging to one of the four spaces $\widetilde{\mathcal H}^\pm_{1/2,m_{1/2}}$.
For this reason, to construct our model, we restrict to $\wtilde{\mathcal{H}}^-_{1/2,1/2}:=\text{span}(\mathcal{G}^+_{1/2,1/2}, \mathcal{G}^-_{1/2,1/2})$, where
\begin{equation}\label{app:eq:lowest-sph-harmonic}
    \mathcal{G}^+_{1/2,1/2}:=\begin{pmatrix}
            \mathcal{E}^+_{1/2,1/2} \\ 0
        \end{pmatrix}
        :=\frac{e^{i\frac{\varphi}{2}}}{2\sqrt{\pi}}
        \begin{pmatrix}
            \cos\frac{\vartheta}{2} \\ -\sin\frac{\vartheta}{2} \\ 0 \\ 0
        \end{pmatrix},\quad
    \mathcal{G}^-_{1/2,1/2}:=\begin{pmatrix}
            0 \\ \mathcal{E}^-_{1/2,1/2} 
        \end{pmatrix}
        :=\frac{e^{i\frac{\varphi}{2}}}{2\sqrt{\pi}}
        \begin{pmatrix}
            0 \\ 0 \\ \cos\frac{\vartheta}{2} \\ \sin\frac{\vartheta}{2}
        \end{pmatrix}.
\end{equation}
Therefore, considering the following  ansatz
\[
    P\Psi : = \Psi_+ \mathcal{G}^+_{1/2,1/2} + \Psi_- \mathcal{G}^-_{1/2,1/2}=\begin{pmatrix}
        \Psi_+\mathcal{E}^+_{1/2,1/2}\\\Psi_-\mathcal{E}^-_{1/2,1/2}
    \end{pmatrix},
\]
and plugging it into \eqref{eq:Dirac-explicit2}, we obtain two coupled equations
\begin{equation*}
    \begin{aligned}
        \Big(\frac{1}{-\nu}\partial_u +\frac{\partial_u\kappa}{2\Omega^2}-\frac1r\Big)\Psi_++ \frac1r\Psi_- &= -i\mathfrak{m}\Psi_-,\\
        \Big(\frac{1}{\kappa}\partial_v -\frac{\partial_v\nu}{2\Omega^2}+\frac{1-\mu}{r}\Big)\Psi_- - \frac1r\Psi_+ &= -i\mathfrak{m}\Psi_+.
    \end{aligned}
\end{equation*}
Performing the gauge invariant reweights
\[
    \psi_+:=r\Psi_+,\quad \psi_-:=\Psi_-,
\]
we finally obtain the \textit{reduced Dirac equations}
\begin{equation}\label{app:eq:final-Dirac}
    \begin{aligned}
        \Big(\frac{1}{-\nu}\partial_u +\frac{\partial_u\kappa}{2\Omega^2}\Big)\psi_+  &= -(1+i\mathfrak{m}r)\psi_-,\\
        \Big(\frac{1}{\kappa}\partial_v -\frac{\partial_v\nu}{2\Omega^2}+\frac{1-\mu}{r}\Big)\psi_-  &=\frac{1}{r^2} (1 -i\mathfrak{m}r)\psi_+.
    \end{aligned}
\end{equation}
Moreover, assuming $\nu<0$, we may define
\[
    \Phi_+ := \kappa^{1/2} \psi_+ ,\quad \Phi_- := r(-\nu)^{1/2}\psi_-,
\]
and find
\begin{equation}\label{app:eq:final-Dirac-bis}
    \begin{aligned}
        \partial_u\Phi_+&=-\frac{\Omega}{r}\Big(1+i\mathfrak{m}r\Big)\Phi_-,\\
        \partial_v\Phi_- &=\frac{\Omega}{r}\Big(1-i\mathfrak{m}r\Big)\Phi_+.
    \end{aligned}
\end{equation}

Let us now proceed to explicitly compute the components of the Dirac energy-momentum tensor $T_{\mu\nu}$.
Recalling that 
\[
    \Psi= P^{-1}\begin{pmatrix}
        \Psi_+\mathcal{E}^+_{1/2,1/2}\\\Psi_-\mathcal{E}^-_{1/2,1/2}
    \end{pmatrix}=
    \frac{e^{i\varphi/2}}{2\sqrt{\pi}}
    \begin{pmatrix}
        \Psi_+ \cos\vartheta/2\\
        -\Psi_-\sin\vartheta/2\\
        \Psi_- \cos\vartheta/2\\
        -\Psi_+\sin\vartheta/2
    \end{pmatrix},
\]
one obtains
\begin{align*}
    T_{uu} &= \frac{i}{2}\big(\overline{\Psi}\uline{\gamma_u} D_u \Psi - D_u \overline{\Psi} \uline{\gamma_u} \Psi\big)
    \\&= \frac{i}{2}\left((\Psi^*\gamma^0) (g_{uv}\uline{\gamma}^v)\Big(\partial_u -\frac{\partial_u\kappa}{2\kappa}\gamma^0\gamma^3\Big)\Psi -\Big( \partial_u\Psi^* \gamma^0 +\frac{\partial_u\kappa}{2\kappa}\Psi^*\gamma^0\gamma^3 \Big)(g_{uv}\uline{\gamma}^v)\Psi\right)
    \\&= \frac{i\nu}{2}\Big[\Psi^*(\text{Id}_4+\gamma^0\gamma^3)\partial_u\Psi - \partial_u\Psi^*(\text{Id}_4+\gamma^0\gamma^3)\Psi \Big]= \frac{-\nu}{2\pi}\Im\big[\oline{\Psi_-}\partial_u\Psi_-\big]= \frac{-\nu}{2\pi}\Im\big[\oline{\psi_-}\partial_u\psi_-\big].
\end{align*}
Similarly, one has
\begin{align*}
    T_{vv} &= \frac{\kappa}{2\pi} \Im\big[ \overline{\Psi_+}\partial_v \Psi_+ \big] = \frac{\kappa}{2\pi r^2} \Im\big[ \oline{\psi_+} \partial_v\psi_+ \big],\\
    T_{uv}&= \frac{-\nu}{4\pi}\Im[\oline{\Psi_-} \partial_v\Psi_-] + \frac{\kappa}{4\pi}\Im[\oline{\Psi_+} \partial_u\Psi_+] = \frac{-\nu}{4\pi}\Im[\oline{\psi_-} \partial_v\psi_-] + \frac{\kappa}{4\pi r^2}\Im[\oline{\psi_+} \partial_u\psi_+],\\
    T_{\vartheta\vartheta}&= \frac{r}{4\pi}\Im\big[\oline{\Psi_-}\Psi_+\big]=\frac{1}{4\pi}\Im\big[\oline{\psi_-}\psi_+\big],\\
    T_{\varphi\varphi}&=\sin^2\vartheta\, T_{\vartheta\vartheta}.
\end{align*}
Using \eqref{app:eq:final-Dirac}, we compute
\[
    T_{uv} %= \frac{\Omega^2}{2\pi r} (\Im \big[\oline{\Psi_-} \Psi_+\big]-mr\Re\big[\oline{\Psi_-}\Psi_+\big]) 
    = \frac{\Omega^2}{2\pi r^2}(\Im \big[\oline{\psi_-}\psi_+\big]-\mathfrak{m}r\Re\big[\oline{\psi_-}\psi_+\big]),
\]
and thus verify Remark \ref{app:rk:trace-T}:
\[
\text{tr}_gT=g^{\mu\nu}T_{\mu\nu}=\frac{\mathfrak{m}}{2\pi r}\Re\big[\oline{\psi_-}\psi_+\big]=\mathfrak{m}\oline\Psi\Psi.
\]

Finally, we recall the Einstein equations in spherical symmetry
\begin{equation}\label{app:eq:final-Einstein}
\begin{aligned}
        \partial_u \partial_v r  &= -\frac{2\Omega^2  m}{r^2} + 4\pi r T_{uv},\\
        \frac{1}{2}\partial_u\partial_v \log \Omega^2 &=\frac{2\Omega^2  m}{r^3} -  4\pi T_{uv}-4\pi\Omega^2 g^{AB}T_{AB},\\
        \partial_u (\Omega^{-2} \partial_u r) &= -\frac{4\pi r}{\Omega^2} T_{uu},\\
        \partial_v (\Omega^{-2} \partial_v r) &= -\frac{4\pi r}{\Omega^2} T_{vv},
\end{aligned}
\end{equation}
where the capital indices denote the spherical variables. 
Using the relation $g^{AB} T_{AB} = \text{tr}_gT+\Omega^{-2}T_{uv}$, one recovers \eqref{eq:wave-eq-r}--\eqref{eq:Raych-v} (in the case when $\mathfrak{m}=0$).
\begin{remark}
A solution $(\psi_+,\psi_-)$ to the reduced Dirac equations \eqref{app:eq:final-Dirac} (or equivalently $(\Phi_+,\Phi_-)$ solving \eqref{app:eq:final-Dirac-bis}) satisfies the following \textit{Bianchi identities}
\begin{equation}\label{app:eq:Bianchi}
\begin{aligned}
    \partial_v(r^2T_{uu})+\partial_u(r^2 T_{uv})&=2 r^2 \Big(T_{uv}\Big( \partial_u\log\Omega - \frac{\nu}{r}\Big) - \frac{\Omega^2 \nu}{r} \text{tr}_g T\Big),\\
    \partial_u(r^2 T_{vv})+\partial_v(r^2 T_{uv})&=2r^2 \Big( T_{uv}\Big(\partial_v\log\Omega-\frac{\lambda}{r}\Big) -\frac{\Omega^2\lambda}{r}\text{tr}_g T\Big).
\end{aligned}
\end{equation}
\end{remark}

\section{Local well-posedness for the Einstein--Dirac system}\label{app:lwp-Einstein-Dirac}
In this section, we briefly trace the argument for local well-posedness of the spherically symmetric Einstein--Dirac system in characteristic slabs. For a treatment in the absence of symmetry, we refer to \cite{zhao2026local}. We include the argument here, since the formulation below is more directly suited to our needs.

As already observed in \cite{chen2022global}, a direct application of the standard local well-posedness theory to the reduced Einstein--Dirac system is complicated by a \textit{potential} loss of derivatives. Indeed, the spin connection coefficients in the reduced Dirac equations contain derivatives of the metric coefficients.
For instance, in the equation for $\psi_-$, one can use the equation for $\partial_u\kappa$ and obtain an equation schematically of the form
\[
    \partial_u\psi_+ = \mathcal N(g,\p g,\psi,\partial_u\psi_-),
\]
which prevents a direct application of the usual quasi-linear theory.
To circumvent this issue, a standard approach consists in considering the second-order equations satisfied by $\psi_\pm$, obtained by applying $\partial_v$ and $\partial_u$ to the equations satisfied by $\psi_+$ and $\psi_-$, respectively, and carefully exploiting the Einstein equations to reduce the system to a quasi-linear hyperbolic system of the form
\begin{align*}
    \p^2\psi &= \mathcal N(g,\p g,\psi,\p\psi),\\
    \p^2g &= \mathcal N(g,\p g,\psi,\p\psi),
\end{align*} 
for which local well-posedness holds by \cite{foures1952theoreme}. One then recovers a solution to the original Einstein--Dirac system by propagating the constraint equations and the Dirac equations from the initial bifurcate null hypersurface.

As this procedure requires many substitutions and involves lengthy computations, we rather opt to formulate the well-posedness problem in terms of $(r,\Omega,\Phi_+,\Phi_-)$, for which the derivative loss can be avoided. The construction requires the additional condition  $\nu<0$ on the \textit{absence of anti-trapped spheres}, which is not restrictive in our setting. Another reason for carrying out this argument here is that, since the dominant energy condition is not satisfied by the Dirac field, we cannot directly rely on the local well-posedness results of \cite{kommemi2013global}. Nevertheless, we follow the same general strategy, while replacing the monotonicity arguments relying on the dominant energy condition by arguments adapted to the Einstein--Dirac system.

For simplicity, let us now recall here Proposition \ref{prop:slab-existence}, which constitutes the main result of this section.
\begin{proposition}[Local existence on slabs]\label{app:prop:slab-existence}
    For any $\mathfrak{m}\ge0$ and $B>0$, there exists a constant $\ve_{\text{slab}}>0$ with the following property.
    Let $(\mathring r,\mathring\Omega,\mathring\psi^{\out}_+, \mathring\psi^{\ing}_-)$ be a characteristic initial data set for the spherically symmetric Einstein--Dirac system on $\mathcal C(U_0,U_1,V_0,V_1)$ as in Definition \ref{def:init-data-set} satisfying
    \[
        \V\log\mathring r\V_{C^2(\C)} + \V\log\mathring \Omega\V_{C^2(\C)} + \V \mathring\psi^\out_+ \V_{C^0(\C_\out)} + \V \partial_v\mathring\psi^\out_+ \V_{C^0(\C_\out)} + \V \mathring\psi^\ing_- \V_{C^0(\uline{\C}_\ing)} + \V \partial_u\mathring\psi^\ing_- \V_{C^0(\uline{\C}_\ing)}\le B.
    \]
    Then there exists a unique smooth solution $(r,\Omega,\psi_+,\psi_-)$ to the spherically symmetric Einstein--Dirac system \eqref{app:eq:final-Dirac-bis}--\eqref{app:eq:final-Einstein} on the slab
    \begin{equation*}
      \D := \R\big(U_0,U_0+\min\{\ve_\mathrm{slab},U_1-U_0\},V_0,V_1\big)\cup\R\big(U_0,U_1,V_0,V_0+\min\{\ve_\mathrm{slab},V_1-V_0\}\big),
    \end{equation*}
    satisfying $\nu<0$ on $\mathcal{D}$ and which extends the initial data. Moreover, the norms 
    \[
        \V\log r\V_{C^k(\D)}, \V\log\Omega\V_{C^k(\D)}, \V\psi_+ \V_{C^k(\D)},\V \psi_- \V_{C^k(\D)}
    \] 
    are finite for any $k$ and can be bounded in terms of appropriate higher order initial data norms.
    \end{proposition}

The proof is a consequence of the following two lemmas. We first establish a local existence result.
\begin{lemma}\label{app:prop:local-exist} For any $\mathfrak{m}\ge0$ and $B>0$, there exists a constant $\ve_0>0$ with the following property. Let $(\mathring r, \mathring\Omega,\mathring\psi^\out_+,\mathring\psi^\ing_-)$ be a characteristic initial data set for the spherically symmetric Einstein--Dirac system on $\C(U_0,U_1,V_0,V_1)$ as in Definition \ref{def:init-data-set}. If $U_1-U_0<\ve_0$, $V_1-V_0<\ve_0$ and
\[
    \V\log\mathring r\V_{C^2(\C)} + \V\log\mathring \Omega\V_{C^2(\C)} + \V \mathring\psi^\out_+ \V_{C^0(\C_\out)} + \V \partial_v\mathring\psi^\out_+ \V_{C^0(\C_\out)} + \V \mathring\psi^\ing_- \V_{C^0(\uline{\C}_\ing)} + \V \partial_u\mathring\psi^\ing_- \V_{C^0(\uline{\C}_\ing)}\le B.
\]
Then there exists a unique smooth solution $(r,\Omega,\psi_+,\psi_-)$ to the spherically symmetric Einstein--Dirac system \eqref{app:eq:final-Dirac-bis}--\eqref{app:eq:final-Einstein} on $\R(U_0,U_1,V_0,V_1)$ satisfying $\nu<0$ on $\R$ and which extends the initial data. Moreover, the norms
\[
    \V\log r\V_{C^k(\R)}, \V\log\Omega\V_{C^k(\R)}, \V\psi_+ \V_{C^k(\R)},\V \psi_- \V_{C^k(\R)}
\]
are finite for any $k$ and can be bounded in terms of appropriate higher order initial data norms.
\end{lemma}
\begin{proof} We follow the approach in \cite{dafermos2005extension} and is based on a standard contraction mapping argument, which requires a nested iteration due to the coupling of the Dirac components.

Since the initial data satisfy $\mathring\nu<0$ on $\C$, we may define the renormalized Dirac components
    \begin{equation}\label{app:eq:renorm-init-data}
        \mathring\Phi^\out_+ := \mathring\kappa^{1/2}\mathring\psi^\out_+,\quad \mathring\Phi_-^\ing := \mathring r(-\mathring\nu)^{1/2}\mathring\psi_-^\ing,
    \end{equation}
on $\C_\out$ and $\uline{\C}_\ing$ respectively,
and the spaces
    \begin{align*}
        \mathcal{A}&:= \{(r,\Omega)\in C^2(\R)\times C^1(\R): \V\log r\V_{C^2(\R)} + \V\log\Omega\V_{C^1(\R)} \le 2B\},\\
        \mathcal{A'}&:= \{(r,\Omega)\in (C^2(\R))^2: \V\log r\V_{C^2(\R)} + \V\log\Omega\V_{C^2(\R)} \le 2B\}\subset \mathcal{A}.
    \end{align*}
For $(r,\Omega)\in\mathcal{A}$, we define the mapping $\Uppsi:(\Phi_+,\Phi_-)\to(\wtilde\Phi_+,\wtilde\Phi_-)$ on 
\[
    \mathcal{B}:=\{(\Phi_+,\Phi_-)\in C^0(\R)^2: \V\Phi_+\V_{C^0(\R)}+\V\Phi_-\V_{C^0(\R)}\le2B\},
\] where
    \begin{equation}
        \begin{aligned}
        \wtilde\Phi_+(U,V) &:= \mathring\Phi_+^\out(V) - \int^U_{U_0} \Big(\frac{\Omega}{r}(1+i\mathfrak{m}r)\Phi_-\Big)(U',V)\,dU',\\ 
        \wtilde\Phi_-(U,V) &:= \mathring\Phi_-^\ing(U) + \int^V_{V_0} \Big(\frac{\Omega}{r}(1-i\mathfrak{m}r)\Phi_+\Big)(U,V')\,dV'.
        \end{aligned}
    \end{equation}
We start by proving that $\Uppsi$ is a contraction. One readily has $\Uppsi:\mathcal{B}\to\mathcal{B}$, as by construction
    \[
        |\wtilde\Phi_\pm|\le B + \ve_0 CB \le 2B,
    \]
for $\ve_0$ small enough, and where $C$ does not depend on $B$ nor on $\ve_0$. Moreover, we consider $(\Phi_+^{(1)},\Phi_-^{(1)})$, $(\Phi_+^{(2)},\Phi_-^{(2)})\in\mathcal{B}$. For $\ve_0$ suitably small, we have
    \[
        |\wtilde\Phi_+^{(1)}-\wtilde\Phi_+^{(2)}|\le\int^U_{U_0}\Big|\frac{\Omega}{r} \Big||\Phi_-^{(1)}-\Phi_-^{(2)}| \,dU'\le \ve_\text{loc}CB\V\Phi_-^{(1)}-\Phi_-^{(2)}\V_{C^0(\R)} < \V\Phi_+^{(1)}-\Phi_+^{(2)}\V_{C^0(\R)} + \V\Phi_-^{(1)}-\Phi_-^{(2)}\V_{C^0(\R)},
    \]
and analogously for $|\wtilde\Phi_-^{(1)}-\wtilde\Phi_-^{(2)}|$, yielding
\[
    \V\wtilde\Phi_+^{(1)}-\wtilde\Phi_+^{(2)}\V_{C^0(\R)} + \V\wtilde\Phi_-^{(1)}-\wtilde\Phi_-^{(2)}\V_{C^0(\R)} < \V\Phi_+^{(1)}-\Phi_+^{(2)}\V_{C^0(\R)} + \V\Phi_-^{(1)}-\Phi_-^{(2)}\V_{C^0(\R)}.
\]
Therefore, given $(r,\Omega)\in\mathcal{A}$ there exists a unique solution $(\Phi_+,\Phi_-)$ solving the Dirac equation 
\begin{equation}\label{app:eq:dirac-lwp}
    \partial_u\Phi_+= -\frac{\Omega}{r}(1+i\mathfrak{m}r)\Phi_-,\quad    \partial_v\Phi_-= \frac{\Omega}{r}(1-i\mathfrak{m}r)\Phi_+,
\end{equation}
on the metric defined by $r$ and $\Omega$, with the prescribed initial conditions \eqref{app:eq:renorm-init-data}. We note that the equations readily imply $|\partial_u\Phi_+|,|\partial_v\Phi_-|\le CB$.
    Furthermore, writing 
    \begin{align*}
        \p_v\Phi_+(U,V) &= \p_v\Phi_+^\out(V)+\int^U_{U_0}(\p_u\p_v\Phi_+)(U',V)\,dU'\\&= \p_v\Phi_+^\out(V)+\int^U_{U_0}\Big(\frac{\Omega}{r^2}(-r\p_v\log\Omega(1+i\mathfrak{m}r)+\lambda)\Phi_- -\frac{\Omega^2}{r^2}(1+\mathfrak{m}^2r^2)\Phi_+\Big)\,dU',\\
        \p_u\Phi_-(U,V) &= \p_u\Phi_-^\ing(U)+\int^V_{V_0}(\p_u\p_v\Phi_-)(U,V')\,dV' \\&= \p_u\Phi_-^\ing(U)+\int^V_{V_0}\Big(\frac{\Omega}{r^2}(r\p_u\log\Omega(1-i\mathfrak{m}r)-\nu)\Phi_+ -\frac{\Omega^2}{r^2}(1+\mathfrak{m}^2 r^2)\Phi_-\Big)\,dV',
    \end{align*}
    we similarly deduce the bounds
    \[
        |\p_v\Phi_+|,|\p_u\Phi_-|\le 2B,
    \]
    for $\ve_\text{loc}$ small enough.
    Hence, given $(\Phi_+,\Phi_-)$ we are now able to define the energy-momentum components $T_{uu}$, $T_{uv}$, $T_{vv}$,
    $T_{\vartheta\vartheta}$, $T_{\varphi\varphi}$, 
    \begin{align*}
        T_{uu}&:=\frac{1}{2\pi r^2}\Im[\oline{\Phi_-}\partial_u\Phi_-],\quad &T_{vv}&:= \frac{1}{2\pi r^2}\Im[\oline{\Phi_+}\partial_v\Phi_+],\quad &T_{uv}&:= \frac{\Omega}{2\pi r^3}(\Im \big[\oline{\Phi_-}\Phi_+\big]-\mathfrak{m}r\Re\big[\oline{\Phi_-}\Phi_+\big]),\\
        T_{\vartheta\vartheta}&:= \frac{1}{4\pi r\Omega}\Im\big[\oline{\Phi_-}\Phi_+\big],\quad
    &T_{\varphi\varphi}&:=\sin^2\vartheta\, T_{\vartheta\vartheta},
    \end{align*}
which can be bounded in view of the previous estimates. Furthermore, thanks to \eqref{app:eq:dirac-lwp}, one can infer proper bounds on $\p_u T_{uv},\p_v T_{uv},\p_u T_{\vartheta\vartheta},\p_v T_{\vartheta\vartheta},\p_u T_{\varphi\varphi},\p_v T_{\varphi\varphi}$.

Next, we define $\Upphi:\oline{\mathcal{A}'}\to \mathcal{A}$, mapping $(r,\Omega)\mapsto(\tilde r,\wtilde\Omega)$, where
    \begin{align*}
        \tilde r(U,V):=& \mathring r(U_0,V)+ \mathring r(U,V_0)-\mathring r(U_0,V_0)+\int^{U}_{U_0}\int^{V}_{V_0}\Big(-\frac{\Omega^2}{r} - \frac{\lambda\nu}{r}+ 4\pi r T_{uv}\Big)(U',V')\,dU'dV',\\
        \log\tilde\Omega(U,V):=&\log\mathring\Omega(U_0,V)+\log\mathring\Omega(U,V_0)-\log\mathring\Omega(U_0,V_0)\\+&\int^{U}_{U_0}\int^{V}_{V_0}\Big(\frac{\Omega^2 }{\tilde r^2} + \frac{\tilde\lambda\tilde\nu}{\tilde r^2}-  4\pi T_{uv}- \frac{4\pi\Omega^2} {\tilde r^2}g_{\mathbb{S}^2}^{AB}T_{AB}\Big)(U',V')\,dU'dV',
    \end{align*}
where $g_{\mathbb{S}^2}$ denotes the metric on the sphere.
Up to taking $\ve_0$ small enough and making use of the previous bounds, one can differentiate the relations defining $\tilde r,\wtilde\Omega$ and find that $\Upphi$ is a contraction on $\mathcal{A}'$ in the norm of $\mathcal{A}$ (we refer to \cite{dafermos2005extension} for more details). Given its fixed point $(r,\Omega)$, we define $(\Phi_+,\Phi_-)$ as before. 
In particular, by using the Bianchi identities \eqref{app:eq:Bianchi}, one can compute explicitly
    \begin{align*}
        \partial_v(r\Omega^2\partial_u (\Omega^{-2} \partial_u r) + 4\pi r^2 T_{uu})&=0,\\
        \partial_u(r\Omega^2\partial_v (\Omega^{-2} \partial_v r) + 4\pi r^2 T_{vv})&=0,
    \end{align*}
hence propagating the constraint equations. Therefore, writing
\[
    \partial_u(\Omega^{-2}\nu) = -\frac{4\pi r}{\Omega^2} T_{uu} = -\frac{2}{r\Omega^2}\Im[\oline{\Phi_-}\p_u\Phi_-],
\]
we have on $\R$
\[
    \Omega^{-2}\nu(U,V) = \mathring\Omega^{-2}\mathring\nu(U_0,V) -\int^{U}_{U_0} \Big(\frac{2}{r\Omega^2}\Im[\oline{\Phi_-}\p_u\Phi_-]\Big)(U',V)\,dU' \le \mathring\Omega^{-2}\mathring\nu(U_0,V) + \ve_0CB <0,
\]
for $\ve_0$ suitably small. Hence, we can reconstruct the original Dirac components $\psi_+ := \kappa^{-1/2} \Phi_+$, $\psi_- := r^{-1}(-\nu)^{-1/2}\Phi_-$ and bound $\V\psi_\pm\V_{C^1(\R)}$ by the previous estimates on $(r,\Omega,\Phi_+,\Phi_-)$. Showing that $(r,\Omega,\psi_+,$ $\psi_-)$ is the unique solution of the equations is now straightforward, while higher regularity norms follow immediately if assumed on initial data.
\end{proof}

The second ingredient we need is the following local spacetime estimate, in the spirit of \cite{kommemi2013global,kehle2024extremal}. We again stress that the additional assumption $\nu<0$ is essential, as it allows us to work with the components $\Phi_\pm$.
\begin{lemma}\label{app:lem:extension-aux} For any $\mathfrak{m}\ge0$ and $B>0$, there exists a constant $C_*<+\infty$ with the following property. Let $(r,\Omega,\psi_+,\psi_-)$ be a solution of the spherically symmetric Einstein--Dirac system on $\R':=\R(U_0,U_1,V_0,V_1)\setminus\{(U,V)\}$. Assume $U_1-U_0\le C_0,V_1-V_0\le C_0$, the initial data estimates 
\begin{equation}\label{app:eq:hyp1-extension}
    \V\log r\V_{C^2(\C)} + \V\log\Omega\V_{C^2(\C)} + \V\psi_+ \V_{C^1(\C)} + \V \psi_- \V_{C^1(\C)}\le C_0,
\end{equation}
and the global estimates
\begin{equation}\label{app:eq:hyp2-extension}
    \iint_{\R'} \Omega^2\,dudv\le C_0,\quad \sup_{\R'}|\log r|\le C_0,\quad \nu<0 \text{ on $\R'$.}
\end{equation}
Then the following estimate holds
\[
    \V\log r\V_{C^2(\R')} + \V\log\Omega\V_{C^2(\R')} + \V\psi_+ \V_{C^1(\R')} + \V \psi_- \V_{C^1(\R')}\le C_*.
\]
\end{lemma}
\begin{proof} In the following, we will use the notation $A\les1$, meaning that $A\le C$ where $C$ is a constant depending only on $\mathfrak{m},C_0$. Moreover, we will make no distinction between $\R$ and $\R'$, even though the integrands are only assumed to be defined only on $\R'$.

As $\nu<0$ on $\R'$, we can express as before the Einstein--Dirac system in terms of the renormalized Dirac components $\Phi_\pm$, which satisfy
\begin{equation}\label{app:eq:Phi_pm_init}
    \V\Phi_+ \V_{C^1(\C)} + \V \Phi_- \V_{C^1(\C)}\le C_0,
\end{equation}
by \eqref{app:eq:hyp1-extension}.
Rewriting the first equation in \eqref{app:eq:final-Einstein} as
\[
    4\pi r^2 T_{uv} = \frac12 \partial_u\partial_v(r^2) + \Omega^2,
\]
we may integrate it over $\mathcal{R}'$ and obtain
\begin{equation}\label{app:eq:T_uv-bound}
    \iint_{\mathcal{R}} |T_{uv}|\,dUdV \les \iint r^2 |T_{uv}|\, dUdV \les \iint_{\mathcal{R}} |\p_u\p_v (r^2)|\,dUdV + \iint_{\mathcal{R}}\Omega^2\,dUdV\les 1,
\end{equation}
by \eqref{app:eq:hyp1-extension}--\eqref{app:eq:hyp2-extension}.
Moreover, rewriting again the same equation as
\[
    \p_u(r\p_vr)= - \Omega^2 + 4\pi r^2 T_{uv},
\]
and integrating in $u$, we obtain
\[
    \sup_{[U_0,U_1]\times\{V\}} |r\p_v r|\les 1 + \int^{U_1}_{U_0} (\Omega^2 + r^2 |T_{uv}|)(U,V)\,dU,
\]
for all $V\in[V_0,V_1]$, yielding
\[
    \int^{V_1}_{V_0} \sup_{[U_0,U_1]\times\{V\}}|\lambda|\,dV\les \int^{V_1}_{V_0} \sup_{[U_0,U_1]\times\{V\}}|r\lambda|\,dV \les 1.
\]
As $\partial_u(r\partial_v r)=\partial_v(r\partial_ur)$, reversing the roles of $u$ and $v$, one finds the analogous bound
\[
    \int^{U_1}_{U_0} \sup_{\{U\}\times[V_0,V_1]}|\nu|\,dU \les 1.
\]
Therefore, estimating
\[
    \iint_{\mathcal{R}} |\lambda\nu|\,dUdV \le \Big(\int^{V_1}_{V_0} \sup_{[U_0,U_1]\times\{V\}}|\lambda|\,dV\Big)\Big(\int^{U_1}_{U_0}\sup_{\{U\}\times[V_0,V_1]}|\nu|\, dU\Big),
\]
we infer a bound on the Hawking mass
\[
    \iint_\mathcal{R} \Omega^2| m|\,dudv\les1.
\]
Moreover, \eqref{app:eq:T_uv-bound} implies
\begin{equation}\label{app:eq:Phi_pm_aux}
    \iint_\R|\Phi_-||\Phi_+|\,dUdV \les \iint_\R \frac{\Omega}{r^2}|\Phi_-||\Phi_+|\,dUdV\les \iint_\R |T_{uv}|\,dUdV\les1,
\end{equation}  
and therefore, from the definition of $T_{\vartheta\vartheta}, T_{\varphi\varphi}$,
\[
    \iint_\R \Omega^2 g^{AB} |T_{AB}|\les 1.
\]
Thus, from the wave equation in $\log\Omega$ in \eqref{app:eq:final-Einstein}, we obtain the pointwise estimate
\[
    \sup_{\R'}|\log\Omega|\les 1 + \Big| \iint_\R \p_u\p_v\log\Omega\,dUdV\Big| \les 1 + \iint_\R  \Omega^2| m| + |T_{uv}| + \Omega^2 g^{AB}|T_{AB}|\,dUdV\les 1.
\]
Now, using \eqref{app:eq:final-Dirac-bis}, we compute 
\[
    \partial_u(|\Phi_+|^2) = 2\Re[\oline{\Phi_+}\partial_u\Phi_+] = -\frac{2\Omega}{r}\Re[(1+i\mathfrak{m}r)\oline{\Phi_+}\Phi_-],
\]
integrate again in $u$ as before, and infer from \eqref{app:eq:Phi_pm_aux} the bound
\[
    \int^{V_1}_{V_0}\sup_{[U_0,U_1]\times\{V\}}|\Phi_+|^2\,dV\les 1,
\]
and similarly,
\[
    \int^{U_1}_{U_0}\sup_{\{U\}\times[V_0,V_1]}|\Phi_-|^2\,dU\les 1.
\]
In the same manner, using the wave equation in $\log\Omega$ in \eqref{app:eq:final-Einstein} again, we may integrate in $u$ and find
\[
    \sup_{[U_0,U_1]\times\{V\}}|\partial_v\log\Omega|\les 1 +\Big| \int^{U_1}_{U_0}\p_u\p_v\log\Omega\,dUdV\Big|,
\]
yielding
\[
    \int^{V_1}_{V_0}\sup_{[U_0,U_1]\times\{V\}}|\partial_v\log\Omega|\,dV\les 1,
\]
and in the same fashion
\[
    \int^{U_1}_{U_0}\sup_{\{U\}\times[V_0,V_1]}|\partial_u\log\Omega|\,dU\les 1.
\]
Therefore, from the initial bounds and \eqref{app:eq:final-Dirac-bis},
\begin{align*}
    \sup_{\R'}|\Phi_+| &\les 1 + \Big|\iint_\R \p_u\p_v\Phi_+\,dUdV\Big| \\&= 1 + \Big|\iint_\R \frac{\Omega}{r^2}\Big(-r\partial_v\log\Omega(1+i\mathfrak{m}r) + \lambda\Big)\Phi_- -\frac{\Omega^2}{r^2}(1+\mathfrak{m}^2 r^2)\Phi_+\,dUdV\Big|
    \\& \les 1 + \Big(\int^{V_1}_{V_0} \sup_{[U_0,U_1]\times\{V\}} (|\partial_v\log\Omega| +|\lambda|)\,dV\Big)\Big(\int^{U_1}_{U_0}\sup_{\{U\}\times[V_0,V_1]}|\Phi_-|\,dU\Big) \\&+ \Big(\int^{V_1}_{V_0}\sup_{[U_0,U_1]\times\{V\}}|\Phi_+|^2\,dV\Big)\Big(\int^{U_1}_{U_0}\sup_{\{U\}\times[V_0,V_1]}\frac{\Omega^2}{r^2}\,dU\Big)\les1,\\
    \sup_{\R'}|\Phi_-| &\les1,
\end{align*}
where the second estimate follows by analogous computations.
These bounds imply a uniform pointwise control also on 
\[
\sup_{\R'}\big(|T_{uv}|+|T_{\vartheta\vartheta}|+|T_{\varphi\varphi}|\big)\les1.
\]
Therefore, by integrating again $\partial_u(r\lambda)=\partial_v(r\nu)$, we obtain
\[
    \sup_{\R'} \big(|\nu|+|\lambda|\big)\les 1,
\]
yielding a pointwise estimate on the Hawking mass $ m$, and consequently
\[
    \sup_{\R'} \big(|\partial_u\log\Omega|+|\partial_v\log\Omega| +|\p_u\p_v\log\Omega|\big)\les 1.
\]
Hence, it follows
\begin{align*}
    \sup_{\R'}|\partial_v\Phi_+|&\les 1+ \Big|\int_{U_0}^{U_1} \p_u\p_v\Phi_+\,dU\Big|\\&\les 1 + \Big|\int_{U_0}^{U_1} \frac{\Omega}{r^2}\Big(-r\partial_v\log\Omega(1+i\mathfrak{m}r) + \lambda\Big)\Phi_- -\frac{\Omega^2}{r^2}(1+\mathfrak{m}^2 r^2)\Phi_+\,dU\Big|\les 1,
\end{align*}
and similarly 
\[
    \sup_{\R'}|\partial_u\Phi_-|\les 1,
\]
implying
\[
    \sup_{\R'}(|T_{uu}|+ |T_{vv}|+ |\p_vT_{uv}|+|\p_u T_{uv}|+|\p_vT_{\vartheta\vartheta}|+|\p_uT_{\vartheta\vartheta}|+|\p_vT_{\varphi\varphi}|+|\p_uT_{\varphi\varphi}|)\les 1.
\]
Thus, by \eqref{app:eq:final-Einstein}
\[
    \sup_{\R'} (|\partial^2_u r|+|\partial^2_v r| +|\p_u\p_v r|)\les 1,
\]
and by commuting the wave equation in $\log\Omega$,
\[
    \sup_{\R'} (|\partial^2_u \log\Omega|+|\partial^2_v \log\Omega|)\les 1.
\]
Recalling $\psi_+ = \kappa^{-1/2} \Phi_+, \psi_- = r^{-1}(-\nu)^{-1/2}\Phi_-$, from the previous bounds we obtain
\[
    \sup_{\R'} \big(|\psi_+|+|\p_v\psi_+|+|\p_u\psi_+|+|\psi_-|+|\p_v\psi_-|+|\p_u\psi_-|\big)\les 1,
\]
concluding the proof.
\end{proof}

\begin{proof}[Proof of Proposition \ref{app:prop:slab-existence}] 
The proof follows the argument of \cite[Proposition 3.17]{kehle2024extremal}, using Lemmas \ref{app:lem:extension-aux} and \ref{app:prop:local-exist} together with a bootstrap argument.
\end{proof}

\addcontentsline{toc}{section}{References}
\bibliographystyle{alpha}
\bibliography{bibliography}

\end{document}